\documentclass{optica-article}

\journal{opticajournal} 

\articletype{Roadmap}

\usepackage{lineno}
\usepackage[normalem]{ulem}
\usepackage{orcidlink}

\usepackage{upgreek}
\usepackage{bm}
            
\newcommand{\usection}[1]{\section*{#1}
\addcontentsline{toc}{section}{\protect\numberline{}#1}}

\providecommand{\pnl}[1]{{\textcolor{black}{(#1)}}}

\linenumbers 

\let\Re\relax 
\DeclareMathOperator{\Re}{Re}
\let\Im\relax 
\DeclareMathOperator{\Im}{Im}

\nolinenumbers

\begin{document}

\title{Roadmap on Nonlocality in Photonic Materials and Metamaterials}


\author{
Francesco Monticone\,\orcidlink{0000-0003-0457-1807}\authormark{1,*}, N.~Asger~Mortensen\,\orcidlink{0000-0001-7936-6264}\authormark{2,3,**},
Antonio~I.~Fern\'{a}ndez-Dom\'{i}nguez\,\orcidlink{0000-0002-8082-395X}\authormark{4,5}, 
Yu Luo\,\orcidlink{0000-0003-2925-682X}\authormark{6,7}, 
Xuezhi~Zheng\,\orcidlink{0000-0002-5301-2908}\authormark{8} 
Christos~Tserkezis\,\orcidlink{0000-0002-2075-9036}\authormark{2},
Jacob~B.~Khurgin\,\orcidlink{0000-0003-0725-8736}\authormark{9},
Tigran~V.~Shahbazyan\,\orcidlink{0000-0001-5139-1097}\authormark{10},
Andr{\'e}~J.~Chaves\,\orcidlink{0000-0003-1381-8568}\authormark{2,11}, 
Nuno~M.~R.~Peres\,\orcidlink{0000-0002-7928-8005}\authormark{2,12,13},
Gino~Wegner\,\orcidlink{0000-0001-6225-5269}\authormark{14},
Kurt~Busch\,\orcidlink{0000-0003-0076-8522}\authormark{14,15},
Huatian Hu\,\orcidlink{0000-0001-8284-9494}\authormark{16}, 
Fabio~Della~Sala\,\orcidlink{0000-0003-0940-8830}\authormark{16,17}, 
Pu~Zhang\,\orcidlink{0000-0002-6253-0555}\authormark{18},
Cristian Cirac{\`i}\,\orcidlink{0000-0003-3349-8389}\authormark{16}, 
Javier~Aizpurua\,\orcidlink{0000-0002-1444-7589}\authormark{19,20,21},
Antton Babaze\,\orcidlink{0000-0002-9775-062X}\authormark{22}, 
Andrei~G.~Borisov\,\orcidlink{0000-0003-0819-5028}\authormark{23},
Xue-Wen Chen\,\orcidlink{0000-0002-0392-3551}\authormark{18},
Thomas~Christensen\,\orcidlink{0000-0002-9131-2717}\authormark{24}, 
Wei Yan\,\orcidlink{0000-0002-8150-1194}\authormark{25,26},
Yi Yang\,\orcidlink{0000-0003-2879-4968}\authormark{27},
Ulrich~Hohenester\,\orcidlink{0000-0001-8929-2086}\authormark{28}, 
Lorenz Huber\,\orcidlink{0009-0000-8321-4896}\authormark{28},
Martijn~Wubs\,\orcidlink{0000-0002-8286-7825}\authormark{24,29}, 
Simone De Liberato\,\orcidlink{0000-0002-4851-2633}\authormark{30,31,32},
P.~A.~D.~Gon\c{c}alves\,\orcidlink{0000-0001-8518-3886}\authormark{2,33},
F.~Javier~Garc{\'i}a~de~Abajo\,\orcidlink{0000-0002-4970-4565}\authormark{33,34},
Ortwin Hess\,\orcidlink{0000-0002-6024-0677}\authormark{35},
Illya~Tarasenko\,\orcidlink{0009-0007-0401-8290}\authormark{35},
Joel~D.~Cox\,\orcidlink{0000-0002-5954-6038}\authormark{2,3},
Line~Jelver\,\orcidlink{0000-0001-5503-5604}\authormark{2},
Eduardo J. C. Dias\,\orcidlink{0000-0002-6347-5631}\authormark{2},
Miguel~S{\'a}nchez~S{\'a}nchez\,\orcidlink{0000-0002-8067-2274}\authormark{36}, 
Dionisios Margetis\,\orcidlink{0000-0001-9058-502X}\authormark{37}, 
Guillermo G{\'o}mez-Santos\,\orcidlink{0000-0002-4048-1481}\authormark{38}, 
Tobias Stauber\,\orcidlink{0000-0003-0983-2420}\authormark{39},
Sergei~Tretyakov\,\orcidlink{0000-0002-4738-9987}\authormark{40},
Constantin Simovski\,\orcidlink{0000-0003-4338-4713}\authormark{40},
Samaneh~Pakniyat\,\orcidlink{0000-0001-9621-4676}\authormark{41},
J. Sebasti{\'a}n G{\'o}mez-D{\'i}az\,\orcidlink{0000-0002-6852-8317}\authormark{41},
Igor~V.~Bondarev\,\orcidlink{0000-0003-0739-210X}\authormark{42}, 
Svend-Age Biehs\,\orcidlink{0000-0002-5101-191X}\authormark{43}, 
Alexandra~Boltasseva\,\orcidlink{0000-0001-8905-2605}\authormark{44,45}, 
Vladimir M. Shalaev\,\orcidlink{0000-0001-8976-1102}\authormark{44},
Alexey V. Krasavin\,\orcidlink{0000-0003-2522-5735}\authormark{46}, 
Anatoly V. Zayats\,\orcidlink{0000-0003-0566-4087}\authormark{46},
Andrea~Al{\`u}\,\orcidlink{0000-0002-4297-5274}\authormark{47,48},
Jung-Hwan Song\,\orcidlink{0000-0001-9502-5718}\authormark{49},
Mark~L.~Brongersma\,\orcidlink{0000-0003-1777-8970}\authormark{50},
Uriel Levy\,\orcidlink{0000-0002-5918-1876}\authormark{51},
Olivia~Y.~Long\,\orcidlink{0000-0003-4636-9463}\authormark{52}, 
Cheng Guo\,\orcidlink{0000-0003-4913-8150}\authormark{53},
Shanhui Fan\,\orcidlink{0000-0002-0081-9732}\authormark{52,53},
Sergey~I.~Bozhevolnyi\,\orcidlink{0000-0002-0393-4859}\authormark{54},
Adam Overvig\,\orcidlink{0000-0002-7912-4027}\authormark{55},
Filipa~R.~Prud{\^e}ncio\,\orcidlink{0000-0002-7073-0987}\authormark{56,57},
M{\'a}rio~G.~Silveirinha\,\orcidlink{0000-0002-3730-1689}\authormark{56},
S.~Ali~Hassani~Gangaraj\,\orcidlink{0000-0003-1818-3215}\authormark{58},
Christos Argyropoulos\,\orcidlink{0000-0002-8481-8648}\authormark{59},
Paloma~A.~Huidobro\,\orcidlink{0000-0002-7968-5158}\authormark{60,61}, 
Emanuele Galiffi\,\orcidlink{0000-0003-3839-8547}\authormark{47}, 
Fan~Yang\,\orcidlink{0000-0002-8648-1858}\authormark{62},
John B. Pendry\,\orcidlink{0000-0001-5145-5441}\authormark{63}, and
David~A.~B.~Miller\,\orcidlink{0000-0002-3633-7479}\authormark{53}
}

\address{\authormark{1}School of Electrical and Computer Engineering, Cornell University, Ithaca, New York 14853, United States} 
\address{\authormark{2}POLIMA---Center for Polariton-driven Light--Matter Interactions, University of Southern Denmark, Campusvej 55, DK-5230 Odense M, Denmark} 
\address{\authormark{3}D-IAS---Danish Institute for Advanced Study, University of Southern Denmark, Campusvej 55, DK-5230 Odense M, Denmark} 
\address{\authormark{4}Departamento de F\'{i}sica Te\'{o}rica de la Materia Condensada, Universidad Aut\'{o}noma de Madrid, E-28049 Madrid, Spain} 
\address{\authormark{5}Condensed Matter Physics Center (IFIMAC), Universidad Aut\'{o}noma de Madrid, E-28049 Madrid, Spain} 
\address{\authormark{6}School of Electrical and Electronic Engineering, Nanyang Technological University, Singapore 639798, Singapore} 
\address{\authormark{7}Key Laboratory of Radar Imaging and Microwave Photonics, Ministry of Education, College of Electronic and Information Engineering, Nanjing University of Aeronautics and Astronautics, Nanjing 211106, China} 
\address{\authormark{8}WaveCoRE Division, Department of Electrical Engineering, KU Leuven, Leuven, B-3001 Belgium}
\address{\authormark{9}Department of Electrical and Computer Engineering, Johns Hopkins University, Baltimore, Maryland 21218, United States} 
\address{\authormark{10}Department of Physics, Jackson State University, Jackson, MS 39217 United States} 
\address{\authormark{11}Department of Physics, Aeronautics Institute of Technology, 12228-900, So Jos dos Campos, SP, Brazil} 
\address{\authormark{12}Centro de F\'{\i}sica (CF-UM-UP) and Departamento de F\'{\i}sica, Universidade do Minho, P-4710-057 Braga, Portugal} 
\address{\authormark{13}International Iberian Nanotechnology Laboratory (INL), Av Mestre Jos\'e Veiga, 4715-330 Braga, Portugal} 
\address{\authormark{14}Max-Born-Institut, 12489 Berlin, Germany} 
\address{\authormark{15}Humboldt-Universit{\"a}t zu Berlin, Institut f{\"u}r Physik, AG Theoretische Optik and Photonik, 12489 Berlin, Germany} 
\address{\authormark{16}Center for Biomolecular Nanotechnologies, Istituto Italiano di Tecnologia, Via Barsanti 14, 73010, Arnesano, Italy} 
\address{\authormark{17}Institute for Microelectronics and Microsystems (CNR-IMM), Via Monteroni, Campus Unisalento, 73100 Lecce, Italy} 
\address{\authormark{18}School of physics, Huazhong University of Science and Technology, Luoyu Road 1037, Wuhan 430074, China} 
\address{\authormark{19}Donostia International Physics Center DIPC, 20018 Donostia-San Sebastián, Spain} 
\address{\authormark{20}Ikerbasque, Basque Foundation for Science, 48009 Bilbao, Spain} 
\address{\authormark{21}Department of Electricity and Electronics, University of the Basque Country, 48940 Leioa, Spain} 
\address{\authormark{22}Department of Applied Physics, University of the Basque Country, 20018 Donostia, Spain} 
\address{\authormark{23}Universit{\'e} Paris-Saclay, CNRS, Institut des Sciences Mol{\'e}culaires d’Orsay, Orsay, France} 
\address{\authormark{24}Department of Electrical and Photonics Engineering, Technical University of Denmark, Kongens Lyngby, Denmark} 
\address{\authormark{25}Key Laboratory of 3D Micro/Nano Fabrication and Characterization of Zhejiang Province, School of Engineering, Westlake University, Hangzhou, Zhejiang Province, China} 
\address{\authormark{26}Institute of Advanced Technology, Westlake Institute for Advanced Study, Hangzhou, Zhejiang Province, China} 
\address{\authormark{27}Department of Physics and HK Institute of Quantum Science and Technology, The University of Hong Kong, Pokfulam, Hong Kong, China} 
\address{\authormark{28}Institute of Physics, University of Graz, Universit{\"a}tsplatz 5, 8010 Graz, Austria} 
\address{\authormark{29}NanoPhoton -- Center for Nanophotonics, Technical University of Denmark, 2800 Kongens Lyngby, Denmark} 
\address{\authormark{30}School of Physics and Astronomy, University of Southampton, Southampton SO17 1BJ, United Kingdom} 
\address{\authormark{31}Sensorium Technological Laboratories, Nashville, Tennessee 37210, United States} 
\address{\authormark{32}Istituto di Fotonica e Nanotecnologie, Consiglio Nazionale delle Ricerche (CNR), 20133 Milano, Italy} 
\address{\authormark{33}ICFO--Institut de Ciencies Fotoniques, The Barcelona Institute of Science and Technology, 08860 Castelldefels (Barcelona), Spain} 
\address{\authormark{34}ICREA--Instituci\'o Catalana de Recerca i Estudis Avan\c{c}ats, Passeig Llu\'{\i}s Companys 23, 08010 Barcelona, Spain} 
\address{\authormark{35}School of Physics and CRANN Institute, Trinity College Dublin, Dublin, 2 Ireland} 
\address{\authormark{36}Instituto de Ciencia de Materiales de Madrid, CSIC, E-28049 Madrid, Spain} 
\address{\authormark{37}Department of Mathematics, and Institute for Physical Science and Technology, University of Maryland, College Park, Maryland 20742, United States} 
\address{\authormark{38}Departamento de F\'{\i}sica de la Materia Condensada, Instituto Nicol{\'a}s Cabrera and Condensed Matter Physics Center (IFIMAC), Universidad Aut{\'o}noma de Madrid, E-28049 Madrid, Spain} 
\address{\authormark{39}Instituto de Ciencia de Materiales de Madrid, CSIC, E-28049 Madrid, Spain} 
\address{\authormark{40}Aalto University, PO Box 15500, FI-00076 Aalto, Finland} 
\address{\authormark{41}Electrical and Computer Engineering Department, University of California, Davis, Davis, California 95616, United States} 
\address{\authormark{42}Department of Mathematics \& Physics, North Carolina Central University, Durham, NC 27707, United States} 
\address{\authormark{43}Institut f{\"u}r Physik, Carl von Ossietzky Universit{\"a}t, 26111, Oldenburg, Germany} 
\address{\authormark{44}Elmore Family School of Electrical and Computer Engineering, Purdue Quantum Science and Engineering Institute, and Birck Nanotechnology Center, West Lafayette, IN 47907, United States} 
\address{\authormark{45}School of Materials Engineering, Purdue University, West Lafayette, IN 47907, United States} 
\address{\authormark{46}Department of Physics and London Centre for Nanotechnology, King’s College London, London, WS2R 2LS, United Kingdom} 
\address{\authormark{47}Photonics Initiative, Advanced Science Research Center, City University of New York, New York, New York 10031, United States} 
\address{\authormark{48}Physics Program, Graduate Center of the City University of New York, New York, New York 10016, United States} 
\address{\authormark{49}College of Design and Engineering, National University of Singapore, Engineering Drive 3, Singapore 117583} 
\address{\authormark{50}Materials Science and Engineering, Stanford University, 496 Lomita Mall Suite 102, Stanford, CA 94305, United States} 
\address{\authormark{51}Department of Applied Physics, The Faculty of Science, The Hebrew University of Jerusalem, Jerusalem, Israel} 
\address{\authormark{52}Department of Applied Physics, Stanford University, Stanford, CA 94305, United States} 
\address{\authormark{53}Ginzton Laboratory and Department of Electrical Engineering, Stanford University, Stanford, CA 94305, United States} 
\address{\authormark{54}Center for Nano Optics, University of Southern Denmark, Campusvej 55, DK-5230 Odense M, Denmark} 
\address{\authormark{55}Department of Physics, Stevens Institute of Technology, Hoboken, NJ 07030, United States} 
\address{\authormark{56}University of Lisbon – Instituto Superior T{\'e}cnico and Instituto de Telecomunica\c{c}{\~o}es, Avenida Rovisco Pais 1, 1049-001 Lisbon, Portugal} 
\address{\authormark{57}Instituto Universit{\'a}rio de Lisboa (ISCTE-IUL), Avenida das For\c{c}as Armadas 376, 1600-077 Lisbon, Portugal} 
\address{\authormark{58}Optical Physics Division, Corning Research and Development, Corning, New York 14831, United States} 
\address{\authormark{59}Department of Electrical Engineering, The Pennsylvania State University, University Park, Pennsylvania 16802, United States} 
\address{\authormark{60}Departamento de F{\'i}sica Te{\'o}rica de la Materia Condensada, Universidad Aut{\'o}noma de Madrid, E-28049, Madrid, Spain} 
\address{\authormark{61}Condensed Matter Physics Center (IFIMAC), Universidad Aut{\'o}noma de Madrid, E-28049, Madrid, Spain} 
\address{\authormark{62}College of Physics and Key Laboratory of High Energy Density Physics and Technology of the Ministry of Education,Sichuan University, Chengdu, Sichuan 610065, China} 
\address{\authormark{63}The Blackett Laboratory, Department of Physics, Imperial College London, London SW7 2AZ, United Kingdom} 

\medskip

\email{\authormark{*}francesco.monticone@cornell.edu} \email{\authormark{**}asger@mailaps.org} 


\begin{abstract*} 
Photonic technologies continue to drive the quest for new optical materials with unprecedented responses. A major frontier in this field is the exploration of nonlocal (spatially dispersive) materials, going beyond the local, wavevector-independent assumption traditionally made in optical material modeling. On one end, the growing interest in plasmonic, polaritonic and quantum materials has revealed naturally occurring nonlocalities, emphasizing the need for more accurate models to predict and design their optical responses. This has major implications also for topological, nonreciprocal, and time-varying systems based on these material platforms. Beyond natural materials, artificially structured materials—metamaterials and metasurfaces--can provide even stronger and engineered nonlocal effects, emerging from long-range interactions or multipolar effects. This is a rapidly expanding area in the field of photonic metamaterials, with open frontiers yet to be explored. In the case of metasurfaces, in particular, nonlocality engineering has become a powerful tool for designing strongly wavevector-dependent responses, enabling enhanced wavefront control, spatial compression, multifunctional devices, and wave-based computing. Furthermore, nonlocality and related concepts play a critical role in defining the ultimate limits of what is possible in optics, photonics, and wave physics. This Roadmap aims to survey the most exciting developments in nonlocal photonic materials, highlight new opportunities and open challenges, and chart new pathways that will drive this emerging field forward--toward new scientific discoveries and technological advancements.
\end{abstract*}


\newpage

\tableofcontents

\newpage

\section[Introduction (Monticone \& Mortensen)]{Introduction}
\label{sec:Introduction}

\author{Francesco Monticone\,\orcidlink{0000-0003-0457-1807} \& N.~Asger~Mortensen\,\orcidlink{0000-0001-7936-6264}}

In the optical sciences--the study of light and light-matter interactions--research primarily focuses on the optical frequency region of the broader electromagnetic spectrum. This field encompasses the exploration and manipulation of electromagnetic waves, which have catalyzed groundbreaking developments over recent decades. These advancements have led to the emergence of numerous transformative concepts in photonics, all deeply rooted in the classical electrodynamics of Maxwell's equations.  

In a sense, all the advancements in the field of optics over the decades and centuries have been based on shaping materials and progressively relaxing assumptions about their optical response. Consider, for example, a simple piece of glass. Its macroscopic electromagnetic properties--at least at the level relevant to, for instance, lens design--are very simple and can be described by linear, time-invariant, isotropic, local, nonmagnetic, passive, constitutive parameters, namely, a scalar frequency-dispersive electric susceptibility and the free-space magnetic permeability. Significant more complexity can be unlocked and accessed by relaxing these assumptions. 

For instance, breaking the assumption of linearity gave rise to the field of nonlinear optics, which gained prominence with the invention of the laser and has had profound implications for both fundamental science and technology. Relaxing the assumption of an isotropic response leads, among other things, to polarization-dependent functionalities, while more complex linear constitutive relations--incorporating effects such as negative constitutive parameters, magneto-electric coupling, chirality, nonreciprocity, and bianisotropy---have been central to the field of metamaterials, enabling unprecedented control over electromagnetic waves. Furthermore, relaxing the assumption of time-invariance is one of the most active areas of research in applied electromagnetics, photonics, and wave physics, allowing for the breaking of frequency/energy conservation and time-reversal symmetry. These examples illustrate how efforts to challenge any of these fundamental assumptions have sparked and motivated entirely new research fields. 

Yet, among these assumptions, spatial locality has arguably been the least explored. This is not to say that nonlocal effects have been overlooked until recently. Early studies date back to the mid-20th century with even earlier seminal contributions by various authors (for further details, see Section~\ref{Sec:Mortensen}). By the early 1960s, the basic physical mechanisms underlying optical nonlocality in natural materials were well understood, as in the notable work by Hopfield and Thomas~\cite{Hopfield:1963} (the former gained recognition for his work on physical computing and artificial neural networks, for which he won the 2024 Nobel Prize in Physics). Part of the Introduction of this paper is worth quoting here for its clarity~\cite{Hopfield:1963}: \emph{In [the local] approximation, the dielectric polarization $P$ within a small volume of radius $r_0$ ($r_0 \ll$ any wavelength involved) depends only on the value of the electric field inside this volume (at the present time and in the past) and is not explicitly dependent on the electric field or other parameters outside the volume under consideration. The term "spatial dispersion" has been used to apply to dielectric behavior for which the local description is not valid. In general, spatial dispersion refers to the wave-vector dependence of the dielectric constant. Implicitly contained in the supposition of local dielectric behavior is the neglect of the transport of energy by any mechanism other than electromagnetic waves. When energy transport by other mechanisms must be considered anomalous (nonlocal) dielectric behavior results, often accompanied by new physical phenomena.}

Although the fundamental ideas of nonlocality were already present in these early works, it was only with the advent of the field of plasmonics and nano-optics that their significance became clearly recognized. In particular, as research in metamaterials and plasmonics started uncovering extreme forms of light-matter interactions, such as ultra-strong field confinement and enhancement in plasmonic nanostructures, it became evident that the traditional local approximation was insufficient. More sophisticated models were required to regularize and accurately describe how these systems respond to tightly confined electromagnetic fields, which contain large wavevector components in their spatial spectrum. This need is even more evident today in the study of nonreciprocal, topological, and time-varying systems, especially those made of continuous media with negative constitutive parameters, whose overall response often depends on the large-wavevector behavior of the system~\cite{Monticone:2020}. As another example, nonlocal response is also important for quantum-fluctuation phenomena~\cite{Intravaia:2019} where spatial correlations in nonequilibrium quantum systems -- often overlooked in local thermal equilibrium approximations -- significantly influence interactions like quantum friction and near-field radiative forces~\cite{Intravaia:2016}. In  many cases,local models can lead to incorrect or even nonphysical predictions, whereas nonlocal models correct the asymptotic response for large wave vectors, ensuring a gradual vanishing of the polarization response. 

In this context, nonlocality is both \emph{a curse and a blessing}. On one hand, it resolves the unphysical issues of local models and predicts new physical effects that would otherwise be missed by the local approximation. On the other hand, it also imposes fundamental limits on how strong and localized light-matter interactions can be. 

In parallel to the exploration of nonlocality in natural materials, the field of metamaterials has enabled the realization of strong \emph{artificial} nonlocality, emerging--at the "effective medium" level--from long-range interactions across multiple unit cells or from the response of multipolar meta-atoms. In this case, nonlocality is clearly a "blessing", offering a powerful new tool to engineer strongly wavevector-dependent functionalities. "Nonlocality engineering" adds wavevector as a new degree of freedom in the metamaterial design toolbox, enabling a new degree of control over the angular response of incident light, strong angular/frequency selectivity, space compression, multifunctional devices, and much more. Notably, nonlocal metamaterials or metasurfaces can even be used to perform mathematical operations on an input wavefront by tailoring their spatial impulse response, using nonlocality for wave-based physical computing (an incidental but intriguing parallel to the career of one of the pioneers of this field, John J. Hopfield.)

The goal of this Roadmap is to review the current state of this emerging frontier in optics and photonics, highlighting both the challenges and opportunities of nonlocality---its \emph{curse and blessing}. By surveying recent developments across various aspects of nonlocality, we also aim to define new directions that will drive this field forward, opening pathways for fundamental scientific discoveries and technological advancements.

Toward these goals, we have assembled a large collection of contributions from many of the leaders of this field. The Roadmap consists of thirty-four sections, structured into five parts:
\begin{itemize} 
\item Part I starts with a general introduction of the concept of nonlocal response in the interaction of electromagnetic waves with matter (Section~\ref{Sec:Mortensen}) and semi-classical hydrodynamics (Section~\ref{sec:Fernandez-Dominguez}), followed by three sections that explore some of the foundational aspects of this field in greater detail (Sections~\ref{sec:Khurgin}, \ref{sec:Shahbazyan}, \ref{sec:Chaves}).

\item Part II introduces the continuum framework for nonlocal plasmonics (Section~\ref{sec:Wegner}), and continues with discussions of quantum hydrodynamics (Section~\ref{sec:Hu}). This is followed by contributions on density-functional theory approaches in the jellium model (Section~\ref{sec:Aizpurua}) and in atomistic representations (Section~\ref{sec:Zhang}). Implications of nonlocality for light-matter interactions are also discussed (Section~\ref{sec:Tserkezis}). Finally, aspects of surface-response formalism are discussed (Sections~\ref{sec:Christensen} and \ref{sec:Hohenester}).

\item Part III highlights significant directions that go beyond the nonlocal response of free-electron bulk metals, ranging from the response of doped semiconductors (Section~\ref{sec:Wubs}) and polar dielectrics (Section~\ref{sec:DeLiberato}) to the nonlocal behavior of 2D plasmons in graphene and noble-metal surface-states (Sections~\ref{sec:Goncalves}, \ref{sec:Hess} and \ref{sec:Cox}), nonlinear and nonlocal effects in 2D materials (Section~\ref{sec:Jelver}) and nonlocal chirality in moir{\'e} multilayer systems (Section~\ref{sec:SanchezSanchez}). 

\item Part IV explores the principles, design strategies, and emerging applications of nonlocal metamaterials and metasurfaces, highlighting their potential to transform modern optics and photonics. Part IV starts with a general discussion of strong spatial dispersion in metamaterials and metasurfaces (Section~\ref{sec:Tretyakov}) and then explores the role of nonlocality and the new opportunities enabled by nonlocality engineering in a wide range of meta-structures for several applications, including hyperbolic metamaterials and metasurfaces (Section~\ref{sec:Pakniyat}), "transdimensional" plasmonic thin films (Section~\ref{sec:Bondarev}), nanorod metamaterials for light management (Section~\ref{sec:Krasavin}) 
, spatio-temporal metamaterials (Section~\ref{sec:Alu}), metasurfaces with high spectral/angular control (Section~\ref{sec:Song}), nonlocal metalenses (Section~\ref{sec:Levy}), metasurfaces for space compression (Section~\ref{sec:Long}), electro-optic spatiotemporal nonlocal metasurfaces (Section~\ref{sec:Bozhevolnyi}), and multi-functional meta-optics (Section~\ref{sec:Overvig}). 

\item Part V delves into the implications of combining nonlocality with other advanced concepts to investigate new physics, extreme scenarios, and fundamental limits in photonics. Topics include the role of nonlocality in topological metamaterials (Section~\ref{sec:Prudencio}) and nonreciprocal plasmonics (Section~\ref{sec:HassaniGangaraj}). This section also features contributions on combining transformation optics and nonlocal models to study singular geometries (Section~\ref{sec:Huidobro}) and it explores the idea of "overlapping nonlocality"---a concept related to, but distinct from, standard nonlocality--and its impact on the ultimate thickness limits of optical systems (Section~\ref{Sec:Miller}).


\end{itemize}

The Roadmap concludes with some brief remarks by the lead authors (Section~\ref{sec:Conclusion}).

\usection{\emph{Part I} --- Fundamental frameworks}

\label{roadmap:Part1}

\section[Linear-response formalism for nonlocal electrodynamics (Mortensen \& Monticone)]{Linear-response formalism for nonlocal electrodynamics}

\label{Sec:Mortensen}

\author{N.~Asger~Mortensen\,\orcidlink{0000-0001-7936-6264} \& Francesco Monticone\,\orcidlink{0000-0003-0457-1807}}

\subsection*{Overview}

The early fundamental contributions to the concepts of nonlocal response and spatial dispersion in condensed matter physics and electromagnetism arose primarily in the mid-20th century. These ideas expanded on classical theories of material response to electromagnetic fields by considering the fact that the response at a given point in a material might depend on the electromagnetic fields at other, nonlocal points~\cite{Landau:1984}. However, this nonlocal dependence is limited to a vicinity of the initial point, with the range governed by the underlying microscopic dynamics of the system.

The numerous independent developments in this field make it challenging to present a chronological or cohesive review that fully honors the significant contributions across quantum physics, condensed-matter physics, plasma physics, and electrodynamics. The study of nonlocal phenomena is inherently tied to the hydrodynamic framework, including the Navier–Stokes equations of classical continuum mechanics. Notably, Madelung’s seminal work established a vital connection between the Hamiltonian formalism of quantum mechanics and the hydrodynamic formalism of classical physics~\cite{Madelung:1927}. Equally important are Bloch’s contributions to the hydrodynamic descriptions of electrons in solids~\cite{Bloch:1933}, Landau’s work on the damping of longitudinal waves in plasmas~\cite{Landau:1946}, Lindhard’s quantum mechanical calculations of the nonlocal dielectric response function of electrons in solids~\cite{Lindhard:1954}, and the development of the Kubo formalism for analyzing the linear response of quantum systems~\cite{Kubo:1966}. Each of these contributions fundamentally addresses the nonlocal response of the underlying electron systems.

\textbf{The common implicit assumption.} As a general theme, these developments have largely relied on the common and often intuitive assumptions of a linear and local response of matter to light fields~\cite{Jackson:1999}. Mathematically, Maxwell's equations are augmented by linearized and spatially local constitutive relations between the relevant fields, such as $\boldsymbol{D}(\boldsymbol{r}) =\varepsilon(\boldsymbol{r})\boldsymbol{E}(\boldsymbol{r})$ for the displacement field $\boldsymbol{D}(\boldsymbol{r})$ arising from an applied electric field $\boldsymbol{E}(\boldsymbol{r})$ in the spatial point $\boldsymbol{r}$. The use of Ohm's law $\boldsymbol{J}(\boldsymbol{r}) =\sigma(\boldsymbol{r})\boldsymbol{E}(\boldsymbol{r}) $ is another such example where the induced current density $\boldsymbol{J}(\boldsymbol{r})$ is arising from an applied electric field $\boldsymbol{E}(\boldsymbol{r})$ in the spatial point $\boldsymbol{r}$. However, there is no fundamental reason to assume that the material's response will be confined to the exact point of perturbation.

\subsection*{Current status}

\begin{figure}
    \centering
    \includegraphics[width=0.6\linewidth]{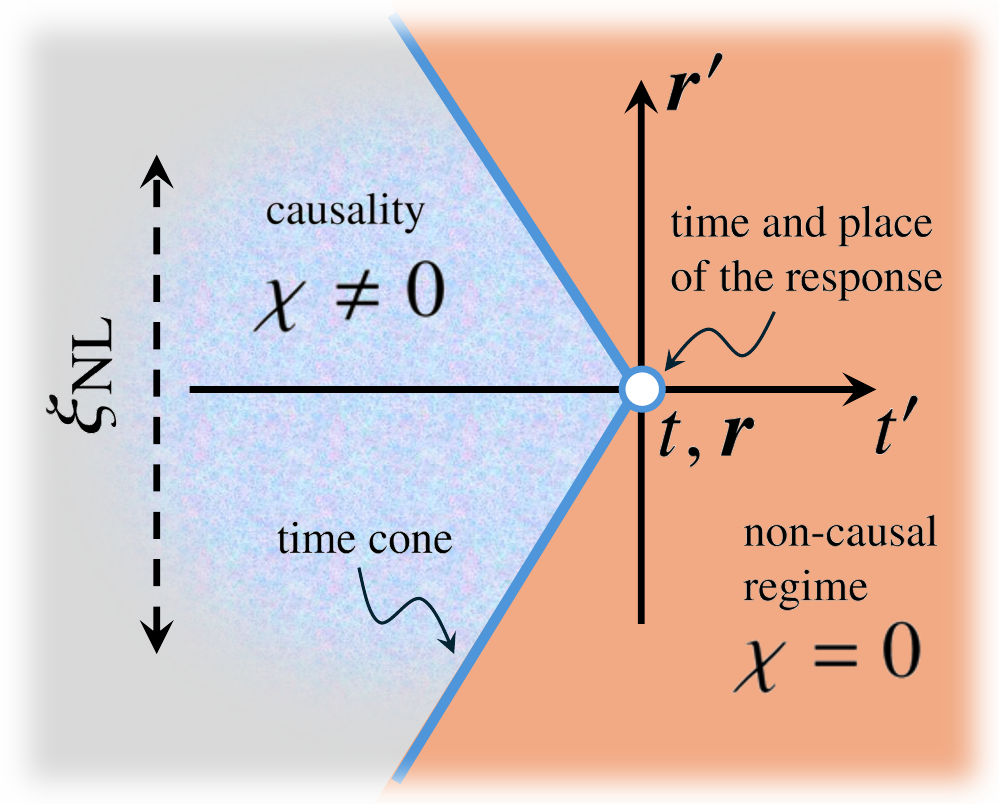}
    \caption{
    Schematic illustration (inspired by Ref.~\cite{Svirko:1998}) of the constraints of causality on the response function $\chi(t- t',\boldsymbol{r},\boldsymbol{r}')$ and the spatio-temporal integrals in Eq.~\eqref{eq:Mortensen:Eq1}. The causal regime (gray-shaded) represents the region where the response function can take non-zero values, while the non-causal regime (orange-shaded) indicates where the response function is strictly zero. The "time cone" is determined by the characteristic speed $v$ of propagation in the medium, ensuring no response occurs from points $\boldsymbol{r}'$ located farther from the origin $(t,\boldsymbol{r})$ than $v \left\vert t' -t \right\vert$. The finite range of the spatially nonlocal response (associated with intrinsic dynamics of the matter) is denoted by $\xi_\mathrm{NL}$.}
\label{fig:Mortensen-Fig1}
\end{figure}

\textbf{Beyond the local-response approximation.} Building on the concepts of Taylor expansions, the Volterra series offers a systematic approach to phenomenologically describe spatio-temporal nonlocal effects. Assuming a system response $\mathscr{Z}$ associated with a system input $\mathscr{F}$, i.e. $\mathscr{Z}[\mathscr{F}]$, the linear contribution in the Volterra series is formally expressed as~\cite{Svirko:1998,Wubs:2016}
\begin{equation}
    \mathscr{Z}(t,\boldsymbol{r})= \int_{-\infty}^t dt'\,\int d\boldsymbol{r}'\, \chi(t-t'; \boldsymbol{r},\boldsymbol{r}') \mathscr{F}(t,\boldsymbol{r}'),
    \label{eq:Mortensen:Eq1}
\end{equation}
where $\chi(t-t'; \boldsymbol{r},\boldsymbol{r}')$ is the spatio-temporal nonlocal response function, with the spatial and temporal integrals extending over all space and time, naturally subject to the restrictions by causality~\cite{Svirko:1998}. Fig.~\ref{fig:Mortensen-Fig1} illustrates the spatio-temporal domain, where the response function is finite (shaded area), while being strictly zero in the noncausal part. The "time cone" is defined by the bounds on the speed $v$ of propagation in the medium, being naturally bounded by the speed of light $c$ in vacuum. Thus, causality influences both the temporal and spatial integrals in Eq.~(\ref{eq:Mortensen:Eq1}). The topics of causality and  Kramers--Kronig (KK) relations are being further discussed in Sec.~\ref{sec:Khurgin}. The nonlocal response function is typically characterized by a finite range $\xi_\mathrm{NL}$ which phenomenologically reflects the inherent microscopic or mesoscopic properties of the material~\cite{Ginzburg:2013,Mortensen:2013,Mortensen:2021a}. Naturally, the phenomenology itself does not include detailed information about these underlying characteristics. Typical examples of nonlocal electrodynamics associated with matter excitations include plasmons, excitons, or phonons~\cite{Fox:2001,Haug:2009,Pekar:1958,Agranovich:1984,Bruus:2014}. Starting from a microscopic Hamiltonian quantum description, the Madelung formalism (see Sec.~\ref{sec:Chaves} of this Roadmap) provides a pathway to a hydrodynamic framework (see Sec.~\ref{sec:Fernandez-Dominguez} of this Roadmap), which can ultimately be expressed as a nonlocal response function $\chi$. 

\textbf{The local-response approximation (LRA).} The common assumption of spatial locality naturally emerges by the approximation
\begin{subequations}
\begin{equation}
    \chi(t-t'; \boldsymbol{r},\boldsymbol{r}') \approx \chi(t-t',\boldsymbol{r}) \delta (\boldsymbol{r}-\boldsymbol{r}'),
    \label{eq:Mortensen:Eq2}
\end{equation}
where $\delta(\boldsymbol{r})$ is the Dirac delta function, which exhibits no long-range behavior; it contributes only when $\boldsymbol{r}'$ is identical to $\boldsymbol{r}$. The spatial integral in Eq.~(\ref{eq:Mortensen:Eq1}) is now performed straightforwardly, giving the common LRA result 
\begin{equation}
\mathscr{Z}(t,\boldsymbol{r})\approx \int_{-\infty}^t dt'\,\chi(t-t',\boldsymbol{r}) \mathscr{F}(t,\boldsymbol{r}).
\label{eq:Mortensen:Eq3}
\end{equation}
\end{subequations}
The remaining temporal integral is associated with the frequency-dispersive response, subject to KK conditions imposed by causality.

\textbf{Translational invariance.} For translationally invariant media, it is useful to Fourier transform Eq.~(\ref{eq:Mortensen:Eq1}) with respect to the temporal and spatial coordinates, resulting in a form that is particularly familiar to the communities of condensed-matter physics and solid-state spectroscopy~\cite{Kuzmany:1998}
\begin{equation}
    \mathscr{Z}(\omega,\boldsymbol{q})=  \chi(\omega,\boldsymbol{q}) \mathscr{F}(\omega,\boldsymbol{q}),
    \label{eq:Mortensen:Eq4}
\end{equation}
with the response function $\chi(\omega,\boldsymbol{q})$ now explicitly exhibiting both its temporal and spatial dispersion through its explicit dependence on both the temporal frequency $\omega$ and the spatial wave vector $\boldsymbol{q}$.

Exploiting light fields with the dispersion relation $\omega(q) = c q$, there has been a natural preference to account for light-matter interactions within the LRA, which in Fourier space amounts to the approximation $\chi(\omega,\boldsymbol{q})\approx \chi(\omega,\boldsymbol{q}\rightarrow 0)$. This limit is well motivated by the fact that commonly $q \xi_\mathrm{NL} =  \omega \xi_\mathrm{NL} /c \ll 1$ for the most encountered optical materials. In many cases, the momentum of optical photons is too low, and the nonlocal length scale is too short for nonlocal effects to have any significant impact. Alternatively, this can be understood as the speed of light $c$ being much greater than any intrinsic velocities $v$ characteristic for excitations in matter. Consequently, in most practical scenarios, common optical materials can be approximated as homogeneous and local in their response to light fields. This assumption is largely justified by the small size of the constituent atoms and their dense arrangement within solids, where all microscopic length scales are much smaller than the wavelength of light.

\textbf{Beyond the low-$\boldsymbol{q}$ regime.} \emph{What would happen if translational invariance were broken or if we could artificially increase the separation between atoms?} In the first scenario, where the system has small but finite dimensions $a$, the relaxation of Noether's conservation laws~\cite{Kosmann-Schwarzbach:2011} would pragmatically allow probing at finite wavevectors
$q\sim 1/a$. In the second scenario, introducing an artificial lattice constant 
$a$, the associated discrete translational invariance would enable \emph{Umklapp} scattering processes~\cite{Aschroft:1976} among photonic Bloch states~\cite{Sakoda:2005}, which also facilitate probing at finite wavevectors 
$q\sim 1/a$ (or higher multiples). Manipulating and engineering a finite $q$-response is no longer merely aspirational --- the advent of photonic nanostructures and metamaterials has turned this idea into reality. In recent decades, advancements have enabled the fabrication of increasingly smaller nanostructures and intricate photonic crystals and metamaterials with engineered lattice constants, along with improved capabilities for exploring light fields at sub-wavelength scales, including various photon and electron-based near-field techniques~\cite{Dunn:1999,Hecht:2000,Chen:2019,GarciadeAbajo:2010,Polman:2019}. As a result, we are now moving beyond the traditional paradigm of probing only the $q\rightarrow 0$ limit when investigating naturally occurring nanostructures and artificially engineered metamaterials. This progress has enabled a deeper fundamental understanding of nonlocal light-matter interactions, including potential quantum effects within the underlying matter systems. Simultaneously, it has opened entirely new possibilities for engineering the complex optical responses of nonlocal metamaterials. 

\textbf{Homogeneous media versus metamaterials.} The nonlocal responses of both homogeneous materials and metamaterials involve spatial dispersion, where field variations influence the optical response. In homogeneous materials, this nonlocality arises from intrinsic microscopic and mesoscopic properties, leading to wavevector-dependent effects. In metamaterials, nonlocality can emerge from the artificial periodicity of the composite with strong interactions between neighboring resonant meta-units, likewise giving rise to wavevector-dependent effects. These two interpretations -- material-induced and resonance-driven nonlocality -- are often discussed separately, yet they share a common underlying principle: both modify wave propagation by introducing additional modes and spatial dispersion at different scales. As noted by Podolskiy during the 2015 Faraday Discussions on nonlocality~\cite{GarciadeAbajo:2015}:  

\begin{itemize}
    \item[] 
\emph{"The drastic difference between nonlocality in homogeneous media studied previously and nonlocality in metamaterials lies in the origin of additional waves. In homogeneous materials, the nonlocal response is attributed to the spatial dispersion of the material itself. In metamaterials, however, nonlocality appears at the 'effective medium' level; the response of every component of the metamaterial may remain local, while the granularity of the composite leads to the spatial dispersion of the effective permittivity."}  
\end{itemize}

Recognizing this dual nature of nonlocality is essential for establishing a unified conceptual framework for nonlocal photonics, as explored throughout this Roadmap.

\textbf{Short-range approximation.} Returning to Eq.~\eqref{eq:Mortensen:Eq1} we finally offer a short-range version, which corresponds to retaining terms up to order $q^2$ in Eq.~\eqref{eq:Mortensen:Eq4}. Initially motivated by Ref.~\cite{Ginzburg:2013}, this is enabled by the introduction of real-space moments of the short-range response function~\cite{Mortensen:2013,Mortensen:2021a}. For simplicity, we consider an isotropic scalar response, and we get the following spatial expression
\begin{equation}
    \mathscr{Z}(\omega,\boldsymbol{r})\approx \left[\chi(\omega, \boldsymbol{r})+ \xi_\mathrm{NL}^2\nabla^2  \right]\mathscr{F}(\omega,\boldsymbol{r}),
    \label{eq:Mortensen:Eq5}
\end{equation}
where the first term on the right-hand side emerges as a result of the LRA in Eq.~\eqref{eq:Mortensen:Eq2} (corresponding to $q\rightarrow 0$), while in the second nonlocal term with the Laplacian operator, $\xi_\mathrm{NL}$ is the range of the nonlocal response, being formally defined as the second spatial moment of the response function~\cite{Mortensen:2013,Mortensen:2021a}.
This scalar expression can be generalized to vectorial fields and non-scalar responses, with the nonlocal hydrodynamic model serving as a key example, where a nonlocal Ohm's law can be formulated in the form~\cite{Toscano:2013}
\begin{equation}
    \mathscr{Z}(\omega,\boldsymbol{r})\approx \left[\chi(\omega, \boldsymbol{r})+ \xi_\mathrm{NL}^2\nabla [\nabla \cdot ]  \right]\mathscr{F}(\omega,\boldsymbol{r}).
    \label{eq:Mortensen:Eq6}
\end{equation}
The change from a Laplacian to a gradient-of-divergence nonlocal correction term introduces the possibility for additional longitudinal waves~\cite{Wubs:2015,Raza:2015}, which is a main difference between nonlocality in homogeneous (conductive) media versus metamaterials (with local constituents).

\subsection*{Concluding remarks}

Having attempted to offer a general introduction to the concept of nonlocal response in the interaction of electromagnetic waves with matter, the remainder of this first part of the Roadmap explores some of these foundational aspects in greater detail, and for further insight, we direct readers to key textbooks~\cite{Landau:1984,Boardman:1982a}, as well as reviews on the nonlocal electrodynamics of metals~\cite{Barton:1979,Pitarke:2007,Raza:2015,Varas:2016,Mortensen:2021a} and plasma~\cite{Manfredi:2018,Moldabekov:2018,Bonitz:2018,Bonitz:2019}.
\section[Nonlocal hydrodynamics (Fern\'{a}ndez-Dom\'{i}nguez \emph{et al.})]{Nonlocal hydrodynamics}

\label{sec:Fernandez-Dominguez}

\author{Antonio~I.~Fern\'{a}ndez-Dom\'{i}nguez\,\orcidlink{0000-0002-8082-395X}, Yu Luo\,\orcidlink{0000-0003-2925-682X},
Xuezhi Zheng\,\orcidlink{0000-0002-5301-2908} 
\& Christos~Tserkezis\,\orcidlink{0000-0002-2075-9036}}

\subsection*{Overview}

\textbf{The hydrodynamic Drude model.} The hydrodynamic Drude model (HDM) constitutes the first and most straightforward realization of a nonlocal permittivity of metals, with a lifetime that is approaching a century~\cite{Bloch:1933}. It is the direct descendant of the standard Drude theory developed 30 years earlier~\cite{Drude:1900}, which describes free electrons in the metal as a gas, whose atoms move freely until they collide with the heavier nuclei (or, with much smaller probability, with each other). The corresponding angular-frequency $\omega$-dependent relative permittivity $\varepsilon$ is then~\cite{Aschroft:1976}
\begin{equation}\label{eq:Fernandez-Dominguez-Eq1}
\varepsilon_{\mathrm{m}} (\omega) =
\varepsilon_{\infty} (\omega) - 
\frac{\omega_p^{2}}
{\omega \left(\omega + \mathrm{i} \gamma\right)}~,
\end{equation}
where $\omega_p = \sqrt{n e^{2}/(m_{\mathrm{e}} \varepsilon_{0})}$ is the plasma frequency (here $\varepsilon_{0}$ is the vacuum permittivity, $e$ is the elementary charge, $m_{\mathrm{e}}$ the mass of electrons and $n$ their density), while $\gamma$ is a phenomenological damping rate accounting for the aforementioned collisions. Any contribution to the permittivity from bound electrons or other mechanisms (e.g. interband transitions) is included via $\varepsilon_{\infty}$, which can be obtained by subtracting the free-electron contribution from experimental data, or by modeling any additional known mechanism as a Lorentzian; for a purely free-electron metal (as in alkali), $\varepsilon_{\infty} = 1$.

\begin{figure}[hb!]
\centering\includegraphics[width=0.9\textwidth]{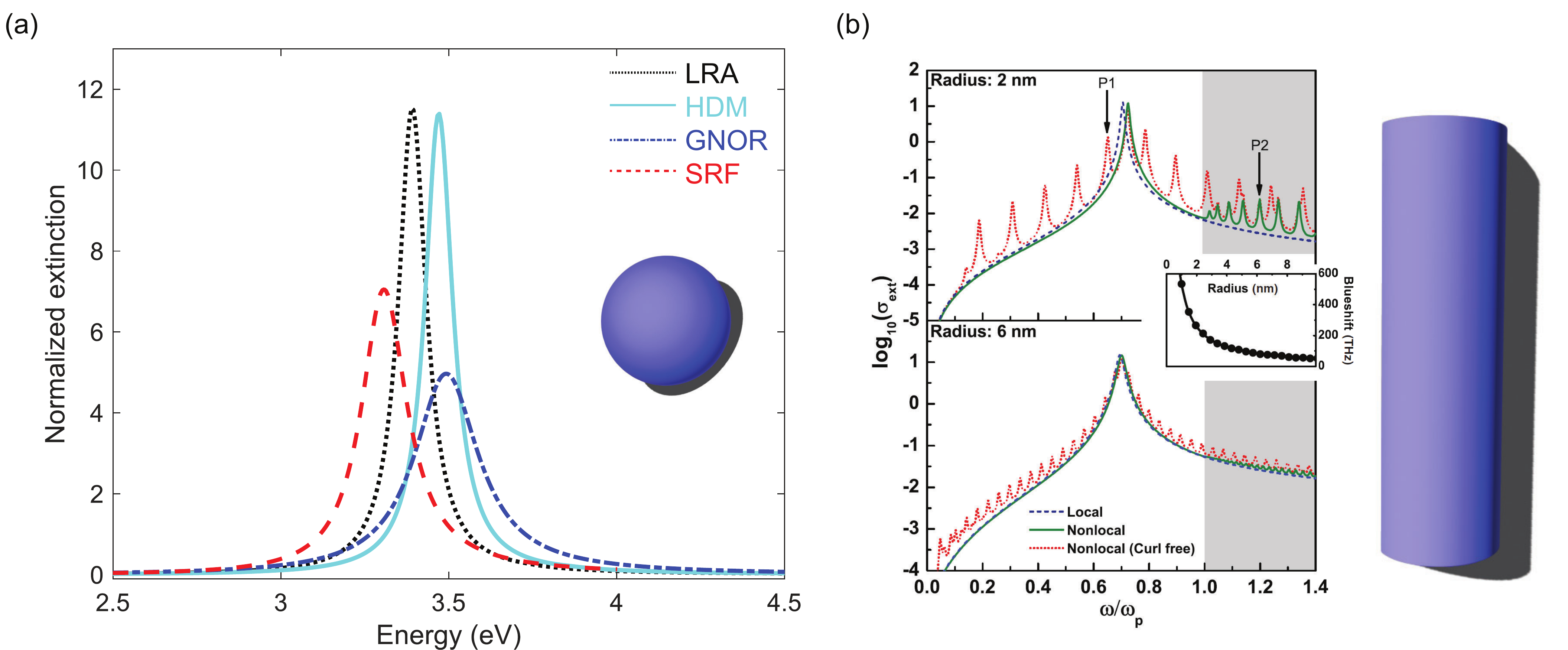}
\caption{Effect of nonlocality to the far-field spectra. \pnl{a} Normalized (to the geometrical cross section) extinction cross section for a Na sphere of radius $R = 5$\,nm, as calculated within LRA (dotted black line), HDM (solid light-blue line), GNOR (dash-dotted dark-blue line), and SRF (dashed red line), in the energy window of the dipolar LSP mode). \pnl{b} Normalized extinction cross section (in logarithmic scale) for an infinitely-long Ag cylinder, within LRA (dashed black lines), HDM (solid green lines), and HDM with a curl-free approach (dotted red lines), for two different cylinder radii. The inset shows the blueshift of the main LSP as a function of radius. Adapted with permission from Ref.~\cite{Raza:2011} (Copyright~\textcopyright~2011 American Physical Society).}
\label{fig:Fernandez-Dominguez-Fig1}
\end{figure}

HDM emerged as a generalization of the above permittivity to reflect, in an effective manner, the quantum character of electrons. As quantum particles, they obey Pauli's exclusion principle, which inevitably turns the electron gas into a \emph{compressible} one~\cite{Bloch:1933}, obeying an equation of motion of the form~\cite{Barton:1979,Pitarke:2007,Raza:2015}
\begin{equation}\label{eq:Fernandez-Dominguez-Eq2}
\frac{\partial\boldsymbol{v}}{\partial t} +
\left(\boldsymbol{v} \cdot \nabla \right)\boldsymbol{v} =
- \frac{e}{m_{\mathrm{e}}} 
\left( \boldsymbol{E} +\boldsymbol{v} \times \boldsymbol{B} \right) -
\gamma\boldsymbol{v} -
\frac{\beta^{2}}{n} \nabla n .
\end{equation}
Here, $\boldsymbol{v} ( \omega, \boldsymbol{r})$ is the hydrodynamic velocity of the free-electron gas (in general, a function of position $\boldsymbol{r}$ and angular frequency, while $\boldsymbol{E}$ and $\boldsymbol{B}$ are the electric and magnetic field of the incident electromagnetic (EM) field, respectively, which is the driving force for the motion of the electron gas. The hydrodynamic parameter $\beta$ is directly proportional to the Fermi velocity $v_{F}$ of the metal; in particular, for a three-dimensional (3D) electron gas, it takes the value $\beta^{2} = v_{F}^{2}/3$ in the low-frequency limit ($\omega \ll \gamma$), while $\beta^{2} = 3 v_{F}^{2}/5$ in the high-frequency (optical) limit ($\omega \gg \gamma$)~\cite{Halevi:1995}. For more details about the derivation of these values, we direct the reader
to~\cite{Raza:2015, Mortensen:2021a, Wegner:2023}. The pressure term $\beta^{2} \nabla n/n$ in Eq.~\eqref{eq:Fernandez-Dominguez-Eq2} can be interpreted as an additional force that tends to homogenize the electron density.

Using $\boldsymbol{J}= e n_{0}\boldsymbol{v}$ to define the current density (where $n_{0}$ is the equilibrium electron density, while the driving EM field induces a perturbation $n_{1})$, the equation of motion Eq.~\eqref{eq:Fernandez-Dominguez-Eq2}, together with the continuity equation, result in the generalized Ohm's law for HDM,
\begin{subequations}    
\begin{equation}\label{eq:Fernandez-Dominguez-Eq3}
\frac{\beta^{2}}
{\omega \left(\omega +\mathrm{i} \gamma \right)}
\nabla
\left[\nabla \cdot \boldsymbol{J}(\omega, \boldsymbol{r}) \right] +
\boldsymbol{J}(\omega, \boldsymbol{r}) =
\sigma (\omega) 
\boldsymbol{E} (\omega, \boldsymbol{r})
,
\end{equation}
where $\sigma$ is the Drude conductivity. This equation is to be solved coupled to the EM wave equation
\begin{equation}\label{eq:Fernandez-Dominguez-Eq4}
\nabla \times \nabla \times \boldsymbol{E} (\omega, \boldsymbol{r}) =
\frac{\omega^{2}}{c^{2}} \varepsilon_{\infty}
\boldsymbol{E} (\omega, \boldsymbol{r}) +
\mathrm{i} \omega \mu_{0}
\boldsymbol{J} (\omega, \boldsymbol{r}) ,
\end{equation}
\end{subequations}
where $\mu_{0}$ is the vacuum permeability, and $c = 1/\sqrt{\varepsilon_{0} \mu_{0}}$ the speed of light in vacuum.

The aforementioned inclusion of quantum pressure has several immediate implications. First of all, it introduces a degree of spatial extent of induced charges, rather than being strictly localized at the metal--environment interface. But this means that the centroid of induced charge is pushed inwards (so-called electron spill-in); thinking in terms of classical harmonic oscillators, this reduced path increases the strength of the restoring force experienced by the electron cloud, leading thus to an increase in resonant frequency or, in terms of wavelength, a blueshift of the LSPs as compared to the predictions of LRA. This is shown in Fig.~\ref{fig:Fernandez-Dominguez-Fig1}\pnl{a}, comparing the black 
dotted line (LRA) with the light-blue solid one (HDM), for a sodium (Na) sphere of radius $R = 5$\,nm, described by the Drude
model in Eq.~\eqref{eq:Fernandez-Dominguez-Eq1} with $\hbar \omega_p = 5.89$\,eV, $\hbar \gamma = 0.1$\,eV, and $\varepsilon_{\infty} = 1$. As the NP size becomes smaller, the deviation between HDM and LRA predictions increases, see, e.g. Fig.~1\pnl{a} in Ref.~\cite{Stamatopoulou:2022}. On the one hand, the LRA spectra converge to the quasistatic approximation ($\omega_p/\sqrt{3}$ for a metal with $\varepsilon_{\infty} = 1$), since the NP behaves increasingly more like a classic point dipole; on the other hand, the nonlocal length and the shift of the centroid of the induced charge become more significant compared to the overall NP size, leading to steadily stronger resonance blueshifts.

The physical origin of the electron spill-in described above can be linked to the optical excitation of longitudinal waves (plasmons) in metal NPs modeled using the HDM. Longitudinal waves are not allowed in metals described by Eq.~\eqref{eq:Fernandez-Dominguez-Eq1}, as the permittivity vanishes only at the plasma frequency, well above the optical range of the EM spectrum. On the contrary, within the HDM, the convective motion of the electron gas induces a spatial dispersion on the longitudinal component of the metal permittivity, while leaving the transverse part as given by Eq.~\eqref{eq:Fernandez-Dominguez-Eq1}. Thus, the dispersion of the longitudinal waves within the metal acquires the form~\cite{Melnyk:1968}
\begin{equation}\label{eq:Fernandez-Dominguez-Eq5}
\varepsilon_{\mathrm{mL}}(\omega, k_{\mathrm{mL}}) \equiv
\varepsilon_{\infty} - 
\frac{\omega_p^{2}}
{\omega \left(\omega + \mathrm{i}  \gamma\right)
- \beta^{2} k_{\mathrm{mL}}^{2} } = 0 ,
\end{equation}
where $k_{\mathrm{mL}}$ is the longitudinal wavenumber in the metal. Equation~\eqref{eq:Fernandez-Dominguez-Eq5} shows that for $\omega<\omega_{p}$, the longitudinal waves are evanescent, decaying exponentially in the radial direction into the NP, giving rise to the aforementioned spill-in. In the limit $\omega\ll\omega_{p}$, the thickness of the spilled-in surface charges can be estimated as~\cite{Fernandez-Dominguez:2012}
\begin{equation}\label{eq:Fernandez-Dominguez-Eq6}
\delta_{\mathrm{mL}}\sim\frac{1}{k_{\mathrm{mL}}}\simeq\frac{\omega_{p}}{\sqrt{\varepsilon_{\infty}}\beta},
\end{equation}
which yields $\delta_{\mathrm{mL}}\sim 0.1$\,nm for realistic parameters (see below), in good agreement with the Thomas--Fermi wavelength in noble metals~\cite{Kittel:2005}. This spatial encoding of the HDM has been exploited for the development of effective, simplified models of nonlocal effects in the optical response of metal nanostructures~\cite{Luo:2013}, which take into account the spill-in of induced charges at their boundaries. Back to Eq.~\eqref{eq:Fernandez-Dominguez-Eq5}, at $\omega>\omega_{p}$, the longitudinal waves become propagating in all directions, leading to longitudinal bulk plasmons resonances, as shown by the green lines in the upper panel of Fig.~\ref{fig:Fernandez-Dominguez-Fig1}\pnl{b} (in this case, for a gold (Au) cylinder with $\omega_p = 8.81$\,eV, $\hbar \gamma = 0.075$\,eV, $\varepsilon_{\infty} = 1$ and $v_{F} =  1.39 \times 10^{6}$\,m/s~\cite{Raza:2011}). These bulk plasmon resonances have been reported already since the 1970~\cite{Ruppin:1973,Ruppin:1975}, but their resulting resonances are substantially weaker (notice the logarithmic scale in the figure) and largely affected by material
losses.

\vskip 2ex
\textbf{Boundary conditions.} Finally, the addition of an extra differential equation, and the existence of longitudinal waves, directly leads  to the need of an additional boundary condition, for the optical response of a nonlocal sphere to be calculated analytically or numerically. Boundary conditions have been discussed in detail in the past, with different choices often leading to different physics, as can be seen for example in the spectra of Fig.~\ref{fig:Fernandez-Dominguez-Fig1}\pnl{b}, and the striking differences between the so-called curl-free model~\cite{McMahon:2010a} (red dotted line) and the implementation following the so-called Sauter boundary condition~\cite{Sauter:1967} (green solid line). This, more natural choice, is based on the clear, physical understanding that the induced-current component normal to the metal--environment interface must be equal to zero, i.e. there is no electron spill-out. For $\varepsilon_\infty=1$ this boundary condition yields naturally the continuity of the normal component of the electric field as well, while for $\varepsilon_\infty\neq1$, differences emerge between imposing current or electric field continuity, which can be directly related to the (to some degree) artificial distinction between bound and free electrons inherent to the Drude model and HDM~\cite{Ciraci:2013}. 

The so-called hard-wall condition introduced above, i.e. the premise that the work function of the metal is so high that the electron-density profile  terminates abruptly at the assumed interface~\cite{Melnyk:1970} (see Ref.~\cite{Mystilidis:2024} for a detailed discussion) has been the cornerstone of criticism towards the traditional HDM, as it fails, by construction, to describe the response of metals where spill-out is considerable, such as Na~\cite{Teperik:2013}. Spill-out shifts the centroid of induced charge outwards, rather than inwards, thus leading to resonance redshifts, as shown in Fig.~\ref{fig:Fernandez-Dominguez-Fig1}\pnl{a} with a dashed red line---here, the spectrum is calculated with the surface-response formalism (SRF) based on Feibelman parameters (see e.g. Ref.~\cite{Goncalves:2020} as well as Secs.~\ref{sec:Christensen} and \ref{sec:Hohenester} of this Roadmap). Ways to relax the hard-wall boundary condition within the framework of HDM have also been proposed, and are currently in use~\cite{Toscano:2015,Ciraci:2016}. Finally, we should also mention that HDM, as described here, also fails to account for the well-known size-dependent broadening of the resonances, already measured experimentally back in the 1970s~\cite{Kreibig:1969,Kreibig:1970}. This is well reproduced by SRF, while in the context of HDM it is dealt with by the generalized nonlocal optical response (GNOR) model~\cite{Mortensen:2014} [see dark-blue dashed-dotted line in Fig.~\ref{fig:Fernandez-Dominguez-Fig1}\pnl{a}], which extends HDM by allowing $\beta$ to become complex, with the imaginary part describing electron diffusion (and thus increased collisions and damping), on top of their convective motion.

\subsection*{Current status, challenges and opportunities}

\begin{figure}[hb!]
\centering
\includegraphics[width=0.90\textwidth]{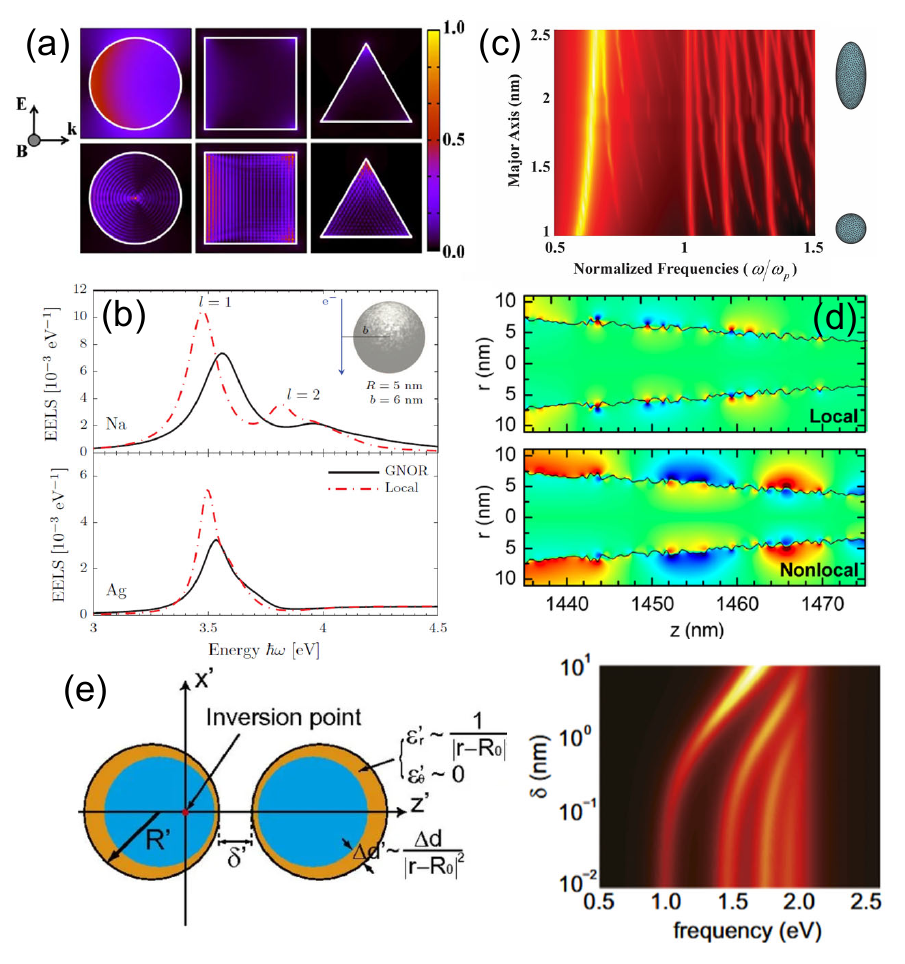}
\caption{Nonlocal modeling of plasmonic devices: \pnl{a} Spurious longitudinal plasmons in NPs of different sizes are absent (present) in local (curl-free HDM) FDTD calculations in the top (bottom) maps. Adapted with permission from Ref.~\cite{McMahon:2010a} (Copyright~\textcopyright~2010 American Physical Society). \pnl{b} EELS spectra obtained through the GNOR and local FEM models of metallic NPs. Adapted with permission from Ref.~\cite{Raza:2015} (Copyright~\textcopyright~2015 Institute of Physics). \pnl{c} BEM-HDM calculations of the extinction spectra of ellipsoid NPs of increasing eccentricity, representing data similar to Ref.~\cite{Zheng:2018}. \pnl{d} FEM calculations that show the nonlocal robustness to surface roughness of the plasmonic modes propagating in metal tips. Reprinted (adapted) with permission from Ref.~\cite{Wiener:2012} (Copyright~\textcopyright~2012 American Chemical Society). Transformation optics HDM description of the absorption spectrum of Au spheres separated by gaps between 10 and 0.01\,nm. The left panel sketches the mapping procedure in the analytical calculations. Adapted with permission from Ref.~\cite{Luo:2014} (Copyright~\textcopyright~2014 National Academy of Sciences).}
\label{fig:Fernandez-Dominguez-Fig2}
\end{figure}

\textbf{HDM solutions and other theoretical models.} From the above discussion, it becomes clear that HDM and its generalizations introduce an additional dimension to the analytical or numerical treatment of electrodynamic problems. The existence of longitudinal waves, and the additional boundary condition that is needed, have to be implemented in
any solver. For canonical architectures such as infinite planes, spheres, and infinitely long cylinders, analytic solutions that describe the reflection, scattering, and absorption of light when spatial nonlocality is taken into account can be derived, and this was done already in the 1970s~\cite{Fuchs:1969,Ruppin:1973,Ruppin:1989}. But, in order to be of practical use to the nanophotonics community, this treatment needed to be generalized to account for more complex particle shapes and configurations, exploited in plasmonics designs. This triggered a rapid growth of nonlocal computational electrodynamics in the late 2000s and early 2010s. McMahon and collaborators introduced the simplified curl-free incarnation of HDM into finite-difference time-domain (FDTD) calculations, essentially by Fourier-transforming the corresponding version of Eq.~\eqref{eq:Fernandez-Dominguez-Eq3} and solving it coupled to the wave equation in the time domain~\cite{McMahon:2010a,McMahon:2010b}. Fig.~\ref{fig:Fernandez-Dominguez-Fig2} displays the squared modulus of the displacement field at the brightest localized plasmon resonance sustained by cylindrical, square, and triangular Au nanowires with size 50\,nm (see the inset sketching the incident fields). Top and bottom panels present nonlocal and local FDTD calculations, respectively, illustrating the optical excitation of spurious longitudinal plasmons below the plasma frequency in the former~\cite{McMahon:2010a}. Toscano \emph{et al.} proceeded  directly in the frequency domain, using the finite-element method (FEM) as their tool for solving coupled differential equations~\cite{Raza:2011,Toscano:2012} using a commercial solver, while Hiremath \emph{et al.} presented the full details of a rigorous formulation based on N\'{e}d\'{e}lec finite elements~\cite{Hiremath:2012}. Fig.~\ref{fig:Fernandez-Dominguez-Fig2}\pnl{b} shows FEM calculations of the energy loss rate experienced by swift electron interacting with Na (top) and Au nanospheres~\cite{Raza:2015}. As discussed in the context of experiments, electron-energy-loss spectroscopy (EELS) allows the characterization of plasmonic resonances with excellent spatial and spectral precision, and thus, it is also an ideal testbed for nonlocal approaches, here comparing GNOR versus local predictions.

Combining the strengths of the two methods, HDM was also introduced in discontinuous-Galerkin time-domain codes~\cite{Huynh:2016,Schmitt:2016} (see also Sec.~\ref{sec:Wegner} of this Roadmap). This method, like FDTD and FEM, is volume-based, which comes with an increased computational cost, since the entire simulation domain needs to be discretized. On the contrary, boundary-element methods (BEM) only need to describe the interface between two different media, and are thus more efficient in terms of computational time and resources. Fig.~\ref{fig:Fernandez-Dominguez-Fig2}\pnl{c} illustrates the implementations of HDM into BEM, presenting the extinction cross-section of ellipsoidal nanoparticles (NPs) with different eccentricity~\cite{Zheng:2018}, revealing the strong dispersion of the bright and dark plasmonic modes that they support. Other BEM descriptions of quantum nonlocal effects in metallic nanostructures can be found in Refs.~\cite{GarciadeAbajo:2010,Trugler:2017,Mystilidis:2023}. Although most of the theoretical research in this context has focused on finite systems comprising nanometric-sized metallic elements (such as antennas or cavities), HDM approaches have also shed light into the optical performance of extended devices. Thus, Fig.~\ref{fig:Fernandez-Dominguez-Fig2}\pnl{d} shows the comparison between local (top) and nonlocal (bottom) predictions for the near-field of a nanotip. These maps reveal that spatial dispersion effectively blurs the structure boundaries, reducing the sensitivity of EM fields to surface roughness with sizes of the order of $\delta_\mathrm{mL}$ in Eq.~\eqref{eq:Fernandez-Dominguez-Eq6}. Thus, as a result of quantum nonlocality, the optical modes at the tip suffer lower scattering losses in their propagation towards its apex.  

In the literature, the versatile and powerful numerical tools introduced above are complemented by analytical efforts to introduce nonlocal refinements to local EM models. One of these exploited the theoretical framework of transformation optics to describe the plasmonic fields sustained by nanostructures in the quasi-static regime~\cite{Aubry:2010}. This approach is illustrated in Fig.~\ref{fig:Fernandez-Dominguez-Fig2}\pnl{e}, which presents the calculation of the absorption cross-section of 30\,nm radius Au spherical dimers separated by nano and subnanometric gaps~\cite{Luo:2014}. The left sketch shows that, through the inversion of two parallel HDM metal surfaces, sphere dimers with a $\beta$ parameter that varies in space is obtained. The asymmetric orange shell renders $\Delta d^\prime\simeq\delta^\prime_\mathrm{mL}\propto\beta^\prime(\mathbf{r^\prime})$~\cite{Fernandez-Dominguez:2012}, as defined in Eq.~\eqref{eq:Fernandez-Dominguez-Eq6} (the primes indicate the dimer coordinate system). The correct treatment of these mapping effects, leveraging the geometric interpretation of HDM in Ref.~\cite{Luo:2013}, yields the absorption spectra in the right contour plot, which reproduces correctly the main features of the nonlocal phenomenology reported experimentally (see later discussion on experiments). Another analytically-inspired approach shed light into the emergence of electron tunneling effects in these plasmonic gaps. This consists in the modeling of the medium between the metallic elements through a permittivity of the form of Eq.~\eqref{eq:Fernandez-Dominguez-Eq1} with a very large $\gamma$, parametrized with electron transmission (\emph{ab initio}) calculations~\cite{Esteban:2012}.  

\begin{figure}[hb!]
\centering
\includegraphics[width=0.9\textwidth]{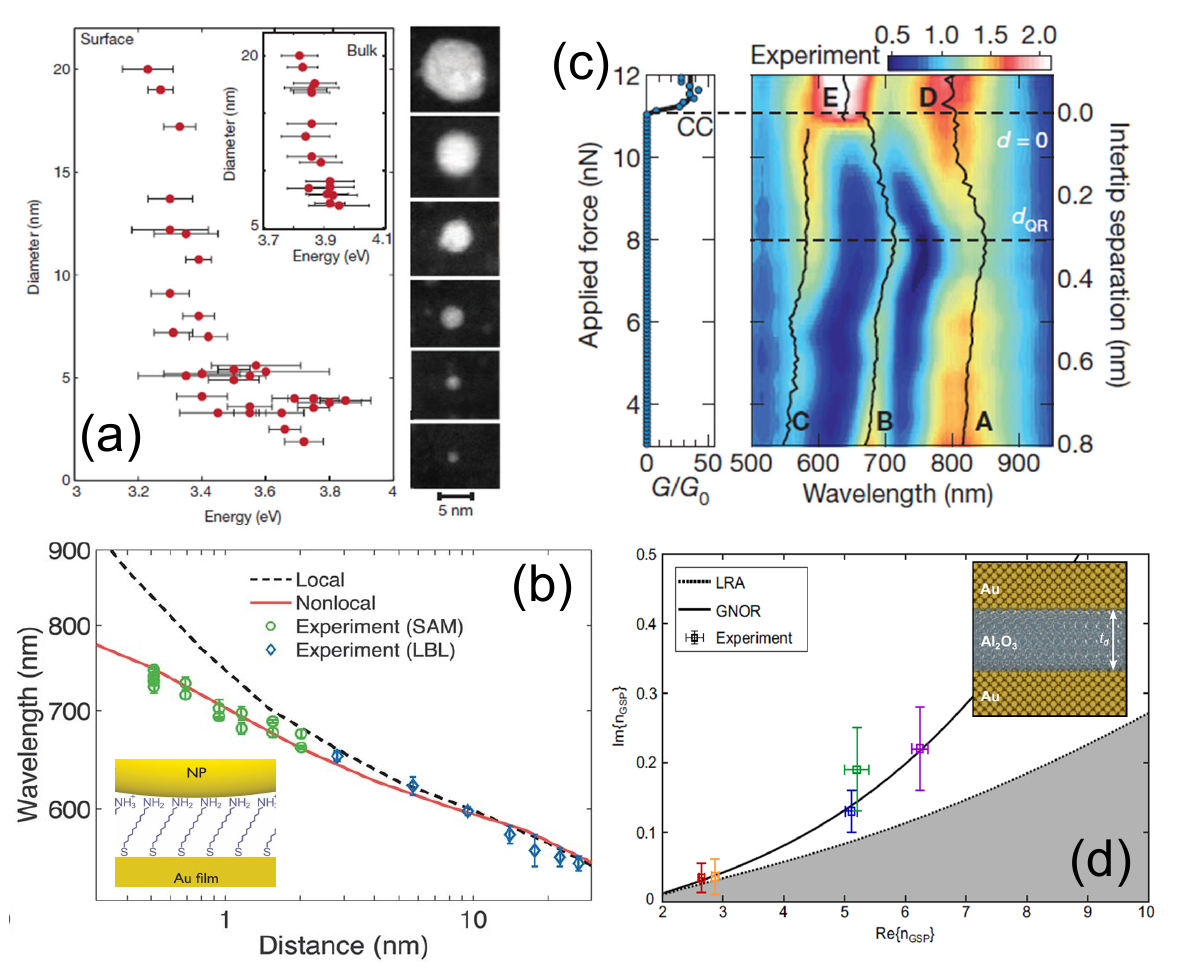}
\caption{Experimental demonstration of quantum nonlocal effects in metal nanostructures: \pnl{a}~Blueshift of the EELS-determined plasmon frequency with decreasing nanoparticle diameter, with the inset depicting bulk resonance energies. Adapted with permission from Ref.~\cite{Scholl:2012} (Copyright~\textcopyright~2012 Nature Springer). \pnl{b}~Deviation from local predictions of the dark-field scattering resonances of a 30\,nm radius NPoM with gap sizes in the nanometric range. Adapted with permission from Ref.~\cite{Ciraci:2012} (Copyright~\textcopyright~2012 American Association for the Advancement of Science). \pnl{c}~Scattering spectra as a function of the distance (measured through the force exerted) between two atomic force microscopy nanotips, revealing the onset of electron tunneling effects. Adapted with permission from Ref.~\cite{Savage:2012} (Copyright~\textcopyright~2012 Nature Springer). \pnl{d}~Imaginary versus real part of the effective index of the gap plasmon modes propagating between Au surfaces, revealing the impact of nonlocal damping. Adapted with permission from Ref.~\cite{Boroviks:2022} (Copyright~\textcopyright~2022 Nature Springer).}
\label{fig:Fernandez-Dominguez-Fig3}
\end{figure}

\vskip 2ex
\textbf{Experiments.} The investigation of nonlocal effects in metal nanostructures has attracted much experimental attention in recent years. Fig.~\ref{fig:Fernandez-Dominguez-Fig3} presents a general perspective on the different optical characterization approaches and sample geometries explored, as well as the nonlocal phenomenology reported in each experimental configuration. EELS has been instrumental for our current understanding of nanophotonic systems~\cite{GarciadeAbajo:2010,Egerton:2011}. It has also been a crucial tool enabling the assessment of surface quantum and nonlocal effects in metal NPs with morphologies presenting sub-nanometric features. Fig.~\ref{fig:Fernandez-Dominguez-Fig3}\pnl{a} shows EELS measurements of the resonant frequency of the dipolar plasmonic mode sustained by silver (Ag) spheres with decreasing diameter~\cite{Scholl:2012}. The EELS data demonstrate a modal blueshift of $\sim 0.5$\,eV from $20$\,nm to $1$\,nm diameter, and a much lower effect on the bulk plasmon resonance at higher frequencies. The authors of this study linked these observations to quantum confinement effects in the NP conduction-electron wavefunctions. Their findings were also completely compatible with a hydrodynamic model of the Ag permittivity as described above, as demonstrated by a subsequent report~\cite{Raza:2013a}. Similar EELS experiments shed light into the disappearance of higher order plasmon modes in isolated Au NPs, and the occurrence of a different nonlocal phenomenology, modal redshift with decreasing junction cross-section, in connected Au bowties~\cite{Wiener:2013}.      

Apart from monolithic structures, nanometric gaps (at the encounter between two metallic elements) have been probably the platform where quantum nonlocality has been investigated most intensively. Although EELS has also been employed in this geometry~\cite{Scholl:2013,Tan:2018}, dark-field spectroscopy has enabled a deep far-field exploration of gap plasmonic modes. Note that the systematic far-field probing of nonlocal effects in nanometric NPs is not possible due to their diminute optical cross-section~\cite{Palomba:2008}. Fig.~\ref{fig:Fernandez-Dominguez-Fig3}\pnl{b} plots the position of the lowest plasmonic resonance sustained by 30\,nm radius Au nanoparticle-on-mirror (NPoM) samples in which the gap length was set through molecular spacers generated by two different techniques: layer-by-layer deposition (from 25 to 3\,nm) and self-assembled monolayers (from 2 to 0.5\,nm). These experiments revealed that, contrary to the continuous redshifting with decreasing gap predicted by local EM calculations, a saturation of the plasmon resonance was observed, in quantitative agreement with a nonlocal hydrodynamic model~\cite{Ciraci:2012} with realistic parameters ($\beta\sim 10^6$\,m/s). These experiments indirectly established the limits of plasmon-assisted field enhancement within plasmonic gaps, and proved that NPoM structures function as far-field rulers with {\AA}ngstr{\"o}m-level precision~\cite{Hill:2012}. 

The investigation of smaller and smaller plasmonic gaps revealed a regime where the quantum nature of conduction electrons in metals manifests in a completely different fashion. When the gap size enters the {\AA}ngstr{\"o}m range and becomes comparable to the electronic spill-out length, the conduction electrons can tunnel between NPs, giving rise to charge transfer phenomena that alter the plasmonic modes at the gap. Fig.~\ref{fig:Fernandez-Dominguez-Fig3}\pnl{c} presents experimental evidence of the onset in this tunneling regime between two Au atomic-force-microscopy tips~\cite{Savage:2012}. The color plot renders dark field scattering versus incident wavelength and force applied to approach the nanotips. The lateral panel plots the electrical conductance through the gap, whose abrupt step reveals the point of contact between the tips. Through the comparison against full quantum and quantum-corrected~\cite{Esteban:2012} classical simulations, the authors of the study set the onset of the tunneling regime at $d_{QR}=0.31$\,nm (8\,nN applied force), where the dispersion of the gap plasmonic modes changes abruptly. At larger gap sizes (lower forces), the modes redshift with decreasing distance in the same way as reported in Ref.~\cite{Ciraci:2012} due to the emergence of strong spatial dispersion in the Au permittivity. At gaps smaller than $d_{QR}$, the modes blueshift with applied force, which was identified as the fingerprint of the occurrence of charge transfer between the tips. Apart from EELS and dark-field scattering, the impact of quantum tunneling in plasmonic gaps has been assessed using nonlinear optical processes. These range from surface enhanced Raman scattering (SERS) from thiophenol monolayers attached to the surface of Au nanodisk dimers~\cite{Zhu:2014} to third-harmonic generation from alkanethiol monolayers at the gap of Au NPoM cavities~\cite{Hajisalem:2014}.  

More recently, advances in nanofabrication techniques have allowed the scrutiny of quantum nonlocal effects in extended systems, which involve interfacing plasmonic surfaces within areas larger than the operating wavelength, orders of magnitude larger than the gap between NPs. Thus, the characterization of the dispersion characteristics of the extremely confined modes propagating within the nanometric gap, $t_d$, between two Au flakes could be carried out using scanning near-field optical microscopy~\cite{Boroviks:2022}. Fig.~\ref{fig:Fernandez-Dominguez-Fig3}\pnl{d} plots the imaginary part of the effective mode index of these gap surface plasmons, $n_\mathrm{GSP}$, versus its real part for $t_{d}$ between 2 and 20\,nm and at an excitation wavelength of 1550\,nm. The local-response approximation (LRA) predictions underestimate the ratio $\Im\{n_\mathrm{GSP}\}/\Re\{n_\mathrm{GSP}\}$. On the contrary, the GNOR (generalized nonlocal optical response) model, which incorporates a diffusive-like, imaginary part in the $\beta$ factor, reproduced the experimental observations accurately. These findings revealed that nonlocal damping, encoded in $\Im\{n_\mathrm{GSP}\}$, has a significant impact in propagating plasmon modes confined at the nanoscale.       

Van der Waals heterostructures provide another platform for achieving extreme optical confinement. Notably, graphene supports surface plasmons at THz frequencies, with naturally transverse decay lengths on the order of a few nanometers. Furthermore, the dispersion characteristics of these plasmon modes can be tuned by adjusting the graphene carrier density. Coupling THz graphene plasmons to metal structures and other two-dimensional materials has enabled the characterization and electrical tailoring of quantum nonlocal effects in these systems~\cite{Lundeberg:2017, AlcarazIranzo:2018}. Recently, the interface between inorganic semiconductors, such as indium antimonide (InSb) or gallium arsenide (GaAs), and metamaterial resonators has also been explored to investigate the emergence of nonlocal effects at THz frequencies~\cite{Rajabali:2021,Aupiais:2023}. While the influence of dielectric spatial dispersion in these systems differs significantly from that observed in nanostructured metals at optical frequencies, these experiments have demonstrated the universal nature of nonlocal EM phenomena.

\subsection*{Future developments to address challenges}

Given that in this Roadmap, the theoretical treatment of quantum nonlocal effects in nanophotonics systems will be discussed in great detail, we provide here only a brief overview of the broader prospects of the field, focusing on general insights beyond the specific topics explored in detail in the following sections. On the theoretical front, a comprehensive framework combining electrodynamics with ab-initio condensed matter physics is still lacking. Such framework will be essential to fully capture the quantum phenomenology found in the experiments briefly introduced above. These were conducted on samples of relevance from an optical perspective, in terms of size, material complexity and geometrical design. Another promising direction is the effective integration of nonlocal electromagnetics with quantum optical models, which would enable the unified description of the quantum nature of electrons and photons on the same footing. However, both of these avenues face significant challenges in terms of numerical complexity, demanding substantial efforts in the years ahead. On the experimental side, cleaner experiments are still needed to efficiently isolate quantum electronic effects in optical measurements, overcoming current constraints on sample design and detection sensitivity to provide unambiguous measurements of their strength in devices optimizing photonic functionalities ranging from light collection and concentration to light-matter interaction and quantum nano-optics.  
\subsection*{Concluding remarks}

The hydrodynamic Drude model, presented here in its linearized form, remains a foundational framework for the investigation and interpretation of nonlocal phenomena. Secs.~\ref{sec:Hu} and \ref{sec:Wegner} delve deeper into the hydrodynamic framework, addressing also physical and computational aspects of self-consistency and nonlinearities in greater detail.

\section[Imaginary  part of nonlocal permittivity: how it localizes absorption and imposes limits on field confinement and enhancement in plasmonics (Khurgin)]{Imaginary  part of nonlocal permittivity: how it localizes absorption and imposes limits on field confinement and enhancement in plasmonics}

\label{sec:Khurgin}

\author{Jacob B. Khurgin\,\orcidlink{0000-0003-0725-8736}}

\subsection*{Overview}

Space-time duality in photonics has been recognized since the formulation of Maxwell's equations, but only in recent years has it been rigorously studied both theoretically and experimentally~\cite{Salem:2013}. The duality between beam diffraction in space and the dispersion of short pulses in fibers has been demonstrated and applied to phenomena such as temporal lensing. When combined with nonlinearity, this duality leads to analogies between spatial and temporal solitons. More recently, space-time duality has been further explored through the introduction of temporal reflection and photonic time crystals, which serve as full analogs to their spatial counterparts~\cite{Caloz:2020}.
Simultaneously, nonlocality -- whether in natural materials like metals or in artificial metamaterials -- has become a subject of  significant interest~\cite{David:2011}. Nonlocality is the spatial counterpart to delayed (non-instantaneous) temporal responses, and in the Fourier domain, it manifests as wavevector-dependent dielectric permittivity $\varepsilon(\omega,\boldsymbol{k})$ -- or susceptibility $\chi(\omega,\boldsymbol{k})$ -- in addition to its ubiquitous frequency dependence. The permittivity of a material is a complex quantity, with the real part describing retardation and the imaginary part associated with decay or absorption. It is well established that the dispersion of the real part of permittivity is linked to the dispersion of the imaginary part via the Kramers--Kronig (KK) relations. Strong, and even anomalous, dispersion occurs near absorptive resonances, as first noted by Kundt in 19th Century~\cite{Kundt:1871}. This suggests that the presence of dispersion implies the existence of absorption somewhere in the frequency domain.
But what is the spatial analog of this relation? The KK relations arise from the causality of susceptibility $\chi(\tau<0)=0$ which is manifested in the Fourier transform being analytic in the upper complex half-plane. In the spatial domain, however, the nonlocal susceptibility behaves differently, $\chi_\omega(-\boldsymbol{\rho})=\chi_\omega(\boldsymbol{\rho})$, and a direct application of the KK transform is not possible. Since the Fourier transform of a symmetric function is real, the real and imaginary parts of complex nonlocal susceptibility $\chi_\omega(\boldsymbol{\rho})=\chi_\omega'(\boldsymbol{\rho})+i \chi_\omega''(\boldsymbol{\rho})$ have their own separate spatial Fourier transforms $X_\omega'(\boldsymbol{k})=F[\chi_\omega'(\boldsymbol{\rho})]$ and $X_\omega''(\boldsymbol{k})=F[\chi_\omega''(\boldsymbol{\rho})]$. Nevertheless, because the susceptibility $X_\omega(\boldsymbol{k})$ still contains frequency dependence and for a given $\boldsymbol{k}$ KK relations hold true in temporal domain, it becomes possible under certain conditions to establish KK-like relationships between the real and imaginary parts in the spatial domain. It follows then that manifestation of strong spatial dispersion in a material that is nominally transparent at small wavevectors is usually accompanied by strong absorption somewhere at higher $\boldsymbol{k}$’s. While the nonlocality of the real part of permittivity has been extensively studied, and several approximations have been developed to describe it effectively, its impact remains relatively limited. Typically, it results in shifts of resonant frequencies in various plasmonic and nanophotonic structures. In contrast, the effects of nonlocality on the imaginary part of permittivity are far more significant. These changes not only contribute to enhanced losses but also impose a limit on the field confinement enhancement in the aforementioned plasmonic and nanophotonic structures, as will be briefly discussed below.

\begin{figure}[ht!]
    \centering
    \includegraphics[width=0.9\linewidth]{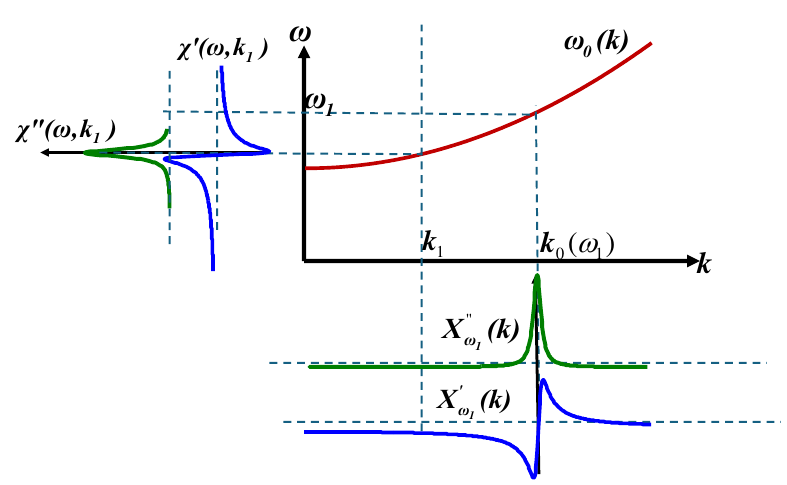}
    \caption{Spatial dispersion of resonant frequency in nonlocal dielectric. On  the left -- temporal dispersion of susceptibility. At the bottom -- spatial dispersion of susceptibility.}
    \label{fig:Khurgin-Fig1}
\end{figure}

Let us begin by examining the simplest scenario of nonlocality in a material with a single resonance frequency $\omega_0(k)= \omega_0(0)+a k^2/2$, where $a$ can be positive, as in the case of a Wannier--Mott exciton~\cite{Hopfield:1963}, or negative, such as in the case of an optical phonon~\cite{Rajabali:2024}. Since $2ak$ is a velocity these examples highlight the fundamental physical origin of nonlocality: the motion of the medium interacting with light. This interaction causes polarization induced at one point $\boldsymbol{r}-\boldsymbol{\rho}$ to propagate to another point $\boldsymbol{r}$ giving rise to nonlocal susceptibility $\chi_\omega(\boldsymbol{\rho})$. 

Consider the scenario of a positive $a$ as shown in Fig.~\ref{fig:Khurgin-Fig1}. For the fixed wavevector $\boldsymbol{k}_1$, the susceptibility exhibits a Lorentzian $\chi_\omega(\boldsymbol{k}_1)\sim \left[\omega_0(\boldsymbol{k}_1) -\omega-i\gamma\right]^{-1}$ profile with a broadening $\gamma$. The  real and imaginary components of susceptibility, visualized on the left side of Fig.~\ref{fig:Khurgin-Fig1}, are clearly interconnected by the KK relation. At the same time, for fixed frequency $\omega_1>\omega_0(0)$, the spatial dispersion of susceptibility takes the form of
\begin{equation}
    X_{\omega_1}(\boldsymbol{k}) \sim \left[\omega_0(0) + a k^2 -\omega_1-i\gamma\right]^{-1}\sim \left[k^2 - k_0^2(\omega_1) -i\gamma/a\right]^{-1},
\label{eq:Khurgin-Eq1}
\end{equation}
where the "resonance" wavevector is $k_0^2=\left[\omega_1-\omega_0(0)\right]/a$. The spatial dispersion of $X'(\boldsymbol{k})$ and $X''(\boldsymbol{k})$ is illustrated at the bottom of Fig.~\ref{fig:Khurgin-Fig1} and the KK-like relation between them is apparent. Note, however, that for $\omega_1 <\omega_0(0)$, the $X_{\omega_1}(\boldsymbol{k})$ remains real for all wavevectors and a KK-like relation does not exist. We may conclude that unlike the temporal domain, the presence of absorption in the spatial domain always implies nonlocal dispersion, but the converse is not necessarily true.  

\subsection*{Current status}

\begin{figure}[ht!]
    \centering
    \includegraphics[width=0.9\linewidth]{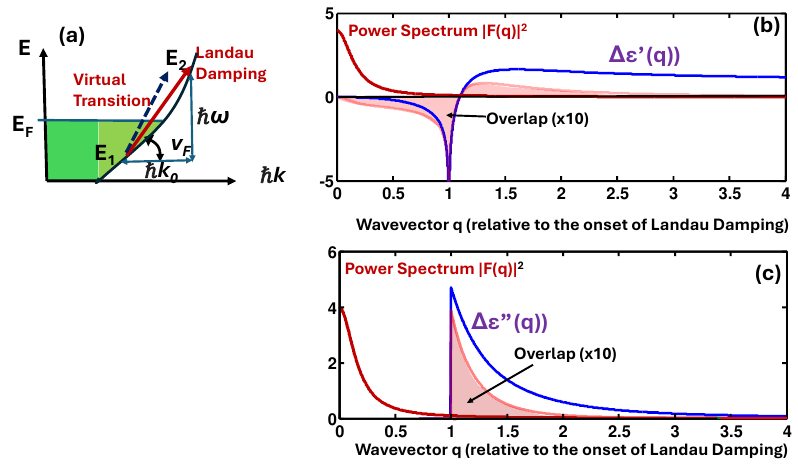}
    \caption{\pnl{a} Real (solid arrow) and virtual (dashed arrow) spatially indirect transitions in metal. \pnl{b} Real and \pnl{c} imaginary parts of the change of permittivity induced by LD and their overlaps with modal power spectrum.
}
    \label{fig:Khurgin-Fig2}
\end{figure}

Consider now the case of a metal such as gold (Au) or silver (Ag), where in the absence of scattering, absorption is forbidden at small wavevectors due to momentum conservation. However, once the wavevector approaches $k_0(\omega)=\omega/v_F$, where $v_F$ is a Fermi velocity, absorption becomes permitted as indicated by the tilted solid arrow in Fig.~\ref{fig:Khurgin-Fig2}\pnl{a}. This phenomenon is called Landau damping (LD), being also further discussed in Sec.~\ref{sec:Shahbazyan}. In one dimension the dispersion of resonant frequency is $\omega_0(k)=v_F k$, and a KK-like relation similar to the one depicted in Fig.~\ref{fig:Khurgin-Fig1} can be established. In three dimensions, the resonance broadens as absorption is now allowed for all $k\geq k_0(\omega)$, and the integration of~\eqref{eq:Khurgin-Eq1} is expected to show that KK-like relation will demonstrate its preservation. Indeed, according to Lindhard the spatial dispersion of dielectric permittivity is given by~\cite{Lindhard:1954} 
\begin{equation}
\varepsilon(\omega,\boldsymbol{k})=\varepsilon_b+\frac{3\omega_p^2}{v_F^2 k^2}\left[1-\frac{\omega}{2v_F k} \ln \frac{\omega+v_F k}{\omega-v_F k}\right],
    \label{eq:Khurgin-Eq2}
\end{equation}
where $\varepsilon_b$ is the dielectric constant due to bound electrons and $\omega_p$ is the plasma frequency. At $k=0$ the permittivity is $\varepsilon(\omega,0)=\varepsilon_b-\omega_p^2/\omega^2$ and we can write for $\Delta \varepsilon(\omega,q)=\varepsilon(\omega,q)-\varepsilon(\omega,0)$ that
\begin{equation}
\Delta\varepsilon'(\omega,q)= \frac{\omega_p^2}{\omega^2}\left[1+\frac{3}{q^2}-\frac{3}{2q^3}\ln \left|\frac{1+q}{1-q} \right|\right],\quad \Delta\varepsilon''(\omega,q\geq 1)= \frac{\omega_p^2}{\omega^2}\frac{3\pi}{2q^3},
    \label{eq:Khurgin-Eq3}
\end{equation}
where $q=k/k_0=v_F k/\omega$ as shown in Fig.~\ref{fig:Khurgin-Fig2}\pnl{b,c}. It is evident that a KK-like relation exists between the two parts, and the onset of LD at $q=1$ is a spatial analog to the onset of band edge absorption in temporal domain. One can interpret the spatial dispersion of the real part of the permittivity as the change in permittivity due to "virtual" LD, i.e., virtual indirect transitions in $k$-space where energy and momentum conservation cannot be simultaneously satisfied. This is illustrated by a dashed line in Fig.~\ref{fig:Khurgin-Fig2}\pnl{a} where $q<1$ is not sufficient to achieve LD but still contributing to the change of real part of permittivity. 

\subsection*{Challenges and opportunities}

To assess the influence of LD on the properties of plasmonic modes, one can employ a straightforward technique. This involves calculating the power density spectrum of the longitudinal electric field component and subsequently determining the effective change in dielectric constant for the entire mode
\begin{equation}
    \Delta\varepsilon_\mathrm{eff}=\frac{\int_0^\infty \Delta\varepsilon(q)\left|F_\parallel(q)\right|^2 \,d^3\boldsymbol{q} }{ \int_0^\infty \left|F_\parallel(q)\right|^2 \,d^3\boldsymbol{q}}.
        \label{eq:Khurgin-Eq4}
\end{equation}
Figure~\ref{fig:Khurgin-Fig2}\pnl{b,c} illustrate the overlap between the mode and the LD-induced permittivity change. As the mode's power spectrum extends to larger wavevectors, this overlap increases. For visible light ($\omega\sim 3\times 10^{15}\,\mathrm{s}^{-1}$) in Au or Ag ($v_F\sim 1.4\times 10^6\,\mathrm{m/s}$) the value of LD offset wavevector $k_0\sim 0.34\,\mathrm{nm}^{-1}$ is significantly larger than the mode  bandwidth of $\left|F_\parallel(q)\right|^2$. Therefore, most of the overlap occurs at large wave vectors in the tail of the $\left|F_\parallel(q)\right|^2$ only due to presence of sharp boundary indicating that LD and nonlocality is largely  surface phenomena. The impact of LD on the real part of the permittivity is relatively limited due to sign change  at $q=1$ primarily resulting in modifications to the resonant frequency, as extensively documented in the literature. In contrast, the impact of alterations in the imaginary component of the permittivity is substantially more pronounced. The increase in absorptive loss not only reduces field enhancement but also influences field confinement as the field distribution tends to avoid high-loss regions.  

In fact, one can evaluate the effective damping rate due to LD as 
\begin{equation}
   \gamma_\mathrm{LD}=\frac{3\pi\omega}{2}\frac{\int_{q>1}^\infty  q^{-3} \left|F_\parallel(q)\right|^2 \,d^3\boldsymbol{q} }{\int_0^\infty \left|F_\parallel(q)\right|^2 \,d^3\boldsymbol{q}}
        \label{eq:Khurgin-Eq5}
\end{equation}
and then simply add it to the bulk damping rate $\gamma_b$ in Drude permittivity 
\begin{equation}
   \varepsilon_\mathrm{eff}(\omega)=\varepsilon_b - \frac{\omega_p^2}{\omega\left(\omega+i\gamma_b + i\gamma_\mathrm{LD}\right)} .
        \label{eq:Khurgin-Eq6}
\end{equation}
As shown in Ref.~\cite{Khurgin:2019} one can evaluate effective LD rate as $\gamma_s = \frac{3v_F}{8 d_\mathrm{eff}}$, where the effective surface to volume ratio is~\cite{Khurgin:2019,Shahbazyan:2016,Shahbazyan:2018}:
\begin{equation}
   d_\mathrm{eff}^{-1}=\frac{\int \varepsilon_0 E_\perp^2 (\boldsymbol{r})\, dS}{\int_\mathrm{metal} \varepsilon_0 E^2 (\boldsymbol{r})\, dV},
        \label{eq:Khurgin-Eq7}
\end{equation}
where $E_\perp^2$ is component of the electric field normal to the surface. This confirms that LD is a surface phenomenon. Its rate, $\gamma_\mathrm{LD}$ can be interpreted as the inverse of the time it takes an average electron, moving along the electric field direction, to reach the surface. For plasmon polaritons propagating along the metal-dielectric interface, the result $\gamma_\mathrm{LD}= \frac{3}{4}\alpha v_F$ where $\alpha$ is the exponential decay constant inside the metal. For spherical nanoparticles of diameter $a$, the result is $\gamma_\mathrm{LD}=\frac{3}{4}v_F/a$ closely matching Kreibig's intuitive interpretation of surface damping~\cite{Kreibig:1995}. Given that in noble metals bulk damping is on the scale of $\gamma_b\sim 10^{14}\,\mathrm{s}^{-1}$, it follows that LD becomes the dominant damping process when the characteristic mode size is less than 10\,nm.

The impact of LD and changes in imaginary part of permittivity is now clear. Concentrating electric fields in SPP modes increases damping, which alters the dielectric permittivity according to \eqref{eq:Khurgin-Eq6}, affecting the mode shape and necessitating reevaluation of the damping constant. This iterative process leads to a self-consistent solution, and the achievable field concentration and enhancement are significantly lower than predicted by models neglecting nonlocal effects or considering only first-order real changes in the real part of permittivity (hydrodynamic models~\cite{Pitarke:2007}). Notably, the mode shape is influenced both inside and outside the metal, with the latter being crucial for applications like nonlinear optics. This is evident in various structures, such as propagating SPPs~\cite{Khurgin:2015a}, nanospheres~\cite{Khurgin:2015b}, and nano-dimers~\cite{Khurgin:2017}, where field enhancement is often desired outside the metal itself. It's worth noting that even including the diffusion term~\cite{Mortensen:2014} in the hydrodynamic model of nonlocality only redistributes the field within the metal, leaving the field outside unaffected. 

Beyond its impact on field concentration and enhancement, LD absorption also plays a crucial role in hot carrier effects in plasmonics. As absorption occurs near the metal surface and the generated hot carriers are preferentially directed normally to the surface, it makes them more likely to be ejected from the metal. This process underpins plasmon-assisted photodetection and photocatalysis~\cite{Khurgin:2019,Khurgin:2020}.

\subsection*{Future research and concluding remarks}

While this summary has highlighted the impact of loss-induced nonlocality on the imaginary part of the permittivity, there remain unexplored aspects of this phenomenon. One intriguing question is how nonlocality affects absorption in the quantum regime, specifically whether the presence of a single photon at a point $\boldsymbol{r}_1$ can influence the absorption of a second photon at another point $\boldsymbol{r}_2$. Additionally, the "reverse LD", or luminescence of metallic structures, warrants further investigation, particularly regarding the spatial and angular distribution of emitted photons and polaritons. It is hoped that future research will uncover more phenomena with potential practical applications.
\section[Landau damping of surface plasmons in metal nanoparticles: the RPA approach (Shahbazyan)]{Landau damping of surface plasmons in metal nanoparticles: the RPA approach}

\label{sec:Shahbazyan}

\author{Tigran V. Shahbazyan\,\orcidlink{0000-0001-5139-1097}}

\subsection*{Overview}

There has been a renewed interest in the role of nonlocal phenomena in optical response of metal-dielectric structures~\cite{McMahon:2009,Ginzburg:2013,Mortensen:2014}. \emph{Landau damping} (LD) of localized surface plasmons (LSP) is one of the earliest manifestations of nonlocal effects observed as broadening of the LSP resonance in optical spectra of small metal nanoparticles (NP)~\cite{Kreibig:1995}. The optically excited LSP decays into single-particle excitations while momentum matching is provided by the electron scattering off the confining potential. For small NPs, this momentum relaxation mechanism can be incorporated, along with the bulk phonon and impurity scattering, into Drude's dielectric function of the metal $\varepsilon(\omega) = \varepsilon_{\infty} - \omega_{p}^{2}/\omega(\omega+i\gamma)$, where $\omega_{p}$ is the plasma frequency, and $\gamma$ is the scattering rate. The latter is presented as the sum $\gamma=\gamma_{0}+\gamma_{s}$ of bulk scattering rate $\gamma_{0}$ and of surface-induced rate
\begin{equation}
\label{eq:Shahbazyan-Eq1}
\gamma_{s}=A\frac{v_{F}}{L},
\end{equation}
where $v_{F}$ is the electron Fermi velocity, $L$ is NP's characteristic size, and $A$ is a phenomenological constant in the range 0.3--1.5 accounting for surface-related effects~\cite{Kreibig:1995}. 

\subsection*{Current status}

\begin{figure}[ht]
\centering
\includegraphics[width=0.75\columnwidth]{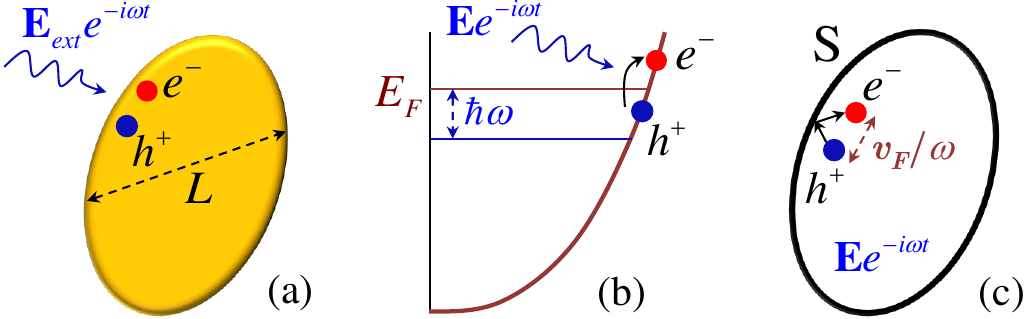}
\caption{
Schematics for surface-assisted excitation of an \emph{e-h} pair with energy $\hbar\omega$. \pnl{a}~An external optical field incident on a metal nanostructure of characteristic size $L$, \pnl{b}~excites LSP that decays into an \emph{e-h} pair, \pnl{c} with momentum matching provided by carrier surface scattering in a region of size $v_{F}/\omega$.}
\label{fig:Shahbazyan-Fig1}
\end{figure}

The scattering rate $\gamma_{s}$ was initially associated with electron's classical scattering (CS) time $\tau_\mathrm{cs}=L/v_{F}$ across the NP~\cite{Genzel:1975,Coronado:2003}, but later it was recognized as intrinsically nonlocal effect. An alternating electric field excites an electron-hole (\emph{e-h}) pair with energy $\hbar\omega$ across the Fermi level $E_{F}$. For typical LSP energies $\hbar\omega\ll E_{F}$, such a process requires momentum transfer $q=\hbar\omega/v_{F}$, facilitated by surface scattering, which defines nonlocal length scale $\xi=\hbar/q=v_{F}/\omega$~\cite{Mortensen:2014}. For common plasmonic metals such as gold (Au) and silver (Ag), this scale is below 1\,nm (e.g., for Au, $\xi\approx 0.5$\,nm at wavelength $\lambda=700$\,nm), implying that \emph{e-h} pair excitation takes place in the close proximity $\xi$ to the metal surface (see Fig.~\ref{fig:Shahbazyan-Fig1}). For a NP with characteristic size $L$, the probability of such process occurring during the optical cycle is $\sim \omega\xi/L$, leading to Eq.~\eqref{eq:Shahbazyan-Eq1}. Despite subnanometer scale of $\xi$, the broadening of optical spectra associated with $\gamma_{s}$ has been observed for NPs of various shapes with sizes up to $\sim 10$ nm. In the presence of surface scattering, the LSP resonance quality factor is $Q=\omega/\gamma\approx Q_{0}/[1+(\xi/L)Q_{0}]$, where $Q_{0}=\omega/\gamma_{0}$, implying that nonlocal effects in plasmonics persist at much larger scale $L\sim \xi Q_{0}\gg \xi$. 

For spherical NPs in the size range transitioning from metal clusters to several nm, calculations within jellium model using time-dependent local-density approximation (TDLDA) highlighted the important role of electron confining potential and electron density spillover beyond the NP classical boundary~\cite{Molina:2002,Lerme:2010}. At the same time, for larger NPs in the size range above several nanometers, TDLDA calculations revealed that the precise shape of confining potential is largely unimportant and the overall magnitude of $\gamma_{s}$, defined by the coefficient $A$, is determined by the electron spillover and dielectric environment effects~\cite{Lerme:2011,Manjavacas:2014}. The reasonable accuracy of Eq.~\eqref{eq:Shahbazyan-Eq1} even for relatively large NPs indicates that, for $L\gg \xi$, the LD rate $\gamma_{s}$ can be obtained as a correction to the Drude dielectric function within a standard analytical approach such as the random-phase approximation (RPA) and the Lindhard dielectric function~\cite{Lindhard:1954}. 

\textbf{RPA calculations} of $\gamma_{s}$ have been carried out in the course of several decades~\cite{Kawabata:1966,Lushnikov:1974,Kraus:1983,Yannouleas:1992,Uskov:2014,Shahbazyan:2016,Shahbazyan:2018,Khurgin:2019}. While in earlier studies, the scattering rate for \emph{spherical} NPs with radius $a$ was derived as $\gamma_\mathrm{sp}=3v_{F}/4a$, the subsequent works focused on obtaining $\gamma_{s}$ for NPs of more complex shapes. The major challenge was to obtain the LD rate for NPs of \emph{arbitrary} shape, even irregular one, for which no numerical calculations are feasible. The condition $L\gg \xi$ implies that the length scales for electronic and LSP excitations are well separated. For hard-wall confining potential, the electronic contribution to $\gamma_{s}$ can be effectively "integrated out" and the LD rate is obtained as a \emph{nonlocal} correction to the Drude decay rate as $\gamma=\gamma_{0}+\gamma_{s}$, where $\gamma_{s}$ has the form \eqref{eq:Shahbazyan-Eq1} with the characteristic size $L$ depending on the field distribution in the metal~\cite{Shahbazyan:2016,Shahbazyan:2018,Khurgin:2019}:
\begin{equation}
\label{eq:Shahbazyan-Eq2}
L=\frac{\int dV\, |\boldsymbol{E}|^{2} }{\int dS\,|E_{n}|^{2}}.
\end{equation}
Here, $\boldsymbol{E}$ is the electric field inside a NP of volume $V$, $E_{n}$ is the field component \emph{normal} to the NP surface $S$, and $A=3/4$. Importantly, $\gamma_{s}$ is sensitive to \emph{polarization} of the LSP field that drives the electrons towards the interface. Such form of $L$ (and $\gamma_{s}$) is valid for NPs in the size range $\xi \ll L\ll \lambda$, where $\lambda$ is the optical wavelength, which includes most plasmonic systems used in the applications.

\textbf{RPA approach to Landau damping in small metal nanoparticles.} Here we briefly outline the RPA approach to LD of LSP in metal NPs following Refs.~\cite{Shahbazyan:2016,Shahbazyan:2018}. An external field excites an LSP in a NP which subsequently decays into an \emph{e-h} pair by promoting, with its alternating field $\boldsymbol{E}e^{-i\omega t}$, a conduction band electron across the Fermi level, while the momentum matching is provided by carriers' surface scattering (see Fig.~\ref{fig:Shahbazyan-Fig1}). The full dissipated power $Q$ in a metal NP is given by 
\begin{equation}
\label{eq:Shahbazyan-Eq3}
Q=\frac{\omega}{2}\Im\int dV\, \boldsymbol{E}^{*}\cdot \boldsymbol{P},
\end{equation}
where $\boldsymbol{P}(\boldsymbol{r})$ is the electric polarization vector (the star stands for complex conjugation). The bulk contribution to $Q$ is obtained by relating, in the \emph{local} limit, the polarization vector to the electric field as $\boldsymbol{P}(\boldsymbol{r})=\boldsymbol{E}(\boldsymbol{r})[\varepsilon(\omega)-1]/4\pi$, yielding the standard expression 
\begin{equation}
\label{eq:Shahbazyan-Eq4}
Q=\frac{\omega\varepsilon''(\omega)}{8\pi} \int dV\, |\boldsymbol{E}|^{2},
\end{equation}
where $\varepsilon''(\omega)$ is the imaginary part of metal dielectric function due to bulk relaxation processes. For small NPs, there is also a surface contribution $Q_{s}$ to the dissipated power arising from momentum relaxation due to surface scattering. The general expression for $Q_{s}$ is obtained by relating $\boldsymbol{P}(\boldsymbol{r})$ to \emph{nonlocal} electron polarization operator $\Pi(\omega;\boldsymbol{r},\boldsymbol{r}')$ via the induced charge density $\rho(\boldsymbol{r})$ as
\begin{equation}
\label{eq:Shahbazyan-Eq5}
\boldsymbol{\nabla}\cdot \boldsymbol{P}(\boldsymbol{r})=-\rho(\boldsymbol{r})=-e\int  dV'\,\Pi(\omega;\boldsymbol{r}, \boldsymbol{r}') \Phi(\boldsymbol{r}').
\end{equation}
Here, the potential $\Phi (\boldsymbol{r})$ is defined as $e\boldsymbol{E} (\boldsymbol{r})=-\boldsymbol{\nabla} \Phi (\boldsymbol{r})$, where $e$ is the electron charge. Using Eq.~\eqref{eq:Shahbazyan-Eq5}, after integrating Eq.~\eqref{eq:Shahbazyan-Eq3} by parts, the dissipated power takes the form
\begin{equation}
\label{eq:Shahbazyan-Eq6}
Q_{s}=\frac{\omega}{2}\mathrm{Im}  \int  dV dV'\, \Phi^{*}(\boldsymbol{r})\Pi(\omega;\boldsymbol{r},\boldsymbol{r}') \Phi(\boldsymbol{r}'),
\end{equation}
where $\Pi(\omega;\boldsymbol{r},\boldsymbol{r}')$ includes only the electronic contribution. Within RPA, $\Pi(\omega;\boldsymbol{r},\boldsymbol{r}')$ is replaced by the polarization operator for noninteracting electrons, yielding 
\begin{equation}
\label{eq:Shahbazyan-Eq7}
Q_{s}=\pi \omega\sum_{\alpha\beta}|M_{\alpha\beta}|^{2}
\left [f(\epsilon_{\alpha})-f(\epsilon_{\beta})\right ]
\delta(\epsilon_{\alpha}-\epsilon_{\beta}+\hbar\omega),
\end{equation}
where $M_{\alpha\beta}=\int dV \psi_{\alpha}^{*} \Phi\psi_{\beta}$ is the transition matrix element of potential $\Phi(\boldsymbol{r})$ calculated from the wave functions $\psi_{\alpha}(\boldsymbol{r})$ and $\psi_{\beta}(\boldsymbol{r})$ of electron states with energies $\epsilon_{\alpha}$ and $\epsilon_{\beta}$ separated by $\hbar\omega$, $f(\epsilon)$ is the Fermi distribution function, and spin degeneracy is included. For NPs of arbitrary shape, a direct numerical evaluation of $M_{\alpha\beta}$ is not possible due to complexity of the electron wave functions. However, for NPs with characteristic size $L\gg \xi$, this issue can be bypassed by extracting the \emph{surface} contribution to the matrix element as~\cite{Shahbazyan:2016}
\begin{equation}
\label{eq:Shahbazyan-Eq8}
M_{\alpha\beta}^{s}= \frac{-e\hbar^{4}}{2m^{2} \epsilon_{\alpha\beta}^{2}} \int dS \,[\nabla_{n}\psi_{\alpha}(\boldsymbol{s})]^{*}E_{n}(\boldsymbol{s}) \nabla_{n}\psi_{\beta}(\boldsymbol{s}),
\end{equation} 
where $\nabla_{n}\psi_{\alpha}(\boldsymbol{s})$ is wave function's derivative normal to the surface, $E_{n}(\boldsymbol{s})$ is the corresponding normal field component, $\epsilon_{\alpha\beta}=\epsilon_{\alpha}-\epsilon_{\beta}$ is the \emph{e-h} pair excitation energy, and $m$ is the electron mass. The integration in Eq.~\eqref{eq:Shahbazyan-Eq8} takes place over the NP surface $S$, while the volume contribution to the matrix element is negligibly small due to near-vanishing overlap of the electron wave-functions in the presence of slowly varying potential $\Phi$. Using this expression for the matrix element, the surface contribution to the dissipated power \eqref{eq:Shahbazyan-Eq7} can be recast as
\begin{equation}
\label{eq:Shahbazyan-Eq9}
Q_{s}=
 \int \int  dS dS'\, E_{n}(\boldsymbol{s})E_{n'}^{\ast}(\boldsymbol{s}') F(\boldsymbol{s},\boldsymbol{s}'),
\end{equation}
where $F(\boldsymbol{s},\boldsymbol{s}')$ is the \emph{e-h} correlation function, which is expressed via normal derivatives of the electron and hole Green functions in a hard-wall cavity (see Ref.~\cite{Shahbazyan:2016}).

Evaluation of $Q_{s}$ hinges on the observation that excitation of an \emph{e-h} pair with energy $\hbar\omega$ takes place in a region of size $\sim \xi\ll L$. Then it can be shown that $F(\boldsymbol{s},\boldsymbol{s}')$ peaks in the region $|\boldsymbol{s}-\boldsymbol{s}'|\sim \xi$ and rapidly oscillates outside of it. Since the electric field is relatively smooth on the scale $L$, the \emph{e-h} correlation function can be approximated by $F(\boldsymbol{s},\boldsymbol{s}')=F_{0}\,\delta(\boldsymbol{s}-\boldsymbol{s}')$, where the coefficient $F_{0}$ is evaluated using single-scattering approximation for the electron Green function as $F_{0}=(3v_{F}\omega_{p}^{2}/32\pi \omega^{2})$~\cite{Shahbazyan:2016}. The final expression for the surface-induced dissipated power has the form
\begin{equation}
\label{eq:Shahbazyan-Eq10}
Q_{s}
=\frac{3v_{F}}{32\pi}\frac{\omega_{p}^{2}}{\omega^{2}} \int  dS\, |E_{n}|^{2}.
\end{equation}
Comparing $Q_{s}$ with the bulk expression \eqref{eq:Shahbazyan-Eq4}, one observes that both contributions can be combined together by adding an imaginary nonlocal correction $\delta \varepsilon_{s}$, where
\begin{equation}
\label{eq:Shahbazyan-Eq11}
\delta \varepsilon_{s} =i\frac{\omega_{p}^{2}\gamma_{s}}{\omega^{3}},
\quad
\gamma_{s}=\frac{3 v_{F}}{4}
 \frac{\int  dS\, |E_{n}|^{2}}{\int  dV \,|\boldsymbol{E}|^{2}},
\end{equation}
to the Drude bulk dielectric function. The above form for $\delta \varepsilon_{s}$ implies that the scattering rate in the Drude dielectric function $\varepsilon(\omega)$ should be modified as $\gamma=\gamma_{0}+\gamma_{s}$. A similar expression for $\gamma_{s}$ was obtained in Ref.~\cite{Khurgin:2019} using a different approach.

\begin{figure}[ht]
\centering
\includegraphics[width=0.6\columnwidth]{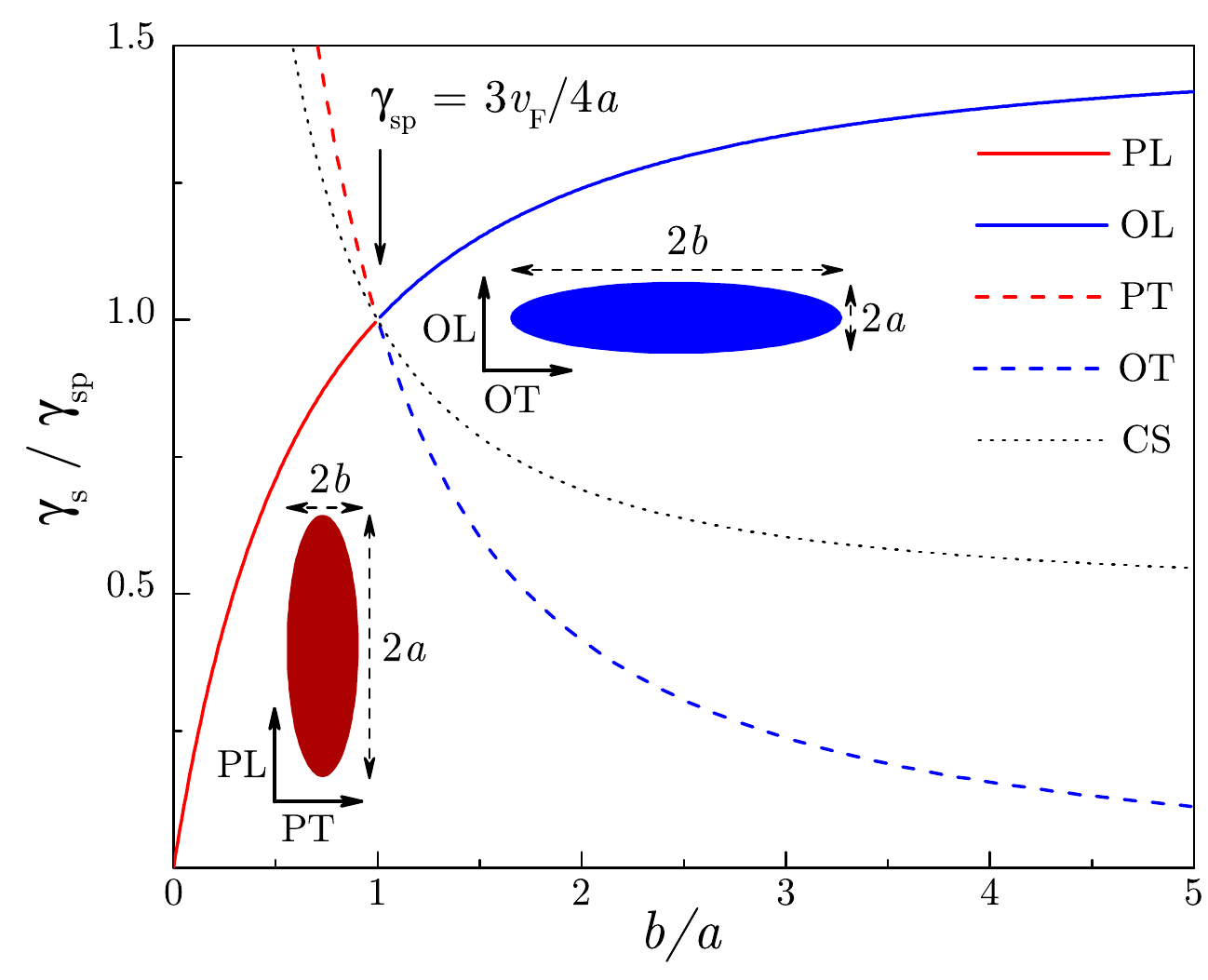}
\caption{
Normalized surface scattering rates for prolate and oblate spheroids along with the CS rate are plotted against aspect ratio $b/a$. Insets: Schematics of LSP modes.}
\label{fig:Shahbazyan-Fig2}
\end{figure}

Although $\gamma_{s}$ is independent of the electric field's overall amplitude, it is highly sensitive to field's polarization that defines the electrons motion relative to the metal-dielectric interface. In the case of asymmetric NPs, such polarization dependence implies that the LD rate can vary significantly for \emph{different} LSP modes excited in the \emph{same} system. This point is illustrated in Fig.~\ref{fig:Shahbazyan-Fig2} for longitudinal (L) and transverse (T) modes in nanorods and nanodisks, modeled by the prolate (P) and oblate (O) spheroids, respectively. For such systems, explicit analytic expressions for the LD rate has been obtained in the form~\cite{Shahbazyan:2016} $\gamma_{s} =\gamma_\mathrm{sp}f_{L,T}$, where 
\begin{align}
\label{eq:Shahbazyan-Eq12}
f_{L}
=\frac{3}{2\tan^{2}(\alpha)}\left [\frac{2\alpha}{\sin (2\alpha)}-1
\right ],
\quad
f_{T}
=\frac{3}{4\sin^{2}(\alpha)}\left [1-\frac{2\alpha}{\tan (2\alpha)}
\right ].
\end{align}
Here, for a prolate spheroid (nanorod), $\alpha=\arccos (b/a)$, where $a$ and $b$ are semi-major and semi-minor axis respectively, while for an oblate spheroid (nanodisk), the LD rates have the same form \eqref{eq:Shahbazyan-Eq12} but with $\alpha=i\mathrm{arccosh}(b/a)$. For comparison, the CS model rate $\gamma_\mathrm{cs}=v_{F}S/4V$, which is independent of mode polarization, is also plotted in Fig.~\ref{fig:Shahbazyan-Fig2}. For visual convenience, all rates are normalized by the LD rate $\gamma_\mathrm{sp}=3v_{F}/4a$ for a spherical NP of radius $a$. At the sphere point $a=b$, the normalized rates continuously transition into each other (e.g., PL to OL and PT to OT), but away from it, the rates for different modes exhibit dramatic difference in magnitude depending on the mode polarization.

\subsection*{Discussion and further developments}

The surface scattering rate Eq.~\eqref{eq:Shahbazyan-Eq11} quantifies LD in "simple" NPs characterized by a single metal surface. Further developments involved more complex plasmonic systems such as thin films~\cite{Yuan:2008}, core-shell hybrid structures~\cite{Li:2013,Kirakosyan:2016,Uskov:2022} and NP dimers~\cite{Khurgin:2017,Tserkezis:2017}. In thin films or metal nanoshells with dielectric core, the surface scattering that accompanies \emph{e-h} pair excitation can take place from \emph{both} the inner and outer metal boundaries. The interference between these processes can lead to \emph{coherent} oscillations (quantum beats) of $\gamma_{s}$ with changing metal thickness $d$. Such oscillations were reported in TDLDA studies of thin Ag films~\cite{Yuan:2008} and later in RPA calculations of $\gamma_{s}$ for spherical metal nanoshells~\cite{Li:2013,Kirakosyan:2016}. Specifically, for metal nanoshells with thickness $d$, the LD rate acquires an interference-induced contribution $\gamma_{s}^\mathrm{int}\approx B\sin (d/\xi)$ where the coefficient $B$ depends on the nanoshell's thickness and overall size~\cite{Kirakosyan:2016}.

In plasmonic dimers, LD has been shown to play a critical role in limiting the field enhancement in the gap between closely spaced NPs~\cite{Khurgin:2017,Tserkezis:2017}. Numerical calculations of the field intensity performed within RPA using the Lindhard dielectric function~\cite{Khurgin:2017} and within the generalized nonlocal optical response (GNOR) model~\cite{Tserkezis:2017} revealed a significant increase in the LD rate that reduced the field enhancement factor in the gap by nearly two orders of magnitude. Qualitatively, the LD rate increase for narrow gaps can be inferred from the characteristic size $L$, given by Eq.~\eqref{eq:Shahbazyan-Eq2}, which depends on the field distribution inside the metal. Note that, in the absence of electron wave-function overlap between two NPs, excitation of an \emph{e-h} pair can take place in either NP, so that the LD rate for a dimer is the sum of individual NP LD rates. In dimers,  the field intensity is highest in the gap between the NPs and is much weaker in the metal, where it is  concentrated in a relatively small volume around the gap. More precisely, the integrated LSP density of states (DOS) $\rho(\omega)$ can be split into contributions from the regions \emph{inside} the metal and \emph{outside} of it as $\rho(\omega)=\rho_\mathrm{in}(\omega)+\rho_\mathrm{out}(\omega)$, with their relative magnitude $\rho_\mathrm{out}(\omega)/\rho_\mathrm{in}(\omega)=|\varepsilon'(\omega)|\gg 1$, where $\varepsilon'(\omega)$ is the real part of metal dielectric function~\cite{Shahbazyan:2023}. With decreasing NP separation, as the LSP resonance redshifts~\cite{Khurgin:2017,Tserkezis:2017}, this ratio increases with $|\varepsilon'(\omega)|$ implying that the electric field is further pushed out from the metal into the gap. As a result, the volume-integrated field intensity in the numerator of Eq.~\eqref{eq:Shahbazyan-Eq2} decreases relative to the surface-integrated intensity in the denominator, leading to a reduced $L$ and, hence, enhanced LD rate for narrow gaps.

The overall magnitude of the LD rate \eqref{eq:Shahbazyan-Eq1} is defined by the coefficient $A$ that depends on the electron confining potential and charge density profile, which, in turn, determine the electric fields near the interface. Recent TDLDA calculations for relatively large (up to 10\,nm) NPs~\cite{Lerme:2011,Manjavacas:2014} indicate that the main impact on the LSP resonance width comes from the electron density spillover and dielectric environment, rather than the electronic states in the cavity. Since the electron spillover is not sensitive to the overall NP shape, the TDLDA value $A\approx 0.32$ for a nanosphere~\cite{Lerme:2011,Manjavacas:2014} is likely shape independent. Note, however, that substantially larger values in the range 0.3--1.5 were reported in the experiment depending on the surrounding dielectric medium~\cite{Kreibig:1995}. On the other hand, the presence of d-band electrons with a nearly step-like density profile in noble-metals gives rise to a thin surface layer, in which the conduction electrons with extended density tail are no longer screened by the d-band electrons, which leads to the field enhancement near the interface~\cite{Liebsch:1995}. In Ag NPs, this effect has been shown to cause a blueshift of the LSP resonance which, in fact, overcompensates the resonance redshift due to the electron density spillover~\cite{Campos:2019}. One can expect that a similar competition between these two nonlocal mechanisms will take place for the LD rate as well and bring the coefficient $A$ closer to the experiment. A related effect has been recently studied for Ag NPs coated with a thin dielectric shell~\cite{Uskov:2022}. In this case, the electron spillover into a medium with a much weaker field screening results in a noticeable LD rate enhancement.

\subsection*{Concluding remarks}

In this contribution, we tried to present a brief outlook on the Landau damping (LD) of surface plasmons in metal nanoparticles (NPs). This phenomenon has attracted a constant interest for over 50 years due to its prominent role in the optical spectra of small NPs. There is a very extensive literature on LD in NPs, and several theoretical and numerical approaches have been developed during this time span. We have focused on recent advances within the RPA approach which resolved a long-standing problem of describing, by means of a relatively simple analytical model, the LD of surface plasmons in metal NPs of arbitrary shape. On the applications side, LD is an efficient source of hot electrons widely used in photochemistry and light harvesting; there are excellent reviews on this topic the readers are referred to.
\section[Madelung hydrodynamics, nonlocal plasmonics, and nonlinear optics (Chaves \emph{et al.})]{Madelung hydrodynamics, nonlocal plasmonics, and nonlinear optics}

\label{sec:Chaves}

\author{Andr{\'e} J. Chaves\,\orcidlink{0000-0003-1381-8568}, N. Asger Mortensen\,\orcidlink{0000-0001-7936-6264} \& Nuno~M.~R.~Peres\,\orcidlink{0000-0002-7928-8005}}

\subsection*{Overview}

Quantum-hydrodynamic models provide a robust framework for exploring the electrodynamic behavior of both three-dimensional (3D) and two-dimensional (2D) electron gases, particularly in the fields of nonlocal plasmonics and nonlinear optics. The Madelung equations, when coupled with Poisson's equation, allow for the analysis of magnetoplasmon spectra, magneto-optical conductivity, and nonlocal Fermi pressure corrections, and are generally relevant for nonlinear and nonlocal light-matter interactions. Applications include plasmonics~\cite{Eguiluz:1976,Ciraci:2013,Yan:2015a,Ciraci:2016,Baghramyan:2021}, nonlinear optics~\cite{Corvi:1986,Brevet:1997,Ginzburg:2015,Zayats:2018}, transistor design~\cite{Crabb:2021}, the analysis of two-dimensional materials~\cite{Fritz:2009,Chaves:2017,Lucas:2018,Fritz:2022}, and semiconductor optics~\cite{Maack:2018}. 
	
The model originates from early developments in quantum mechanics, incorporating both the statistical interpretation of the wave function and the hydrodynamic formulation. In 1926, Born introduced the statistical interpretation of the wave function~\cite{Born:1926,Pais:1982,Bernstein:2005}. Shortly thereafter, Madelung reformulated the single-particle time-dependent Schr{\"o}dinger equation in hydrodynamic form~\cite{Madelung:1927,Wyatt:2005}, linking quantum mechanics to some kind of fluid dynamics.
	
The resulting Madelung equations~\cite{Madelung:1927} resemble the continuity and Euler equations in fluid dynamics but initially lacked mechanisms to account for many-body effects, such as statistical pressure~\cite{Bonitz:2018}. Over time, the framework was expanded to include these effects, such as Fermi pressure and exchange-correlation terms~\cite{Ding:2017a}. The relationship between the hydrodynamic model and the Schr{\"o}dinger equation is bidirectional; starting from the hydrodynamic formulation, an effective Schr{\"o}dinger equation can also be derived~\cite{Manfredi:2001,Alves:2020,Alves:2022}. This approach facilitates the study of complex quantum systems using established numerical methods.
	
When studying electronic fluid dynamics under electric and magnetic fields~\cite{Fetter:1985,Cui:1991,Halperin:1992,Dulikravich:1997,Sano:2021}, the self-consistent determination of these fields becomes necessary, often requiring that the hydrodynamic model to be solved alongside Maxwell's equations. In particularly within plasmonics~\cite{Barnes:2016}, the non-retarded approximation~\cite{Locarno:2023} simplifies these calculations by replacing Maxwell's equations with Poisson's equation, applicable under specific conditions.
	
The original Madelung equations, however, do not account for many-body quantum phenomena~\cite{Khan:2014}. This limitation can be addressed by either reformulating the approach using a many-body wave function~\cite{Janossy:1968, Khan:2014} or by deriving the equations from the moments of the Wigner distribution function~\cite{Manfredi:2001}. The latter approach naturally incorporates elements such as statistical pressure, the Bohm term, and additional contributions from the energy potential~\cite{Manfredi:2001}, thus extending the formalism to include many-body effects.

\textbf{Madelung hydrodynamics.} The time-dependent Schr{\"o}dinger equation is given by
\begin{equation}\label{eq:Peres-Eq1}
    i\frac{\partial\Psi(\boldsymbol{r},t)}{\partial t} = H\Psi(\boldsymbol{r},t),
\end{equation}
where $H$ is the system's Hamiltonian describing a particle of mass $m$ subjected to a potential $U$, and $\Psi(\boldsymbol{r},t)$ is the complex-valued wave function. By expressing the wave function as $\Psi(\boldsymbol{r},t) = \sqrt{n(\boldsymbol{r},t)}\exp[iS(\boldsymbol{r},t)]$, where $ n(\boldsymbol{r},t)$ and $S(\boldsymbol{r},t)$ are real-valued functions, and substituting into Eq.~\eqref{eq:Peres-Eq1}, two equations emerge:
\begin{subequations}
\begin{equation}\label{eq:Peres-Eq2a}
    \frac{\partial n}{\partial t} + \boldsymbol{\nabla} \cdot (n\boldsymbol{v}) = 0,
\end{equation}
\begin{equation}\label{eq:Peres-Eq2b}
    \frac{\partial\boldsymbol{v}}{\partial t} + \frac{1}{2}\nabla\boldsymbol{v}^2 = -\frac{1}{m}\boldsymbol{\nabla}U + \frac{\hbar^2}{2m^2} \boldsymbol{\nabla} \left( \frac{1}{\sqrt{n}} \nabla^2 \sqrt{n} \right).
\end{equation}
\end{subequations}
Here, $\boldsymbol{v} = \hbar\boldsymbol{\nabla}S/m$ is the velocity field. Eq.~\eqref{eq:Peres-Eq2a} resembles the continuity equation, while \eqref{eq:Peres-Eq2b} has the form of the Euler equation for fluid motion. The term $-m^{-1}\boldsymbol{\nabla}U$ accounts for external forces, and the final term, known as the quantum (or Bohm) potential, arises purely from quantum mechanics, vanishing as $\hbar \to 0$. This quantum potential introduces spatial dispersion, leading to nonlocal effects in systems like plasmons.
When $U(\boldsymbol{r},t)$ is the self-consistent electrostatic potential $V_\mathrm{sc}(\boldsymbol{r},t)$, the system is supplemented by Poisson's equation:
\begin{equation}
    \nabla^2 V_\mathrm{sc}(\boldsymbol{r},t) = -\frac{q}{\epsilon_0}n(\boldsymbol{r},t),
\end{equation}
where $q$ is the particle charge and $\epsilon_0$ is the vacuum permittivity. In cases where retardation is significant, Maxwell's equations replace Poisson's equation.
	
\textbf{Many-body effects.}	The derived Euler equation excludes many-body effects because it originates from the single-particle Schr{\"o}dinger equation. Many-body effects can be incorporated by introducing a phenomenological term or by deriving equations from a many-body wave function.

In many-body systems, the Euler equation acquires an additional term, the Fermi force, given by
\begin{equation}
    \boldsymbol{\mathcal{F}}_F = -\frac{\boldsymbol{\nabla}p_F[n(\boldsymbol{r},t)]}{mn},
\end{equation}
where $p_F$ is the Fermi pressure, which depends on the system's dimensionality and particle dispersion. For specific cases, it is given by
\begin{equation}
    p_F =
	\begin{cases}
		\frac{\hbar^2 n}{5m}(3\pi^2n)^{2/3}, & \text{(3D electron gas)} \\
		\pi\frac{\hbar^2 n^2}{2m}, & \text{(2D electron gas)} \\
		\frac{\hbar v_F}{3\pi}(\pi n)^{3/2}, & \text{(graphene)}
	\end{cases}
\end{equation}
where $v_F$ is the Fermi velocity for graphene. Both the Bohm potential and the Fermi pressure contribute to the $k$-dependence of the dielectric function, differing in the powers of $k$, i.e., $k^2$ for Fermi pressure and $k^4$ for the Bohm potential.

\textbf{Wigner function approach.}
The Madelung equations can also be derived from the Wigner distribution function, which preserves information about both position and momentum while maintaining the interference properties characteristic of quantum mechanics. The Wigner distribution approach assumes a distribution function of the following form (in 2D)
\begin{equation}
		f(\boldsymbol{r},\boldsymbol{p},t)=\frac{1}{(2\pi)^{2}\hbar^{2}}\sum_{\alpha}P_{\alpha} \int d\boldsymbol{r^{\prime}}\psi_{\alpha}^{\ast}
(\boldsymbol{r}+\boldsymbol{r}^{\prime}/2,t)\psi_{\alpha}(\boldsymbol{r}-\boldsymbol{r}^{\prime}/2,t)e^{i\boldsymbol{p}\cdot\boldsymbol{r}^{\prime}/\hbar}, 
\label{eq:Peres-Eq7}
\end{equation}
where $\psi_{\alpha}(\boldsymbol{r},t)$ is a single-particle wave function obeying the time-dependent Schr{\"o}dinger equation~\eqref{eq:Peres-Eq1}. The system is assumed to be in a mixed state with a density matrix $\rho$ of the form $\rho(\mathbf{r},\mathbf{r}^\prime, t)=\sum_{\alpha}P_{\alpha}\psi_{\alpha}^\ast (\mathbf{r}^\prime,t) \psi_{\alpha}(\mathbf{r},t)$, where the probabilities $P_{\alpha}$ obey the sum rule $\sum_{\alpha}^{N_{s}}P_{\alpha}=1$, where $N_{s}$ is the number of states in the mixture.
The time derivative of $f(\boldsymbol{r},\boldsymbol{p},t)$ can be obtained using the time-dependent Schr{\"o}dinger equation~\eqref{eq:Peres-Eq1}. Introducing the following two moments of the Wigner distribution
\begin{equation}
		n(\boldsymbol{r},t)=\int d\boldsymbol{p}\,f(\boldsymbol{r},\boldsymbol{p},t), \quad
		n(\boldsymbol{r},t)\boldsymbol{v}(\boldsymbol{r},t)=\frac{1}{m}\int d\boldsymbol{p}\,\boldsymbol{p}\,f(\boldsymbol{r},\boldsymbol{p},t),
\end{equation}
the continuity equation arises from the first moment, while Euler's equation arises from the second moment, with the latter incorporating both statistical and quantum (Bohm term) pressures.
	
\subsection*{Current status}
	
\textbf{Magnetoplasmon spectrum and optical conductivity.} The hydrodynamic formulation of quantum mechanics has been effectively applied to problems such as second-harmonic generation in 2D and 3D electron gases and the calculation of plasmon dispersion, both in the presence and absence of a magnetic field (magnetoplasmons). Including a magnetic field in the Madelung's equation amounts to changing the right-hand side of Euler's equation by including the term
\begin{equation}
		-\frac{q}{m}\boldsymbol{\nabla}V-\frac{q}{m}\frac{\partial\boldsymbol{A}}{\partial t}+\frac{q}{m}{ \boldsymbol{v}}\times(\boldsymbol{\nabla}\times\boldsymbol{A}),
\end{equation}
where $V$ and $\boldsymbol{A}$ are the scalar and vector potential, respectively. This does not include the Fermi pressure, which can be incorporated in an \emph{ad hoc} manner. Additionally, it is assumed that a weak magnetic field allows for the use of the previously provided expressions for the Fermi pressure. When an external electromagnetic field is introduced, the magnetic field component must be accounted for to ensure a consistent calculation of the system's nonlinear optical properties. With all terms included, Euler's equation reads
\begin{equation}
		\frac{\partial{ \boldsymbol{v}}}{\partial t}+({ \boldsymbol{v}}\cdot\boldsymbol{\nabla}){ \boldsymbol{v}}=-\frac{\pi\hbar}{m^{2}}\boldsymbol{\nabla}n-\frac{q}{m}\boldsymbol{\nabla}V_\mathrm{sc}+\frac{q}{m}{\boldsymbol{v}}\times\boldsymbol{B},
\end{equation}
where $V_{\mathrm{sc}}$ is the self-consistent potential. By linearizing the problem, the calculation of the magnetoplasmon spectrum arises from the simultaneous solution of the continuity, Euler's, and Poisson's equations, giving~\cite{Akbari:2013}
\begin{equation}		\omega=\Omega_{\boldsymbol{k}}=\sqrt{\omega_{c}^{2}+ak+\beta^{2}k^{2}}.\label{eq:Peres-Eq10}
\end{equation}
Here, $a=q^{2}n_{0}/(2m\epsilon_{0})$, $\omega_{c}=\vert qB\vert/m$ is the cyclotron frequency, and $\beta^{2}=v_{F}^{2}/2$, where $\beta$ corresponds to the speed of the first sound in a 2D electron gas~\cite{Mazarella:2009}. The previous result is valid in the quasi-static approximation, that is, at low frequencies compared to the electron-electron collisions rate $\gamma_{ee}=(k_{B}T)^{2}/(\hbar E_{F})$, where $k_{B}$, $T$, and $E_{F}$ are the Boltzmann constant, the temperature, and the Fermi energy, respectively. For the 3D electron gas we have $\beta^{2}=v_{F}^{2}/3$.
	
Applying an external electromagnetic field rather than a self-consistent electrostatic potential induces a charge current $\boldsymbol{J}_{1}=qn_{0}\boldsymbol{{v}}_{1}$, which  leads to the optical conductivity tensor
\begin{equation}
		\overleftrightarrow{\sigma}=-\frac{q^{2}n_{0}}{D(\boldsymbol{k},\omega)}\left[\begin{array}{cc}
			im\omega & -qB\\
			qB & im\omega
		\end{array}\right], 
\end{equation}
where $D(\boldsymbol{k},\omega)=q^{2}B^{2}+m^{2}\beta^{2}k^{2}-m^{2}\omega^{2}$. When nonlocal corrections are neglected~\cite{Aires:2011}, this result agrees with calculations based on Boltzmann's kinetic equation (see also Refs.~\cite{Andreeva:2020,Roldan:2013}).
	
\textbf{Nonlinear effects.} We can extend the formalism for the calculation of nonlinear optical properties of the electron gas \cite{Tokman:2019,Zhou:2022,Jelver:2024}. Up to second order, the nonlinear current is defined as $\boldsymbol{J}_\mathrm{NL}=qn_{0}\boldsymbol{v}_{2}+qn_{1}\boldsymbol{v}_{1}$, where $\boldsymbol{v}_{2}$ is the new term resulting from the expansion of the velocity field up to second order. Proceeding as in linear response, and assuming a zero static magnetic field, we find that
\begin{equation}
		\boldsymbol{J}_{\mathrm{NL},x}  =\frac{3}{2}k_{x}\frac{e^{3}v_{F}^{2}}{4\pi\hbar^{2}\omega^{3}}E_{\omega,x}^{2}e^{-i2\omega t}
		-\frac{3}{2}k_{x}\frac{e^{3}v_{F}^{2}}{4\pi\hbar^{2}\omega^{3}}E_{\omega,x}^{*}E_{2\omega,x}e^{-i\omega t}+\text{ c.c.}\,.
\end{equation}
This result agrees with the literature~\cite{Manzoni:2015,Cox:2018} and where we have used that $q=-e<0$, with $e$ being the elementary charge. We emphasize that $\boldsymbol{E}_{\omega}(\boldsymbol{r})$ corresponds to the \emph{total field} and \emph{not} to the external field. Note that we have assumed that the total electric field has a finite in-plane momentum component $k_x$. For normal incidence, $k_x=0$ and the second order response vanishes, as it should according to inversion-symmetry arguments.

\textbf{Plasmon-enhanced second-harmonic generation.} The calculation of second-harmonic generation assisted by plasmons~\cite{Cox:2017} can be performed by including a potential of the form $V_{\mathrm{ext}}+V_\mathrm{ind}$ in Euler's equation, where $V_{\mathrm{ext}}$ corresponds to the external potential and $V_\mathrm{ind}$ represents the induced potential from plasmon excitation, which is accounted for via Poisson's equation.
  
Proceeding in the same manner as for the determination of the magnetoplasmon spectrum, we find, including nonlocal corrections, that~\cite{Cardoso:2024}
\begin{multline}
         \phi_{2\omega}^{\text{ind}}  =  -\frac{e^{3}k^{3}n_{0}}{2\epsilon_{0}m^{2}} \frac{3\omega^2 - \gamma^2 +\beta^2k^2}{\left(\beta^2k^2 -\omega(\omega+i\gamma)\right)\left(4\beta^2k^2 - 2\omega(2\omega+i\gamma)\right)} \\\times \frac{1}{(\omega+i\gamma)^{2}} \frac{1}{\hat{\epsilon}(2\boldsymbol{k},2\omega)}\frac{1}{[\hat{\epsilon}(\boldsymbol{k},\omega)]^{2}}(\phi_{\omega}^{\text{ext}})^{2},\label{eq:Peres-Eq13}   
\end{multline}
where $\phi_{\omega}^{\mathrm{ext}}$ is the external field. The denominators of $\phi_{2\omega}^{\mathrm{ind}}$ can be easily expressed in terms of the dielectric function with damping and reads as follows:
\begin{equation}
	\hat{\epsilon}(\boldsymbol{k},\omega)=1 - \frac{\omega_k^2}{\omega(\omega+i\gamma) - \beta^2k^2}.
\end{equation}
The longitudinal excitations of the gas are determined from $\hat{\epsilon}(\boldsymbol{k},\omega)=0$, implying the spectrum $\omega(k)=\sqrt{\omega_{k}^{2}+\beta^{2}k^{2}}$. Therefore, the SHG is enhanced by plasmon excitation.
	
As opposed to the case studied previously for zero magnetic field, where nonlocal corrections were  ignored, the finite value of $\phi_{2\omega}^{\mathrm{ind}}$ requires a finite in-plane wave vector.

\begin{figure}
\centering
\includegraphics[width=0.5\textwidth]{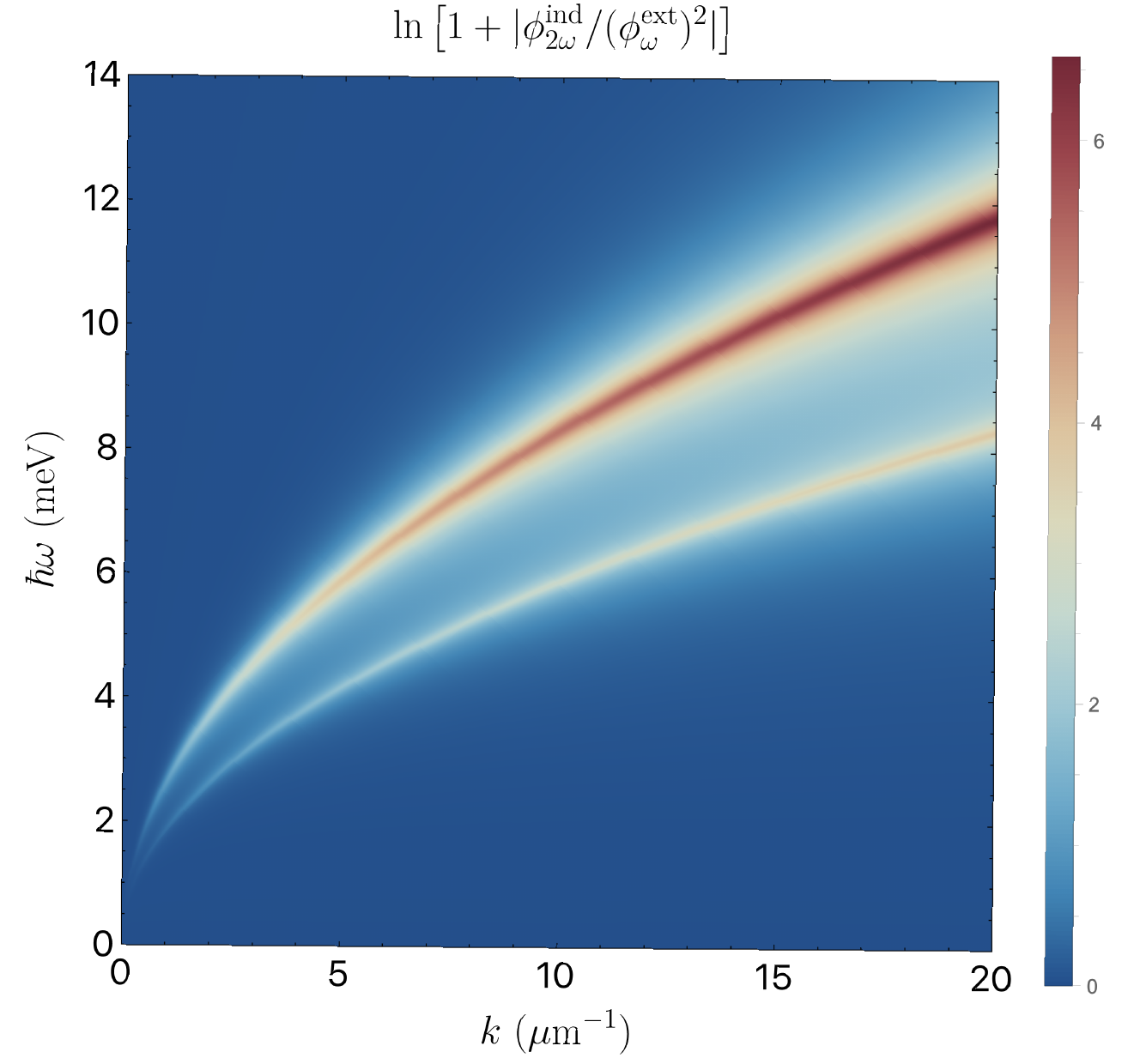}
\caption{Second-harmonic generation assisted by plasmons. The two bright lines correspond to the fundamental and second harmonic. The parameters are $n_{0}=10^{12}$\,cm$^{-2}$ and $\hbar\gamma=0.25$\,meV, and using for $m$ the free electron mass.
\label{fig:Chaves-Fig1} }
\end{figure}

In Fig.~\ref{fig:Chaves-Fig1} we depict the function $\phi_{2\omega}^{\text{ind}}$  as function of the plasmons wave number and energy $\hbar\omega$. The two harmonics are clearly visible, whose dispersion follows the plasmon curve of the 2D electron gas.

\subsection*{Challenges and opportunities}

The primary advantage of the hydrodynamic model lies in its simplicity compared to other methods that incorporate quantum mechanical effects. This simplicity proves advantageous in two distinct contexts: \emph{i)} deriving straightforward analytical results for relevant problems and \emph{ii)} numerically solving complex systems where the hydrodynamic model effectively captures quantum effects. Below, we briefly discuss how these features can motivate the application of the hydrodynamic model to recent and intriguing problems.

First, the connection between the hydrodynamic and Schr{\"o}dinger equations, as demonstrated in the previous section, enables their parallel use alongside Maxwell's or Poisson's equations. This approach has been previously explored to study electron spill-out~\cite{Alves:2022}, but could be further applied to various plasmonic problems using numerical methods from quantum mechanics.

The Wigner transformation has been applied in the context of nanoscale electronic transport~\cite{Nedjalkov:2011}, with the possibility of incorporating electron-phonon~\cite{Nedjalkov:2006} and electron-impurity interactions~\cite{Querlioz:2009}. Similar approaches could be employed to incorporate the effects of phonons and impurities in the hydrodynamic model, aiding in the understanding of their role in plasmon propagation, spectral broadening, and plasmon-phonon hybridization. Furthermore, the hydrodynamic equation can account for viscosity~\cite{Conti:1999,Principi:2016}.

As nanoparticle size decreases, light-matter interactions deviate from their bulk counterparts due to quantum confinement~\cite{Fitzgerald:2016}. The hydrodynamic model can incorporate quantum effects, such as the Fermi pressure term, while remaining compatible with most numerical approaches in nano-optics, such as the finite-element method~\cite{Ciraci:2013}. Thus, the hydrodynamical equation can be used to investigate the plasmonic properties of nanostructures, including novel 2D van~der~Waals materials.

Although this study focused on second-harmonic generation in graphene, the procedure can be extended to higher-order nonlinear responses, such as third-harmonic generation~\cite{Hajisalem:2014}, four-wave mixing~\cite{Simkhovich:2014}, plasmon self-modulation~\cite{DeLeon:2014}, and solitons~\cite{Feigenbaum:2007}. 
In the case of self-modulation phenomena driven by nonlinearity, we expect a renormalization of the plasmon dispersion and an amplification of nonlocal effects due to nonlinearity. Using the hydrodynamic formalism, an analytic expression for the renormalized spectra can, in principle, be derived.
Furthermore, rather than analyzing the Fourier decomposition of the fields for each order, one could first solve the hydrodynamic equations in the time domain and subsequently decompose the fields in the frequency domain.

This work considered the Schr{\"o}dinger Hamiltonian corresponding to a Fermi liquid described by an effective mass Hamiltonian. However, in the realm of quantum materials, ranging from graphene to topological insulators~\cite{Keimer:2017}, one can expect phenomena such as flat bands, strongly correlated systems, and nontrivial Bloch curvature. Other material excitations, like excitons, phonons, and Cooper pairs, strongly couple with light~\cite{Basov:2021}. A relevant question, then, is whether a hydrodynamic model akin to the one presented here can be developed for these excitations.

\subsection*{Future developments to address challenges}

Generalizations of the quantum hydrodynamic model to address light-matter interactions in anisotropic systems, such as phosphorene, and strongly correlated systems, such as magnetic systems, are urgently needed. Finally, a formulation based on the Wigner distribution  is a possible development to address the mentioned challenges.

Developing models that go beyond the hydrodynamic model while retaining similar simplicity is a challenging yet rewarding endeavor. One such relevant problem is the study of plasmons in magic-angle twisted bilayer graphene, which exhibits flat bands~\cite{Cao:2018,Kuang:2021}.

\subsection*{Concluding remarks}

The quantum-hydrodynamic framework, derived from the Madelung equations, has evolved into a versatile tool for analyzing nonlocal and nonlinear effects in photonic materials. Its capability to incorporate many-body interactions, magnetic fields, and self-consistent electrostatic potentials enables rich insights into electron dynamics. By bridging quantum mechanics and hydrodynamics, this approach paves the way for advancing applications of advanced materials.

\usection{\emph{Part II} --- Nonlocal effects in free-electron metals and plasmonic nanostructures}
\label{roadmap:Part2}

\section[Continuum framework for plasmonic systems (Wegner \& Busch)]{Continuum framework for plasmonic systems} 

\label{sec:Wegner}

\author{Gino Wegner\,\orcidlink{0000-0001-6225-5269} \& Kurt Busch\,\orcidlink{0000-0003-0076-8522}}

\subsection*{Current status}
	
The continuum framework for plasmonic systems deals with the underlying many-body dynamics while, at the same time, embracing only a small set of degrees of freedom, thereby providing access to a large class of analytically and numerically tractable observables.
This approach is particularly suited to handle the collective conduction electron dynamics, i.e. the protagonists of plasmonic excitations. In fact, besides bare plasmons, these plasmon excitations also include hybrid light-plasmon excitations commonly referred to as plasmon polaritons. While plasmons occur both, in the bulk and near surfaces, the finite penetration depth of electromagnetic fields in metals confines plasmon polaritons to surfaces (surface plasmon polaritons, SPPs) or small nano-particles (localized surface plasmon polaritons, LSPs). Over the past few decades, these plasmonic excitations have been studied extensively both, theoretically and experimentally, and find numerous applications in nano-photonic devices and plasmon-driven chemistry. 
	
The continuum framework rests on local averaging schemes (in the multi-scale modeling also known as coarse-graining), which yield balance equations for the bulk of the form
\begin{equation}
	   \partial_t \mathcal{F}^{(\alpha)}(\boldsymbol{r}, t) 
	   =   
	   - \nabla \cdot \mathcal{F}^{(\alpha+1)}(\boldsymbol{r}, t) 
	   + \mathcal{S}^{(\alpha)}(\boldsymbol{r}, t),
\end{equation}
thereby connecting consecutive-order moments $\mathcal{F}^{(\alpha)}$ of increasing tensorial rank while introducing different source/sink terms $\mathcal{S}^{(\alpha)}$. Here, a position $\boldsymbol{r}$ specifies a volume element $dV$ which is small compared to the whole continuum, but large compared to the typical distance of constituents~\cite{Landau:1987}.
Contemporary bulk models in plasmonics describe the dynamics of the charge density $\rho$, which, in the absence of external electromagnetic fields, is assumed to be spatially homogeneous. In turn, the charge density is coupled to the current density $\boldsymbol{J}$ by the continuity equation~\cite{Jackson:1999}
\begin{equation}
	    \partial_t \rho(\boldsymbol{r}, t) 
	    = 
	    - \nabla \cdot \boldsymbol{J} (\boldsymbol{r}, t) \,.
	    \label{eq:Wegner-Eq1}
\end{equation}
Similarly, the current balance reads 
\begin{equation}
	  \partial_t  \boldsymbol{J} (\boldsymbol{r}, t) 
	  = 
	  - \nabla \cdot \underline{\underline{\Pi}}(\boldsymbol{r}, t) + \mathcal{S}^{(1)} (\boldsymbol{r}, t) \,,
	  \label{eq:Wegner-Eq2}
\end{equation}
and is tied to momentum balance by the ubiquitous relation $\boldsymbol{J} = (\rho/m)\boldsymbol{p}$ where $\boldsymbol{p}$ denotes the momentum.
	
\begin{figure}[hb]
\centering
\includegraphics[width=0.95\textwidth]{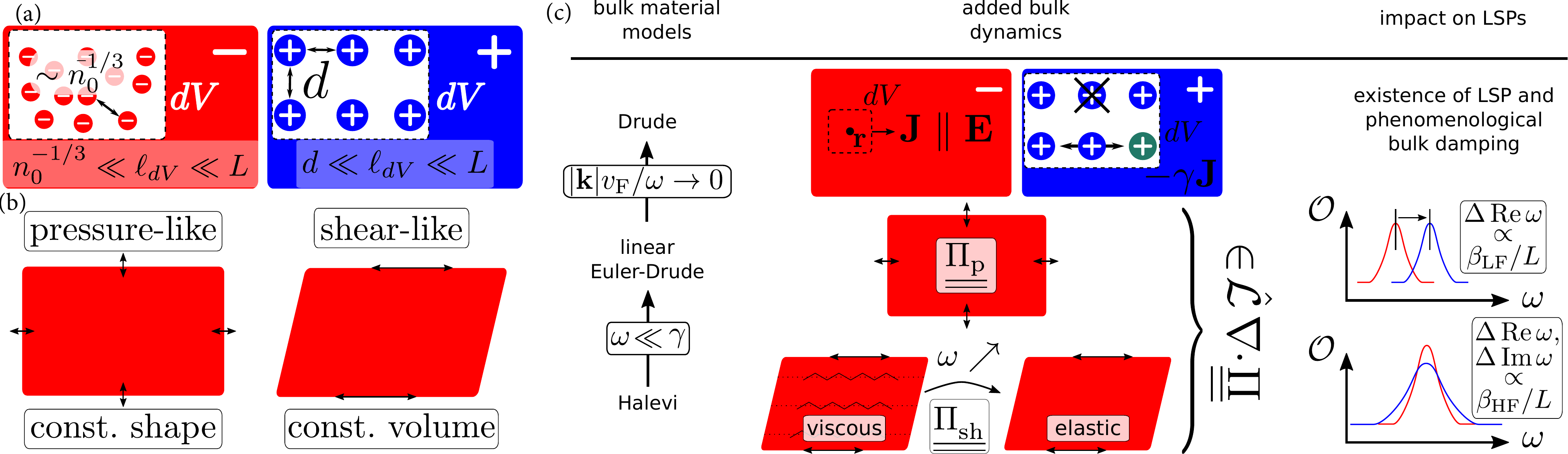}
\caption{Illustration of a continuum framework in plasmonics. \pnl{a} Connection of the micro- and macroscopic scales of the conduction electron (left) and the ionic background (right) continuum, where $\ell_{dV}$ and $L$ denote the extent of a continuum parcell and a linear dimension of the respective continuum, respectively. The microscopic scales $n_0^{-1/3}$ and $d$ correspond to the mean electronic distance and lattice constant, respectively. \pnl{b} Continuum deformations that are considered. \pnl{c} Selected bulk material models, their (hierarchical) interrelation in Fourier-space (with wavevector $\boldsymbol{k}$ and frequency $\omega$), correspondingly added bulk dynamics and associated impact on LSPs. $\hat{\mathcal{L}}$ denotes the linear projection operator, while $\mathcal{O}$ refers to an arbitrary observable used to study the spectral properties of LSPs. Blue (red) lines denote nonlocal (local) realizations of the observable. Nonlocal models are combined with hard-wall conditions throughout.
\label{fig:Wegner-Fig1}}
\end{figure}

Different continuum models differ in the form of $\underline{\underline{\Pi}}$ which, in linear response, describes the stress tensor~\cite{Landau:1987} and is typically split into pressure- and shear-like contributions, i.e. $\underline{\underline{\Pi}}=\underline{\underline{\Pi}}_\mathrm{p} +\underline{\underline{\Pi}}_\mathrm{sh}$. The stress tensor describes locally the macroscopic response due to short-ranged (classical and quantum) interactions induced by deformations of the continuum~\cite{Landau:1970} [see Fig.~\ref{fig:Wegner-Fig1}\pnl{a,b}]. In practice, it is often expressed by equations of state~\cite{Haas:2011} via $\rho$ and/or $\boldsymbol{J}$ -- introducing nonlocality which originates in the conduction electron's Fermi--Dirac statistics and scales with the Fermi velocity $v_{F}$.
In the Drude model~\cite{Drude:1900}, the charge density is assumed to be incompressible and constant $\underline{\underline{\Pi}}\equiv \underline{\underline{0}}$ and the current sink/source is given by a phenomenological damping term via a relaxation rate $\gamma$ and an external field, which adds to the internal Coulomb field. The ionic background is treated as a rigid continuum for overall charge-neutrality and a restoring force~\cite{Aschroft:1976}. As a general feature, the Drude model allows for the existence of plasmons, SPPs, LSPs and non-radiative dissipation. 
Within the Euler--Drude model~\cite{Wegner:2023} (i.e., the linearized version of the HDM discussed in Sec.~\ref{sec:Fernandez-Dominguez} of this Roadmap) attributed to Bloch~\cite{Bloch:1933}, the Drude model is augmented by a hydrostatic pressure term, such that
$\nabla \cdot \underline{\underline{\Pi}}_\mathrm{p} = \beta_\mathrm{LF}^2\nabla\rho$, where $\beta_\mathrm{LF}=v_{F}/\sqrt{3}$ is the characteristic velocity of compressional waves. When combined with hard-wall boundary conditions, this model is known to qualitatively reproduce the size-dependent nonlocal blueshift of LSPs in noble metals. In fact, the hard-wall boundary condition prohibits the spill-out of the charge density which is implied by the large values of these systems' work function~\cite{Mortensen:2021a} (see also Sec.~\ref{sec:Fernandez-Dominguez} of this Roadmap). 
Further, the viscoelastic model~\cite{Conti:1999} introduces shear. It interpolates between the viscous shear at small frequencies of the incident field compared to the rate $\gamma$ and non-dissipative elastic shear~\cite{Landau:1970} at large frequencies.
The dissipative nature of the former limit~\cite{Jewsbury:1981,Landau:1987} [corresponding to a Navier--Stokes(--Drude) model] is tied to a damping of the internal stresses on the time scale $\sim 1/\gamma$.
The viscoelastic shear stress is reminiscent of highly viscous fluids and evolves according to~\cite{Tokatly:1999,Landau:1970}
\begin{equation}
 \left[\partial_t + \gamma \right] \nabla \cdot \underline{\underline{\Pi}}_\mathrm{sh}(\boldsymbol{r} , t) 
        = 
        - \frac{4}{3}\beta^2_\mathrm{el}\nabla \left[ \nabla \cdot \boldsymbol{J} (\boldsymbol{r} , t)\right] 
        + \beta^2_\mathrm{el} \nabla \times \nabla \times \boldsymbol{J} (\boldsymbol{r} , t),
\end{equation}
where $\beta_\mathrm{el}=v_{F}/\sqrt{5}$ denotes the characteristic velocity of transverse plasma waves due to elastic shear~\cite{Wegner:2024}. The double-curl-operator introduces transverse nonlocality, while the other term, together with the pressure embodies longitudinal nonlocality.
    
As a last model, we introduce a contribution due to Halevi~\cite{Halevi:1995}, which describes the longitudinal projection {($\hat{\mathcal{L}}$)} of the viscoelastic model. As such, the nonlocal correction for plasmonic frequencies, where $\omega\gg\gamma$, is given by $\hat{\mathcal{L}}\nabla\cdot \left[\underline{\underline{\Pi}}_\mathrm{p}+\underline{\underline{\Pi}}_\mathrm{sh}\right](\boldsymbol{r}, \omega) \approx \left[\beta^2_\mathrm{TF} + (4/3)\beta^2_\mathrm{el}\right] \nabla \rho(\boldsymbol{r}, \omega)$. Therein, the often introduced (squared) velocity  $\beta^2_\mathrm{HF} = \beta^2_\mathrm{TF} + (4/3)\beta^2_\mathrm{el} $ appears due to joint action of pressure and shear\cite{Wegner:2023} (see also Sec.~\ref{sec:Fernandez-Dominguez} of this Roadmap). It increases the expected LSP blueshifts. In fact, the Halevi model gives rise to a unified (qualitative) description of line shifts and broadenings of LSPs in noble metals~\cite{Wegner:2023}. In Fig.~\ref{fig:Wegner-Fig1}\pnl{c}, we illustrate the hierarchical interrelation in Fourier space of the above-described bulk continuum models along with the correspondingly added dynamics and concomittant effects on LSPs.
We emphasize, that only two bulk parameters, $\omega_{p}$ and $\gamma$, have to be determined (e.g. via fits to experimental refractive indices that exclude surface effects~\cite{Wegner:2023,Pfeifer:2023,Mortensen:2021a}). Specifically, the Fermi velocity represents a third parameter which occurs in those models that feature a stress tensor and it can be derived from the metal's equilibrium electron density $n_0$, {using} $\omega_{p}=\omega_{p}(n_0)$.

Besides a bulk material description, a complete material model for plasmonic scatterers requires an appropriate treatment of the surface. 
In the simplest case, a description of the metal by a local bulk response (e.g. Drude) is carried all the way up to a discontinuous transition to a dielectric environment with spatially-constant bulk response (which includes vacuum)~\cite{Mortensen:2021a} leading to a sharp boundary. 
A more general framework is based on a treatment of Maxwell's equations and the introduction of a sub-wavelength transition layer (also called selvedge~\cite{Sipe:1980}). This layer's response deviates from that of the corresponding bulk medium~\cite{Bedeaux:2002}. In the spirit of continuum models, the charges, currents and fields in this microscopic surface layer are integrated out (in surface-normal direction) to yield surface excess quantities. These govern the (dis)continuities of the bulk fields at material interfaces~\cite{Bedeaux:2002}, thereby leading to corrections of the treatment of sharp boundaries above. 
The (dis)continuities represent a formal set of boundary conditions including, as special cases, the classical Maxwell~\cite{Jackson:1999} as well as the mesoscopic boundary conditions discussed in Secs.~\ref{sec:Fernandez-Dominguez} and \ref{sec:Hohenester} of this Roadmap. 
In principle, these excesses can be derived from the charges and currents induced in the surface layer of a scatterer with nonlocal continuum bulk models, assuming the latter to be equipped with additional boundary conditions (ABCs)~\cite{Pekar:1958}.
In particular, the surface charges and currents using nonlocal continuum models can then be used to derive surface-response functions (e.g., Feibelman parameters~\cite{Feibelman:1982}), for use in the new boundary conditions (such as, e.g., in the mesoscopic treatment as displayed in~Ref.\cite{Mortensen:2021a}). 
In such scattering problems, the lower bound of resolved bulk scales is lifted, such, that typically a local description of the metal's bulk is used. This scheme is analytically more tractable and simplifies time-domain simulations~\cite{Mortensen:2021a}, while allowing for a successively more elaborate surface-treatment via continuum-derived excesses (see e.g. the Euler--Drude-derived Feibelman parameter listed in Ref.~\cite{Mortensen:2021a}). In turn, the latter would enrich the desired database of Feibelman parameters, discussed in Sec.~\ref{sec:Hohenester} of this Roadmap. We would like to stress, that the reduction to a local bulk model removes confined bulk plasmons, e.g. the standing waves in spheres employing the Euler--Drude model~\cite{Christensen:2014}. 
	
\subsection*{Challenges and opportunities}

Naturally, with increasing complexity of a continuum model and/or scatterer geometry, the underlying set of differential equations (completed with initial and boundary conditions) may not anymore be amenable for analytical treatment so that numerical Maxwell solvers have to be employed. 
In this context, discontinuous-Galerkin finite-element methods represent a natural choice as they are tailored for the spatial discretization of balance equations in the presence of material interfaces.
In a nodal discontinuous-Galerkin time-domain (DGTD) approach to Maxwell's equations~\cite{Hesthaven:2008,Busch:2011}, the electromagnetic and matter fields are discretized on unstructured finite-element meshes for adaptive discretization of complex geometries, equipped with element-local interpolating nodal basis functions of adjustable order for efficient sampling of the relevant bulk length scales.
In order to obtain a global solution, the fields in neighboring elements are connected via numerical fluxes which facilitates a flexible incorporation of boundary conditions. 
The spatial discretization leads to a set of coupled ordinary differential equations in time which, owing to the element-local discretization, can be solved with high-order explicit time-stepping schemes (e.g. low-storage Runge--Kutta schemes) that allow to take full advantage of the efficient spatial discretization.
Besides pulsed plane-wave excitations (e.g. for computing spectra and field distributions via on-the-fly Fourier transforms~\cite{Wegner:2023,Kanehira:2023,Hille:2016}), the DGTD approach has been extended to point-dipole sources (which allow, among others, to compute Green's tensors, e.g. for Casimir--Polder calculations~\cite{Kristensen:2023}) and electron beams (for determing energy-electron loss and cathodoluminescence spectra~\cite{Stamatopoulou:2024,Schroeder:2015,Matyssek:2011}). In addition, the numerical scheme allows the treatment of cavities and open systems as well~\cite{Busch:2011}. 

On the analytical level, a number of Mie solutions with successively refined continuum bulk models exist~\cite{Wegner:2023,Mortensen:2021a,Raza:2013b,Wegner:2024,Aden:1951}. 
In fact, with increasing order of the spatial derivatives in Eq.~\eqref{eq:Wegner-Eq2} more field modes appear (satisfying separate Helmholtz equations~\cite{Wegner:2024}) each with its own wavenumbers $k$ via the implicit longitudinal and transverse dispersion relations~\cite{Ruppin:2001}, $\varepsilon_\mathrm{L}(k_\mathrm{L}, \omega)=0$ and $k^2_\mathrm{T}= \varepsilon_\mathrm{T}(k_\mathrm{T}, \omega) \omega^2 / c^2_0$, that result from the corresponding models' longitudinal and transverse nonlocal dielectric functions $\varepsilon_\mathrm{L}$ and $\varepsilon_\mathrm{T}$. With an increasing number of modes, the problem becomes analytically intractable. Independent of this and since the particular choice of ABCs significantly impacts the interplay of these modes, it is highly desirable to develop a general framework to derive ABCs for nonlocal continuum models (beyond typical hard-wall~\cite{Mortensen:2021a} and no-slip conditions~\cite{Hannemann:2021}).
    
\subsection*{Future developments to address challenges}
	
In order to discuss future developments of the continuum framework for plasmonic systems it is, again, instructive to distinguish analytical and numerical pathways. 
	
In the analytical realm, the focus lies on a systematic refinement of the bulk models, where connections to kinetic equations for the statistical dynamics in phase-space have been suggested~\cite{Wegner:2023,Tokatly:1999,Haas:2011,Conti:1999} and relations to orbital-free approximations of dynamical equations appearing in TDDFT have been pointed out~\cite{DellaSala:2022}. 
Eventually the refined bulk models should also consider the influence of the Fermi--Dirac statistics on electron-electron interaction. As an early precursor, we would like to mention the incorporation of the Dirac exchange term in Eq.~\eqref{eq:Wegner-Eq2}~\cite{Jensen:1937} as a function of the density gradient. When keeping same-spin electrons apart, this term counteracts the hydrostatic pressure, thereby reducing the LSP lineshifts in noble metals. 
However, since this respresents an extension of the dynamical equation suggested by Bloch, the need for modifications towards plasmonic frequencies still has to be discussed. Moving beyond noble metals, refined bulk models should address the modifications to the Drude model that are required to accurately describe transition metals~\cite{Wolff:2013}.
While all the above treatments of the bulk refer to the conduction electrons, a complete model of the optical response of metals also necessitates the (on the linear level additive~\cite{Darwin:1934,Darwin:1943}) incorporation of the response of bound electrons due to the significant absorption stemming from interband transitions (IBTs) as well as due to the (often d-band) screening of the conduction electrons away from the IBT range~\cite{Liebsch:1993} using few parameters with clear physical meaning~\cite{Pfeifer:2023}.  

Regarding the surface response, derivations of Feibelman parameters based on TDDFT-informed continuum bulk models would be very interesting, since eventually Feibelman parameters are often determined from or checked against TDDFT simulations (see Secs.~\ref{sec:Christensen} and \ref{sec:Hohenester} of this Roadmap). Further, this would allow the separation of the approximations that are tied to bulk continuum models from those that are tied to use of Feibelman parameters.
Further, we suggest to examine the applicability of the excess formalism also to balance equations beyond the level of charge conservation. This will allow the connection to the bulk nonlocal response and should lead to a generalized set of ABCs. Specifically, assuming a planar surface in the $xy$-plane, the continuity equation has led to the general result~\cite{Bedeaux:2002}
\begin{equation}
		J_\mathrm{n}(x,y,z\to0^-) -J_\mathrm{n}(x,y,z\to0^+) 
		= 
		\partial_t \rho_\mathrm{s}(x,y) 
		+ \nabla_\parallel \cdot \boldsymbol{J}_{\mathrm{s},\parallel}(x,y)
\end{equation}
where $\nabla_\parallel= (\partial_x, \partial_y, 0)$ is the gradient with surface-parallel components only and the subscript $\textrm{s}$ denotes an excess.

In the numerical realm, the focus lies on the implementation of the analytical forms of the above-discussed bulk and surface response.
On the bulk level, for DGTD a particular difficulty emerges with higher-order spatial derivatives related to the stress tensor such as when introducing the von~Weizs{\"a}cker potential~\cite{Parr:1989}.
As a time-domain formulation, the DGTD allows to treat also nonlinear response. So far, the convective nonlinearity~\cite{Landau:1987} of the  full Euler--Drude model has been treated and this could serve as a starting point for similar developments regarding more refined models. 
The geometrical flexibility inherent in finite-element approaches such as boundary-element methods (see Sec.~\ref{sec:Hohenester} of this Roadmap) or DGTD should further be leveraged by considering different boundary conditions in order to reflect on the subtleties of the transition layer also for irregular shapes, including e.g. surface roughness. Here we stress, that some surface-response functions such as the Feibelman parameters obtained from the Euler--Drude model with hard-wall boundary conditions~\cite{Mortensen:2021a}, are usually derived in frequency domain. Consequently, their inclusion in boundary-element methods is quite advanced (see Sec.~\ref{sec:Hohenester} of this Roadmap) but, so far, is largely absent for DGTD.

As regards the analytical and numerical studies as well, modern plasmonic studies are also focused on the interaction with external resonances, e.g. in core-shell particles, where excitons of dye-shells (like J-aggregate layers) couple to the plasmons in the metallic core. While aiming for more realistic descriptions, it has been realized that the choice of model for each bulk material determines the nature and also amount of resonances of the core-shell system (see, e.g., Ref.~\cite{Stete:2023}). We also suggest to explore the role of metallic bulk nonlocality and surface roughness on the matching of plasmonic and excitonic resonance as well as the losses, determining wether strong coupling prevails. 
Further, the applicability of continuum bulk models in other materials, such as doped semiconductors~\cite{DeCeglia:2018} or meta-materials, has to be studied, similar to the suggestion on such attempts for surface response functions in Sec.~\ref{sec:Christensen} of this Roadmap.
	
\subsection*{Concluding remarks}

Overall, the continuum framework for plasmonic systems provides an efficient and extensible description of the collective dynamics of conduction electrons. Depending on the resolved scales and, thus, accuracy, either nonlocal bulk models are matched with additional boundary conditions or local models are equipped with mesoscopic boundary conditions, which can be derived from the charges and currents originating in the former case. The established plasmonic continuum models have demonstrated the necessity to distinguish between longitudinal and transverse bulk nonlocality, with analytical solutions being available for selected combinations of material models and geometries. To increase the class of tractable material models and geometries the DGTD has proven itself as a flexible and resource-efficient Maxwell solver.    

\section[Quantum hydrodynamic model (Hu \emph{et al.})]{Quantum hydrodynamic model}

\label{sec:Hu}

\author{Huatian Hu\,\orcidlink{0000-0001-8284-9494}, Fabio Della Sala\,\orcidlink{0000-0003-0940-8830}, Pu Zhang\,\orcidlink{0000-0002-6253-0555} \& Cristian~Cirac{\`i}\,\orcidlink{0000-0003-3349-8389}}

\subsection*{Overview}

Optical spatial nonlocality, a spatially dispersive effect where local polarization depends on nearby fields, is an omnipresent phenomenon in natural materials. In fact, beyond the temporal nonlocality that leads to the well-known frequency dispersion, spatial dispersion depending on the momentum is often neglected in conventional photonics involving bulk materials. This effect is instead very relevant in the context of nanophotonics, especially plasmonics, where light can be concentrated into a deep-subwavelength region. In this case, the near field of the light can resolve the medium's microscopic spatial graininess, which violates the conception of homogeneous constitutive parameters in the bulk materials with an averaged field. Once the scale of light-matter interaction is confined down to nanometer-scale, quantum effects play a dominant role and the dielectric constant becomes nonlocal.

\subsection*{Current status}

\begin{figure}[htb]
\centering\includegraphics[width=0.7\textwidth]{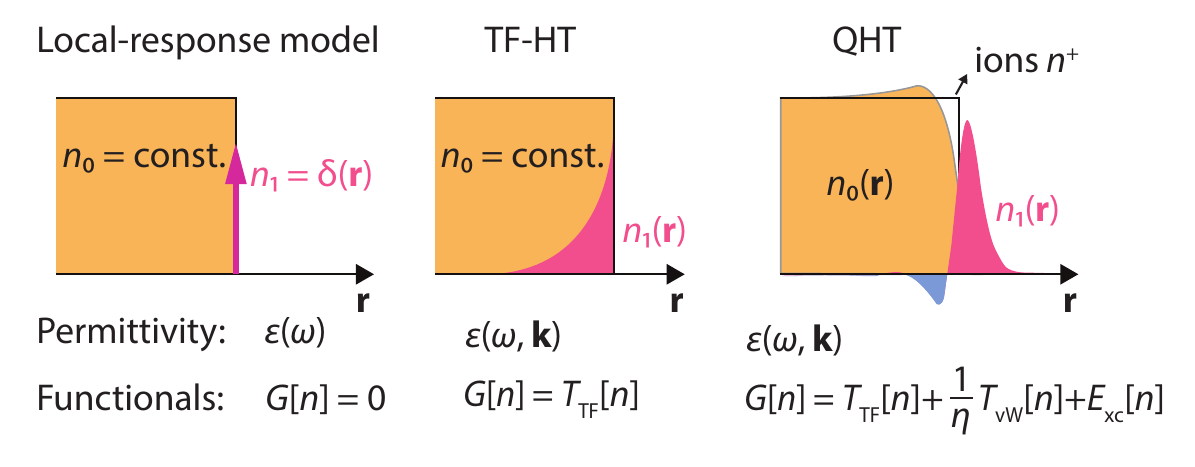}
\caption{A summary of three different response models of a free-electron gas. In the local-response model, the induced charge density is a Dirac delta function on the metal surface. The background equilibrium density $n_0$ is uniform and confined in the ion boundary; in the Thomas--Fermi hydrodynamic theory (TF-HT), $n_0$ is uniform as local-response model while the induced electrons accumulate near the metal surface without being able to escape (hard-wall condition); in the quantum hydrodynamic theory (QHT) the equilibrium density is exponentially decaying across the metal surface and the induced charges are smeared both inside and outside the metal-air interface. Different functionals used are summarized as below.}
\label{fig:Hu-Fig1}
\end{figure}

The Kohn--Sham (KS) density-functional theory (DFT) approach and its time-dependent extension (TD-DFT) is a powerful method for accurately describing quantum effects~\cite{Ullrich:2012} (see also Sec.~\ref{sec:Aizpurua} of this Roadmap). However, its computational cost is prohibitively high for mesoscopic plasmonic nanostructures containing billions of electrons. Such systems are predominantly governed by the collective behavior of electrons rather than individual single-particle properties, allowing for a trade-off between accuracy and the computational demand of calculating each electron's orbital. Orbital-free density functional theory (OF-DFT) methods~\cite{Witt:2018,DellaSala:2022,Mi:2023}, which approximate the kinetic energy functional by assuming an effective orbital, provide an efficient alternative for addressing these large-scale systems. They are less accurate but much faster than KS methods.
Building on this principle, the quantum hydrodynamic theory (QHT)~\cite{Yan:2015a, Toscano:2015, Ciraci:2016,DellaSala:2022} provides a self-consistent framework that incorporates microscopic quantum effects--such as nonlocality, electron spill out, and quantum tunneling--while seamlessly integrating with full-vector Maxwell’s equations. This approach effectively predict both near- and far-field properties of mesoscopic systems. It borrows the concept of electron density $n(\mathbf{r},t)$ from DFT, together with another macroscopic variable -- the electron fluid velocity $v(\mathbf{r},t)$, enabling the dynamics of the electron fluid to be expressed as:
\begin{equation}\label{eq:Hu-Eq1}
m_e \left( \frac{\partial}{\partial t} + \mathbf{v} \cdot \nabla + \gamma \right) \mathbf{v} = -e \left( \mathbf{E} + \mathbf{v} \times \mathbf{B} \right) - \nabla \frac{\delta G[n]}{\delta n}.
\end{equation}
where all quantum effects are collected into the functional $G[n]$. The functional $G[n]$ can be expressed as
\begin{equation}
    G[n]=T[n]+E_\mathrm{XC}[n]
    \label{eq:Hu-Eq2}
\end{equation}
i.e. as the sum of non-interacting kinetic energy, $T[n]$, and exchange-correlation potential energy $E_\mathrm{XC}[n]$, based on the level of approximation that is considered. Notably, if the energy functionals $G[n]$ are neglected, in the linear regime Eq.~\eqref{eq:Hu-Eq1} will reduce to the Drude model, i.e., local-response approximation (LRA), as shown in Fig.~\ref{fig:Hu-Fig1}. 

The Fermi theory of a free-electron gas introduces electron-electron interactions leading to a collective repulsion of electrons due to the Pauli exclusion principle. As a result, when electrons are forced toward the material boundary by an external electric field, they spread (see the induced density $n_1$ of middle panel of Fig.~\ref{fig:Hu-Fig1}). This approximation, denoted as $G[n]=T_\mathrm{TF}[n]$, is referred to as Thomas--Fermi hydrodynamic theory (TF-HT). To properly describe the boundary behavior, additional conditions--so-called hard-wall boundary condition (HW-BC)--must be imposed, ensuring that the normal component of the polarization vanishes. Physically, the HW-BC imply that electrons cannot cross the boundary. TF-HT has been shown to successfully account account for the blueshift of plasmonic resonances and the decrease of the field enhancement in confined regions~\cite{Ciraci:2012, Ciraci:2013}. 

The TF-HT framework is not always accurate because it often considers the background equilibrium density ($n_0$) to be constant and strictly confined within the material boundaries, neglecting the quantum effects such as electron spill out and tunneling. To address these limitations QHT~\cite{Yan:2015a, Toscano:2015, Ciraci:2016} incorporates a more advanced KE approximation by employing a fraction ($\lambda$) of the von~Weizsäcker (vW) kinetic energy functional, namely
\begin{equation}
T[n]=T_\mathrm{TF}[n]+\lambda T_\mathrm{vW}[n]
\label{eq:Hu-Eq3}
\end{equation}
together with the exchange-correlation with local-density approximations (LDA). We note that XC functionals beyond LDA can also be employed. However, given the approximations inherent in the leading term--the kinetic energy (KE) functional--it is not particularly beneficial to use more sophisticated XC functionals in this context.
The KE functional in Eq.~\eqref{eq:Hu-Eq3} is also widely used for ground-state calculations of finite systems within OF-DFT.
The OF-DFT ground-state calculation in the jellium model is equivalent to solve the zero-order equation on the QHT, which can be easily coupled to the Poisson's equation to account for the electrostatic potentials~\cite{Ciraci:2017}. 
Additionally, the near-field potential from the ionic lattice and the electron affinity of the interfacing dielectric become relevant when considering electron spill-out and its tunability at the interface between, for example, a noble metal and a dielectric~\cite{Zhou:2023b}.
Thus the spatially-varying equilibrium electron density $n_0 (\boldsymbol{r})$ can be calculated [Fig.~\ref{fig:Hu-Fig1}\pnl{right}), for metallic structures of any shape. Structures such as dimers and other nanogap configurations exhibiting electronic quantum tunneling effects can be accurately modeled using this framework, eliminating the need for additional boundary conditions. More precisely, in the context of QHT, the concept of well-defined boundary conditions, such as HW-BC of TF-HT, is no longer applicable since electrons can cross boundaries and decay in space, resulting in continuous fields throughout the system. However, realistic simulations require the physical domain to have finite dimensions, and the choice and truncation of the simulation region might significantly influence the accuracy of the results.
Greater accuracy and stability in QHT can be achieved by improving the KE functional, incorporating for example Laplacian-level kinetic energy contributions~\cite{Baghramyan:2021}.
Moreover, numerical schemes for modeling the nonlinear optical response of QHT have been proposed~\cite{Takeuci:2022}, along with efficient simulation approaches that can handle full three-dimensional systems~\cite{Vidal-Codina:2023}.

\subsection*{Opportunities and Challenges}

QHT is a versatile multiscale numerical tool well-suited for a broad range of applications in systems that require quantum-mechanical modeling of electrons alongside the ability to account for macroscopic optical interactions. 

\textbf{Applicability beyond metals}. In recent decades, QHT has been successfully applied to describe optical properties of alkali, e.g., sodium (Na), and noble metals, e.g., gold (Au) and silver (Ag)~\cite{Zhou:2023b}. The study of alkali metals holds fundamental significance as it validates QHT by benchmarking its predictions against DFT calculations. On the other hand, research on noble metals has its practical relevance as these materials are the most used in conventional plasmonic applications. However, extending the application of QHT to other material systems has been limited. More recently, significant efforts have focused on exploring alternative materials that support plasmons, including degenerate electron systems such as transparent conduction oxides (TCO), e.g., indium tin oxide (ITO) in the near-infrared, graphene, and heavily doped semiconductors, e.g., indium gallium arsenide (InGaAs) and indium phosphide (InP), in the mid-infrared. These materials exhibit exceptional linear and nonlinear optical properties, as well as electrical tunability. TF-HT has been employed to characterize the linear and nonlinear optical responses of ITO~\cite{Scalora:2020} and doped semiconductors~\cite{DeLuca:2021}.
Intersubband transitions in heavily doped semiconductor systems, or metallic quantum wells (QWs), present particularly intriguing phenomena since their properties can be tuned externally, and they exhibit exceptionally strong optical nonlinearities~\cite{Lee:2014}.
When doping levels are sufficiently high, such that the Fermi level can cover the conduction band and a few QW states, the intersubband transitions become \emph{plasmonic} since the potential has a much lower contribution to the Hamiltonian.
Organized into nanometer-scale layers, these heavily doped semiconductor or metallic QWs can support remarkably strong optical nonlinearities~\cite{Qian:2016, Yu:2022}.
Conventionally, QW properties are solved using Schr{\"o}dinger-Poisson equations~\cite{Silberberg:1992} and other phenomenological models~\cite{Zanotto:2014}, but their nonlinear susceptibilities remain complex to predict. Preliminary work based on TF-HT has been undertaken to address the strong nonlinearity of such QWs~\cite{Hu:2024}. However, more precise predictions of QHT applied to these and other emerging material systems remains a promising yet largely unexplored avenue for future research.

\textbf{Nonlinear nano-optics empowered by QHT}. Among the various applications of QHT, those focused on nonlinear phenomena are particularly compelling. Contrarily to effective-response approaches, such as the use of Feibelman $d$-parameters~\cite{Feibelman:1982} (see also Secs.~\ref{sec:Christensen} and \ref{sec:Hohenester} of this Roadmap), or quantum-corrected methods~\cite{Esteban:2012,Yan:2015b}, which are often restricted to linear response regimes, QHT inherently provides a comprehensive, fully dynamical framework. Moreover, while nonlocal and electron spill-out effects are traditionally regarded as "unwanted" side effects that increase broadening and degrade field enhancement, from a nonlinear perspective, these effects can instead enhance the free-electron response~\cite{Khalid:2020}.
Low-electron-density Drude materials such as TCOs, and heavily doped semiconductors can indeed support high optical nonlinearity, as their effective nonlinear susceptibilities scale inversely with some power (depending on the order) of the electron density~\cite{DeLuca:2021} as well as their effective electron mass~\cite{DeLuca:2022b}. Although TF-HT has been successfully applied to describe heavily doped semiconductors~\cite{Rosetti:2025}, ITO~\cite{Scalora:2020}, and silicon (Si), these approaches rely on the assumption that spill-out effects can be neglected. By contrast, QHT --accounting for spatially varying equilibrium electron densities--becomes particularly relevant for configurations where geometry dimensions are comparable with the typical scale of electron spill-out. Examples include QWs and atomic scale protrusions supporting extremely localized optical modes (see Sec.~\ref{sec:Zhang} in this Roadmap). Additionally, the local spatially varying electron density can indeed have a significant impact either directly on the nonlinear optical properties~\cite{Khalid:2020,DeLuca:2022a} or on the linear response in a nonlinear manner through externally controlled $n_0$~\cite{Li:2022a,Zurak:2024}. A self-consistent QHT proves very suitable for both scenarios.

\textbf{Longitudinal bulk plasmons}. A longitudinal bulk plasmon (LBP) is an electronic density wave within the bulk material, enabled by nonlocal effects in the medium's constitutive relations. LBPs have been extensively studied, both theoretically and experimentally, for decades in free-electron gas materials with TF-HT~\cite{Lindau:2007,Anderegg:1971,Ruppin:1973,Ruppin:2001,DeCeglia:2018}. Yet, their application scenarios remain largely unexplored, primarily due to the challenges associated with efficiently exciting them in conventionally sized structures.
Recent advances in material science and nanofabrication have however have begun to unlock new possibilities. Unlike localized surface plasmons, which occur at the material boundaries at lower energy, LBPs resonate above the plasma frequency in the bulk region of the material. This distinction leads to unique opportunities: Firstly, the resonance of LBPs lies above the plasma frequency, which correspond to the epsilon-near-zero (ENZ) region commonly associated with TCOs. This allows LBPs in TCOs to possibly operate in the near-infrared or even visible spectrum, hereby providing a valuable complement to silicon photonics. Secondly, since the LBP optical mode profile reaches into the bulk regions, they offer larger active interaction volumes compared to surface-bound modes. This larger volume enables higher nonlinear efficiencies. Recent studies using TF-HT have revealed that LBPs in plasmon-QW hybrid structures can exhibit exceptionally high Kerr nonlinearities and low-power-threshold optical bistability~\cite{Hu:2024}.
Looking forward, the application of QHT is expected to unveil new physics in this domain. By incorporating a more accurate description of spatially varying electron densities and dynamical responses, QHT can provide deeper insights into LBP behavior and open pathways for innovative applications in nonlinear and quantum photonics.

\textbf{QHT as a constitutive description of nonlocal media}. QHT couples with Maxwell's equations to form a closed description for plasmonic responses, it effectively serves as a constitutive relation of spatially dispersive metals and semiconductors. Building upon the full-wave solution obtained by solving the coupled Maxwell-QHT equations, a more formal and comprehensive electromagnetic theory for nonlocal media emerges. 
This theory, grounded in QHT, is particularly useful for analyzing the optical responses and bridging gaps with other models.
Notably, QHT, as a continuum theory, has recently been integrated into the generalized Lorentz model (GLM)~\cite{Zhou:2021}.
Within the theoretical framework of GLM electromagnetic theorems such as Poynting theorem and Lorentz reciprocity are generalized to nonlocal media. This extension further enables the development of a QHT-based quasinormal mode (QNM) theory (see Sec.~\ref{sec:Zhang} of this Roadmap). The QNM theory not only provides modal analysis, as evidenced with the classical QNM theory, it promises a route to field quantization through the QNM expansion of Green's function~\cite{Richter:2019}. The elevation of QHT to a quantized theory is yet still an unaccomplished feat.

\subsection*{Future developments to address challenges}

While QHT can incorporate functionals to capture both linear and nonlinear optical properties of degenerate electron systems, there remain considerable challenges to be solved. 

\textbf{Full self-consistency}. Due to the mathematical complexity of QHT equations, these are often solved using a perturbation approach where the equilibrium (ground-state) electron density $n_0$ is calculated independently by solving the zeroth-order equation~\cite{Ciraci:2017}. Then, one assumes that the sum of all higher-order perturbations remains much smaller than $n_0$. However, this assumption becomes less straightforward near boundaries, where $n_0$ exhibits exponential decay, where a different asymptotic decay velocity might render this condition false for any arbitrary small perturbation. 
In the context of quantum systems, one must ensure that $n= n_0+ \delta n>0$, while this is straightforward when $\delta n>0$ it might be problematic when $\delta n<0$. From a foundational perspective, this highlights the need for a time-dependent, fully self-consistent model to address QHT and eliminate the perturbative shortcomings~\cite{Covington:2021,Shao:2021}. A non-perturbative, time-dependent treatment would account for saturation effects and allow for the full exploration of plasmon dynamics, including modulation effects driven by changes in $n_0$.

\textbf{Accuracy of QHT}. The accuracy of QHT is strongly influenced by the choice of KE functionals, far more so than by than the exchange-correlation energy. A detailed discussion on the subject can be found in Ref.~\cite{DellaSala:2022}. Two main pathways for enhancement can be identified:

The first approach involves better capturing the nonlocality of the KE functional. In the TF-HT framework the KE functional is entirely local, although the use of Eq.~\eqref{eq:Hu-Eq3} introduces a semilocal description. Laplacian-level (LL) KE functionals, such as those described in Ref.~\cite{Baghramyan:2021}, further increase nonlocality, but they also introduces significant complexity in practical implementations. This is particularly problematic for nonlinear applications, where expansions in LL-QHT become quite cumbersome. In solid-state physics, fully nonlocal KE functionals based on the Lindhard function~\cite{Lindhard:1954}, which describes the exact response of the homogeneous electron gas, have been extensively studied\cite{Witt:2018,Mi:2023}. Unlike LL-KE functionals, these nonlocal functionals are not derived from high-order derivatives of the electron density but instead use convolutions of the electron density with the Lindhard function. 
These nonlocal KE functionals are, however, typically defined in the reciprocal space and thus limited to periodic systems and/or finite systems inside a periodic cell. For applications in nano-optics, where far-field scattering calculations are of interest, KE functionals must be defined in the real-space. Recently, fully nonlocal functionals in real-space have emerged\cite{Sarcinella:2021}.
These functionals involve the calculation of the screened Coulomb potential and its functional derivatives, which can be implemented within the QHT framework for nano-optics. Thus, such functionals can represent an interesting direction to follow to increase the QHT accuracy for diverse systems.

The second pathway is to address the frequency dependence--or non-adiabaticity--of the KE functional, as the exact $G[n]$ functional also depends on time. In the frequency-domain, using Eq.~\eqref{eq:Hu-Eq3}, the KE contributions to the QHT are static and real-valued. However, the exact KE contributions are frequency-dependent and complex-valued~\cite{DellaSala:2022}, with the imaginary part accounting for the broadening~\cite{Ciraci:2017}. This issue have been discussed in Refs.~\cite{Neuhauser:2011,White:2018,Jiang:2021}. For just two electrons, the (static) vW KE functional is exact and the QHT linear response exactly coincides with the reference TD-DFT results~\cite{DellaSala:2022}. For many electron systems, a frequency dependent KE is required\cite{Jiang:2021}. However, to date only nonlocal approximations derived from the Lindhard function exists~\cite{Neuhauser:2011,White:2018,Palade:2018,Jiang:2021}, which cannot be directly applied for real-space nano-optics QHT implementation. Finding a spatially nonlocal and frequency-dependent KE functional in real-space is a challenging interesting direction for future developments.

For both the above development directions machine-learning can be a valuable tool~\cite{Seino:2018,Meyer:2020,Imoto:2021,Remme:2023}.

\subsection*{Concluding Remarks}

QHT is a rapidly evolving and versatile tool for modeling quantum effects in nanophotonics. By incorporating quantum corrections into macroscopic optical models, QHT enables the accurate prediction of both linear and nonlinear optical responses, bridging the gap between quantum mechanics and classical electrodynamics. Despite its successes, significant challenges remain, particularly in expanding its applicability to nonlinear and more complex systems.

Future advancements in QHT hinge on two critical fronts: the development of more accurate and versatile KE functionals and the adoption of fully self-consistent, time-dependent numerical solvers. Addressing the nonlocality of KE functionals will be essential for extending QHT's utility to a broader range of materials and geometries.

QHT’s ability to describe intricate quantum phenomena such as electron spill-out, quantum tunneling, and longitudinal bulk plasmons positions it as a cornerstone for the future of quantum-enabled nanophotonics. By addressing its current limitations and leveraging advancements in computational techniques, QHT can pave the way for transformative innovations in the study of light-matter interactions at the nanoscale.

\section[The jellium model within time-dependent density-functional theory to address quantum effects in nanoplasmonics (Aizpura \emph{et al.})]{The jellium model within time-dependent density-functional theory to address quantum effects in nanoplasmonics}

\label{sec:Aizpurua}

\author{Javier~Aizpurua\,\orcidlink{0000-0002-1444-7589}, Antton Babaze\,\orcidlink{0000-0002-9775-062X} \& Andrei G. Borisov\,\orcidlink{0000-0003-0819-5028}}

\subsection*{Current status}

The collective response of a free-electron gas confined by a metal surface or by the boundaries of a nanoparticle has been one of the main exponents of light control and manipulation at the nanoscale during the last decades. The capacity of a finite structure to sustain localized surface plasmons provides one of the most effective mechanisms to localize light below the diffraction limit in extremely small effective mode volumes, associated with a very large electromagnetic near field. Classical approaches to calculate the interaction of plasmonic nanoparticles with external electromagnetic field are typically based on solving Maxwell’s equations within the linear-response theory, where materials are described by local frequency-dependent dielectric functions. These approaches have been very successful to address optical properties in field-enhanced spectroscopy and microscopy. However, nowadays the building blocks in nanoplasmonics reach the realm of the atomic scale, such as in ultranarrow plasmonic gaps, in metallic clusters, in emitters closely attached to metallic surfaces, or in picocavity configurations. Strong quantum effects thus emerge, derived from the inhomogeneous electron density distribution at the boundaries as well as from nonlocal dynamical screening~\cite{Zhu:2016,Babaze:2023,Teperik:2013}.

A powerful method to account for quantum effects in the optical response of metal surfaces or metal nano-objects relies on the time-dependent density-functional theory (TDDFT)~\cite{Marques:2004}. Different flavors of TDDFT have been developed in condensed-matter physics and in quantum chemistry, depending on the properties to be addressed. The most widely used approach is based on the Kohn--Sham (KS) scheme. The electron density $n(\boldsymbol{r},t)$ is represented using effective single-particle orbitals $\Psi(\boldsymbol{r},t)$, which evolve in time under the action of the KS potential $v_\mathrm{KS} (n; \boldsymbol{r},t)$. This potential comprises the Hartree, exchange–correlation, and external potential, as well as (pseudo) potentials representing the interaction of electrons with nuclei. 

The solution of the KS equations within TDDFT in space and time allows for obtaining key quantities that characterize the linear and nonlinear optical response of a nanosystem, and is a state-of-the-art full \emph{ab initio} approach at the forefront of modern theoretical capabilities to address such response. This approach fully accounts for quantum effects such as \emph{i)} \emph{nonlocal dynamical screening}, \emph{ii)} \emph{finite-size effects and electron spill-out}, and \emph{iii)} \emph{charge transfer} in close-contact nanoparticle surfaces~\cite{Zhu:2016,Babaze:2023,Teperik:2013}. In a metal–dielectric or metal–insulator–metal configuration, the TDDFT properly describes multiphoton and optical-field induced photoemission, as well as optically assisted electron tunneling from one metal to another. This allows for addressing challenging goals in nanoscale transport related to the engineering of petahertz (PHz) electronic devices~\cite{Ludwig:2020}.

\subsection*{Challenges and opportunities}

Successful applications of TDDFT within a full atomistic description of nanoparticles have paved the way towards understanding the ultimate limits of field localization in nanoplasmonics, as well as describing the nonlocal and band structure effects related to the size dependence of plasmon resonances in (noble) metal clusters~\cite{Barbry:2015,Chaudhary:2024}. However, the high computational cost of the method prevents studies on large-scale plasmonic systems with a wide parameter variation, often necessary to gain physical intuition and "nail down" the main nonlocal plasmonic effects. This challenge is met with the free-electron (jellium) model of TDDFT (JM-TDDFT). In this method, the ionic cores of the metal are not treated explicitly, but represented by a uniform charge density, $n_+=(3/4\pi) r_s^{-3}$, where $r_s$ is the Wigner--Seitz radius. Despite its simplicity, JM-TDDFT has proven to be a reliable and adequate many-body description of the dynamics of the conduction band electrons in a metal under optical excitation. Results of this approach have been pivotal for the prediction of quantum phenomena in plasmonics, such as the effect of electron tunneling across a plasmonic gap and electronic coupling between a quantum emitter and a plasmonic nanoantenna. Moreover, the nonlocal optical response of metallic nanoparticles and nanogaps obtained within the JM-TDDFT has been a key reference for benchmarking other theories~\cite{Zhu:2016,Babaze:2023,Teperik:2013,Ludwig:2020}.

\begin{figure}[htb]
    \centering
    \includegraphics[width=0.9\linewidth]{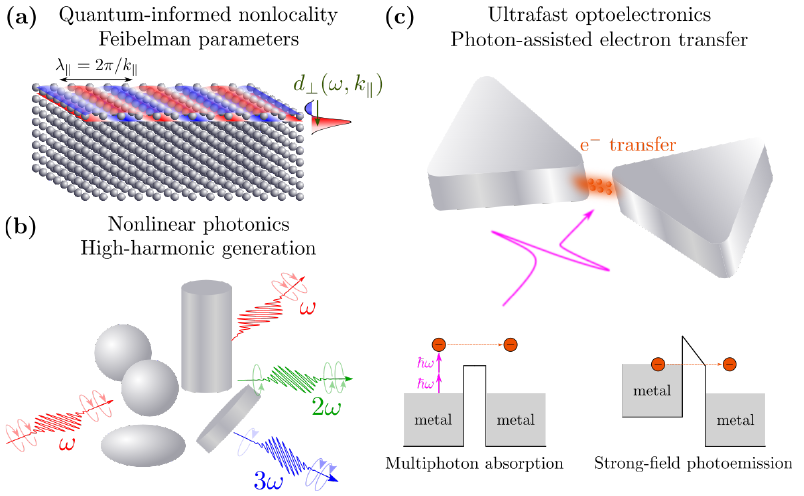}
    \caption{Areas of development of nonlocal nanoplasmonics within the JM-TDDFT approach. \pnl{a} The calculation of Feibelman parameters requires the description of the metal to be extended to include the effect of localized electron bands and atomistic crystallographic facets. \pnl{b} The jellium description of the electron gas allows for performing first-principles calculations of its intrinsic nonlinear optical response together with the effect of the nanoantenna geometrical boundaries, as in symmetry-induced nanoplasmonic response to structured light. \pnl{c} Ultrafast single cycle optical pulses induce photocurrents in nanogaps, determined by the dynamics of sub-fs electron bursts, as described by jellium-based electron models. }
    \label{fig:Aizpurua-Fig1}
\end{figure}

The results of the properties of plasmon modes of metal nanoantennas within the JM-TDDFT, as well as those regarding the Green’s function of a point-like dipolar emitter in proximity to a plasmonic dimer highlight the good performance of the dispersive surface-response formalism which incorporates nonlocality also in the direction parallel to the metal–vacuum interface~\cite{Babaze:2023}. This formalism is based on the implementation of Feibelman parameters~\cite{Feibelman:1982} (see also Secs.~\ref{sec:Christensen} and \ref{sec:Hohenester} of this Roadmap), $d_\perp(\omega, \boldsymbol{k}_\parallel)$ and $d_\parallel(\omega, \boldsymbol{k}_\parallel)$, into modified boundary conditions of Maxwell’s equations~\cite{Yang:2019a,Goncalves:2020} to capture extreme nonlocality at interfaces. The real part of the Feibelman parameter $d_\perp(\omega,\boldsymbol{k}_\parallel)$ accounts for the position of the centroid of the induced charge at a given frequency $\omega$ and at a given parallel momentum transfer $\boldsymbol{k}_\parallel$, and the imaginary part determines the probability of creating electron–hole pairs in the surface region. Crucially, it has been demonstrated that using Feibelman parameters that explicitly depend on $\omega$ and $\boldsymbol{k}_\parallel$ is significantly more accurate than using Feibelman parameters obtained within the long-wavelength approximation~\cite{Babaze:2023}. These Feibelman parameters $d_\perp(\omega,\boldsymbol{k}_\parallel)$ can be obtained using JM-TDDFT for a flat metal surface [Fig.~\ref{fig:Aizpurua-Fig1}\pnl{a}]. 

One of the main challenges of the surface-response formalism based on Feibelman parameters is the limited \emph{availability of quantum-informed parameters} for the different pairs of metal–dielectric interfaces commonly used in experimental realizations of plasmonics. So far, the jellium model of metals has been implemented to generate quantum-informed Feibelman parameters, which allows for addressing nonlocality of free-electron plasmonic nanoparticles [aluminum (Al) and alkali metals such as sodium (Na)] in vacuum. Such a description is not adequate for silver (Ag) and gold (Au), routinely used in spectroscopy and microscopy experiments. In these metals, the localized d-band electrons strongly impact the optical response. Furthermore, the dielectric environment also affects the Feibelman parameters by changing the spill-out of the electron density and the dynamical screening. To address these effects, a full-atomistic \emph{ab initio} TDDFT approach beyond JM-TDDFT needs to be applied (see also Sec.~\ref{sec:Zhang} of this Roadmap), which makes the calculations extremely challenging numerically.

On the other hand, the lighter and more versatile JM-TDDFT opens the door to tackle other nonlocal and, more generally, quantum effects of light–matter interaction in nanoplasmonics, where the underlying physics is determined by the conduction electrons. Thus, the intrinsic \emph{nonlinear electromagnetic response} of conduction electrons leading to frequency conversion in nanoplasmonic configurations can be directly obtained within the jellium scheme [Fig.~\ref{fig:Aizpurua-Fig1}\pnl{b}], enabling the identification of different sources of nonlinearity at the microscopic level. Furthermore, JM-TDDFT allows one to describe the field-induced electron emission from metallic surfaces and tips triggered by single-cycle optical pulses. In particular, when a metallic surface is facing another one in close proximity [Fig.~\ref{fig:Aizpurua-Fig1}\pnl{c}], the bursts of photoemitted electrons exhibit an intriguing and complex \emph{femtosecond dynamics} that depends on the waveform of the incident transient pulse~\cite{Ludwig:2020}. This opens up perspectives for designing nanoscale PHz optoelectronic devices. 

\subsection*{Future developments to address challenges}

Both short- and long-term developments will be required in the implementation of many-body calculations of light–matter interaction in the context of nanoplasmonics. The calculation of Feibelman parameters for interfaces of non-free-electron materials will require the development of atomistic \emph{ab} initio calculations of dispersive Feibelman parameters. This will enable the direct implementation of nonlocality in a set of materials such as gold (Au), silver (Ag), copper (Cu), platinum (Pt), palladium (Pd), etc., interfacing dielectrics, of high interest in field-enhanced catalysis and chemical reactivity. The role of the crystallographic facets and of surface-localized electronic states could thus be addressed in this way. 

In nonlinear nano-optics, specific combinations of the intrinsic nonlinear response of the electron gas with specific geometrical symmetries shown by certain nanoantennas configurations could be exploited in nonlinear chiral nanoplasmonics, with the full nonlocal and nonlinear response directly accounted for by the jellium description of the electron gas.
Finally, the fs photocurrents produced by intense single-cycle optical pulses can be studied in the context of electron transport in STM junctions, with the ultimate goal of providing theoretical support to understand and design devices that combine atomic-scale spatial and sub-fs temporal resolution.

\subsection*{Concluding remarks}

The jellium model (JM-TDDFT) has been a powerful tool for addressing conduction electron dynamics and nonlocal effects in metallic nanoparticles during the last years, offering a computationally efficient yet accurate framework for exploring fundamental quantum phenomena in nanoplasmonics. Moreover, it is evident that this methodology holds significant potential for future advancements, particularly in deepening our understanding of nonlinear nano-optics and ultrafast optoelectronics. At the same time, the development of more sophisticated atomistic descriptions beyond the jellium model will be crucial for advancing in the development of quantum-informed nonlocality. Together, these developments will not only expand our fundamental knowledge of light-matter interaction but also contribute to the design of applications in quantum technologies and ultrafast photonics.
\section[Atomic scale nonlocal plasmonics (Zhang \& Chen)]{Atomic scale nonlocal plasmonics}  

\label{sec:Zhang}

\author{Pu Zhang\,\orcidlink{0000-0002-6253-0555}  \& Xue-Wen Chen\,\orcidlink{0000-0002-0392-3551}}

\subsection*{Overview}

Nanoplasmonics is continuously pursuing ever tighter light confinement. With the conceptual innovations and advances in nanofabrication, the compressed plasmonic field inevitably encounters the eventual graininess of material, or reaches the atomic scale. Studying the impact of the atomic scale structure on the plasmonic response and related applications constitutes atomic scale plasmonics. Holding the promise of achieving the ultimate light confinement and enormous enhancement of light-matter interaction, this research direction opens up an avenue for the plasmonics realm. In this contribution we start with a brief account of atomic scale plasmonics from its commencement to the current status, necessarily reflecting the personal view of the authors. Then we identify some opportunities and challenges faced by this flourishing field. Emphasis is placed on the nonlocal effects and theoretical analysis of the extremely localized field.

\begin{figure}[hb]
    \centering
    \includegraphics[width=0.95\linewidth]{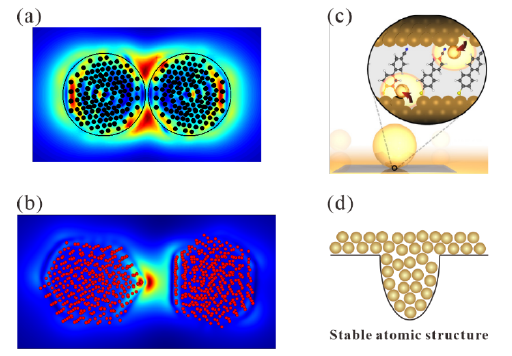}
    \caption{The representative developments and trend of atomic scale plasmonics. \pnl{a}~An early study on the plasmonic response of atomic clusters via an ab-initio approach reveals the impact of the atomic structure on the local field pattern and strength~\cite{Zhang:2014a}. \pnl{b}~The atomic cluster dimer in the tip-to-facet configuration is predicted with TD-DFT to give rise to atomic scale lightning rod effect~\cite{Barbry:2015}. \pnl{c}~Experimental report on "picocavities" based on the observation of bright fluctuational flashes in the real-time SERS spectra~\cite{Benz:2016}. \pnl{d}~The beyond-single-adatom atomic structure is a preferred configuration for atomic scale plasmonics owing to the stability at room temperature.}
    \label{fig:Zhang-Fig1}
\end{figure}

\subsection*{Current status}

Spatial confinement of light is one of the most important strategies to boost light-matter interaction, which lies at the core of the majority, if not all, of light technologies. Conducting material-based nanostructures with abundant and freely moving electrons serve as the driving force for light confinement, as they are able to leverage both the non-resonant lightning rod effect from morphology induced electron accumulation and the resonant augment from the localized surface plasmon resonance. With the typical nanoplasmonic constructs such as sharp tips, nanogap structures and cascaded assemblies, the localized optical field reaches the nanoscale and even smaller. Going a step forward nanoplasmonics encounters the graininess of material.
Atomic scale plasmonics was initiated with theoretical exploration. From the theoretical perspective, classical electromagnetic theory with the local response approximation, or the description of material by the permittivity $\varepsilon(\omega)$, is normally applied in plasmonics, but starts to fail for plasmonic field squeezed into nanoscale spaces because of the emerging nonlocal effects. The alternatives include continuum nonlocal theories (see Sec.~\ref{sec:Fernandez-Dominguez} of this Roadmap), which effectively provide a description based on a spatially dispersive permittivity $\varepsilon(\omega,\boldsymbol{k})$, and \emph{ab initio} approaches. Being a representative of the latter, time-dependent density-functional theory (TD-DFT) encompasses all the non-classical effects of electrons, and is originally developed to treat atomic structures. Using TD-DFT beyond the jellium approximation (see Sec.~\ref{sec:Aizpurua} of this Roadmap), the impact of the atomic structure on the plasmonic near field was first investigated and highlighted in a study on the sodium (Na) cluster dimers consisting of over 600 atoms in total~\cite{Zhang:2014a}. As shown in Fig.~\ref{fig:Zhang-Fig1}\pnl{a}, the near-field pattern clearly has the imprint of the detailed atomic structure. The atoms at the cluster surface shape the localized field into irregular distributions. Intense field even appears in the atomic voids and crevices. The hot spot in the plasmonic near field is of particular importance. Much attention was focused on the formation of hot spots in similar atomic cluster dimers. In particular, an asymmetric dimer in the tip-to-facet configuration illustrated in Fig.~\ref{fig:Zhang-Fig1}\pnl{b} was successfully predicted with TD-DFT to generate hot spots of down to 0.4\,nm$^2$ lateral area~\cite{Barbry:2015}. The underlying mechanism is recognized as the non-resonant lightning rod effect.
The first experimental observation relevant to atomic scale plasmonics was reported in low-temperature time-series surface-enhanced Raman spectroscopy (SERS)~\cite{Benz:2016}. Fluctuational bright flashes were spotted in the real-time spectra besides persistent signals. The experiment was interpreted with the introduction of the "picocavity" in Fig.~\ref{fig:Zhang-Fig1}\pnl{c}, where the plasmonic hot spot at a gold (Au) adatom activated by laser irradiation couples with an individual molecular bond. The interpretation was supported by the experimental observations including the extreme enhancement of the optomechanical coupling and alteration of SERS selection rules by strong optical field gradient. The picocavity has a Raman localization volume below 1\,nm$^3$. Apart from the enormously enhanced Raman scattering, this highly localized hot spot boasts great potentiality in various types of ultrasensitive detection and sensing. For instance, a silver (Ag) tip decorated with an Ag cluster has been shown able to realize sub-nanometer resolution single-molecule photoluminescence imaging~\cite{Yang:2020b}. In this case the atomic structure, or the Ag cluster, is transferred to the tip apex during tip indentations, instead of activated by laser stimulation. Nevertheless, neither methods can deterministically produce stable atomic protrusions. The beyond-single-adatom stable atomic protrusion sketched in Fig.~\ref{fig:Zhang-Fig1}\pnl{d} is being actively sought for~\cite{Paschen:2023,Taneja:2024}.

\begin{figure}[ht]
    \centering
    \includegraphics[width=0.99\linewidth]{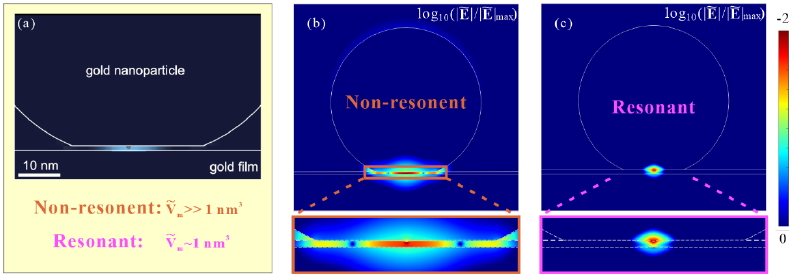}
    \caption{Field confinement at an atomic scale protrusion. \pnl{a}~The schematic of the protrusion in the nanoparticle-on-mirror (NPoM) configuration. The logarithmic-scale normalized magnitude of the modal electric field around the protrusion at the \pnl{b}~non-resonant and \pnl{c}~resonant frequencies calculated with quantum hydrodynamic theory based quasinormal mode theory.}
    \label{fig:Zhang-Fig2}
\end{figure}

\subsection*{Challenges and opportunities}

The vibrantly progressing research area of atomic scale plasmonics certainly faces prominent challenges, which in the meanwhile signify great opportunities. In this section we name a few outstanding ones in hope that they could inspire more research interest in the area.
The most fundamental question would be "What is the ultimate limit of light confinement?". Before any meaningful discussion on the topic, what needs to be resolved in the first place is how to characterize quantitatively the degree of light confinement with a consistent standard. While mode volume is routinely used for the purpose, the way it is evaluated varies a lot from case to case. For instance, Ref.~\cite{Benz:2016} obtained the 1\,nm$^3$ Raman localization volume referring to the Raman scattering enhancement by the picocavity. A more widely accepted approach is the mode volume according to the mode theory of optical microcavity. However, the nonlocal effects including electron nonlocality, electron spill out and nonlocal damping, manifested in atomic scale plasmonics, call for a non-classical theory. Quantum hydrodynamic theory (QHT) elaborated on in Sec.~\ref{sec:Hu} is a viable option. The quasinormal mode theory (QNM) has recently been generalized to QHT, so that a mode theory-based mode volume can be defined in atomic scale plasmonics~\cite{Li:2021,Zhou:2021}. Therein the nonlocal effects are encapsulated into the restoring force $\hat{\Theta}$ and damping $\hat{\Gamma}$ operators in the generalized Lorentz model (GLM)~\cite{Zhou:2021}
\begin{equation}
    i\rho\varepsilon_0\omega_p^2 \boldsymbol{E}-i\rho \hat{\Theta}\boldsymbol{P}-i \rho\hat{\Gamma}\boldsymbol{J}=\omega \boldsymbol{J}
\end{equation}
which constitutes a general theoretical framework for any continuum linear response theory and facilitates the establishment of the corresponding QNM theory. According to the QHT-QNM theory, picocavity working under the non-resonant lightning rod mechanism has a several tens of nm3 mode volume. Plotted in logarithmic scale the local field still spreads across the whole nanogap as illustrated in Fig.~\ref{fig:Zhang-Fig2}\pnl{b}. In contrast, the QHT-QNM theory helps discover an extremely localized mode (ELM) supported by the atomic protrusion~\cite{Li:2021}. At the resonant frequency its mode volume turns out about 1\,nm$^3$, with the modal field in Fig.~\ref{fig:Zhang-Fig2}\pnl{c} extremely localized at the protrusion. Moreover, the ELM can be made very radiative with a 30\% efficiency when coupled with a host antenna. The resonance of the protrusion was also pointed out through the local response calculations~\cite{Wu:2021,Griffiths:2021}. Nevertheless, whether 1\,nm$^3$ mode volume represents the ultimate light confinement is still an open question.
When an optical field is localized on the atomic scale, or commensurate with the extent of the electron wave function, it is naturally expected susceptible to the ambient conditions. Apart from serving for ultrasensitive detection, tuning the optical response with another physical factor in a controlled manner is a promising avenue. At an atomic scale protrusion, the minute number of electrons responsible for the optical response implies that the tuning could be ultrafast and energy efficient. In fact, by using a QHT tamed for noble metals and the corresponding QNM theory, electro-optic modulation of the ELM has been proposed~\cite{Zhou:2023b,He:2019}. A 20\,fs response time and below 100\,aJ per-bit energy consumption were predicted. We note electrical control is far from the only way to modulate the atomic scale plasmon. It is tempting to explore, both theoretically and experimentally, such modulation via e.g. thermal and mechanical stimuli in multi-physics systems.
Since the debut of atomic scale plasmonics, the interactions with quantum processes like molecular vibration~\cite{Benz:2016}, photoluminescence~\cite{Yang:2020a} and exciton~\cite{He:2019} are of particular importance for fundamental interest and potential applications. The tightly confined optical field is able to remarkably expedite the processes and as well to unlock originally forbidden ones by breaking the selection rules with the strong field gradient~\cite{Benz:2016}. There’s still much room left to tap the ELM for boosting light-matter interaction, but it is accompanied with intricate issues. Currently the typical atomic scale protrusions are generated dynamically through laser activation, and become destroyed equally fast at room temperature. The probabilistic nature severely limits the practical application of picocavity enhanced processes. In addition, the lack of controllable realization of an atomic scale protrusion prevents matching the protrusion’s resonant frequency with the intrinsic frequencies of the quantum processes. Almost all the existing studies only leverage the non-resonant lightning rod effect of the protrusion.

\subsection*{Future developments to address challenges}

Facing the challenges we then discuss future developments hopeful for answering the questions. Light confinement needs to be characterized with the consideration of the working wavelength. In the visible and near infrared range, light confinement has been reported theoretically to reach 1\,nm$^3$, but the limit was not addressed. Since the electron response to optical excitation is at the bottom of the mechanisms of light confinement, the length scale of the ultimate confinement, if it exists, should be related to that of the electrons in the material, or the extent of the electron wave function. What role the latter plays in determining the confinement is to be explored. Full quantum or some sophisticated approximate atomistic approaches would be required for the task. On the other hand, there are other known strategies for light localization than those leading to the ELM. Especially applying them to the judicious design of the host of the atomic scale structure is expected to further compress the field.
To exploit the potential of using atomic scale plasmon for modulation, multi-physics scenarios become highly relevant. The existing theories and numerical models for plasmonic response must coherently incorporate the controlling physical factor of interest to characterize its nonlinear coupling with the plasmonic field. In this regard, the QHT is a convenient theoretical platform, as it directly deals with the foundational quantities of electron and current densities. The influences of the controlling factors on the densities are usually easy to introduce in a first-principles fashion as in electro-optic modulation~\cite{Li:2022a}. A comprehensive modulation study necessarily involves inspecting the dynamic tuning process, and thus a time-domain theory is indispensable. At least for QHT, however, the time-domain implementation is challenging (see Sec.~\ref{sec:Hu} of this Roadmap), let alone one generalized to multi-physical systems. Aside from used for information processing, modulating the atomic scale plasmon enables tuning its resonance to match with quantum processes, so that the enhancement is maximized.
Deterministic preparation of a stable atomic scale structure poses an obstacle to the practical application. While beyond-single-adatom stable atomic scale structures are alternatives and have been attempted experimentally in literature, they are still too tiny to be mass fabricated. An effective approach to circumvent the difficulty is to work with highly doped semiconductors instead of metals. Except the working frequency is shifted to the infrared range, almost all the intriguing features are realizable with structures correspondingly scaled to decananometers, which are readily feasible with the current nanofabrication techniques. The extremely localized field is retained considering the mode volume normalized with respect to the wavelength cubed. Dramatic plasmon modulation is also within reach owing to the relatively large screening length in semiconductors. Adapting the non-classical theories to semiconductors is however in need for the new expedition.

\subsection*{Concluding remarks}

Atomic scale plasmonics is a natural development of nanoplasmonics. In this contribution we briefly discussed the current status of this frontier, the challenges and opportunities, and envisioned the future development. Although the picocavity and extremely localized mode represent actively studied topics of atomic scale plasmonics, more diverse plasmonic field localized in atomic scale structures are worthwhile to look into. Plasmonic field squeezed into atomic scale slots was introduced to explain transient "flare" emission observed in plasmonic cavities~\cite{Baumberg:2023}. "picophotonics" was also coined for the study on the microscopic optical waves in the atomic lattice~\cite{Bharadwaj:2022}.

\section[Nonlocal limitations to light--matter interactions (Tserkezis)]{Nonlocal limitations to light--matter interactions}

\label{sec:Tserkezis}

\author{Christos Tserkezis\,\orcidlink{0000-0002-2075-9036}}

\subsection*{Current status}

Probably the biggest appeal of plasmonic nanostructures is related to the impact they can have on enhancing light--matter interactions.
Individual plasmonic nanoparticles (NPs), and even more so plasmonic cavities, provide a dramatic increase in the local density of optical states (LDOS), allowing to control the properties of quantum emitters (QEs) placed near them. This potential manifests in both the so-called weak- and strong-coupling regimes. Under weak coupling conditions, the plasmonic nanostructure is treated as a dielectric environment that alters both the excitation and the emission rate of QEs via the Purcell effect, allowing e.g. to control molecular fluorescence, at the single-molecule level~\cite{Anger:2006,Kuhn:2006,Kinkhabwala:2009}.
In the strong-coupling regime, the QE hybridizes with the optical modes of the cavity, and the two components enter a coherent and reversible exchange of energy, that results in the formation of half-light--half-matter entities, called
polaritons~\cite{Torma:2015,Tserkezis:2020a,Basov:2021}.
In the case of cavities, the local response approximation (LRA) predicts a diverging field enhancement as the gap between individual components shrinks~\cite{Romero:2006}. This unphysical situation reflects the structural weaknesses of LRA. Nonlocality imposes more realistic limits to the performance of nanostructures, especially plasmonic ones; the spatial extent of induced charges, together with the increased Landau damping as a result of collisions with the nanostructure boundaries, regularize the plasmonic near-field enhancement, as discussed already in Ref.~\cite{Ciraci:2012}.
Such limitations have been verified ever since for individual NPs~\cite{Mystilidis:2023} and coupled structures~\cite{Yan:2024}, probed with a variety of sources including swift electrons and point dipoles~\cite{Wiener:2013,Christensen:2014,Goncalves:2023}.

\begin{figure}[htbp]
\centering\includegraphics[width=1.0\textwidth]{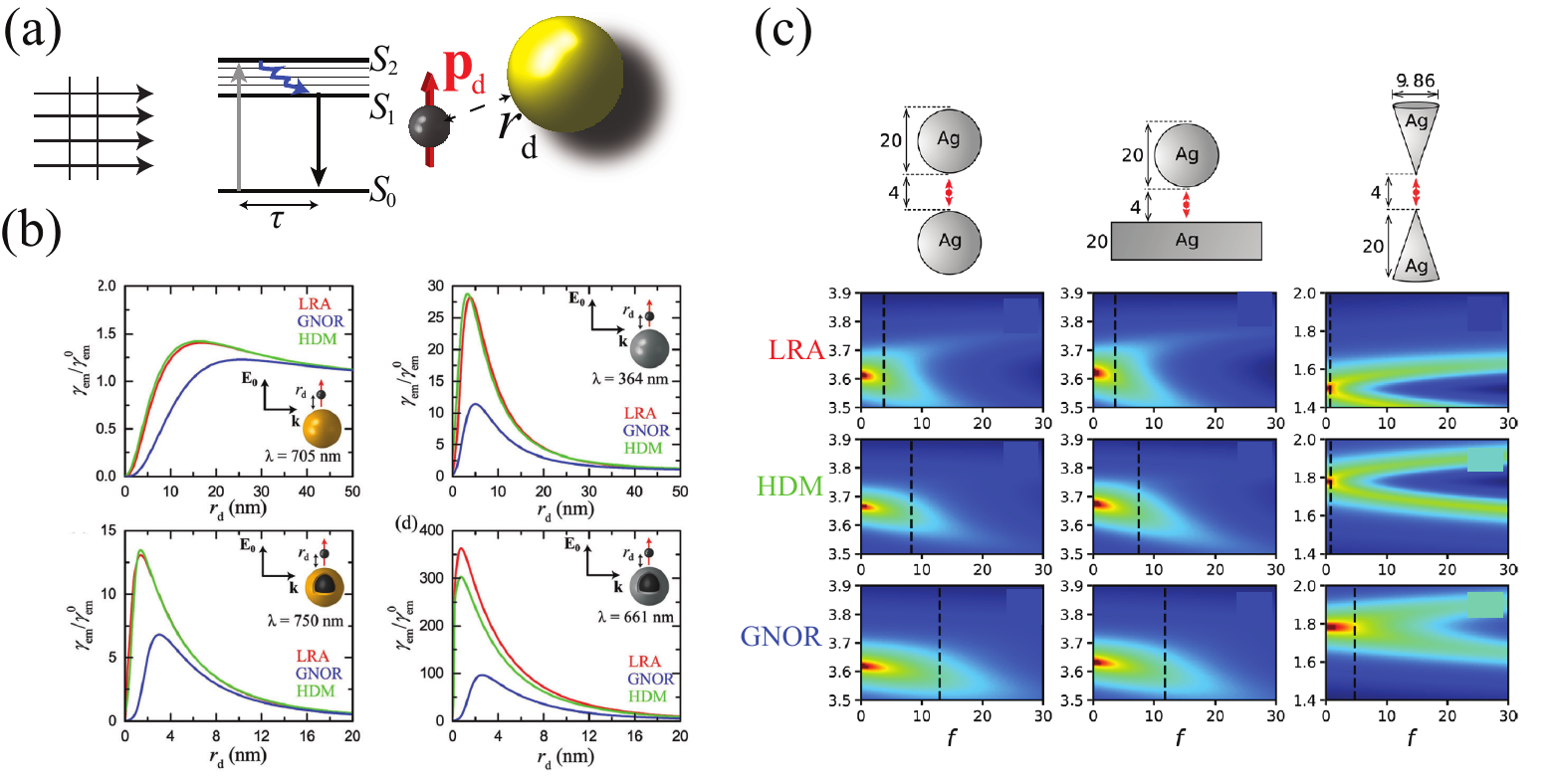}
\caption{Nonlocality in QE--plasmon interactions.
\pnl{a}~Fluorescence near a plasmonic NP; a QE with dipole moment
$\boldsymbol{p}_{\mathrm{d}}$ is placed at distance $r_{\mathrm{d}}$
from the surface of the NP, and both are illuminated by an external
plane wave. The QE goes from the ground state $S_{0}$ to an
excited state $S_{2}$, quickly relaxes to a state $S_{1}$, and then
decays to the ground state after time $\tau$.
\pnl{b}~Modification of fluorescence enhancement near gold (Au) or silver (Ag)
nanospheres and nanoshells as a function of distance $r_{\mathrm{d}}$,
within LRA, HDM, and GNOR. Figure adapted with permission from Ref.~\cite{Tserkezis:2016} (Copyright~\textcopyright~2016 Royal Society of Chemistry).
\pnl{c}~Modification of strong coupling of a single QE in different Ag
plasmonic cavities, as a function of the dipole oscillator strength $f$,
within LRA, HDM, and GNOR. Reprinted (adapted) with permission from Ref.~\cite{Jurga:2017} (Copyright~\textcopyright~2017 American Chemical Society).
}
\label{fig:Tserkezis-Fig1}
\end{figure}

\textbf{Weak coupling.} The hydrodynamic Drude model (HDM) was employed already in the 1980s to study the emission of a point dipole near a flat metallic surface~\cite{Fuchs:1981}, while NPs were more actively considered in the late 2000s~\cite{Chang:2006}. Detailed studies of fluorescence enhancement followed a few years later~\cite{Kosionis:2012,
Tserkezis:2016,Jurga:2017,Gongora:2017,Tserkezis:2017,Kupresak:2022}, and Purcell factors have been calculated for NPs of various shapes~\cite{Filter:2014,Girard:2015}, or for stratified media~\cite{Intravaia:2015}, employing both HDM and the more recent generalized nonlocal optical response (GNOR) theory~\cite{Mortensen:2014}. These studies highlighted that fluorescence can be affected in two additive ways by nonlocality. At first, as sketched in Fig.~\ref{fig:Tserkezis-Fig1}\pnl{a}, the QE has to make the transition from the ground, $S_{0}$, to the excited, $S_{2}$ state, a process that is governed by the $\boldsymbol{p}_{\mathrm{d}} \cdot \boldsymbol{E}$ interaction, where $\boldsymbol{p}_{\mathrm{d}}$ is the dipole moment of the QE, while $\boldsymbol{E}$ is the total field experienced by it, maximized at plasmonic resonances. The frequency shifts predicted by HDM become thus important for the efficient excitation of the QE, as they detune plasmonic resonances from the transition energy. In the second process, after being excited to $S_{2}$, the QE quickly relaxes to a metastable state $S_{1}$ at a close energy, from which it then decays spontaneously to the ground state after a characteristic time $\tau$. When placed near plasmonic NPs, the QE can couple to the radiative dipolar mode, but also to higher-order multipoles, which are non-radiative. Depending on the QE--NP separation, one of the two couplings prevails, as quantified by the ratio of radiative over total decay rates. An optimum QE--NP separation can thus be identified, and it is strongly modified once nonlocal damping is taken into account [see e.g. Fig.~\ref{fig:Tserkezis-Fig1}\pnl{b} for GNOR calculations; Landau damping dominates the process, much of the energy is absorbed by the NP, and the QE needs to be placed farther away from the NP surface for an efficient radiative coupling~\cite{Tserkezis:2016}]. This situation is slightly different when the QE is placed within a plasmonic cavity, because the hybrid plasmon modes they sustain are more radiative.
Different types of plasmonic cavities have been considered within the context of HDM and GNOR~\cite{Tserkezis:2017}, with the main conclusion being that nonlocality tends to regularize the maximum enhancement that is possible for decreasing gaps, bringing theoretical predictions closer to experimental measurements~\cite{Kinkhabwala:2009}.

\textbf{Strong coupling.} In the strong-coupling regime, single QEs, or QE collections supporting collective excitonic modes, interact with a cavity mode, leading to the formation of hybrid polaritonic states~\cite{Bellessa:2004,Gonzales:2013,Delga:2014}; the energy difference between the hybrid modes is known as the Rabi splitting. The main question regarding nonlocality is: can it affect the coupling strength, e.g. by weakening the interacting EM fields, or by detuning the plasmonic mode? In a first attempt to answer these questions, Ref.~\cite{Tserkezis:2018} considered plasmon--exciton coupled systems in a spherical
core--shell geometry. Taking into account that in any experiment the plasmonic resonances will be characterized by the measured frequencies and linewidths, with any nonlocal effects intrinsically included, anticrossing spectra were plotted renormalized to the plasmonic resonances predicted within each model, for different nonlocal models, showing an almost perfect superposition of anticrossings; both hybrid modes shift, as a result of nonlocality, by the same amount, leading to negligible differences in the width of the Rabi splitting. The damping introduced by the nonlocal models is small enough to ensure that no significant broadening occurs, and the main strong-coupling criterion of the Rabi splitting being larger than the average damping rate of the uncoupled modes~\cite{Torma:2015} still holds.
A similar response has been observed with electron-beam excitation as well~\cite{Zouros:2020}. With a different starting point, Ref.~\cite{Jurga:2017} placed QEs with adjustable oscillator strength inside different plasmonic cavities, and observed that the threshold for entering the strong-coupling regime would shift depending on the applied model (see Fig.~\ref{fig:Tserkezis-Fig1}\pnl{c}, for three different cavities, within LRA, HDM, and GNOR). The main difference between Refs.~\cite{Tserkezis:2018} and~\cite{Jurga:2017} is exactly in the oscillator strength of the excitonic material; while the former assumed a collective emitter layer that would always have a large enough dipole moment to enter the strong-coupling regime, the latter explored the conditions for a single QE to achieve this.

\subsection*{Challenges and opportunities}

The above discussion is more relevant in noble-metal NPs and cavities, with high work functions and negligible electron spill-out. The picture is expected to change if spill-out dominates, as in alkali metals. This has only recently started being considered; for instance, in Ref.~\cite{Babaze:2022} the self-interaction Green's function was calculated for sodium nanospheres and nanosphere dimers, both within the surface-response formalism (SRF) based on Feibelman parameters, and within time-dependent density functional theory (TDDFT), showing that the results deviate considerably from the predictions of LRA, already for QE--NP distances or interparticle gaps as large as $2.5$\,nm. Interestingly, the semiclassical SRF treatment was in excellent agreement with TDFFT down to $\sim 0.7$\,nm. Nevertheless, deviations were observed for even smaller distances; these were attributed to the fact that the standard SRF is based on \emph{local} Feibelman parameters, usually obtained for flat metal--dielectric interfaces. Eventually the curvature of the structure needs to be included~\cite{Yan:2015b,Christensen:2017};
this was done by the introduction of \emph{nonlocal} Feibelman parameters~\cite{Babaze:2023}, which is, however, still a rather challenging task, both in terms of obtaining those parameters, and implementing them in computational electrodynamics.
NP dimers were studied in Ref.~\cite{Baghramyan:2022}, focusing on the role of the electron tunneling that comes as a result of spill-out in two sodium nanospheres separated by sub-nm gaps. This study, performed with the quantum hydrodynamic (QHT) model~\cite{Ciraci:2016}, also known as self-consistent hydrodynamic Drude model (SC-HDM)~\cite{Toscano:2015}---a generalization of HDM that relaxes the hard-wall boundary condition---predicted considerable fluorescence quenching inside the cavity, with results that are qualitatively and quantitatively different from those of LRA and HDM; similarly to the case of noble-metal NPs, nonlocality seems to provide the upper limits to the system's performance. In the strong-coupling regime, Ref.~\cite{Bundgaard:2024} found that spill-out, and the plasmonic redshifts that accompany it, do not drastically change Rabi splittings.
In a similar study for flat sodium--air interfaces, strong coupling was indeed retained, also accompanied by a drastic effect on the QE dynamics, which acquires a strong non-Markovian character~\cite{Karanikolas:2021}. On the other hand, Ref.~\cite{Ciraci:2019} observed that the
oscillator-strength threshold for achieving strong coupling changes within each model. In particular, using QHT to account for electron spill-out, a considerable increase in the oscillator threshold was predicted, also related to the fact that coupling to higher-order multipoles occurs already for larger QE--NP distances due to the larger effective NP size. Finally, it is worth mentioning a similar investigation related to the Mollow triplets in QE--NL coupled systems~\cite{Ge:2013}; using GNOR as the nonlocal model, Ref.~\cite{KamandarDezfouli:2017} observed that the side bands become stronger and with narrower linewidths, compared to the predictions of LRA.

Historically, nonlocality in plasmonics has been associated with "bad news", i.e. limitations to the performance of an architecture or a device. While such studies are tremendously important from a fundamental point of view, to establish the physical mechanisms involved, they also tend to become slightly unattractive, by predicting less impressive results than those of LRA. There have only been few situations where nonlocality promised to improve a performance, as e.g. when it was shown to smooth the surface roughness of a metallic tip~\cite{Wiener:2012}.
In the context of light--mater interactions, a recent study~\cite{Goncalves:2020} showed that nonlocality and electron spill-out have the potential to enhance dipole-forbidden
transitions~\cite{Cuartero-Gonzalez:2018} or plasmon-mediated energy transfer, eventually showing that nonlocality is not always a limiting factor, but a possibility of harnessing it for practical applications does exist.

\subsection*{Future developments to address challenges}

From the above discussion, it becomes clear that nonlocality can have a strong, and sometimes unpredictable impact on light--matter interactions. Nevertheless, current literature is large; restricted to plasmonics, and almost exclusively on a theoretical level. Furthermore, most studies have only focused on exploring the role of corrections in the response of the plasmonic component, while the QEs themselves are not considered. There is therefore still plenty of room for future explorations. First of all, it will be important to consider, both in theory and in experiment, the impact of nonlocality on other types of light--matter interacting architectures.
For instance, Ref.~\cite{Alpeggiani:2014} has made a first effort to address semiconductors, focusing on the radiative and nonradiative decay rates of multi-subband plasmons in quantum wells. Magnetoplasmonic systems and Landau polaritons were studied in Ref.~\cite{Rajabali:2021}, which managed to set quantitative limits to the miniaturization of polaritonic devices and to the enhancement of polariton gaps. Nonlocality in phonon-polariton resonators was studied in Ref.~\cite{Gubbin:2020}, albeit only on the level of far-field extinction and reflection properties.

At the same time, one is justified to ask, to what extent it is appropriate to continue refining the description of photonic nanostructures, while retaining the simple point-dipole approximation for the QE. It is almost inevitable that the next generation of nonlocal studies will have to introduce both a quantum-electrodynamic treatment of the problem~\cite{Varas:2016,Hapuarachchi:2018}, and also more elaborate models for the QEs, either entering the domain of computational quantum chemistry~\cite{Flick:2018}, or with semiclassical models as suggested e.g. in Ref.~\cite{Ciraci:2019}.
Strong coupling of a QE with a film described by a hydrodynamic or a Lindhard permittivity has been explored in Ref.~\cite{Vagov:2016}, but retaining again some level of approximation. And this only makes sense, as quantization of the "cavity modes" for open, dispersive and lossy cavities, can be tricky~\cite{Tserkezis:2020a}. Macroscopic cavity quantum electrodynamics (QED) appears to be the most promising route right now, but obviously no treatment can be seen as the unique, conclusive answer yet. This will require a combination of further theoretical work, possibly with the introduction of new numerical methods for mesoscopic electrodynamics~\cite{Hohenester:2022a,Yang:2022}, but more importantly, realization of fine experiments that can  probe the small modal shifts and damping involved, and unambiguously attribute them to nonlocality.

\subsection*{Concluding Remarks}

Nonlocality plays an important role in light--matter interactions, by introducing detuning and additional damping that can modify the coupling strength between plasmonic environments and QEs.
It is, however, important to broaden the horizon of such studies by also departing from the point-dipole description of the QE, while more efficient computational methods might still be needed.
On the other hand, nonlocality is much less explored when QE environments other than plasmonic are concerned, and this promises to be an active direction in the near future.

\section[Quantum surface-response functions for nanoplasmonics: Feibelman parameters (Christensen \emph{et al.})]{Quantum surface-response functions for nanoplasmonics: Feibelman parameters}

\label{sec:Christensen}

\author{Thomas Christensen\,\orcidlink{0000-0002-9131-2717}, Wei Yan\,\orcidlink{0000-0002-8150-1194} \& Yi Yang\,\orcidlink{0000-0003-2879-4968}}

\subsection*{Current status}

The study of quantum-response effects in nanoplasmonics has often centered on the incorporation of specific physical effects otherwise omitted in the piecewise-constant, local-response outset that underlies conventional (or "classical") electromagnetic modeling (CEM).
Focus has especially centered on the incorporation of one of the three key omissions in CEM, namely
(1)~bulk nonlocal response (i.e., the finite range of the dielectric tensor $\bm{\varepsilon}(\mathbf{r},\mathbf{r}')$ with respect to $|\mathbf{r}-\mathbf{r}'|)$, 
(2)~the non-abrupt and smooth variation of the electron density near material surfaces (e.g., Friedel oscillations and electron spill-out), and 
(3)~absorption processes involving non-negligible change of quasiparticle momentum (e.g., surface-enabled Landau damping).
While such single-effect explorations have yielded valuable physical and conceptual insights, they are challenged by the reality that no single beyond-CEM effect strongly dominates the others in general and, moreover, that they may act in opposition.
In extreme cases, including only one effect may \emph{decrease} predictive power, as in the case of hydrodynamic accounts of nonlocality for simple metal surface plasmons.
Observations of beyond-CEM deviations should therefore generally be expected as arising from the interplay of multiple, physically distinct effects.

\begin{figure}[b!]
    \centering
    \includegraphics[width=0.99\linewidth]{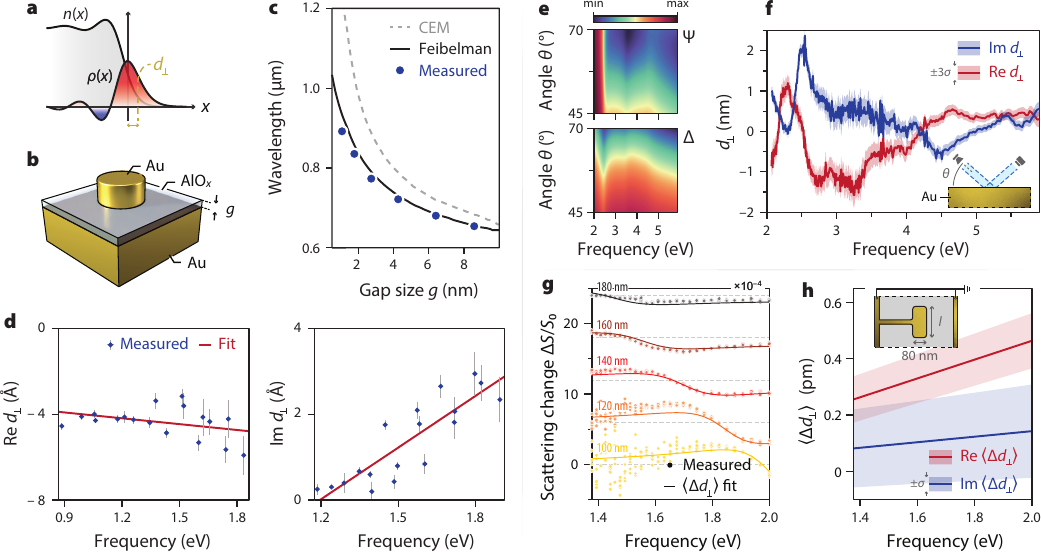}
    \caption{Measuring Feibelman parameters.
        \pnl{a}~%
        The $d_\perp$ and $d_\parallel$ parameters are the frequency-dependent centroids of the induced charge $\rho(x)$ (shown) and the normal derivative of the interface-tangential current $\partial_y J_y(x)$ (not shown).
        Both $\rho(x)$ and $J_y(x)$ are induced quantities due to an incident plane wave, appearing as perturbations to the static equilibrium density $n(x)$.
        \pnl{a--d}~%
        Comparing observations of large deviations between CEM and experiments in gapped film-coupled nanodisks \pnl{b,c} with perturbation theory, Eq.~\eqref{eq:Christensen-Eq4}, can enable experimental inference of the Feibelman parameters \pnl{d}, here of $d_\perp$ for at a gold (Au) aluminum oxide (AlO\textsubscript{$x$}) interface~\cite{Yang:2019a}.
        \pnl{e,f}~%
        The angle- ($\theta$) and frequency-dependent amplitude contrast $\Psi$ and phase difference $\Delta$ of the complex reflection ratio $r_{\textit{p}}/r_{\textit{p}} = e^{i\Delta}\tan \Psi$ between {\textit{p}}- and {\textit{s}}-polarized light are conventionally used to obtain a material's permittivity by fitting to the classical Fresnel reflection coefficients $r_{\textit{p},\textit{s}}$.
        By using the Feibelman-corrected reflection coefficients instead, sufficiently precise measurements of $\Psi$ and $\Delta$ can enable inference of the interface's Feibelman parameters \pnl{f}, here of $d_\perp$ at an Au(111)--air interface~\cite{Chen:2024b}.
        \pnl{g,h}~%
        An external gate voltage of 10\,V, applied over an Au nanoresonator, leads to a frequency-dependent shift $\Delta S$ of the nanoresonator's scattering signal, presented relative to its ungated signal $S_0$ \pnl{b}.
        Measurements over different nanoresonator lengths $l$ are labeled, shifted vertically, and differentiated by color.
        The shift of the scattering signal may be interpreted as due to a gate-tunable (interface-averaged) shift $\langle \Delta d_\perp\rangle$ of the $d_\perp$-parameter at the Au--air interface \pnl{h}, obtained by fitting to experimental data~\cite{Zurak:2024}.
        Inset in \pnl{g} schematically illustrates the nanoresonator, its dimensions, and gating configuration.
        Panels~\pnl{a,c,d} adapted with permission from Ref.~\citenum{Yang:2019a} (Copyright~\textcopyright~2019 Springer Nature).; \pnl{e,f} adapted from Ref.~\citenum{Chen:2024b}; \pnl{g,h} adapted from Ref.~\citenum{Zurak:2024} (American Association for the Advancement of Science, CC BY 4.0).
    }
    \label{fig:Christensen-Fig1}
\end{figure}

Feibelman identified this challenge in the early eighties and introduced his eponymous $d_\perp$ and $d_\parallel$ parameters as a possible remedy~\cite{Feibelman:1981,Feibelman:1982}, further developed with subsequent contributions from Apell, Liebsch, and others~\cite{Apell:1981, Apell:1982, Liebsch:1997}.
In particular, the CEM implies that induced charge density is accumulated only at interfaces, and consequently amounts to a zeroth-order (monopole) approximation of a multipole expansion of a continuously varying induced charge density.
$d_\perp$ parameterizes the omitted first-order (dipole) term of this expansion and $d_\parallel$ parameterizes the discrepancy between the classical and quantum-mechanical monopole terms (i.e., total charge); together, they capture the leading-order corrections to CEM~\cite{Yan:2015b,Christensen:2017}.
Considering a planar interface (residing at, say, $x=0$), $d_\perp$ and $d_\parallel$ are the frequency-dependent centroids of the induced charge density $\rho(x, \omega)e^{i k y}$ [Fig.~\ref{fig:Christensen-Fig1}\pnl{a}] and the normal derivative of the tangential current density $J_y(x, \omega)e^{i k y}$ due to an incident plane wave propagating in the $xy$-plane at frequency $\omega$~\cite{Feibelman:1982, Liebsch:1997, Apell:1981}:
\begin{equation}\label{eq:Christensen:Eq1}
    d_\perp(\omega) = \frac{\int x \rho(x,\omega) \,\mathrm{d}{x}}{\int \rho(x,\omega) \,\mathrm{d}{x}},
    \quad\qquad
    d_\parallel(\omega) = \frac{\int x \partial_x J_y(x,\omega) \,\mathrm{d}{x}}{\int \partial_x J_y(x,\omega) \,\mathrm{d}{x}}.
\end{equation}

As centroids of induced quantities, $d_\perp$ and $d_\parallel$ have units of length with complex amplitudes in the {\AA}ngstr{\"o}m-range for typical plasmonic materials.
They represent surface-response functions of the interface (i.e., intrinsic functions of the material composition across the interface) that describe the tendency of interfaces to polarize, analogously to how the bulk permittivity $\varepsilon$ describes the tendency of a bulk material to polarize.
The associated surface polarization is~\cite{Yang:2019a}:
\begin{equation}\label{eq:Christensen-Eq2}
    \boldsymbol{\pi} 
    =
    \varepsilon_0 d_\perp
    \llbracket E_\perp \rrbracket\hat{\mathbf{n}}
    -
    \varepsilon_0 d_\parallel \llbracket \mathbf{D}_\parallel/\varepsilon_0 \rrbracket,
\end{equation}
with $\llbracket \mathbf{f} \rrbracket \equiv \mathbf{f}(\mathbf{r} + 0^+\hat{\mathbf{n}}) - \mathbf{f}(\mathbf{r} + 0^-\hat{\mathbf{n}})$ denoting the discontinuity of a field $\mathbf{f}$ across interface with normal $\hat{\mathbf{n}}$ and with $f_\perp = \hat{\mathbf{n}}\cdot \mathbf{f}$ and $\mathbf{f}_\parallel = (\mathbf{I}-\hat{\mathbf{n}}\hat{\mathbf{n}}^{\mathrm{T}})\mathbf{f}$ denoting the field's normal and tangential parts.
Unlike the bulk polarization, sourced by the $\mathbf{E}$-field \emph{amplitude}, the surface polarization is sourced by the interface-\emph{discontinuities} $\llbracket E_\perp\rrbracket$ and $\llbracket \mathbf{D}_\perp\rrbracket$.
Compellingly, the corrections to CEM are thus driven by CEM's failure to achieve field continuity.

While Feibelman's original formulation focused only on  planar interfaces, the surface polarization formulation---and the recognition that any surface-curvature dependence of $d_{\perp,\parallel}$ is negligible for leading-order beyond-CEM considerations~\cite{Apell:1982, Christensen:2017}---generalizes the scope to arbitrary photonic structures~\cite{Yan:2015b, Yang:2019a}.
Moreover, as discovered three and a half decades after the Feibelman parameters' introduction, the surface polarization can in fact be abstracted away through a modification of the CEM boundary conditions~\cite{Yang:2019a}:
\begin{subequations}\label{eq:Christensen-Eq3}
\begin{alignat}{3}
    &\llbracket D_\perp \rrbracket
    = 
    d_\parallel \bm{\nabla}_\parallel \cdot \llbracket \mathbf{D}_\parallel \rrbracket,&
    \qquad\qquad
    &\llbracket B_\perp \rrbracket
    =
    0,&
    \\
    &\llbracket \mathbf{E}_\parallel \rrbracket
    =
    - d_\perp {\nabla}_\parallel \llbracket E_\perp \rrbracket,& 
    \qquad\qquad
    &\llbracket \mathbf{H}_\parallel \rrbracket
    =
    I \omega d_\parallel \llbracket \mathbf{D}_\parallel \rrbracket\times \hat{\mathbf{n}}.&
\end{alignat}
\end{subequations}

While Eqs.~\eqref{eq:Christensen-Eq3} are amenable to direct incorporation in full-wave computational frameworks, the general smallness of beyond-CEM corrections naturally suggests perturbative approaches.
Such approaches have been pursued in both nonretarded~\cite{Christensen:2017} and retarded~\cite{Yang:2019a} settings, suitable for deep-subwavelength resonators and multi-scale structures respectively.
In the latter setting, a quasi-normal mode perturbation theory~\cite{Lalanne:2018} gives the leading-order shift $\omega_1$ of the complex resonance frequency $\omega = \omega_0 + \omega_1 + \ldots$ from the CEM frequency $\omega_0$~\cite{Yang:2019a}:
\begin{equation}\label{eq:Christensen-Eq4}
    \omega_1
    =
    \omega_0(\kappa_\perp d_\perp + \kappa_\parallel d_\parallel),
\end{equation}
with mode-, shape- and scale-dependent perturbation amplitudes $\kappa_{\perp,\parallel}$ (inverse length units), obtainable from CEM quasi-normal modes, and with $d_{\perp,\parallel}$ evaluated at $\omega_0$. 
The perturbative expression cleanly separates the mode-, shape-, and scale-dependence ($\kappa_{\perp,\parallel}$) from the material-dependence ($d_{\perp,\parallel}$) of beyond-CEM shifts.
In the nonretarded limit and for a pure plasma (plasma frequency $\omega_p$) adjacent to vacuum, analytical expressions can e.g., be obtained for the resonance frequency $\omega^{\textsc{sp}}_{k}$ of a surface plasmon wave vector $k$ at a planar interface ($\kappa_\perp = -\kappa_\parallel = -\tfrac{1}{2}k$) and for the $l$th order localized surface plasmon frequency $\omega^{\textsc{lsp}}_{l}$ of a nanosphere of radius $R$ ($\kappa_\perp = -\kappa_\parallel = -\tfrac{l+1}{2}R^{-1}$)~\cite{Liebsch:1997, Christensen:2017, Goncalves:2020}:
\begin{equation}
    \omega^{\textsc{sp}}_{k}
    =
    \tfrac{1}{\sqrt{2}}\omega_{\mathrm{p}}
    \big(1 - \tfrac{1}{2}d_{\text{eff}}k\big),
    \qquad
    \omega^{\textsc{lsp}}_{l}
    =
    \tfrac{1}{\sqrt{2+l^{-1}}}\omega_{\mathrm{p}}
    \big(
    1
    -\tfrac{l+1}{2}d_{\text{eff}}R^{-1}
    \big),
\end{equation}
with $d_{\text{eff}} = d_\perp-d_\parallel$ evaluated at $\omega_0$.
The incorporation of Feibelman parameters lifts the characteristic scale-invariance of the nonretarded CEM by introducing the length-scale of $d_{\perp,\parallel}$ to compete with the geometric or modal length-scale $L$, resulting in corrections of order $\mathcal{O}(d_{\text{eff}} L^{-1})$.
These simple examples also reveal the physical interpretations of the complex parts of $d_{\text{eff}}$:
\emph{i)}~$\Re d_{\text{eff}}$ contributes to a shift of the frequency center,
\emph{ii)}~the sign of $\Re d_{\text{eff}}$ determines the shift’s direction (blue-shifting for charge "spill-in", $\Re d_{\text{eff}}<0$, and red-shifting for "spill-out", $\Re d_{\text{eff}}>0$), and
\emph{iii)}~$\Im d_{\text{eff}}$ contributes size- or momentum-dependent broadening.

\subsection*{Challenges and opportunities}

While the introduction Feibelman parameters now lies more 4 decades in the past, the framework's application to modern nanoplasmonic-response questions remains in an early stage and several open, unsolved challenges remain.
Below, we highlight opportunities and challenges in experimental and theoretical determination of the Feibelman $d$-parameters.

\textbf{Experimental measurement and tabulation.} A key strength of the Feibelman parameters is that they provide a physics-agnostic framework for incorporating beyond-CEM effects.
No specific physical effects are included or excluded: the physics in the underlying microscopic response quantities $\rho$ and $\mathbf{J}$ -- equivalently, $\boldsymbol{\varepsilon}(\mathbf{r},\mathbf{r}')$ -- on which $d_{\perp,\parallel}$ depends is automatically incorporated.
Specific models of $d_{\perp,\parallel}$ can therefore usually encompass (to leading order) existing "single-effect" approaches, including e.g., hydrodynamic accounts of nonlocality~\cite{Feibelman:1982}.
This freedom and versatility, however, brings a concomitant challenge: the values and frequency-dependence of $d_{\perp,\parallel}(\omega)$ are not prescribed by the framework itself.
For real materials, they must be obtained by experimental measurements or other sophisticated theoretical modeling.

An analogous problem arises for the bulk permittivity $\varepsilon(\omega)$, which has been met extensive experimental tabulation efforts~\cite{Johnson:1972}.
By comparing observations of large beyond-CEM deviations in AlO\textsubscript{$x$}--Au thin-film-gapped Au-disks [Fig.~\ref{fig:Christensen-Fig1}\pnl{b,c}], Yang~\emph{et al.}~\cite{Yang:2019a} demonstrated that equivalent efforts are feasible for the Feibelman parameters, by inferring $d_\perp$ of an Au--AlO\textsubscript{$x$} interface by comparison with perturbation theory [Fig.~\ref{fig:Christensen-Fig1}\pnl{d}].
Near-field measurements of highly confined thin-film gap surface plasmons at an Au--Al\textsubscript{2}O\textsubscript{3} interface~\cite{Boroviks:2022} demonstrate the feasibility more broadly, as do measurements of beyond-CEM deviations in graphene--metal heterostructures~\cite{Lundeberg:2017,Goncalves:2021}.

While feasible, these approaches often require multiple parametrically varied geometries and are intrinsically limited to the frequency range of the resonant plasmonic feature.
Measurements of the Feibelman parameters across broad frequency ranges is significantly more challenging, especially above the screened plasma frequency, contending with increased smallness of beyond-CEM shifts away from resonance and worsened signal-to-noise ratios.
Recent efforts have made strides towards broadband characterizations [Fig.~\ref{fig:Christensen-Fig1}\pnl{e,f}]~\cite{Chen:2024b}, by incorporating the Feibelman parameters into ellipsometric analysis via Feibelman-corrected Fresnel reflection coefficients and, e.g., demonstrating that while Au spills in at typical surface plasmon frequencies ($\Re d_\perp < 0$), it spills out above the plasma frequency ($\Re d_\perp < 0$).

\textbf{Theory and first-principles modeling.} State-of-the-art computational modeling of the Feibelman parameters arguably remains closely aligned with techniques first developed and applied in the 1980s~\cite{Feibelman:1982, Liebsch:1995}.
These techniques leverage time-dependent density functional theory (TD-DFT), accounting for the dynamics of the highest-lying conduction electrons and treating the remaining electrons through an effective jellium, i.e., uniform potential (see Sec.~\ref{sec:Aizpurua} of this Roadmap).
This captures the essential physics of simple metals well, since they exhibit only negligible screening from lower-lying electrons, and correctly predicts  spill-out ($\Re d_\perp > 0$) and beyond-CEM red-shifts ($\omega_1 < 0$), consistently with measurements~\cite{Tsuei:1989}.
Noble metals, however, are strongly screened by bound electrons, and exhibit beyond-CEM blue-shifting ($\omega_1 > 0$) associated with spill-in ($\Re d_\perp < 0$), contrary to jellium predictions.
Even semi-classically screened generalizations~\cite{Liebsch:1995, Christensen:2017} do not substantially remedy this mismatch.
These facts call for beyond-jellium calculations of the Feibelman parameters, explicitly incorporating both bound electrons and ionic potentials, e.g., by leveraging modern TD-DFT software.

First steps have been made through the incorporation of non-uniform atomic potentials which has e.g., shed light on the influence of crystallographic facet terminations~\cite{RodriguezEcharri:2021a}.
Many important questions remain unaddressed, however, including:
\begin{enumerate}
    \item Influence of dielectric screening for metal--dielectric interfaces~\cite{Jin:2015}.
    From semiclassical considerations, $d_\perp$ is expected to scale roughly inversely dielectric permittivity for noble-metal--dielectric heterostructures~\cite{Yang:2019a}.
    No similar expectation approximate rules arise for simple metals, however, since their free electrons spill into the dielectric cladding, and a general, predictive, and accurate modeling scheme has not been developed.
    \item Impact of residual chemical elements or few-layer intercalates at interfaces, including the wetting layers that form under ambient conditions.
    Their potential influence is crucial for the interpretation and assignment of experimentally obtained Feibelman parameters.
    \item While $d_\parallel$ vanishes in charge-neutral jellium calculations~\cite{Apell:1981} and has been assumed small in experimental treatments, the presence of bound-electron screening~\cite{Liebsch:1995}, surface roughness~\cite{Apell:1984}, and electronic interface states~\cite{Feibelman:1993, RodriguezEcharri:2021a} can lift this requirement.
    An understanding of the relative amplitudes of $d_\parallel$ and $d_\perp$ in general settings  remains absent.
    \item Tunability of $d_{\perp,\parallel}$ by external means, e.g., applied voltages or strain.
    Recent experiments by Zurak \emph{et al.}~\cite{Zurak:2024} used differential scattering response under voltage gating of a plasmonic nanoresonator [Fig.~\ref{fig:Christensen-Fig1}\pnl{g}] to estimate the associated tunability of the Feibelman parameters, finding sensitivities on the order of $10^{-1}\,\text{pm}/\text{V}$ for $d_\perp$ at Au--air interfaces [Fig.~\ref{fig:Christensen-Fig1}\pnl{h}].
    A full theoretical understanding of these experimental results as well as whether larger tunability is attainable by other external controls are exciting open questions.
    \item Jellium TD-DFT's failure to model $\Re d_{\perp,\parallel}$ (frequency shift) also limits its ability to model $\Im d_{\perp,\parallel}$ (frequency broadening) due to Kramers--Kronig relations.
    The impact of bound-electron screening on beyond-CEM broadening is thus still poorly understood.
    One open question is whether some plasmonic materials might be naturally "quantum low-loss" (i.e., small $\Im d_{\perp,\parallel}$) or "quantum high-loss" (i.e., large $\Im d_{\perp,\parallel}$), which may constrain outlooks for low-loss ultra-confined plasmons and hot-electron generation, respectively.
    \item The electron--phonon coupling and phonon-assisted transitions play a significant role in the bulk losses of materials~\cite{Brown:2015}.
    Surprisingly little, however, is known about the electron--phonon coupling's impact on surface damping. 
    E.g., the contribution to $\Im d_\perp$ from phonon-assisted transitions is not currently understood, nor is its relative magnitude compared to intraband electron--hole pair transitions (surface-enabled Landau damping).
\end{enumerate}
Answering these questions will likely necessitate new computational and algorithmic developments for TD-DFT approaches of thin-film and interfacial optical response.

\vskip 2ex
In both theoretical and experimental pursuits of the Feibelman parameters' values, two facts are helpful to validate obtained results and ensure physical consistency.
First, since the Feibelman parameters are response functions, they are causal and consequently obey Kramers--Kronig relations, which equips them with several useful sum rules~\cite{Persson:1983}.
Second, passivity---i.e., that linear passive materials do no net work on the electromagnetic fields---also constrains the Feibelman parameters, just as it constrains the local permittivity by $\Im(\omega \varepsilon)\ge 0$.
In particular, passivity requires that $\Im (\omega d_\perp \llbracket \varepsilon^{-1} \rrbracket) \ge 0$ and  $\Im (\omega d_\parallel \llbracket \varepsilon \rrbracket) \le 0$ for all frequencies ($\Im\omega > 0$)~\cite{Chen:2024b}.

\subsection*{Future developments to address challenges}

While the noted opportunities largely fall within Feibelman's original scope, i.e., anchored in plasmonics, future developments may well be driven most efficiently by pursuing an expanded scope and generalizations of the underlying ideas of surface-response functions.

One particularly apt and simple example of such possible generalizations of scope is doped semiconductors, which are excellent candidates for realizing large beyond-CEM corrections due to long Fermi wavelengths and pronounced nonlocal response~\cite{Maack:2017}.
As are generalizations to synthetically nonlocal meta-materials.
In particular, we anticipate that both phenomenological and semiclassical modeling of the Feibelman parameters may be sufficient to match experiments, since these materials afford a clearer separation between the involved beyond-CEM length-scales than metals.
In turn, this may inform and revitalize phenomenological modeling approaches for Feibelman parameters of metals.

More broadly, several other material and quasiparticle platforms support ultra-confined optical modes, which may correspondingly host substantial beyond-CEM corrections.
Polar materials such as hexagonal boron nitride, for instance, supports highly-confined surface phonon polaritons~\cite{Dai:2014}.
Even ordinary high-index dielectrics, such as silicon, can support ultra-confined, nanometer-scale optical modes in the gaps of bowtie nano-antennae~\cite{Albrechtsen:2022}.
Lacking mobile charge carriers, the physical mechanisms for beyond-CEM effects must involve phonons in either material platform, rather than electrons alone: accordingly, the momentum-dependence of the electron--phonon coupling appears likely to play a key role.
Importantly, regardless of the underlying physical mechanism, the Feibelman-parameter framework will be applicable due to its physics-agnostic formulation.
Recent observations of super-thermal evaporation at water surfaces~\cite{Chen:2023, Chen:2024a}, suggest that even broader material classes may exhibit beyond-CEM effects associated with nonzero Feibelman parameters, with physical mechanisms that are wholly unrelated to the Fermi liquid perspective of plasmons.
Even the classical effects of surface roughness can be abstracted into the framework~\cite{Apell:1984}, a fact that is important to hold in view for experimental inference of Feibelman parameters.

The existence of nonlocality, even the mere existence of an interface, is also known to enable nonlinear effects otherwise forbidden in a local-response bulk.
This includes second-harmonic generation in centrosymmetric materials.
Generalizations of Feibelman parameters to the nonlinear domain, e.g., for capturing hydrodynamic accounts of nonlinearity~\cite{Ginzburg:2014}, would consequently be of substantial interest for describing surface-enabled nonlinear phenomena.
The fact that the second-harmonic nonlinear susceptibility of an interface contains terms proportional to the interface-discontinuity of the electric field~\cite{Guyot-Sionnest:1986} is highly encouraging in this regard.

Taken together, these outlooks suggest the possibility of more far-reaching generalizations to physical problems beyond electromagnetism.
At their core, the Feibelman parameters amount to a leading-order remedy of the field discontinuity imposed by the simple but unphysical assumptions of piecewise constant, and local response.
Such assumptions are widespread in physics, spanning e.g., heat transport, acoustics, and semiconductor physics.
These facts suggest that many notions of surface- and interface-response functions might find generalization and unification in the ideas underlying the Feibelman parameters.

\subsection*{Concluding remarks}

The Feibelman parameters provide a natural and elegant framework for incorporating the salient beyond-classical effects in electromagnetic modeling. 
While the physics of Feibelman parameters for jellium plasma have been well-understood for decades, many questions remain open for the real materials of interest to contemporary polaritonics, e.g., metals, doped semiconductors, polar materials, and high-index dielectrics.
In addition to these questions, there are broad opportunities for generalizations and extensions of the underlying ideas to physics beyond linear electromagnetism.
We hope this overview provides an interesting and illustrative selection of such questions and opportunities, and that it may serve to inspire future contributions to the field.

\section[Computational electrodynamics with Feibelman surface-response formalism\\ (Hohenester~\emph{et al.})]{Computational electrodynamics with Feibelman surface-response formalism}

\label{sec:Hohenester}

\author{Ulrich Hohenester\,\orcidlink{0000-0001-8929-2086}, Lorenz Huber\,\orcidlink{0009-0000-8321-4896} \& Javier~Aizpurua\,\orcidlink{0000-0002-1444-7589}}

\subsection*{Current status}

Metallic nanoparticles can focus light down to extreme sub-wavelength volumes through excitation of surface plasmons, these are coherent surface charge oscillations of electrons at the metal surface~\cite{Hohenester:2020}. The question of how to account for quantum surface effects around sharp particle features, small gap regions, or for incoming fields with high spatial variations such as those produced by quantum emitters in the vicinity of a particle, has puzzled researchers for several decades. A particularly appealing solution is due to Feibelman~\cite{Feibelman:1982} (see also Sec.~\ref{sec:Christensen} of this Roadmap), who devised an approach where all quantum modifications at the boundary are lumped into two dispersive so-called Feibelman parameters, which provide a measure of the distance over which the electric fields deviate from those for a sharp interface. Feibelman parameters were first brought to the field of plasmonics in~\cite{Teperik:2013} for the explanation of confinement-induced plasmon shifts, and further developed to include surface-induced nonlocal effects in the optical response of arbitrarily-shaped plasmonic nanoparticles~\cite{Yan:2015b}. More recently~\cite{Yang:2019a,Goncalves:2020}, it has been suggested that Feibelman parameters can be directly incorporated into classical Maxwell solvers by introducing modified "mesoscopic" boundary conditions. By comparing results from modified Mie theory including Feibelman parameters with those of \emph{ab initio} simulations based on time-dependent density-functional theory (TDDFT)~\cite{Babaze:2023}, it has been shown that for electromagnetic field variations on the nanometer scale one should rather consider Feibelman parameters that are nonlocal in the directions tangential to the boundary, in accordance to the original work of Feibelman who considered for a planar slab geometry a parameter dependence on the wavevector components parallel to the interface.
Despite the strong recent interest in Feibelman parameters and mesoscopic boundary conditions, there are surprisingly few attempts to implement them into computational Maxwell solvers. In the seminal paper of~\cite{Yang:2019a} the authors proposed two methods for dealing with such boundary conditions in the COMSOL Multiphysics software based on the finite-element method (FEM): either by introducing an additional scalar auxiliary potential together with a surface current, or by computing the quasinormal modes of the system under study and accounting for the modifications due to the Feibelman parameters in lowest order of perturbation theory. Both approaches lead to simulations that are slower in comparison to those using normal boundary conditions.
As an alternative approach, in Refs.~\cite{Hohenester:2022a,Huber:2025} we have developed the methodology for directly introducing mesoscopic boundary conditions into a Maxwell solver based on the boundary element method (BEM), using either local or nonlocal Feibelman parameters, and have implemented them in our software toolbox nanobem~\cite{Hohenester:2022b}. Agreement with analytic results based on a modified Mie theory has been demonstrated for single and coupled spheres, together with the expected computational convergence when increasing the number of boundary elements. Further developments are needed to bring our software to a form that can be used flexibly by other users, and to set up a database for Feibelman parameters for various material combinations.

\begin{figure}[htb]
    \centering
    \includegraphics[width=0.99\linewidth]{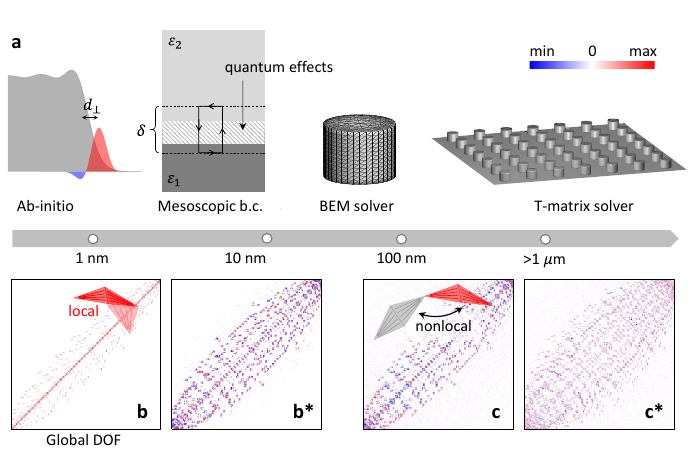}
    \caption{Schematics of BEM simulations including Feibelman parameters. \pnl{a}~We start by computing local or nonlocal Feibelman parameters within an \emph{ab initio} approach, which can be submitted to the mesoscopic boundary conditions and implemented in a BEM solver. The latter provides us with $T$-matrices that could be used for the simulation of particle clusters or particle arrays. \pnl{b,c}~The lower panels show the matrices used in our BEM approach for matching the electromagnetic fields at the boundary, with a representation of the tangential electromagnetic fields in terms of Raviart--Thomas shape elements (see inset) as global degrees of freedom (DOF). \pnl{b}~For local Feibelman parameters the matrix is spare, however, due to the technical details described in the text and in Refs.~\cite{Hohenester:2022a,Huber:2025}, the \pnl{b*} matrix actually used in the BEM approach is already relatively dense. \pnl{c,c*}~For nonlocal Feibelman parameters the matching matrices are equally dense. As the inversion of the matching matrices is computationally expensive, the performance of our BEM approach is not too different for local and nonlocal Feibelman parameters.}
    \label{fig:Hohenester-Fig1}
\end{figure}

\subsection*{Challenges and opportunities}

For computational Maxwell solvers incorporating local or nonlocal Feibelman parameters, the main challenges concern the methodology and the numerical implementation. To understand the challenges, we have to briefly dwell on some of the technicalities. In BEM, the Calderon identities, see Eq.~(9.55) of Ref.~\cite{Hohenester:2020}, relate the tangential electromagnetic fields at the particle boundary to the incoming fields, for instance those produced by plane waves or quantum emitters placed in the vicinity of the particle. Together with the boundary conditions, one can use the identities to compute the unknown tangential fields and, in turn, the fields everywhere else. When considering Feibelman parameters, one can start from the same Calderon identities, and only the field matching at the boundary has to be modified. Here one has to be careful about the tangential derivative $\nabla_\parallel (d_\perp \circ \llbracket E_\perp \rrbracket)$ in the mesoscopic boundary conditions~\cite{Yang:2019a}, where the electric field component normal to the boundary $E_\perp$ additionally has to be related in BEM to the tangential magnetic field $H_\parallel$ through Ampère’s law. This leaves us with two tangential derivatives in the mesoscopic boundary condition. Fortunately, they can be handled in the same manner as the hypersingular operator in the Galerkin implementation of BEM, see Eq. (11.40) of Ref.~\cite{Hohenester:2020}, namely by swapping the derivatives through integration by parts to the Raviart--Thomas shape elements $f_\nu(r_\parallel)$ that are used for the representation of the tangential electromagnetic fields at the boundary. Furthermore, the derivative $\nabla\cdot \hat{n}\times f_\nu(r_\parallel)$ originating from this partial integration, with $\hat{n}$ being the outer surface normal, has to be related to $\nabla_\parallel \cdot f_\nu(r_\parallel)$ using a suitable translation procedure~\cite{Hohenester:2022a}. The important point for our discussion here is that through this translation not only neighbor but also distant shape elements become directly coupled when matching the tangential fields at the boundary, even for local Feibelman parameters, as can be seen in Fig.~\ref{fig:Hohenester-Fig1}\pnl{b,b*} for the matching matrix before and after translation. Quite generally, dense matrices play a central role in BEM~\cite{Hohenester:2020} and the inversion of an additional matrix constitutes no major bottleneck in the computational approach. The same conclusion holds for nonlocal Feibelman parameters, where from the outset distant shape elements are coupled, see Fig.~\ref{fig:Hohenester-Fig1}\pnl{c}, and the translation procedure leads to matrices that are equally dense than those for local Feibelman parameters. With the exception of the filling of these matrices, the consideration of nonlocal Feibelman parameters thus represents no computational drawback.
The implementation of local and nonlocal Feibelman parameters to solve Maxwell’s equations within the BEM requires the availability of a reliable set of these parameters associated with the boundaries separating two different media, which can be obtained from a quantum calculation of the optical response of a flat interface separating the same pair of media. To that end, when it comes to describe an interface between a dielectric and a free-electron gas of s-like electrons, such as in sodium (Na) or aluminum (Al), a TDDFT where the metallic medium is considered to be a jellium turns to be very adequate to describe the surface response~\cite{Babaze:2022}. Interfaces involving noble metals such as gold (Au) or silver (Ag), with a strong effect of d-like electrons on the optical response require more sophisticated approaches.

\subsection*{Future developments to address challenges}

In the future, the computation of Feibelman parameters and the development of Maxwell solvers incorporating mesoscopic boundary conditions should go hand in hand. Only in combination they provide a powerful machinery for considering quantum surface effects in classical Maxwell solvers, and can serve the community as a flexible simulation toolkit.
The current BEM implementations of the mesoscopic boundary conditions~\cite{Hohenester:2022a,Huber:2025} are somewhat experimental, and are neither optimized for flexibility nor speed. In contrast to state-of-the-art BEM solvers, such as nanobem~\cite{Hohenester:2022b}, the additional computation of the matrices for field matching must properly account for the novel features of a spatial cutoff for the nonlocal Feibelman parameters $d(r_\parallel)$ and a sufficiently fine integration grid to resolve the spatial variations of $d(r_\parallel)$. A first step in this direction has been undertaken in Ref.~\cite{Huber:2025}, where we have suggested a separation of the Feibelman parameters into a local contribution and a remainder, accounting for the most salient short- and long-range features. Feibelman parameters are expected to be of importance for sharp particle features, gap structures, or quantum emitters nearby particles, thus calling for a boundary discretization that is fine in the vicinity of the critical points and coarser further away. This requires refined integration routines that can efficiently deal with boundary elements of unequal size. Since the consideration of the boundary matching matrices is computationally equally expensive for local and nonlocal Feibelman parameters, one must assure that the filling of these matrices does not become the slowest part of the BEM simulations. nanobem is a solver particularly suited for small-scale problems. However, with the recent developments of the toolbox it is now possible to compute $T$-matrices, which can incorporate the Feibelman parameters, and to directly load them into $T$-matrix solvers such as SMUTHI or treams. With this it is possible to simulate clusters, arrays, or metasurfaces.
Mesoscopic boundary conditions can be also implemented in other Maxwell solvers. The approach developed in Refs.~\cite{Hohenester:2022a,Huber:2025} can be probably adapted without too many modifications for finite element method (FEM) solvers using a Galerkin scheme. As for other solvers, such as those based on the finite-difference time domain (FDTD) or discrete-dipole approximation (DDA) method, it has to clarified how the spatial derivatives in the tangential directions can be handled.
Regarding the challenges in the calculation of Feibelman parameters, extensive efforts have been invested in developing jellium TDDFT simulations to obtain local and nonlocal parameters of interfaces involving a purely free electron gas, however, the implementation of such a nonlocal scheme of parameters for noble metal crystallographic interfaces is also a need in nowadays nano-plasmonics. Gold and silver interfaces, for instance, are commonly used in practical optical spectroscopy implementations, where strong band-structure effects are in place induced by the d-band of more localized-electrons, and a strong presence of electronic image states also modifies the spectral dependence of the parameters at lower energies. All of these effects enforce the implementation of atomistic \emph{ab initio} calculations of the surface response in periodic supercells, capable to address these issues.

\subsection*{Concluding remarks}

Mesoscopic boundary conditions incorporating Feibelman parameters have been recently introduced as a versatile tool to account for quantum surface effects at metal-dielectric interfaces. However, computational Maxwell solvers incorporating Feibelman parameters are still scarce. One exception is the boundary element method approach, for which Feibelman parameters that are either local or nonlocal in the directions tangential to the interface have been implemented successfully. In the future, we plan to render our software suitable for an easy-to-use and flexible open-source simulation toolbox, and to supplement it with Feibelman parameters computed with state-of-the-art \emph{ab initio} Schrödinger solvers for various material systems.

\usection{\emph{Part III} --- Nonlocal effects beyond free-electron bulk metals}
\label{roadmap:Part3}

\section[Nonlocal response in doped semiconductors (Wubs)]{Nonlocal response in doped semiconductors}

\label{sec:Wubs}

\author{Martijn Wubs\,\orcidlink{0000-0002-8286-7825}}

\subsection*{Current status}
Doped semiconductors are a vast and active research field, see Ref.~\cite{Guo:2023} for a recent review. It is textbook knowledge that the Drude model~\cite{Drude:1900} can describe free-electron systems, not only metals but also doped semiconductors. In doped semiconductors the response can be due to combinations of electrons, and light and heavy holes. Free electrons typically have much smaller reduced masses than holes, so they dominate the response if present. Electron reduced masses $m^{*}$ of small-bandgap semiconductors are smaller than in metals. Free-carrier densities in doped semiconductors are much smaller than in metals. The net effect is that semiconductor plasma frequencies $\omega_p = \sqrt{n e^2/(\varepsilon_0 m^{*})}$ occur in the infrared to THz regimes, compared to the ultraviolet frequencies for metals.

For nanostructured metals, it is known that the Drude model ceases being accurate because of nonlocal response and because of "spill-out" across geometric boundaries  [see Sec.~\ref{sec:Fernandez-Dominguez} of this Roadmap]. Typical sizes (particle radii, thin-film thicknesses) should be well below ten nanometers to see nonlocal effects in metals. Signatures of nonlocal response in metallic nanoparticles are the spectral blueshift as particles get smaller, and longitudinal resonances above the plasma frequency arising in confined geometries, and hydrodynamic surface nonlinearities. It is therefore interesting whether similar nonlocal effects occur in semiconductors.

\begin{figure}[hb!]
\centering\includegraphics[width=0.99\columnwidth]{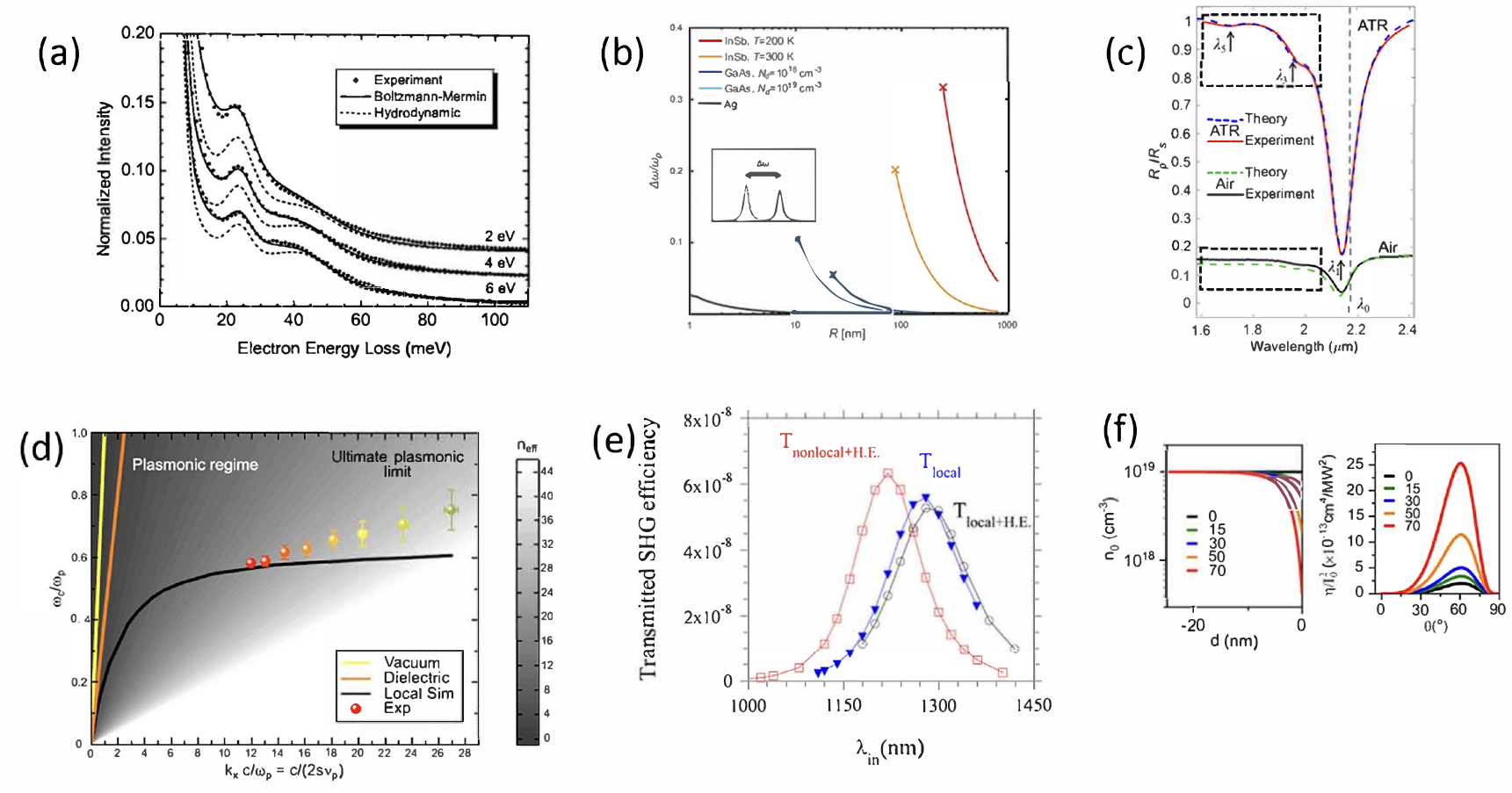}
\caption{Examples of theory and experiments on nonlocal response in semiconductors. \pnl{a} Comparison of hydrodynamic and Boltzmann--Mermin theories with electron energy loss spectroscopy experiments on InSb. Reproduced with permission from Ref.~\cite{Bell:2006} (Copyright~\textcopyright~2006 American Physical Society). \pnl{b} Prediction of large relative nonlocal blueshifts of nanospheres made of doped GaAs or thermally excited InSb, as compared to Ag. Reproduced with permission from Ref.~\cite{Maack:2017} (Copyright~\textcopyright~2017 Europhysics Letters Association). \pnl{c} Observation of nonlocal blueshift and additional thickness-dependent resonances above the bulk plasma frequency in an In:CdO thin film (here: thickness 20\,nm). Reproduced with permission from Ref.~\cite{DeCeglia:2018} (Copyright~\textcopyright~2018 Nature Springer). \pnl{d} Plasmonic dispersion of compact and tunable InSb-based THz surface plasmon cavities. Data points, obtained by thermal tuning, deviate from the local-response curve. Reproduced with permission from Ref.~\cite{Aupiais:2023} (Copyright~\textcopyright~2023 Nature Springer). \pnl{e} Nonlocal effects and increased electron gas temperature tend to pull the transmitted SHG efficiency resonance(s) in opposite directions. Reproduced with permission from Ref.~\cite{Rodriguez-Sune:2020} (Copyright~\textcopyright~2020 AIP Publishing). \pnl{f}~Surface charge depletion and corresponding free-electron third-harmonic efficiency $\eta$ for various static external fields (in V/$\upmu$m). Reproduced with permission from Ref.~\cite{DeLuca:2022a} (Copyright~\textcopyright~2022 American Physical Society).
} 
\label{Fig:Wubs-Fig1}
\end{figure}

In Ref.~\cite{Zhang:2014b}, few-nm sized doped semiconductor nanocrystals are studied in a range of doping concentrations, showing features ranging from size-quantization to classical plasmonic behavior. In Ref.~\cite{Maack:2017} it was argued that the hydrodynamic Drude model (HDM) should apply and nonlocal behavior should be observable in larger doped semiconductor particles, of tens of nanometers in size (considerably larger than for metals), described by the longitudinal dielectric function $\varepsilon(\omega,k) = \varepsilon_{\infty} - \omega_{p}^{2}/[\omega^2 + i \gamma \omega - \beta^2 k^2]$. The surface plasmon resonance of a nanosphere in air then has a nonlocal blueshift of $\Delta\omega = \beta/(\sqrt{2} R)$. In small-bandgap semiconductors [such as indium antimonide (InSb) with $E_g = 0.17$\,eV] electrons are thermally excited to the conduction band, and at low temperatures the nonlocal parameter becomes $\beta = 3 k_{\rm B}T/m_{e}^{*}$~\cite{Bell:2006,Maack:2017}. For $n$-doped semiconductors on the other hand, the $\beta$ is given by $\sqrt{3/5}(\hbar/m^{*})(3 \pi^2 n)^{1/3}$, where $n$ is the electron density~\cite{Bell:2006,Maack:2017}. Thus the same hydrodynamic Drude response function can describe microscopically quite different situations. In both cases, exciton effects can be neglected because of large screening. And in both cases, the relative nonlocal blueshift $\Delta\omega/\omega_p$ for semiconductors -- doped gallium arsenide (GaAs) or thermally excited InSb -- is much larger than for metals~\cite{Maack:2017}, e.g., silver (Ag), see Fig.~\ref{Fig:Wubs-Fig1}\pnl{b}. For the doped semiconductor, $\Delta\omega/\omega_p \propto n^{-1/6} (m^{*})^{-1/2}/R$, which shows that, interestingly, both the lower carrier densities and smaller reduced masses in semiconductors contribute to larger relative nonlocal blueshifts.

In linear optics, convincing hydrodynamic nonlocal behavior in semiconductors has been observed in planar geometries: infrared optical nonlocalities due to conduction electrons have been observed in indium (In)-doped cadmium oxide (CdO) thin films~\cite{DeCeglia:2018}, both a nonlocal blueshift of the main plasmon resonance and two additional longitudinal resonances at higher energies, for film thicknesses of 20\,nm, see Fig.~\ref{Fig:Wubs-Fig1}\pnl{c}. In a recent preprint, heavily n-doped indium arsenide antimonide (InAsSb) thin films are also reported to exhibit nonlocal resonances, even for film thicknesses up to 200\,nm~\cite{Moreau:2024}. Also, compact plasmonic-based THz cavities, with thermal activation of carriers across the small bandgap of InSb, have been shown to operate until the emergence of nonlocality and Landau damping~\cite{Aupiais:2023}, see Fig.~\ref{Fig:Wubs-Fig1}\pnl{d}.
 
Besides this progress in nonlocality in linear optics, there are interesting developments in the nonlinear response of doped semiconductors as well. Ref.~\cite{Rodriguez-Sune:2020} described both the second- and the third-harmonic response from an indium tin oxide (ITO) nanolayer, modeled as a combination of bulk and hydrodynamic nonlinear response, see Fig.~\ref{Fig:Wubs-Fig1}\pnl{e}. Ref.~\cite{DeLuca:2022a} proposed to use surface modulation of the equilibrium charge density of heavily doped semiconductors via an external static potential $n_0$, see Fig.~\ref{Fig:Wubs-Fig1}\pnl{f}. It combines the insights that the hydrodynamic nonlinear response is a surface rather than a bulk response, and that the third-order nonlinear polarization is proportional to $1/n_{0}^2$, so that a reduction of $n_0$ at the surface in response to a static potential can make the nonlinear response orders of magnitude stronger. A recent preprint~\cite{Rosetti:2025} attributes the strong third-harmonic response of n-doped indium gallium arsenide (InGaAs) thin layers to free-electron nonlocal response, with measurements agreeing much better with hydrodynamic than with local-response theory.

\subsection*{Challenges and opportunities}

\textbf{Quantum-confined versus hydrodynamic nonlocality.} While for planar geometries, accurate agreement of hydrodynamic theory and experiment have been found~\cite{DeCeglia:2018,Moreau:2024}, for nanoparticles and for other geometries their local response has been harder to identify as being hydrodynamic. There can be challenges to make spherical particles, and a more fundamental challenge is to distinguish hydrodynamic from size-quantization effects that also produce blueshifts as sizes are reduced~\cite{Zhang:2014b}. And thus non-classical phenomena, as observed in Ref.~\cite{Schimpf:2014} for example, may be due to either. It is difficult to draw a hard boundary, but generally in larger systems, single-particle size quantization effects become less likely and before we reach larger sizes where classical plasmonics applies, the hydrodynamic Drude model is expected to apply. While "spill-out" on the atomic scale is less important for the larger doped semiconductor structures than for metals that exhibit nonlocal response, charge depletion and inhomogeneous doping levels are novel possible features -- and opportunities -- for the semiconductors. 

\textbf{Embedded eigenstates.} Recently, it was argued that taking nonlocal response into account does not destroy embedded eigenstates~\cite{Prudencio:2021}. On the contrary, nonlocal nanospheres feature infinitely many such states, and the nonlocality relaxes the size restrictions of local theories. These do not couple to light, but could be observed instead with electron beams. As nonlocal effects show up in larger systems in semiconductors, as discussed above, doped semiconductors may be the preferred system to observe nonlocal embedded eigenstates, probed by electrons. 
 
\textbf{Two-fluid hydrodynamic response.} The optical response of semiconductors may be produced by more than one type of charge carrier, for example by both heavy and light holes, each with their corresponding plasma frequency. In Ref.~\cite{Zhang:2017}, a hybridization theory for two-component carrier plasmas was proposed, with an antibonding mode where the two components move in phase depends sensitively on the doping densities, and the low-energy bonding mode with opposite charge alignment. The antibonding mode agreed well with experiment. A hydrodynamic two-fluid theory was also proposed for doped semiconductors~\cite{Maack:2018,Golestanizadeh:2019}. Observing the predicted low-energy hybridized hydrodynamic mode is still an open challenge. More generally, in semiconductors it can be a challenge that plasmon and phonon resonances have similar energies. 

\textbf{Complex nonlocal parameter.} In the standard hydrodynamic Drude model, the $\beta$ parameter is real-valued with $\beta^2 = 3 v_{F}^2/5$ and describes quantum pressure convection, while in the GNOR extension it is complex-valued~\cite{Mortensen:2014}, with $\Im\beta$ interpreted as representing induced charge diffusion and accounting for size-dependent broadening of arbitrary plasmonic structures, including semiconductors. In the theory for semiconductors of Ref.~\cite{DeCeglia:2018}, an imaginary part of $\beta$ is due to the viscosity of the electron gas. Likewise in the theory for semiconductors in Ref.~\cite{Moreau:2024}, a second viscosity term appears exactly as the imaginary part of $\beta^2$ in the standard hydrodynamic model. A complex $\beta$ is also obtained in Halevi's formula $\beta^2(\omega) = (3\omega/5+i\gamma/3)v_{F}^2 /(\omega+ i \gamma)$~\cite{Halevi:1995}, where $\beta$ becomes a running coupling constant interpolating between the $\omega \ll \gamma$ and $\omega \gg \gamma$ limits, see Refs.~\cite{Halevi:1995,Bell:2006,Wegner:2023}. Each of these microscopic mechanisms can be the cause of size-dependent damping, i.e. the commonly observed broadening of spectra for smaller plasmonic structures. Yet size-dependent damping could also be caused by surface roughness, or by a combination of these causes. This makes identifying the actual causes of damping in a given experiment challenging. For example, in the intriguing quest for signatures of viscous electron plasmas, as reviewed in Ref.~\cite{Polini:2020}, many causes will need to be suppressed or ruled out before size-dependent plasmonic damping can be uniquely attributed to viscosity of the electron gas, in semiconductor plasmonics and elsewhere.

\textbf{Beyond hydrodynamic theory.} The hydrodynamic Drude model, or the simple extension provided by the GNOR model~\cite{Mortensen:2014}, are robust in the sense that their nonlocal dielectric function accurately describes many optical experiments where nonlocality shows up~\cite{Moreau:2024,DeLuca:2022a}. A more general dispersion relation for plasmons would be
\begin{equation}
\omega^2 + i \omega \left( \gamma_0 + \sum_{n=1} a_{2n} k^{2n}\right) = \omega_{0}^{2} + \sum_{n=1} b_{2n} k^{2 n},
\end{equation}
with spatial dispersion both in the frequency and in the damping constants and with constants $a_{2n}$ and $b_{2n}$. This general form reduces to $\omega^2 = \omega_{p}^2 /\varepsilon_{\infty}$ for the local Drude model and to $\omega^2 + i \gamma \omega = \omega_{p}^2 /\varepsilon_{\infty} + \beta^2 k^2$ for the HDM and GNOR models, in each case found by putting the corresponding epsilon to zero. So in the hydrodynamic theory, both series in powers of $k^2$ truncate beyond $k^2$. One reason why higher-order terms would be masked in experiments is the damping, where the Drude damping plus quadratic damping $\propto k^2$ may be strong enough to broaden and flatten plasmonic spectra before higher-order dispersion comes in sight for larger $k$. By contrast, in an electron energy loss spectroscopy experiment on InSb~\cite{Bell:2006}, i.e. with electrons rather than light, the measurements did not agree well with the hydrodynamic Drude model, but remarkably well with a Boltzmann--Mermin model. So challenges are to explore this boundary between theories, to find out whether other electron energy loss spectroscopy (EELS) experiments can be found where the HDM is accurate, and vice versa to find examples in linear and nonlinear optics where the HDM ceases to be accurate.

\textbf{Applications.} It has been shown that plasmonic cavities of a few microns in size of low-carrier density InSb can operate in the nonlocal regime of THz plasmonics~\cite{Aupiais:2023}. These tunable cavities may become preferable to metallic ones, with useful modes of operation even in the nonlocal regime. The recent progress in nonlinear free-electron response of doped semiconductors~\cite{Rodriguez-Sune:2020, DeLuca:2022a,Rosetti:2025}, especially their tunable and efficient operation at low carrier densities, has put nonlocal hydrodynamic theory in the spotlight and may lead to novel schemes and compact semiconductor devices for second- and third-harmonic generation. Here further progress depends on material science and design based on demanding simulations of systems that combine both nonlocal and nonlinear response. 

\subsection*{Future developments to address challenges}

\textbf{Combine linear and nonlinear optics.} Evidence is mounting that the nonlinear second- and third-harmonic response of doped semiconductors can largely be attributed to free-carrier hydrodynamic response. This makes doped semiconductors interesting materials for tunable on-chip nonlinear devices that may lead to novel applications. On the fundamental side, as discussed above, the size-dependent linear optics effects of doped semiconductor nanoparticles have been explained either as due to quantum confinement or as hydrodynamic. In the future, combining nonlinear with linear measurements may help to identify the hydrodynamic nature (or not) of the linear response, by studying whether the nonlinear response shows a hydrodynamic nature.

\textbf{Comparing optical with conductivity measurements.} The relation between Drude conductivity and corresponding dielectric function is well known, and that dielectric functions can be studied both by optical and electron transport experiments. Some researchers in optics have attributed the imaginary part of the nonlocal parameter $\beta$ for semiconductors to viscosity of the electron gas~\cite{DeCeglia:2018,Moreau:2024}, while solid-state physicists claim that viscous electron transport has been elusive and was first observed in graphene, see Ref.~\cite{Polini:2020}. In the future, it would be interesting to do both electron transport and optical measurements on the same tunable plasmonic semiconductor systems, since the viscosity of the electron gas should affect both.

\textbf{Comparing optical with EELS measurements.} Above we discussed that the hydrodynamic Drude model could be used to explain nonlocal response in measurements on thin semiconductor films, while in some EELS measurements the HDM was inadequate, clearly not as accurate as a Boltzmann--Mermin model~\cite{Bell:2006}. It would be worthwhile to explore the limits of the HDM, by doing combinations of optical and EELS measurements on the same planar structures.

\subsection*{Concluding remarks}
Here we have given an overview of some predictions and observations of nonlocal effects in plasmonic semiconductors, both in linear and nonlinear optics, and listed some challenges, opportunities and suggestions for future research. 

\section[Nonlocality in polar dielectrics (De Liberato)]{Nonlocality in polar dielectrics} 

\label{sec:DeLiberato}

\author{Simone De Liberato\,\orcidlink{0000-0002-4851-2633}}

\subsection*{Current status}

\begin{figure}[hb!]
    \centering
    \includegraphics[width=0.9\linewidth]{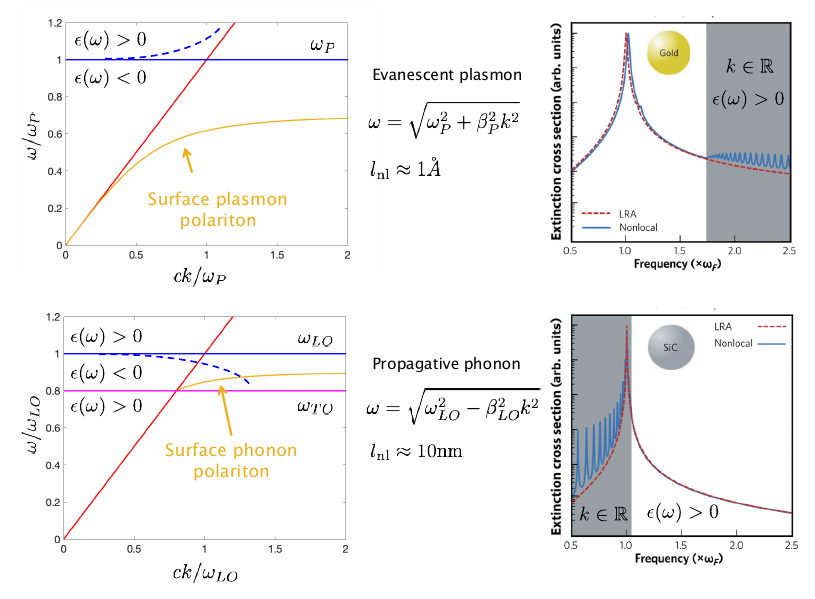}
    \caption{Comparison of nonlocality in plasmon-polariton (top) and phonon-polariton (bottom) systems. On the left column we show a sketch of the surface polariton dispersions (yellow solid line) with highlighted the regions in which the Drude (top) and Lorentz (bottom) dielectric functions change sign. The red solid lines are the light line and the horizontal lines mark the plasma frequency (P) and the LO and TO phonon frequencies. The dashed blue lines sketch the dispersions of the longitudinal modes, written in the quadratic approximation in the central column for each case. On the right column we show the Fr{\"o}hlich resonance, of frequency $\omega_F$, for a gold (top) and SiC (bottom) nanosphere, with the extinction cross section calculated using a local (red dashed line) and nonlocal (blue solid line) theory. The shaded region corresponds to the frequency region spanned by the dispersive mode for real values of the wavevector $k$. Only for phonon-polaritons, due to the red-dispersion of the LO phonon, such a region overlaps with the region of negative dielectric function in which the Fr{\"o}hlich resonance can exist, leading to resonant energy exchange and to a substantially larger nonlocal length scale $\ell_\mathrm{nl}$. Data adapted from Ref.~\cite{Gubbin:2020}.}
    \label{fig:Di-Liberato-Fig1}
\end{figure}

Phonon polaritons, hybrid modes formed by the strong coupling of electromagnetic fields with optical phonons in polar crystals, have emerged as a powerful platform for sub-diffractional mid-infrared (mid-IR) photonics. Unlike plasmons, which rely on free carrier oscillations and can suffer from high losses, phonon polaritons benefit from the intrinsic low-loss nature of lattice vibrations. As such, they have found application in a range of areas including sensing, nonlinear optics and thermal emitters~\cite{Gubbin:2022a}. The standard approach to modeling polar crystals relies on a local dielectric approximation, which implicitly assumes a negligible momentum dependence of the phonon response. This local treatment works well at larger feature sizes, as in bulk transverse optical (TO) and longitudinal optical (LO) phonons have well-defined frequencies at optical wavelengths, separated by the Reststrahlen band. However, at the nanoscale, extreme light confinement can probe the phonon dispersion well inside the Brillouin zone, phonons acquire a non-negligible wavevector dependence, and their propagation within a confined geometry cannot be neglected. Under these conditions nonlocal effects, where the material response at a given point depends on fields in a finite surrounding region, become significant. Studying the electrodynamics of the system requires then taking into account the phononic, mechanical degrees of freedom on-par with the electromagnetic ones. This boils down to the Poynting vector, describing energy fluxes, acquiring a second term, proportional to the stress tensor of the lattice and describing energy propagating in the form of elastic energy in the solid~\cite{Gubbin:2022b}. Imposing the continuity of such a generalized Poynting vector allows one to determine boundary conditions on the mechanical fields that, together with the usual Maxwell boundary conditions, make it possible to fully solve the coupled light-matter problem. One important feature of these boundary conditions is that they mix longitudinal and transverse degrees of freedom, leading to a modified optical response and the emergence of novel excitations known as longitudinal-transverse polaritons (LTPs), which mix both transverse and longitudinal character. 
While nonlocal effects are known to be present also in plasmonic systems, and some consequences of nonlocal phenomenology are similar in plasmonic and phononic systems, the origin and consequences of nonlocality in those two leading nanophotonic platforms is fundamentally different, as schematically shown in Fig.~\ref{fig:Di-Liberato-Fig1}~\cite{Gubbin:2020}. Plasmons disperse toward the blue and longitudinal plasmons, which only exist above the plasma frequency, cannot thus hybridize with electromagnetically localized plasmon-polariton modes, which instead exist only below the plasma frequency where the dielectric function of the metal is negative. Their interaction is necessarily an evanescent phenomenon, leading to nonlocal phenomenology in plasmons being limited to an {\AA}ngstr{\"o}m-thin skin depth. Optical phonons disperse instead toward the red, making the coupling between longitudinal and transverse degrees of freedom a resonant, propagative phenomenon, leading to the emergence of LTPs, and enhancing by of orders of magnitude the length scales on which nonlocal phenomena can be observed and exploited. 
The existence of LTPs have been theoretically predicted and experimentally observed in silicon carbide (SiC) nanopillar arrays~\cite{Gubbin:2019} and in crystal hybrids, polar superlattices with nanometric-sized layers in which standard dielectric modeling fails to reproduce even the qualitative features of experimental reflectance~\cite{Gubbin:2020}. This success highlights that nonlocality in phonon polariton systems is not a theoretical curiosity but a practical concern that can affect device design and operation. Other predicted impacts of nonlocality on phonon polariton systems are an increased linewidth of Fr{\"o}hlich resonances in dielectric nanoparticles~\cite{Gubbin:2020} and the reduction of field confinement in nanogap resonators~\cite{Gubbin:2020a}, as in both cases energy can leak out via emission of LO phonons. 

\subsection*{Challenges and opportunities}

As the length scales of photonic devices approach the phonon propagation length, phonon nonlocality introduces several fundamental and practical challenges. The first and perhaps most direct challenge lies in accurate theoretical modeling. Traditional modeling approaches rely on local dielectric functions fitted to bulk optical constants, which fail to capture the momentum-dependent response of optical phonons. As a result, standard simulation tools and design methodologies yield inaccurate predictions of device resonance frequencies, confinement factors, and field enhancements. This discrepancy complicates the design of ultra-thin and deeply subwavelength structures where the effects of nonlocality are most pronounced. Multiple approaches have been explored, taking explicitly the phononic field into account either in finite-element method (FEM) simulations~\cite{Gubbin:2020a} or in scattering-matrix codes~\cite{Gubbin:2020b}, relying on effective medium theories~\cite{Yu:2023}, or developing analytical mappings to well-known quantum optics models~\cite{Gubbin:2022c}. Still, these methods have been applied to relatively simple materials and interfaces, with the most complex case a single asymmetric Reststrahlen band in Ref.~\cite{Gubbin:2020b}. For more complex and potentially highly anisotropic materials, in which multiple phonon modes are present in overlapping Reststrahlen bands, the resulting set of coupled boundary conditions could be substantially more challenging to solve. Even for simple materials current approaches rely on approximations which limit the theory’s overall domain of applicability. For a start, all the theoretical development cited rely on a quadratic approximation of the phonon dispersion. This means that the frequency of each optical phonon mode is written as a function of the wavevector as $\omega(k)=\sqrt{\omega^2(0) -\beta^2k^2 }$, where $\omega(0)$ and $\beta$ are the zone-center frequency and velocity parameter of the mode. This can be problematic when considering interfaces between materials with Reststrahlen bands with broad overlaps, in which the frequencies provided by the quadratic approximation can diverge significantly from their physical values. While a completely numerical approach that takes as input a parametrization of the phonon dispersion across all the Brillouin zone could in-principle be used to solve the problem in a more general way, this remains an open problem. Moreover, for extremely small nanolayers (1--2\,nm), the microscopic crystal structure starts to play a role, and the thickness of the each layer, used a parameter in the dielectric continuum modeling, is not precisely defined. This adds a material-dependent adjustable parameter to the theory which could be fixed (once each material interface) by \emph{ab initio} simulations~\cite{Gubbin:2020}.
A second challenge is related to experimental characterization. Probing nonlocal phonon effects requires techniques capable of exploring few-nanometer scales, where these phenomena manifest. While various mature approaches to near-field microscopy exist, the dependency of the nonlocal effects on the exact dimensions of the objects requires samples with extremely high uniformity, or the capability to probe a single nano-object, which remains to-date challenging.
Despite these hurdles, the opportunities that nonlocal phonon effects unlock are significant. Nonlocality offers a pathway to create novel types of hybrid modes like LTPs which combine the radiative nature of transverse fields and the strong interactions with electrical currents of longitudinal phonons, potentially bridging the gap between electrical excitations and far-field mid-IR emission. Such a capability could enable direct electrical pumping of mid-IR phonon polariton modes, offering a route to electrically driven emitters without the need for engineered electronic transitions such as those in quantum cascade structures. In such a scheme, the Fr{\"o}hlich interaction, responsible for LO-phonon emission in polar dielectrics, can become dressed by the strong interaction crating the LTPs. The scattering of a conduction electron would then resonantly create an LTP which could subsequently decay emitting mid-infrared light in the far-field. While initial theoretical analysis shows this mechanism to be potentially intense enough for practical applications~\cite{Gubbin:2023}, the intrinsically multi-scale nature of the problem makes it difficult to obtain precise results and an experimental verification of the emission mechanism is still missing. Only recently some more advanced non-equilibrium results obtained with a non-equilibrium Green function approach have appeared~\cite{Gubbin:2024} and still only for simple structures and low values of the applied bias. 
Additionally, nonlocal responses could be used to enhance control over optical anisotropy and hyperbolicity in polar materials. By carefully designing nanostructured heterostructures, superlattices, and patterned nanophotonic elements, it may be possible to engineer the dispersion of phonons and thus the properties of the resulting polaritons. This could lead to enhanced sensing capabilities, ultra-confined modes with engineered dispersion, and new opportunities for nonlinear optical processes.

\subsection*{Future developments to address challenges}

To effectively harness phonon nonlocality, the next years will likely see a concerted effort along two main axes: theoretical modeling and experimental characterization.
From a theoretical standpoint, developing comprehensive and user-friendly computational frameworks that include phonon dispersion and satisfy both electromagnetic and mechanical boundary conditions is a priority. Existing macroscopic models, some already demonstrated in simplified geometries, must be extended and made accessible to a broad community. Incorporating nonlocality into commercial electromagnetic simulation packages will be an important milestone, facilitating widespread exploration. Such models should be flexible enough to handle complex anisotropic materials, arbitrary nanostructures, and large-scale integrated photonic devices. On a fundamental level, more intuitive analytical approaches to interpret nonlocal behavior and guide rational design strategies will also prove valuable. Advanced multi-scale modeling of LTP-based electroluminescent devices will have to advance, allowing to better guide the design of prototype devices to gain a first experimental proof of this novel electroluminescence emission channel.
Experimentally, there is a need for the characterization of more LTP materials, which will play the double role of both verifying and pushing forward theoretical developments. Spectroscopic characterization of monodisperse polar nanosphere of decreasing dimension could provide evidence for some of the yet unobserved predictions of the nonlocal theory as well as clarifying the behavior of materials below critical nanometer dimensions. Experiments are necessary to try and observe LTP electroluminescence. Only after a first unequivocal observation of this emission channel an effort to optimize the emission and collection efficiency could take place, which in turn will be necessary to ascertain whether LTP electroluminescence is at most an interesting curiosity or if it has the potential to empower a novel generation of mid-infrared optoelectronic devices.
Finally, synergy with other fields—such as ultrafast optics, quantum photonics, and topological photonics—may lead to novel concepts and applications. For instance, the interplay of nonlocal phonon effects with coherent control schemes or coupling to quantum emitters (like defects in diamonds, molecular vibrations or quantum wells) could yield unprecedented control over mid-IR light–matter interactions. Similarly, integrating nonlocal phononic elements into topological platforms might provide robust, loss-resistant channels for infrared light.
In short, the future developments needed to address the challenges posed by phonon nonlocality involve building a comprehensive toolkit encompassing theory, experiment, and device engineering. These advances will enable the community to fully realize the potential of nonlocal phonon polaritons in creating next-generation mid-IR photonic technologies.

\subsection*{Concluding remarks}

Phonon nonlocality represents both a theoretical and technological frontier in mid-infrared photonics. By moving beyond the local dielectric approximation, we acknowledge the true complexity of polar materials and unlock new phenomena such as longitudinal-transverse polaritons. Although this introduces modeling challenges and demands more refined experimental techniques, it could open up the possibility of creating more compact, efficient, and versatile mid-IR devices.
The ability to couple electrical currents directly to radiative modes, leverage nonlocal effects for enhanced field confinement, and engineer the phonon dispersion through nanostructuring and heterostructures suggests a rich landscape of future innovations. As theoretical tools mature and integrate with experimental platforms, and as fabrication techniques continue to improve, phonon-based nonlocal photonics will likely play an increasingly important role. Ultimately, this may enable a new generation of mid-IR emitters, detectors, sensors, and nonlinear optical devices that harness the full power of phonon physics and broaden the scope of photonic materials and metamaterials research.

\section[Nonlocal effects in graphene plasmonics (Gon\c{c}alves \& Garc{\'i}a de Abajo)]{Nonlocal effects in graphene plasmonics}

\label{sec:Goncalves}

\author{P. A. D. Gon\c{c}alves\,\orcidlink{0000-0001-8518-3886} \& F. Javier Garc{\'i}a de Abajo\,\orcidlink{0000-0002-4970-4565}}

\subsection*{Current status}

Plasmons in graphene and graphene-based nanostructures possess extraordinary properties, featuring deeply subwavelength field confinement ($\lambda_p \ll \lambda_0$, where $\lambda_p$ and $\lambda_0$ denote the graphene-plasmon and photon wavelengths, respectively), relatively low losses, and active tunability through electrostatic gating~\cite{Goncalves:2016,Garciadeabajo:2014} and exposure to magnetic fields~\cite{Crassee:2012}. Graphene plasmonics evolved into a field of its own, fueled by the demonstration of gate-tunable plasmons that combine high confinement with low losses~\cite{Woessner:2015,Ni:2018}, an exceptional performance as ultrasensitive biochemical sensors~\cite{Rodrigo:2015}, strong nonlinearities promoted by graphene plasmons~\cite{Cox:2019}, and potential for the design of quantum optoelectronic circuitry~\cite{AlonsoCalafell:2019}, just to mention a few. Additionally, and arguably just as important, the accelerated growth of graphene plasmonics played a pivotal role in paving the way for the broader field of two-dimensional (2D) polaritonics in atomically thin materials and related van der Waals (vdW) heterostructures~\cite{Basov:2016}.

In addition to their remarkable plasmonic properties, graphene plasmons (GPs) can exhibit strong signatures of nonlocal effects (i.e., arising from the material's $q$-dependent response, where $q$ is the in-plane plasmon wave vector), in contrast to plasmons in three-dimensional (3D) or even ultrathin noble metals, where nonlocality is inherently weak and challenging to probe---often requiring nanostructures with characteristic length scales in the few-nanometer range~\cite{Zhu:2016,Barbry:2015,Campos:2019}, nanometric graphene--emitter separations~\cite{Goncalves:2020}, or the ability to measure large momentum transfers~\cite{Batson:1983}. 
The unparalleled susceptibility of GPs to nonlocal effects can be largely attributed to two key properties:  
\begin{itemize}
    
 \item[1)] \emph{Two-dimensional character and unique band structure.}~Graphene is a 2D semimetal in which charge carriers behave as massless Dirac fermions, following a linear energy–momentum dispersion. This enables on-demand control of the carrier density $n$ (allowing for a complete on/off switch of graphene's plasmonic response) and, consequently, control over the electronic Fermi wave vector via $k_{F} = \sqrt{\pi n}$. 
 For moderate doping, the associated Fermi wavelength $\lambda_{F}=2\pi/k_{F}$ can take values of tens of nanometers (Table~\ref{tab:Goncalves-Tab1}), which also defines the length scale (e.g., for the plasmon wavelength) below which nonlocal effects are important. Moreover, in finite-sized nanostructures, $\lambda_{F}$ can become comparable to the system size $D$, making both nonlocal and atomistic effects appreciable. 
 
 \item[2)] \emph{Large optical confinement.}~Plasmons in graphene-based systems can achieve extremely large wave vectors $q=2\pi/\lambda_{p}$, which can reach a substantial fraction of the Fermi wave vector $k_{F}$, ultimately approaching the $q/k_{F} \simeq 1$ regime in graphene--dielectric--metal heterostructures, thereby enhancing the role of nonlocality.  
\end{itemize}
Therefore, the combination of large Fermi wavelengths and highly confined optical modes renders graphene plasmons a unique platform for exploring the material's nonlocal response. Beyond its significance for accurately modeling plasmonic resonances in graphene structures~\cite{Thongrattanasiri:2012,Garciadeabajo:2014,Christensen:2014b}, nonlocal effects can also be harnessed to uncover many-body interactions encoded in the nonlocal optical conductivity of this carbon material~\cite{Lundeberg:2017}.

\begin{table}[hb]
\centering
\includegraphics[width=0.7\textwidth]{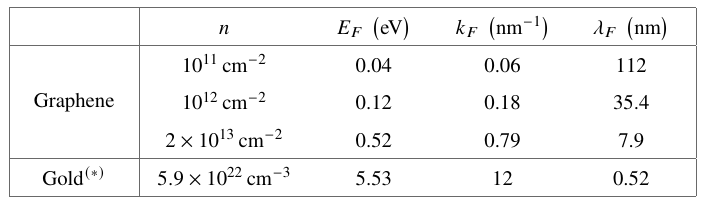}
\caption{Key physical quantities underpinning the electronic properties of graphene (for three different carrier densities) and gold. ${}^{(*)}$Taking the electron density from Table~ 1.1 in Ref.~\cite{Aschroft:1976} and assuming an effective electron mass $m = 9.109\times 10^{-31}$\,kg.}
\label{tab:Goncalves-Tab1}
\end{table}

\begin{figure}[hb]
\centering
 \includegraphics[width=0.95\textwidth]{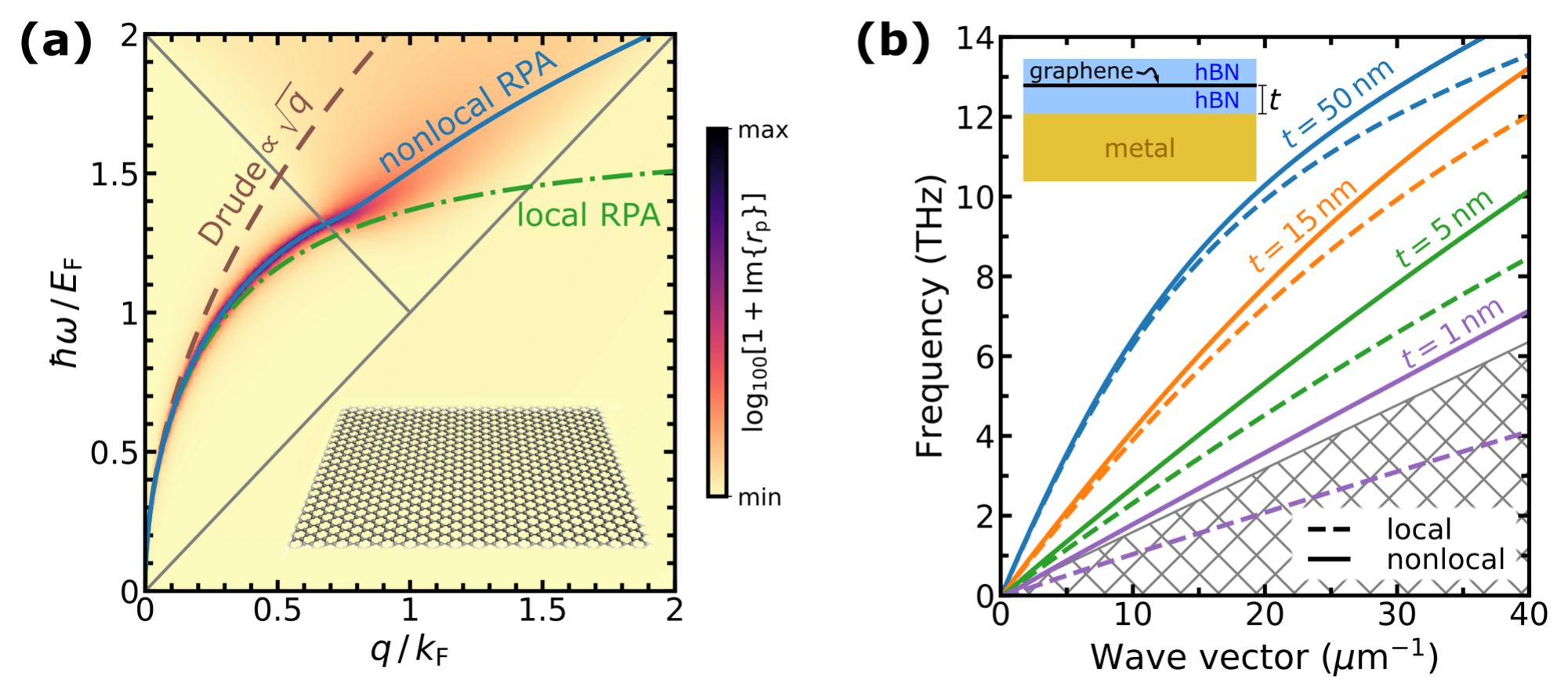}
 \caption{Plasmons in extended graphene systems.
 \pnl{a}~Plasmon dispersion in free-standing graphene obtained using different conductivity models, namely, the Drude approximation as well as the local and nonlocal versions of the random-phase approximation (RPA). \pnl{b}~Local- and nonlocal-RPA dispersion of acoustic graphene plasmons in hBN-encapsulated graphene placed on a metal substrate (which behaves as a perfect conductor in the THz range). Landau damping is active in the hatched region ($\omega < q v_{F}$). The thickness of the top hBN slab is fixed at 10\,nm, whereas the graphene--metal separation $t$ (controlled by the thickness of the bottom hBN film) is varied as indicated in the figure. hBN is modeled as a uniaxial crystal with $\varepsilon_{xx}^{\mathrm{hBN}} = \varepsilon_{yy}^{\mathrm{hBN}} = 6.7$ and $\varepsilon_{zz}^{\mathrm{hBN}} = 3.56$. For graphene [in both \pnl{a} and \pnl{b}], the carrier density and relaxation time are $n = 10^{12}\,\mathrm{cm}^{-2}$ (corresponding to $E_{F} =0.12$\,eV) and $\tau =500$\,fs, respectively.}
\label{fig:Goncalves-Fig1}
\end{figure}

Figure~\ref{fig:Goncalves-Fig1}\pnl{a} shows the dispersion of plasmons in a free-standing graphene monolayer. In the long-wavelength limit (i.e., within the local-response approximation), graphene plasmons exhibit a $\omega_\mathrm{GP} \propto \sqrt{q}$ dispersion, just like any homogeneous 2D electron gas (2DEG) (although with a different dependence on the carrier density, that is, $n^{1/4}$ instead of the $n^{1/2}$ scaling in conventional 2DEGs)~\cite{Goncalves:2016,Hwang:2007}. 
As the plasmon wave vector approaches a significant fraction of $k_{F}$, nonlocal effects induce a blueshift in the dispersion. At very large wave vectors, the plasmon mode eventually enters the electron--hole continuum, enabling Landau damping [observed as a pronounced broadening of the plasmon band in Fig.~\ref{fig:Goncalves-Fig1}\pnl{a}]. 

In extended graphene, the large optical confinement provided by graphene plasmons can be pushed to the ultimate limit in graphene--dielectric--metal (GDM) heterostructures~\cite{Lundeberg:2017,AlcarazIranzo:2018,Epstein:2020}. In this configuration, screening from the metal at small graphene--metal separations significantly alters the plasmon spectrum, reshaping it into a nearly linear dispersion [Fig.~\ref{fig:Goncalves-Fig1}\pnl{b}]~\cite{Alonso-Gonzalez:2017,Lundeberg:2017,Goncalves:2021}. As a result, these so-called acoustic graphene plasmons (AGPs) attain even larger wave vectors than those of conventional (i.e., unscreened) graphene plasmons at the same frequency. This, together with the substantial slow down of the plasmon group velocity, makes AGPs particularly susceptible to nonlocal effects [Fig.~\ref{fig:Goncalves-Fig1}\pnl{b}]. 
Notably, Lundeberg \emph{et al.}~\cite{Lundeberg:2017} measured the AGP dispersion using scattering-type scanning near-field optical microscopy (s-SNOM), revealing not only significant nonlocal effects but also electron–electron interactions, including Fermi-velocity renormalization and compressibility corrections. Furthermore, AGPs are capable of squeezing electromagnetic radiation down to one-atom-thick regions~\cite{AlcarazIranzo:2018,Epstein:2020}.

The significance of nonlocality in graphene plasmonics is not limited to spectroscopic measurements of nonlocal spectral shifts and broadening. 
The interaction between emitters and graphene plasmons, mediated by near-fields with large wave vectors, presents a promising avenue for probing nonlocal phenomena through their influence on emitter dynamics and emission spectra. In particular, graphene plasmons have been predicted to enhance otherwise "forbidden" light--matter interactions---e.g., multipolar transitions, two-plasmon spontaneous emission, and singlet-to-triplet transitions---by several orders of magnitude~\cite{Rivera:2016}. 
For instance, in the presence of large confinement factors ($\lambda_0/\lambda_p$ or $q/k_0$) achievable with AGPs, an atomic transition involving a change in angular momentum of $\Delta l = 5$ (a $5^\mathrm{th}$-order electric multipole transition) can occur at a rate exceeding that of a dipole-allowed transition ($\Delta l = 1$) for the same atom in free space~\cite{Rivera:2016}. In this regime, accounting for nonlocal effects is not just important, but essential for rigorously describing plasmon--emitter interactions mediated by strongly confined graphene plasmons~\cite{Rivera:2016,Petersen:2017}.

Additionally, the unconventional Dirac-cone electronic dispersion of graphene gives rise to a large intrinsic nonlinear response. Combined with the strong field confinement and enhancement associated with graphene plasmons, this makes graphene an attractive platform for nonlinear optics~\cite{Cox:2019}. This potential can be leveraged to develop all-optical nanophotonic devices and implement quantum gates~\cite{AlonsoCalafell:2019,Calajo:2023}. Incidentally, although graphene is a centrosymmetric material and, thus, not expected to exhibit second-order nonlinearities, the tightly confined graphene plasmons can give rise to spatially nonlocal nonlinear optical interactions. These interactions make the second-order response finite, with the corresponding coupling strength scaling as $\sim (q/k_{F})^{7/4}$~\cite{Manzoni:2015}.

\begin{figure}[hb]
\centering
 \includegraphics[width=1.0\textwidth]{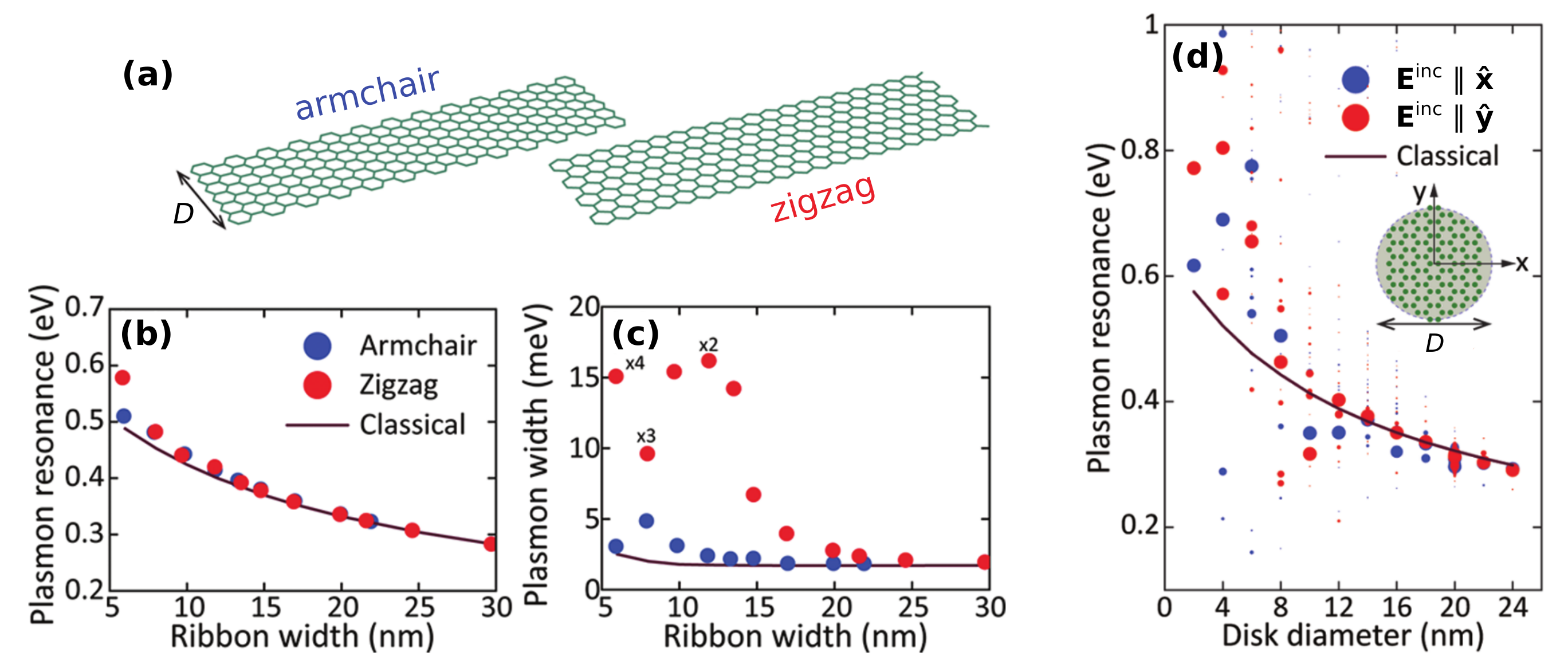}
 \caption{Plasmons in graphene nanostructures. \pnl{a}~Illustration of the two standard types of edge termination in graphene ribbons. \pnl{b}~Dipole plasmon resonances in graphene nanoribbons featuring either armchair or zigzag edge terminations (for normaly incident light polarized across the ribbon). The circles denote atomistic calculations (see Ref.~\cite{Thongrattanasiri:2012} for details), while the solid line indicates the classical result using the bulk conductivity of homogeneous, bulk graphene. \pnl{c}~Linewidths associated with the plasmon resonances in \pnl{b}.
 \pnl{d}~Plasmon resonances in graphene nanodisks (see inset). The circles correspond to quantum-mechanical atomistic calculations for incident plane waves with different polarizations (see labels) under normal illumination. The circles are scaled according to the strength of the resonances. The solid line indicates the dipole resonance energy obtained using a classical electromagnetic framework. 
 In all cases, the graphene nanostructures are considered to be free-standing, the Fermi energy is $E_{F}= 0.4$\,eV, and the relaxation time is $\tau = 411$\,fs. Reprinted (adapted) with permission from Ref.~\cite{Thongrattanasiri:2012} (Copyright~\textcopyright~2012 American Chemical Society).}
\label{fig:Goncalves-Fig2}
\end{figure}

In graphene nanostructures---such as ribbons and disks---nonlocal effects can be particularly pronounced due to the additional localization provided by the finite lateral size (i.e., the structure acts as a resonant cavity). This results in plasmon resonances with wavelength comparable to the characteristic system size $\lambda_{p} \sim D$, and introduces an \emph{effective} wave vector $q \sim D^{-1}$. As mentioned earlier, and as shown in Table~\ref{tab:Goncalves-Tab1}, the Fermi wavelength can satisfy $\lambda_{F} \sim \{\lambda_{p},D\}$ for commonly used carrier densities. At low carrier concentrations, where $\lambda_{F} \gtrsim 100$\,nm, nonlocal effects are observable even in relatively large graphene structures. In small nanostructures with significant edge-to-bulk ratios, nonlocality and quantum finite-size effects become closely intertwined, as localized plasmon resonances can probe the length scales associated with electronic edge states. Specifically, the type of edge termination---zigzag (ZZ) or armchair (AC)---leads to distinct plasmonic responses in nanostructured graphene (Fig.~\ref{fig:Goncalves-Fig2})~\cite{Thongrattanasiri:2012}. Figures~\ref{fig:Goncalves-Fig2}\pnl{b} and \ref{fig:Goncalves-Fig2}\pnl{c} show the plasmon energies and linewidths obtained through RPA calculations using tight-binding eigenstates for different ribbon widths. 
The classically predicted plasmon energies agree well with those obtained from quantum-mechanical calculations down to $D \approx 10$\,nm. For smaller ribbon widths, the atomistic theory predicts an increasing blueshift of the plasmon resonances as the width decreases. Note, however, that while the resonance position is almost identical for both ZZ- and AC-terminated ribbons (except for extremely narrow widths), the corresponding linewidths are significantly different: the plasmon linewidths in AC ribbons closely follow the classical result, whereas those in ZZ ribbons exhibit increased broadening for $D \lesssim 17$\,nm. This broadening arises from edge states forming a weakly dispersive band near the Dirac point, enabling plasmon decay into electron--hole pair excitations for plasmon energies above $E_{F}$~\cite{Thongrattanasiri:2012}. 

Unlike nanoribbons, graphene nanodisks cannot be constructed with purely ZZ or AC terminations, inevitably resulting in mixed edge types  (see inset of Fig.~\ref{fig:Goncalves-Fig2}). Consequently, their plasmonic response becomes polarization-dependent, as the termination varies along the perimeter. This is clearly shown in Fig.~\ref{fig:Goncalves-Fig2}\pnl{c}, together with the observed blueshift of the plasmon resonances when compared with the classical description. Additionally, the figure reveals several weaker resonances, attributed to higher-order multipoles as well as hybridizations with interband transitions (involving electronic edge states from ZZ-terminated regions). 

For systems comprising more than one nanostructure, the Coulomb interaction between neighboring elements leads to a richer plasmonic response~\cite{Thongrattanasiri:2013,Silveiro:2015}.

In passing, we emphasize that, while in Fig.~\ref{fig:Goncalves-Fig2} the classical electromagnetic description accurately predicts plasmon resonances in graphene ribbons and disks with $D \gtrsim$10--17\,nm, this threshold is not universal, as it strongly depends on carrier density. For the cases considered in Fig.~\ref{fig:Goncalves-Fig2}, with $E_{F} =0.4$\,eV ($n \approx1.2\times 10^{13}\,\mathrm{cm}^{-2}$), the corresponding Fermi wavelength is $\lambda_{F} \approx 10$\,nm, which aligns well with this threshold. 
Hence, at lower carrier densities, nonlocal and quantum finite-size effects can manifest in larger graphene nanostructures.

\subsection*{Challenges, opportunities, and future directions}

Graphene plasmonics is now a well-consolidated field from both theoretical and experimental standpoints. However, nonlocal and quantum size effects in graphene nanoplasmonics remain less explored, despite their significance for exploring the fundamental limits of plasmon-enhanced light–matter interactions in graphene and for designing truly nanometer-scale graphene devices.

On the experimental side, many theoretical proposals have yet to be realized, including the unambiguous observation of quantum size effects in small graphene nanostructures (down to "molecular-sized" graphene~\cite{Manjavacas:2013}) and the exploitation of ultraconfined graphene plasmons to infer quantum nonlocal effects in metals~\cite{Goncalves:2021} and in strongly correlated materials~\cite{Costa:2021}. 
While the latter could be already pursued using current state-of-the-art s-SNOM techniques~\cite{Lundeberg:2017,Ni:2018,Hesp:2024}, the fabrication of high-quality, pristine nanographenes on dielectric substrates suitable for plasmonics remains a challenge, with stable systems still limited to crystallographically well-defined metallic surfaces.

On the theoretical front, several promising directions remain open. In extended graphene, further investigation is needed in the hydrodynamic-transport regime---the opposite of the commonly studied quasi-collisionless scenario---where charge carriers behave as a viscous fluid, giving rise to intriguing transport phenomena such as negative nonlocal resistance and current whirlpools~\cite{Bandurin:2016,Palm:2024}. The implications of these effects for graphene plasmons remain largely unexplored. 
Moreover, conventional plasmon modes cannot propagate at velocities below $v_{F}$ [i.e., the plasmon dispersion cannot enter the electron--hole continuum, as shown in Fig.~\ref{fig:Goncalves-Fig1}\pnl{b}]. However, in the hydrodynamic regime, plasmon-like energy waves ("demons") can exist and propagate with velocities as low as $v_{F}/\sqrt{2}$~\cite{Sun:2016,Torre:2019}, further enhancing nonlocal effects. 
In finite-sized graphene, many open questions remain in relation to the influence of edge reconstructions---which can stabilize the edges both energetically and mechanically~\cite{Koskinen:2008}---as well as disorder and edge-passivating adatoms, all of which constitute uncharted territory. Additionally, graphene nanoribbons can exhibit spin-polarized edges and edge magnetism~\cite{Carvalho:2014}, suggesting the exploration of spin-polarized plasmons~\cite{Agarwal:2014}.

In parallel, the growing interest in twisted moiré quantum materials is expected to extend into plasmonics in moiré graphene. This field is already gaining attention, with the first experimental observations reported~\cite{Hesp:2021} alongside theoretical advancements~\cite{Lewandowski:2019,Cavicchi:2024}. Here too, nonlocal effects could not only shed light into the mechanisms governing the complex phases arising in moir{\'e} materials but also open new coupling pathways where quasiparticles can exchange momentum with the twist-induced superlattice reciprocal vectors.

\subsection*{Concluding remarks}

Nonlocal effects in graphene plasmonics arise from the unique interplay between highly confined plasmons and large Fermi wavelengths, making graphene an exceptional platform for exploring nonlocal phenomena. While often considered detrimental due to their association with increased broadening, nonlocal effects can also be leveraged to access otherwise inaccessible phenomena---such as nonlocality-induced nonlinearities and new coupling pathways---or to provide additional insights into electronic correlations intrinsically linked to nonlocal properties. 
Given the advancements and future prospects outlined in this section, the outlook for nonlocal graphene plasmonics is promising.

\section[Nonlocal quantum gain, loss compensation, and plasmon amplification in graphene hyberbolic metamaterials (Hess \& Tarasenko)]{Nonlocal quantum gain, loss compensation, and plasmon amplification in graphene hyberbolic metamaterials}

\label{sec:Hess}

\author{Ortwin Hess\,\orcidlink{0000-0002-6024-0677} \& Illya Tarasenko\,\orcidlink{0009-0007-0401-8290}}

\subsection*{Current Status}

\begin{figure}[hb]
\centering
\includegraphics[width=0.8\linewidth]{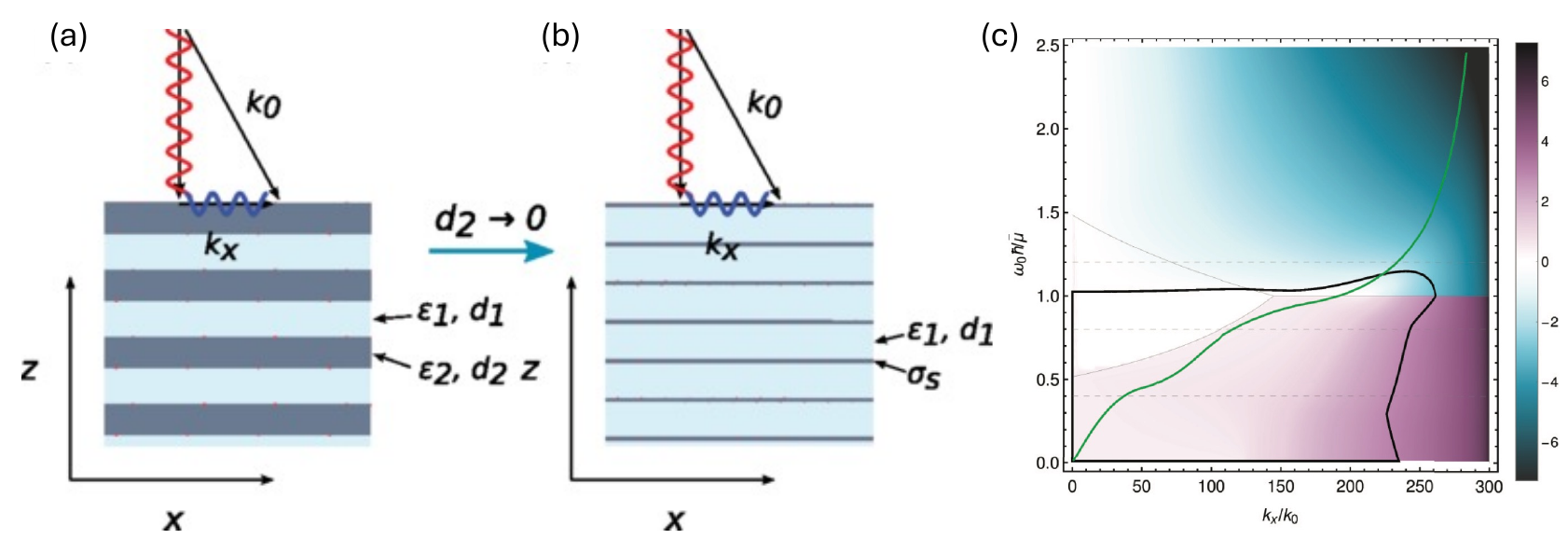}
\caption{Graphene-based hyperbolic metamaterials (GHMMs). \pnl{a} Traditional hyperbolic metamaterials (HMMs) use conductive sheets of finite thickness, where an incident transverse magnetic (TM) wave couples to carrier oscillations in both the $x$- and $z$-directions. \pnl{b} Replacing these metal layers with single sheets of graphene results in graphene-based hyperbolic metamaterials, where carrier oscillations are confined strictly to the plane of the graphene layer. \pnl{c} Density plot, showing the imaginary part of the Bloch wave vector for an infinite Graphene-dielectric periodic structure, shown in \pnl{b}. Here graphene is optically pumped to create inversion. Positive (red) values correspond to amplification, negative (blue) values correspond to attenuation. Black curve bounds the hyperbolic regime of operation, where $\varepsilon_{x-\mathrm{eff}}<0$. Green line is the single sheet plasmon dispersion. Reproduced with permission from Ref.~\cite{Tarasenko:2019} (Copyright~\textcopyright~2019 American Physical Society).}
\label{fig:Hess-Fig1}
\end{figure}

Insight into nonlocal quantum gain represents a pivotal advancement in the development of graphene-based hyperbolic metamaterials (GHMMs), addressing longstanding challenges in loss compensation and plasmon amplification. GHMMs are a class of hyperbolic metamaterials (HMMs) that leverage the exceptional properties of graphene, a two-dimensional material known for its high carrier mobility, tunable optical conductivity, and strong confinement of electromagnetic fields~\cite{Tarasenko:2019,Tarasenko:2020, Shung:1986, Yan:2012b, Ferrari:2015}. Traditional HMMs use metal layers to achieve hyperbolic dispersion [Fig.~\ref{fig:Hess-Fig1}\pnl{a}], but these materials suffer from significant Ohmic losses and limited tunability. In contrast, GHMMs replace metallic layers with single or multiple graphene sheets [Fig.~\ref{fig:Hess-Fig1}\pnl{b}], enabling dynamic control over optical properties while minimizing losses~\cite{Iorsh:2013, Ferrari:2015}.

Hyperbolic dispersion is a key feature of these materials, allowing propagation of high-wave-vector modes and enabling subwavelength confinement of light. These properties are essential for applications in imaging, sensing, and nanolithography~\cite{Wunsch:2006, Hwang:2007}. However, conventional models for graphene conductivity, such as Drude and Kubo formulas, assume locality and linearity. These approximations break down for excitations with large wave vectors, particularly near the Dirac cone, leading to inaccuracies in describing plasmonic phenomena. The introduction of nonlocal quantum conductivity models, derived using the random-phase approximation (RPA), has addressed these limitations. These models account for both interband and intraband transitions, providing an accurate description of the optical response of graphene~\cite{Dai:2015, Pendry:2000, Pyatkovskiy:2009}. This enables the calculation of regions of plasmonic amplification of waves, propagating in active GHMMs, as illustrated in [Fig.~\ref{fig:Hess-Fig1}\pnl{c}].

The chapter highlights the transformative potential of nonlocal quantum gain in GHMMs. To theoretically grasp loss compensation and amplification in GHMM requires a nonlocal quantum conductivity model~\cite{Page:2015, Page:2018, Tarasenko:2019, Tarasenko:2020}
\begin{equation}\label{eq:Hess-Eq1}
\sigma_s(k_x, \omega) = i \omega e^2 \frac{\Pi(k_x, \omega)}{k_x^2}. 
\end{equation}
Here, $\Pi(k_x, \omega)$ is the propagator of the electron-hole pairs. In RPA first order and expressed in Lindhard format~\cite{Lindhard:1954}, the polarization function of the arbitrary nonequilibrium graphene sheet can, in turn, be expressed in the form~\cite{Page:2015, Page:2018, Tarasenko:2019}
\begin{equation}\label{eq:Hess-Eq2}
\Pi(n) = \Pi\big|_{T=0,\mu=0} +
\int_{0}^{\infty} d\epsilon \left[ \frac{\partial \Pi(k_x, \omega)\big|_{T=0,\mu=\epsilon}}{\partial \epsilon} n(\epsilon) \right]
+ \int_{0}^{\infty} d\epsilon \left[ \frac{\partial \Pi(k_x, \omega)\big|_{T=0,\mu=\epsilon}}{\partial \epsilon} \bar{n}(\epsilon) \right],
\end{equation}
where $n(\epsilon)$ and $\bar{n}(\epsilon) = 1 - n(-\epsilon)$ are the distribution functions for electrons and holes, respectively.

Optical pumping of graphene introduces gain by creating an inverted electron-hole plasma, facilitating stimulated recombination that amplifies surface plasmons~\cite{Soukoulis:2010a, Soukoulis:2010b, Page:2015, Page:2018, Basov:2016}. This gain mechanism not only compensates for inherent losses but also enables a shift from loss-dominated to gain-dominated regimes of operation. Such dynamic tunability opens up new possibilities for GHMM applications, including sub-diffraction imaging, efficient light-matter interaction platforms, and quantum photonics~\cite{Koppens:2014, Reserbat-Plantey:2021}.

Recent studies have demonstrated the potential of GHMMs to achieve low-loss imaging and high-resolution performance in nanophotonic devices. For instance, GHMMs have been integrated into subwavelength optical filters, tunable modulators, and plasmonic waveguides. Their ability to dynamically switch between hyperbolic and elliptic regimes further enhances their versatility~\cite{Low:2017, Liu:2021}. Despite these advances, challenges such as gain stability, fabrication precision, and scalability continue to hinder the practical implementation of GHMMs. The field is now focused on overcoming these limitations to fully exploit the capabilities of nonlocal quantum gain.

\subsection*{Challenges and opportunities}

While nonlocal quantum gain has unlocked new capabilities in GHMMs, significant challenges remain that must be addressed to realize their full potential. One of the primary challenges is achieving stable gain mechanisms. Optical pumping of graphene introduces nonlocal quantum gain by creating an inverted electron-hole plasma, but this process can lead to instability, noise, and mode competition~\cite{Koppens:2014}. These effects arise when the gain exceeds a critical threshold, resulting in uncontrolled amplification. Stabilizing gain requires precise control over doping levels, chemical potentials, and excitation regimes. Without careful optimization, gain mechanisms may inadvertently degrade device performance.

Thermal and collision losses also present a significant challenge. Graphene's Fermi level smearing at elevated temperatures and carrier scattering at high doping levels introduce energy dissipation, limiting the effectiveness of loss compensation. These effects are particularly detrimental in applications requiring high field confinement or long propagation lengths. Mitigating these losses will require innovative material design and system integration~\cite{DasSarma:1982, Falkovsky:2007, Geim:2007}.

Fabrication challenges further complicate the development of GHMMs. The precise layering of graphene and dielectric materials at nanoscale dimensions is technically demanding and prone to inconsistencies. Current fabrication techniques, such as chemical vapor deposition and atomic-layer deposition, show promise but require further refinement to ensure reproducibility and scalability. Achieving uniformity across large-scale devices is essential for translating laboratory advances into practical applications.

Despite these challenges, the opportunities presented by nonlocal quantum gain are immense. Amplified plasmonic modes in GHMMs enable enhanced light-matter interactions, which are critical for applications in quantum optics, sensing, and nanophotonics~\cite{Yeh:1988}. For example, GHMMs can serve as platforms for developing efficient quantum light sources, entangled photon generation, and nanoscale waveguides. The ability to dynamically switch between hyperbolic and elliptic regimes also makes GHMMs ideal for reconfigurable optical devices, including modulators, filters, and frequency-selective surfaces.

Another exciting opportunity lies in extending the operational bandwidth of GHMMs. Nonlocal quantum gain enables the propagation of high-wave-vector modes, which can be harnessed for broadband plasmonic devices. This is particularly relevant in applications such as super-resolution imaging, where GHMMs can overcome the diffraction limit and achieve unparalleled spatial resolution. Moreover, hybrid systems combining GHMMs with complementary materials, such as transition metal dichalcogenides (TMDs) or hexagonal boron nitride (hBN), can further enhance device performance and introduce new functionalities~\cite{Basov:2016,  Reserbat-Plantey:2021}.

\subsection*{Future Developments to Address Challenges}

To address the challenges facing GHMMs, significant advancements are needed in material engineering, device design, and fabrication techniques. One of the most promising avenues is the optimization of gain mechanisms. This involves tailoring doping levels, chemical potentials, and excitation regimes to maximize electron-hole recombination while minimizing noise and instability~\cite{Orlov:2011}. Advances in laser excitation techniques and chemical functionalization could enable precise control over carrier dynamics, enhancing the stability and efficiency of plasmon amplification.

Thermal and collision losses can be mitigated by exploring new materials and hybrid structures. For example, integrating graphene with high-thermal-conductivity materials, such as hBN, can help dissipate heat more effectively. Similarly, introducing nanostructured dielectric layers or low-scattering materials into GHMMs can reduce energy dissipation and enhance performance under realistic conditions~\cite{Soukoulis:2010b, Ferrari:2015}.

Innovations in fabrication techniques will also be crucial. Emerging technologies such as 3D nanoscale printing, machine learning-assisted design, and layer-by-layer assembly hold promise for improving precision, reproducibility, and scalability. These approaches could enable the mass production of GHMMs with consistent optical and electronic properties, facilitating their integration into existing photonic and electronic platforms~\cite{Reserbat-Plantey:2021}.

Hybrid systems represent another promising direction for future development. Combining GHMMs with complementary materials, such as perovskites or TMDs, can address losses while introducing new functionalities, such as enhanced light absorption or tailored optical responses. These hybrid systems could leverage the unique properties of each material to create devices with superior efficiency and versatility~\cite{Koppens:2014, Basov:2016}.

In the longer term, GHMMs are expected to play a transformative role in quantum technologies. By leveraging nonlocal quantum gain, researchers can develop advanced platforms for single-photon generation, entangled photon sources, and low-loss quantum waveguides. These capabilities will be critical for advancing quantum communication, sensing, and computing.

Ultimately, the goal is to develop energy-efficient, ambient-compatible plasmonic devices that can operate reliably across diverse applications. Overcoming the current limitations will require a multidisciplinary approach, combining advances in materials science, device engineering, and theoretical modeling. By addressing these challenges, GHMMs could become foundational components of next-generation photonic and optoelectronic technologies~\cite{Tarasenko:2019, Polman:2019}.

\subsection*{Concluding remarks}

The advancements in our understanding leading to non-local quantum gain have established graphene-based hyperbolic metamaterials (GHMMs) as a promising solution to overcome the intrinsic limitations of traditional hyperbolic metamaterials. By leveraging the unique properties of graphene and incorporating precise models of non-local quantum conductivity, GHMMs have demonstrated their potential to address critical challenges such as loss compensation and plasmon amplification. These innovations have enabled significant breakthroughs in nanophotonics, quantum technologies, and advanced imaging systems~\cite{Tarasenko:2019, Yan:2012b, Basov:2016}.

The key takeaway from the discussed work is the role of nonlocal quantum gain in mitigating losses and amplifying plasmonic modes, thereby enhancing the overall performance of GHMMs. This is achieved through the optical pumping of graphene, which facilitates stimulated electron-hole recombination and unlocks new operational regimes. As a result, GHMMs offer tunable and dynamically reconfigurable properties, making them suitable for a wide range of applications, including quantum emitters, nanoscale waveguides, and low-loss photonic devices~\cite{Wunsch:2006,  Reserbat-Plantey:2021}.

Despite these advances, the field faces several challenges, such as achieving stable gain, reducing thermal and collision losses, and overcoming fabrication limitations. Addressing these issues will require multidisciplinary efforts in materials science, device engineering, and theoretical modeling. The development of hybrid systems, integration with complementary materials, and the use of advanced fabrication techniques are promising pathways to mitigate these challenges~\cite{Koppens:2014, Ferrari:2015, Liu:2021}.

Looking ahead, GHMMs are poised to play a transformative role in next-generation technologies. Their ability to dynamically manipulate light at the nanoscale opens new frontiers in quantum communication, high-resolution imaging, and reconfigurable photonic systems. Moreover, the integration of GHMMs into existing photonic and electronic platforms will enable the development of energy-efficient, scalable, and versatile devices~\cite{Low:2017, Soukoulis:2010b}.

In conclusion, the realization of nonlocal quantum gain in GHMMs has paved the way for substantial advancements in the field of nanophotonics. While challenges remain, the opportunities presented by these materials far outweigh the limitations. With continued research and development, GHMMs are set to redefine the boundaries of photonic and optoelectronic technologies, offering solutions to some of the most pressing demands of modern science and engineering.

\section[Surface electrodynamics of crystalline noble metals (Cox \& Mortensen)]{Surface electrodynamics of crystalline noble metals}

\label{sec:Cox}

\author{Joel D. Cox\,\orcidlink{0000-0002-5954-6038} \& N.~Asger~Mortensen\,\orcidlink{0000-0001-7936-6264}}

\subsection*{Current status}

\begin{figure}[hb!]
    \centering
    \includegraphics[width=0.8\linewidth]{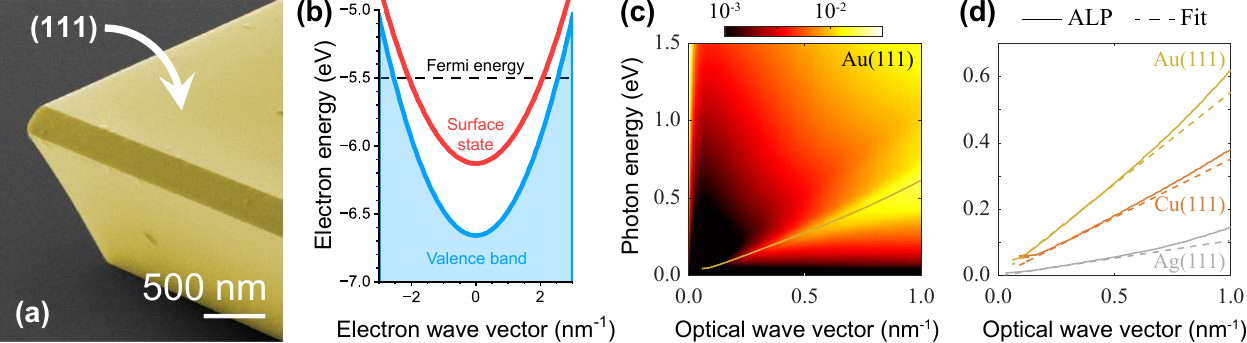}
    \caption{\pnl{a} SEM image of gold crystalline metal flake with indication of the (111) surface facet. Reproduced with permission from Ref.~\cite{Boroviks:2018} (Copyright~\textcopyright~2018 Optica Publishing Group). \pnl{b} Schematic energy-momentum electron dispersion relation for the TS surface states and the corresponding valence band in Au(111). \pnl{c} Plot of the loss function at an Au(111) surface and \pnl{d} acoustic-like surface-plasmon energy-momentum dispersion relation associated with the TS surface states on (111) facets of noble metals. Reproduced with permission from Ref.~\cite{RodriguezEcharri:2021a} (Copyright~\textcopyright~2021 Optica Publishing Group).}
    \label{fig:Cox-Fig1}
\end{figure}

The emergence of crystalline metals in plasmonics~\cite{Huang:2010}, which are often considered superior to their polycrystalline counterparts~\cite{McPeak:2015}, has stimulated substantial efforts in the synthesis~\cite{Hoffmann:2016,Mejard:2017,Cheng:2019} and optical characterization~\cite{Boroviks:2018,Boroviks:2021} of these pristine samples to enable demanding experiments~\cite{Spektor:2017,Frank:2017,Boroviks:2022,Pres:2023}.
However, crystalline metals are also turning notable in relation to their surfaces. In particular, surface science has long established~\cite{Inglesfield:1982} that (111) noble-metal surfaces [Fig.~\ref{fig:Cox-Fig1}\pnl{a}] host in-plane conductive surface states -- often referred to as Tamm--Shockley (TS) surface states~\cite{Tamm:1932,Shockley:1939} -- governed by a free-electron-like parabolic energy-momentum dispersion relation [Fig.~\ref{fig:Cox-Fig1}\pnl{b}] and hosting a two-dimensional electron gas (2DEG) localized to the very surface.
The TS surface state is formally characterized by a surface conductivity $\sigma_\parallel(\omega,\boldsymbol{q}_\parallel)$, exhibiting both frequency and spatial dispersion. The impact of the TS 2DEG can be seamlessly incorporated into electrodynamics by modifying the boundary conditions to account for the surface conductivity. This approach establishes a connection to the Feibelman surface-response function (see also Sec.~\ref{sec:Christensen} in this Roadmap), as $d_\parallel \propto \sigma_\parallel$~\cite{Feibelman:1982}.

The surfaces of crystalline noble-metal particles are typically comprised of multiple facets corresponding to the intrinsic crystal planes of the solid~\cite{Quan:2013}. The facets are in turn the cause for intriguing morphology-dependent and polarization-dependent plasmonic resonances~\cite{Myroshnychenko:2008,Yoon:2019,Elliott:2022,RodriguezEcharri:2021a}. In contrast, large planar flakes are predominantly characterized by (111) surfaces [Fig.~\ref{fig:Cox-Fig1}\pnl{a}], although their edges may naturally exhibit other facets~\cite{Boroviks:2018}. The conductive properties of the (111) facet in crystalline flakes are now renewing interest in TS surface states, particularly in the realm of their electrodynamic response.

\subsection*{Challenges and opportunities}

The electrodynamic response of the homogeneous 2DEG (possibly embedded in general dielectric media), and in particular its supported plasmon excitations exhibiting a square-root dependence of the plasmon energy on the wave vector~\cite{Ando:1982}, are theoretically well explored.
A straightforward approach is to derive $\sigma_\parallel(\omega, \boldsymbol{q}_\parallel \rightarrow 0)$ using a Drude model within the phenomenological relaxation-time approximation, employing a 2D electron density consistent with the common Fermi energy of the bulk and TS surface states. A key challenge remains in gaining a deeper understanding of the relaxation rate in the 2DEG and how it may differ from its bulk counterpart.

Turning to the plasmon excitations of the TS surface states, strong screening by the adjacent 3DEG of the bulk influences the plasmon dispersion~\cite{Echenique:2001}, resulting in an acoustic-like relation $\omega(\boldsymbol{q}_\parallel) \simeq v_\phi q_\parallel$ [Fig.~\ref{fig:Cox-Fig1}\pnl{c}]. This is analogous to the 2D mirror plasmons observed in graphene-on-metal structures~\cite{Lundeberg:2017}, but represents the extreme case where the separation between the 2DEG and the metallic "mirror" approaches zero. This makes the phase velocity $v_\phi$ even lower, leading to longer wave vectors at the same frequency [Fig.~\ref{fig:Cox-Fig1}\pnl{c}], posing a significant challenge for experimental observation in near-field setups.
Angle-resolved electron energy loss spectroscopy (EELS) has identified 2D plasmons on similar surfaces~\cite{Diaconescu:2007,Politano:2015}, but with even lower phase velocities $v_\phi$. This makes them even more tightly confined to the surface than graphene plasmons are to a graphene monolayer~\cite{Lundeberg:2017,Autore:2019}.

In the context of the plasmonic response of the bulk 3D electron gas (3DEG), the surface-response functions are modified by the presence of the TS 2DEG. In addition to the Feibelman perpendicular parameter $d_\perp(\omega)$, there is now also a non-vanishing parallel component $d_\parallel (\omega)\propto \sigma_\parallel (\omega, \boldsymbol{q}_\parallel\rightarrow 0)$~\cite{Feibelman:1982}. Thus, contrary to any other facet where $d_{\parallel} = 0$~\cite{Feibelman:1982,Liebsch:1997}, the (111) surface of noble metals is characterized by $d_{\parallel} \neq 0$~\cite{RodriguezEcharri:2021a}. A compelling question arises: are there any plasmonic signatures that could differentiate (111) surfaces~\cite{Feibelman:1994,Politano:2008} from other surfaces of the same metal?
For faceted noble-metal nanoparticles, the $d_\parallel$ thus varies among its facets, which is predicted to reflect polarization-dependent responses for even the dipole resonances~\cite{RodriguezEcharri:2021a}. The definitive experimental investigation of this phenomenon remains challenging, requiring precise control and characterization of particle morphology and orientation on the surface, along with accurate measurement of its polarization response, for instance, using dark-field microscopy.

Probing surface-response functions remains an experimental challenge~\cite{Yang:2019a}, as their extraction often relies on comparing computational spectra with and without surface effects, which in turn critically depends on accurate experimental knowledge of morphology on the nanoscale. An alternative approach involves electrically gating a single nanostructure, where strong bulk screening confines added electrons to the surface, effectively altering the surface response while leaving the bulk response unchanged. This method recently enabled the first modulation of surface response in a single plasmonic nanoresonator~\cite{Zurak:2024}. While simple models qualitatively capture these results~\cite{Zurak:2024,Mortensen:2021b}, a more precise theoretical and computational description remains elusive. In addition to rigorous determination of $d_\perp(\omega)$ and $d_\parallel(\omega)$ Feibelman parameters for charge-neutral (111) surfaces and other facets with the aid of \emph{ab initio} methods that account for d-band screening, additional key challenges include understanding the mutual equilibrium distribution of added electrons between TS surface states on (111) facets and nearby bulk states on all facets of a nanostructure. While atomistic models and the random-phase approximation (RPA) have been used to study surface-response functions of (111) surfaces~\cite{RodriguezEcharri:2021a}, further investigations with advanced methods of charged systems (contrary to the common charge-neutral scenario), such as time-dependent density-functional theory, are needed. Consequently, the experiments by Zurak \emph{et al.}~\cite{Zurak:2024} outpace the current theoretical foundation.

\subsection*{Future developments to address challenges}

To address the outlined challenges, future developments could focus on several key areas. First, there is a need to refine theoretical models to gain a more precise understanding of the relaxation rates for TS states and their differences from bulk counterparts. This potentially involves the study of the coupling to phonons in surface-terminated systems. By enhancing these theoretical frameworks, researchers can better capture the complexities of TS surface-state dynamics and interactions with bulk states.

Second, advancing experimental techniques for probing surface-response functions is essential. This includes innovating experimental techniques that improve the accuracy of measurements. The highly confined acoustic-like plasmons of the TS 2DEG can be probed using scanning near-field optical microscopy (SNOM) and electron spectroscopy techniques, though both methods present technical challenges, particularly due to low signal-to-noise ratios.
In this context, recent ellipsometry-based characterization of the surface response offers an interesting alternative~\cite{Chen:2024b}. The excitation of acoustic plasmons based on nonlinear wave mixing constitutes another appealing possibility that exploits nonlocal and nonlinear light-matter interactions (see Sec.~\ref{sec:Jelver} in this Roadmap). Here, the energy- and momentum-matching of impinging light to highly confined polaritons could be achieved by interfering counter-propagating beams in a second-order difference-frequency generation scheme, such as that used to probe graphene plasmons from free space by Constant \emph{et al.}~\cite{Constant:2016}. Additional optical momentum can be provided by coupling light from dielectric waveguides to highly confined acoustic plasmon modes. The polarization dependence of the second-order nonlinear response at (111) surfaces~\cite{Boroviks:2021} and plasmonic field enhancement~\cite{RodriguezEcharri:2021b} can provide additional insight to characterize the TS 2DEG.
For nanoparticles, achieving greater control over nanoparticle morphology, orientation, and polarization response characterization will be critical for obtaining reliable data on the effects of surface response and in particular the electrodynamic effects of TS surface states.

Another crucial area is the integration of computational and experimental insights. Future research should focus on designing computational models that can explain experimental results with high fidelity, particularly in the context of gated plasmonic nanostructures. These models and \emph{ab initio} descriptions must address the equilibrium distribution of added electrons between TS surface states and nearby bulk states, potentially allowing theoretical predictions to better align with observed phenomena.

Additionally, further development of electrically gated single-nanostructure experiments will enhance the understanding of TS surface responses. Exploring methods to dynamically modulate and tune plasmonic properties in these nanostructures will yield new insights into the behavior of plasmonic systems under varying conditions. In this context, the combination of electrical gating with the AC lock-in technique employed by Zurak \emph{et al.}~\cite{Zurak:2024} is a promising method to extract weak spectral changes otherwise buried in noise.

Moreover, investigating the plasmonic signatures unique to (111) noble-metal surfaces compared to other facets is vital. Identifying these distinct polarization-dependent resonances can provide valuable information about the fundamental interactions at play in these materials. Incidentally, for delafossite metals, recent reports of nonlocal electrodynamic effects in the hexagonally-symmetric conducting palladium (Pd) planes of palladium cobalt oxide (PdCoO$_2$) reveal the importance of facets in the Fermi surface, suggesting analogous dependencies on Fermi-surface facets in reciprocal space~\cite{Baker:2024}.

Furthermore, there is a significant opportunity to explore transdimensional plasmonics~\cite{Boltasseva:2019} (see also Sec.~\ref{sec:Bondarev} in this Roadmap), particularly the transition from three-dimensional (3D) to two-dimensional (2D) plasmonic behavior in few-atom thin films subjected to out-of-plane quantum confinement. In these circumstances, where distinguishing between surface and bulk properties becomes less straightforward, it may be interesting to revisit Sipe's selvage considerations~\cite{Sipe:1980}.

Lastly, the similarities between graphene plasmonics and TS surface states could be advantageous, as graphene plasmons may serve as a tool to probe and reveal quantum phenomena in the 2DEG linked to TS surface states. For instance, Gon\c{c}alves \emph{et al.}~\cite{Goncalves:2021} proposed extending graphene-on-metal experiments~\cite{Lundeberg:2017} to investigate the interplay between nonlocal 2D plasmons (from both graphene and TS types) and their screening by the adjacent 3D electron gas of the bulk metal, which also responds nonlocally~\cite{Dias:2018,RodriguezEcharri:2019}. While impressive experiments have utilized polycrystalline metal films~\cite{Lundeberg:2017,AlcarazIranzo:2018}, the change to crystalline flakes with (111) surfaces may present a new paradigm when interfaced with graphene. The hybridization of graphene and TS-type plasmons could pave the way for new explorations of quantum effects in highly confined electron gases.

\subsection*{Concluding remarks}

Crystalline noble metals, particularly their (111) facets with Tamm--Shockley surface states, present opportunities to explore surface-specific phenomena, including nonlocal response and refined surface-response formalisms. Progress requires advancing theoretical models for relaxation rates, nonlocal effects, and their distinctions from bulk properties, as well as improving experimental methods to probe surface responses with high precision. Greater control over nanoparticle morphology and the integration of computational and experimental approaches, especially in gated nanostructures, are key. Investigating surface-state plasmonic properties, including their nonlocal behavior, and hybrid interactions with graphene plasmons could pave the way for deeper insights into surface electrodynamics.

\section[Concerted nonlinear and nonlocal electrodynamics in 2D materials (Jelver \emph{et al.})]{Concerted nonlinear and nonlocal electrodynamics in 2D materials}

\label{sec:Jelver}
\author{Line Jelver\,\orcidlink{0000-0001-5503-5604}, Eduardo J. C. Dias\,\orcidlink{0000-0002-6347-5631} \& Joel D. Cox\,\orcidlink{0000-0002-5954-6038}}

\subsection*{Current status}

Nonlocal electrodynamic phenomena in nanophotonics have been extensively studied in the regime of linear optics, where a one-to-one correspondence is maintained between the energy and momentum carried by photons and the polarization they induce in matter. However, spatially-extended nonlocal light-matter interactions become inherently more complex in the realm of nonlinear optics, where the self-interaction of intense optical fields comprised of many photons is mediated by their mutual strong coupling with matter. For instance, sharp variations in the spatial distribution of near-fields can significantly impact the nonlinear response of individual nanostructures~\cite{Wolf:2016}, while engineered long-range interactions among subwavelength scatterers in metasurfaces can give rise to a large collective nonlinear optical response~\cite{Krasnok:2018, Sharma:2023, Kolkowski:2023}.

\begin{figure}[hb]
    \centering
    \includegraphics[width=.8\linewidth]{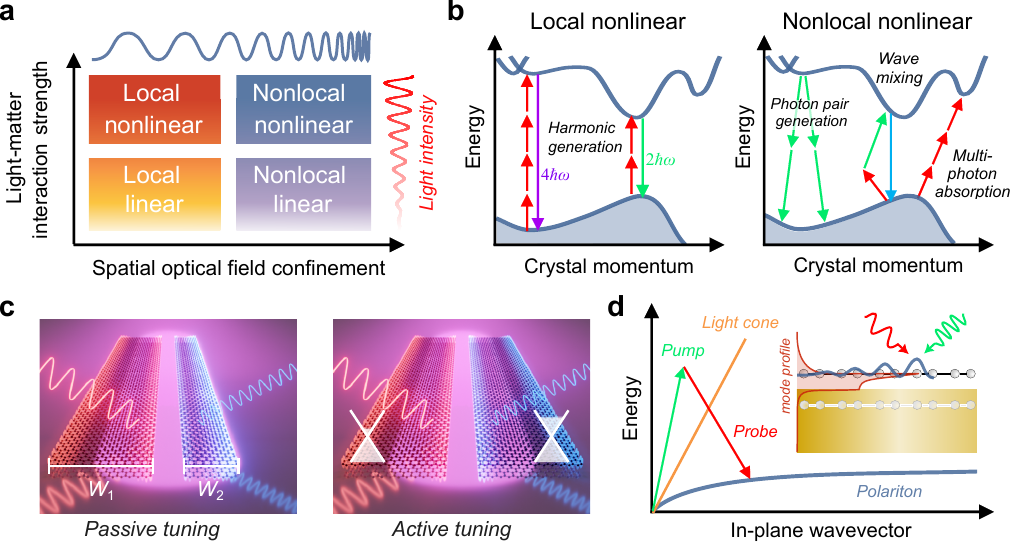}
    \caption{\pnl{a} Map of nonlocal and nonlinear light-matter interaction regimes according to interaction strength and field confinement.
    \pnl{b} Examples of local nonlinear optical processes, where electronic transitions are independent of photon momentum (left), contrasted with momentum-dependent nonlinear light-matter interactions (right). \pnl{c}~Illustration of plasmon-driven second harmonic generation in graphene nanoribbon heterostructures enabled by passive (nanostructuring, left) and active (charge carrier doping, right) mechanisms that shape near-fields to break inversion symmetry. \pnl{d}~Proposed scheme to excite highly confined polaritons, e.g., image polaritons in a 2D vdW material/metal heterostructure (inset), from free space through nonlinear wave mixing.}
    \label{fig:Jelver-Fig1}
\end{figure}

The regimes of nonlocal and nonlinear electrodynamics overlap when high-intensity light is confined on extremely small length scales, as summarized in Fig.~\ref{fig:Jelver-Fig1}\pnl{a}. This situation is uniquely embodied by polaritons---quasiparticles formed by the hybridization of light with collective dipole-carrying matter excitations---in two-dimensional (2D) atomically thin or van~der~Waals (vdW) materials. Here, reduced screening due to the vanishing thickness of 2D materials pushes the near-field confinement of their supported polaritons to extreme levels~\cite{Basov:2016, Alonso-Gonzalez:2017, Low:2017}, leading to enhanced spatial dispersion and an intrinsically nonlocal response. In tandem, the concentrated electromagnetic near-fields associated with 2D polaritons can effectively drive nonlinear light-matter interactions that exhibit an involved dependence on both frequency and spatial dispersion (see Fig.~\ref{fig:Jelver-Fig1}).

Although significant effort has been dedicated to studying the nonlocal behavior of polaritons in 2D materials (see Sec.~\ref{sec:Goncalves} in this Roadmap), such as through scanning near-field optical microscopy (SNOM) imaging~\cite{Li:2015, Lundeberg:2017, Ma:2018, Ni:2018}, most investigations have focused on the linear response regime. On the other hand, studies of nonlinear polariton-driven light-matter interactions in 2D materials have thus far been mainly restricted to graphene plasmonics~\cite{Cox:2014, Manzoni:2015, Constant:2016, Kundys:2018, Jiang:2019, AlonsoCalafell:2021, Li:2022b}. Owing to its linear electronic band structure and atomic thickness, graphene plasmon resonances that strongly confine electromagnetic near-fields can be induced and actively tuned by electrostatic gating. The Dirac cone electronic dispersion relation of graphene also ensures that its free charge carriers undergo highly anharmonic motion, leading to a large intrinsic nonlinear optical response. These properties combined have motivated intensive research efforts in \emph{nonlinear graphene plasmonics} to leverage the highly confined and actively tunable 2D plasmons for applications in nonlinear nanophotonics~\cite{Cox:2019}. However, graphene plasmon resonances are generally limited by achievable charge doping levels to terahertz (THz) and infrared (IR) frequencies, and, albeit longer-lived than their noble metal counterparts, suffer from substantial losses~\cite{Ni:2018}.

Beyond isolated 2D materials, nonlocal light-matter interactions become particularly important in vdW heterostructures. In particular, interfacing a vdW material with a metal film behaving as a perfect conductor gives rise to \emph{image polaritons}, formed by the self-interaction of 2D polaritons with their mirror image in the metal~\cite{Menabde:2022a}. The extreme confinement of the associated near-fields is characterized by an acoustic-like plasmon dispersion relation that extends well beyond the light cone, making these so-called acoustic plasmon modes extremely sensitive to nonlocal effects. In the case of metal/graphene hybrids, the deviation of their experimentally observed behavior from predictions based on a local description allows these polaritons to probe the nonlocal response of the combined system~\cite{Dias:2018}. Although the strong near-fields associated with image polaritons hold great promise for nonlinear optics, studies of these excitations have mainly focused on the linear response regime. 

\subsection*{Challenges and opportunities}

In centrosymmetric crystals, the leading contribution to the second-order nonlinear optical response is proportional to the gradient of the electromagnetic field that breaks inversion symmetry, and thus is inherently nonlocal~\cite{Manzoni:2015,Butet:2015}. While this symmetry dependence impedes second-order nonlinear optical effects (e.g., second-harmonic generation, sum- or difference-frequency generation, and optical rectification) in bulk crystals, it presents an opportunity to control the nonlinear response by engineering spatially-extended light-matter interactions. This concept has been theoretically explored in graphene plasmonics, where the resonant excitation of plasmons in inversion-symmetric ensembles of graphene nanoribbons is predicted to drive a large second-order nonlinear response when the inversion symmetry is broken by tuning the ribbon doping charge densities independently~\cite{Rasmussen:2023}, as illustrated in Fig.~\ref{fig:Jelver-Fig1}\pnl{c}.

Meanwhile, the extreme near-field confinement associated with image polaritons in graphene-metal heterostructures should give rise to an intense optical nonlinearity that is highly sensitive to nonlocal effects. Preliminary results have been achieved with heterostructures of graphene interfacing noble metal ribbons, where the broken translational symmetry of the metal ribbons facilitates in- and out-coupling to the far-field in experiments on harmonic generation~\cite{AlonsoCalafell:2021, AlonsoCalafell:2022}. However, the large volume of metal in these systems presents challenges when attempting to distinguish the metal and graphene contributions, and is problematic due to the large Ohmic losses in noble metals.

Nonlinear optical effects can also facilitate the excitation of propagating polaritons in extended systems. For instance, the difference frequency and wave vector produced by the second-order mixing of intense counter-propagating electromagnetic fields can be tailored to couple with the polariton dispersion extending beyond the light cone, as schematically represented in Fig.~\ref{fig:Jelver-Fig1}\pnl{d}. This concept has been experimentally demonstrated for graphene plasmons, enabling the excitation of plasmons by free space illumination~\cite{Constant:2016}, as well as the optoelectronic tuning of multiple waveguided plasmon modes sustained by a graphene sheet embedded in an optical waveguide~\cite{Li:2022b}.

The combined nonlocal and nonlinear interactions of waveguided polaritons offers exciting prospects for developing integrated quantum photonic devices. In particular, propagating quantized plasmons in graphene nanoribbons are predicted to exhibit unity-order coherent nonlinear interactions, including two-plasmon absorption~\cite{AlonsoCalafell:2019} and $\pi$-phase shifts induced by the collision of counter-propagating plasmons~\cite{Calajo:2023}. However, these schemes impose demanding structural and electronic quality requirements on 2D sheets and nanostructures, presenting a significant challenge to their experimental realization. Although recent advancements have enabled atomically precise control over the width and edge configurations of graphene nanoribbons~\cite{Wang:2021, Jiang:2023, Lyu:2024}, as well as in the fabrication of nanoscale transition metal dichalcogenide (TMD) ribbons and tubes~\cite{Fu:2023, Nakanishi:2023}, scalable and robust production methods for polariton-based devices remains a critical barrier. Addressing this challenge will require further development of scalable fabrication techniques to fully unlock the potential of 2D materials in optoelectronic applications.

\subsection*{Future developments to address challenges}

The unique optoelectronic properties of atomically thin materials open numerous avenues for applications in nanophotonics, while the strong field confinement of polaritons in 2D materials beyond graphene can help overcome current limitations in graphene plasmonics. For instance, the lack of crystal inversion symmetry in monolayer TMDs allows even-ordered nonlinear optical processes to occur in extended samples, which can be enhanced by excitons with valley-dependent degrees of freedom~\cite{Liu:2020}. Although the low group velocities exhibited by exciton-polaritons limit their ability to propagate optical signals, nonlinear and nonlocal light-matter interactions in engineered heterostructures could alleviate this constraint while leveraging the high exciton coherence for emerging technologies~\cite{Kurman:2018}. The directional dependence of propagating polaritons in anisotropic 2D materials---such as plasmons in doped phosphorene or in emerging vdW materials like palladium phosphoselenide (PdPSe) and arsenic trisulfide (As$_2$S$_3$)---offers additional spatial control that can influence polariton-driven nonlinear optical phenomena. Notably, As$_2$S$_3$ features a record high birefringence~\cite{Sortino:2025}, further expanding the design possibilities for advanced nanophotonic systems. 

The greatest challenges for the realization of devices that harness concerted nonlinear and nonlocal effects lie in the need for highly precise fabrication methods that produce defect-free and well-aligned nanostructures while ensuring seamless integration with existing technologies. On the theory side, advanced condensed matter and electrodynamic models that incorporate spatial dispersion in the nonlinear response functions of extended 2D materials must be developed to accurately capture nonlocal effects~\cite{Rostami:2017}. In mesoscopic 2D systems, atomistic simulations offer a powerful framework to simultaneously incorporate nonlocal and quantum-mechanical effects in the polariton-driven nonlinear optical response~\cite{Karimi:2021, Jelver:2023b}.

The creation of heterostructures that integrate graphene with other 2D materials offers another promising direction for next-generation photonic and optoelectronic technologies. Such heterostructures offer highly confined polaritons, exhibiting enhanced in-plane dispersion and strong associated near-fields, and can leverage the high intrinsic nonlinear response and optoelectronic tunability of graphene to explore nonlocal and nonlinear optical phenomena. Future research could focus on graphene-hBN heterostructures, where hybrid plasmon-phonon modes enable ultrafast nonlinear switching and low-loss waveguiding in the mid-IR~\cite{Golenic:2024}. The exploration of graphene-phosphorene heterostructures will assess their potential for reconfigurable and anisotropic photonics in broadband signal processing~\cite{Pogna:2024}. Furthermore, hybrid structures comprised of hyperbolic 2D materials open new paths to engineer polariton dispersion, unlocking multifunctional photonic platforms~\cite{Voronin:2024,Wang:2024}. Finally, advances in vertical integration and interlayer coupling could lead to more efficient nonlinear optical devices with dynamic control of both spectral and spatial responses.

\subsection*{Concluding remarks}

While low-dimensional materials are ubiquitous in nanophotonics, studies of spatially-extended nonlocal optical phenomena have largely been limited to the linear response regime. The intense near-field concentration associated with polaritons in 2D materials presents unique opportunities to explore the interplay of nonlocal and nonlinear light-matter interactions. To this end, novel theory frameworks that incorporate optical nonlocality in the response functions of 2D materials are needed to describe the complex energy- and momentum-dependent nonlinear interactions of their supported polaritons. On the experimental front, advanced nanofabrication techniques are required to create high-quality 2D material heterostructures capable of probing and leveraging spatial and frequency dispersion in integrated nonlinear photonic devices.

\section[Nonlocal chirality in twisted multilayer (S\'anchez S\'anchez \emph{et al.})]{Nonlocal chirality in twisted multilayer}
\label{sec:SanchezSanchez}
\author{Miguel S\'anchez S\'anchez\,\orcidlink{0000-0002-8067-2274}, Dionisios Margetis\,\orcidlink{0000-0001-9058-502X}, Guillermo~G\'omez-Santos\,\orcidlink{0000-0002-4048-1481} \& Tobias Stauber\,\orcidlink{0000-0003-0983-2420}}

\subsection*{Current status}

Etymologically, the word chirality comes from the Greek word for "hand" and is thus related to a three-dimensional object whose mirror image does not match the original by any translation or rotation in 3D space. In reverse, this means that purely two-dimensional systems cannot be genuinely chiral, since the object itself would serve as a mirror plane.

Chiral objects define two enantiomers which are identical in all scalar properties such as density or eigenfrequencies. Nevertheless, in their interaction with other chiral objects, opposite enantiomers can be detected. Circularly polarized (CP) light can thus distinguish left or right-handed samples by measuring the different absorption cross-sections of left and right CP light giving rise to circular dichroism (CD)~\cite{Barron:2004}. Since linearly polarized light is the superposition of left and right CP light, there will also be a rotation of the polarization plane. This has been observed in neutral twisted bilayer graphene (TBG) for certain resonant frequencies related to transitions around the van Hove singularities~\cite{Kim:2016}. 

Polarization rotation and CD can usually be observed by breaking either time-reversal or rotational/mirror symmetries \cite{Barron:1984}. In TBG, both symmetries are conserved, which makes the observation of CD remarkable since Maxwell's equations are reciprocal in unbiased two-dimensional systems and there should be no dependence on the direction of the CP light. However, there is a slight breaking of the two-dimensionality of the system as the two layers are separated by a distance $a=3.4$\AA. In fact, the chiral response in TBG is due to the {\em non-local} correlation between the current-density of layer 1 in say $x$-direction and the current density of layer 2 in the (transverse) $y$-direction:
%
\begin{align}\label{eq:SanchezSanchez-Eq1}
\chi^\mathrm{chiral}(\omega)=-\frac{i}{\hbar}\int_0^\infty dt\,\ e^{i\omega t}\left\langle\left[ j^1_x(t),j^{2}_{y}(0)\right]\right\rangle,
\end{align}
where $j^\ell_\nu(t)$ is the $\nu$-directed current operator ($\nu=x,y$) in layer $\ell$ in the interaction picture, and $\langle \cdot \rangle$ is the equilibrium average. Eq.~\eqref{eq:SanchezSanchez-Eq1} provides the average current in layer $1$ due to the electric field in layer $2$, and could be interpreted as the system response to the "gradient" of the electric field along the third dimension~\cite{Landau:1984}. 

The chiral response only depends indirectly on the distance between the two layers, $a$, via the equilibrium average. However, typical chiral observables are directly proportional to $a$ and thus depend on $a\chi^\mathrm{chiral}$~\cite{Stauber:2018}.

The nonlocal response between perpendicular current-directions of different layers is not the only requirement to observe CD, because for electron-hole symmetric systems the chiral response is identically zero at charge neutrality. Only due to small electron-hole symmetry breaking, electron-like and hole-like transitions are not compensated and the enhanced density of states around van Hove singularities can give rise to an observable CD~\cite{SuarezMorell:2017,Stauber:2023}. In fact, two main resonances are observed in Ref.~\cite{Kim:2016}, related to transitions around two different van~Hove singularities. 

\subsection*{Challenges and opportunities}

To calculate the CD, one needs to go beyond the dipole approximation $e^{i k z}\approx 1+i k z$, again emphasizing that chirality is a three-dimensional phenomenon. This introduces the dimensionless scale $ka$, where $a$ is the typical extension of the chiral object and $k=2\pi/\lambda$ the wave number of light. This scale is usually small making chiral effects hard to use in typical nanoscale devices. 

Field-theoretically, the chirality of an electromagnetic field with electric field $\boldsymbol{E}$ and magnetic field $\boldsymbol{B}$ in a dielectric medium with relative dielectric permittivity $\varepsilon$ and magnetic permeability $\mu$ is defined by
\begin{align}
\mathcal{C}=\frac{\varepsilon\varepsilon_0}{2}\boldsymbol{E}\cdot(\nabla\times\boldsymbol{E})+\frac{1}{2\mu\mu_0}\boldsymbol{B}\cdot(\nabla\times\boldsymbol{B}),
\end{align}
which can further be related to a chiral flux via a continuity equation~\cite{Barron:2004}. For CP light, the chirality is proportional to its frequency $\omega$ and intensity $|\boldsymbol{E}_0|^2$. For fixed wavelengths, the chirality can thus only be modified by changing the amplitude which can be achieved by confining light. In fact, decay into evanescent modes gives rise to fluorescence quenching which has been used to detect chiral molecules with the help of TBG~\cite{Moreno:2024}.

Also, surface-plasmon polaritons could enhance the chiral coupling and may enhance the chirality up to 4000 times the one of corresponding propagating CP light~\cite{Stauber:2020}. Cavities composed of TBG should thus offer tremendous opportunities as plasmon-induced near-field chiralities may enhance photocatalytic processes for enantiomer-selectivity. 

Quasi two-dimensional moir\'e structures can host plasmons which are inherently chiral~\cite{Stauber:2018}. Especially, many organic molecules are intrinsically chiral mainly due to the chemical structure of carbon with its four valence electrons. In fact, all amino acids are chiral as only glycine is achiral. Moreover, the RNA aptamer or Spiegelmer is chiral as well as hexahelicene and pentahelicene. All these may be synthesized in an enantiomer-selective way giving rise to new functionalities.

\begin{figure}[hb]
\centering
\includegraphics[width=0.7\columnwidth]{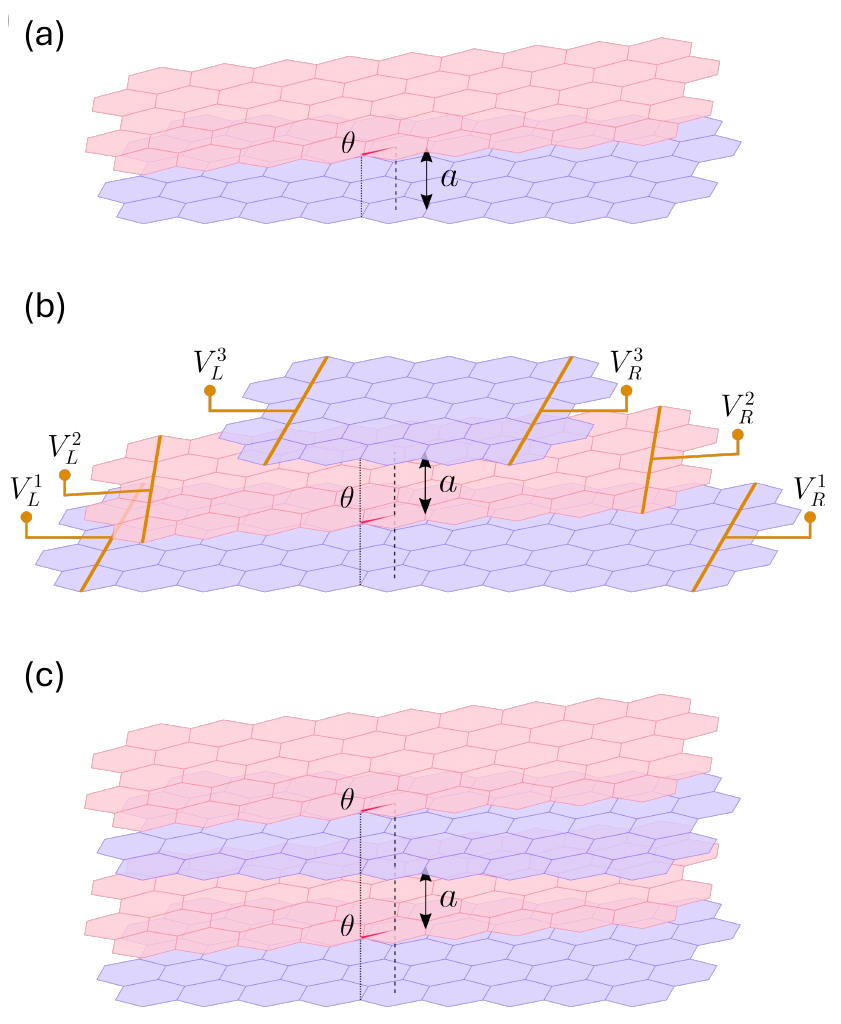}
\caption{Multilayer moir\'e systems with \pnl{a} $N=2$, \pnl{b} $N=3$, and \pnl{c} $N=4$ layers. The structures with alternating twist angles and an even layer number are inherently chiral, whereas those with an odd layer number are achiral as they display a mirror plane at the central layer.}
\label{fig:SanchezSanchez-Fig1}
\end{figure}

\subsection*{Future developments to address challenges}

The crucial requirement for the chiral response in moir\'e systems is the spatial separation between the layers and thus the nonlocal correlations between the sheet currents. However, increasing the layer distance would result in weakening the quantum mechanical coupling. Hence, there is a trade-off between optimal coupling and large layer separation, probably best met by transition-metal dichalcogenides such as tungsten diselenide (WSe$_2$).

Another route to enhance the inherent chirality is to stack several van~der~Waals materials on top of each other~\cite{Mannix:2022,Ji:2024}. The chiral response of multilayers with width $d$ should thus be increased by the geometric factor $d/a=N$,
where $N$ is the number of layers. However, the moir\'e unit cell grows exponentially with the number of layers even for commensurable twist angles, and for incommensurable structures it is difficult to obtain the electromagnetic response~\cite{Cances:2017}.

For the special case of alternating twisted multilayers, the analysis can be considerably simplified as it can be mapped to independent twisted bilayers and one single layer in case of an odd number of layers~\cite{Khalaf:2019}. In Fig.~\ref{fig:SanchezSanchez-Fig1}, the multilayer moir\'e systems with $N=2$, $N=3$, and $N=4$ layers are shown. The structures with an even layer number are inherently chiral, whereas those with an odd layer number are achiral as they display a mirror plane at the central layer. Nevertheless, even for $N=3$ the nonlocal "chiral" correlations that involve the perpendicular current densities of adjacent layers give rise to vertical gradients of the magnetic moment as a response to an electric in-plane field~\cite{Margetis:2024}. This layer nonlocality in nominally "achiral" systems could be detected by layer-discriminated contacts, as shown in Fig.~\ref{fig:SanchezSanchez-Fig1}\pnl{b} and also discussed in Ref.~\cite{Bahamon:2024}. 

Let us note that we can estimate the chiral response for moir\'e systems with \emph{arbitrary} twist angles between the layers using the decoupling procedure proposed in Ref.~\cite{Khalaf:2019}. For that, we rely on the fact that decreasing (increasing) twist angles can be mimicked by properly increasing (decreasing) the interlayer tunnel amplitude~\cite{Bistritzer:2011}. Neglecting effects of non-commensurability, we can thus map any multilayer moir\'e system onto a set of independent twisted bilayers and one single layer in case of an odd number of layers. 

\subsection*{Concluding Remarks}

The nonlocal coupling between sheet-currents flowing the perpendicular directions gives rise to a chiral response for neutral multilayer moir\'e systems without a mirror plane and slight electron-hole asymmetry. Also for multilayer moir\'e systems with a mirror plane, nonlocal "chiral" correlations can be detected by layer-discriminating contacts. These effects can be considerably enhanced by non-radiative (plasmonic) modes leading the path to chiral cavities composed of multilayer moir\'e systems with improved enantiomer-selective capabilities. The functionality can be further optimized by tuning the twist angles and doping levels.

\usection{\emph{Part IV} --- Engineered nonlocal responses in metamaterials and metasurfaces}
\label{roadmap:Part4}

\section[Strong spatial dispersion in metamaterials and metasurfaces (Tretyakov \& Simovski)]{Strong spatial dispersion in metamaterials and metasurfaces}

\label{sec:Tretyakov}

\author{Sergei Tretyakov\,\orcidlink{0000-0002-4738-9987} \& Constantin Simovski\,\orcidlink{0000-0003-4338-4713}}

\subsection*{Overview}

Metamaterials (MMs) are commonly defined as arrangements of artificial structural elements, designed to achieve advantageous and unusual electromagnetic properties~\cite{Capolino:2009}. This definition implies that metamaterials behave as effectively homogeneous media. This property distinguishes MMs from photonic crystals and other arrays that cannot be homogenized and modeled by constitutive relations for macroscopic fields. Structural elements of MMs, usually called \emph{meta-atoms} (MAs), as a rule, are small compared with the wavelength $\lambda$ and are arranged with the interparticle distance which is also much smaller than $\lambda$. However, some arrays of large or even infinitely extent MAs (thin wires in a dielectric ambient or thin metal layers alternating with thin dielectric layers), can be classified as MMs (\emph{hyperbolic} MMs) because these arrays can be homogenized if their MAs are arranged with a subwavelength period~\cite{Capolino:2009}. 

Among many parameters of meta-atoms that can be varied to design and optimize metamaterials and metasurfaces, the key role is played by their sizes and shapes. Shape effects on electromagnetic response of objects become significant when the object size becomes comparable (not negligibly small) with the wavelength. Dependence of the material response on the wavelength is called \emph{spatial dispersion}. SD means non-locality of the polarization response. Understanding SD and using its implications is in the core of MM science and technologies. The same refers to metasurfaces (MSs) which are the 2D analogues of MMs -- thin composite layers~\cite{Glybovski:2016}. Instead of bulk material parameters, MSs are characterized either by their surface impedance, or by effective electric, magnetic, and magneto-electric collective polarizabilities, or susceptibilities~\cite{Glybovski:2016}, which are in most cases spatially dispersive. SD, being a spatial non-locality, is manifested in both MMs and MSs as the dependence of the material parameters (bulk or surface ones) on the wavevector $\boldsymbol{k}$. In the absence of spatial dispersion, metasurfaces cannot realize such important functionalities as, for example, full absorption or reflection phase control, so that the possible applications are basically limited to frequency-selective surfaces and some polarizers. 

Weak SD is inherent to MMs composed by small complex-shaped MAs. It corresponds to weak $\boldsymbol{k}$-dependence of the permittivity tensor $\varepsilon(\omega,\boldsymbol{k})$ which obviously depends also on the frequency $\omega$. The first-order terms in the Taylor expansion of $\varepsilon(\omega,\boldsymbol{k})$, linear in $\boldsymbol{k}$, define such effects of weak SD as chirality and omega coupling~\cite{Capolino:2009,Serdyukov:2001}. Among the effects of the second-order SD the most important one is artificial magnetism~\cite{Capolino:2009,Serdyukov:2001}. Both first- and second-order SD effects are observed also in MSs~\cite{Glybovski:2016}. For MMs artificial magnetism results in the relative permeability being different from unity in the absence of natural magnetics. For MSs this effect results in the nonzero magnetic surface susceptibility and (in the alternative model) in magnetic collective polarizability. Chirality in both MMs and MSs leads to polarization transformation and conversion~\cite{Capolino:2009,Glybovski:2016,Serdyukov:2001}. Omega coupling results in asymmetry of reflection (for both MS and MM layers in homogeneous space)~\cite{Serdyukov:2001}. More information about weak SD in MMs can be found in Refs.~\cite{Capolino:2009,Serdyukov:2001,Simovski:2018}. Weak SD in MSs is reviewed in Refs.~\cite{Asadchy:2018,Asadchy:2017}. Strong SD corresponds to the case when the $\boldsymbol{k}$-dependence of $\varepsilon(\omega,\boldsymbol{k})$ is resonant~\cite{Simovski:2012}, or its Taylor expansion poorly converges, including too many terms~\cite{Capolino:2009}. Strong SD is inherent to hyperbolic MMs~\cite{Capolino:2009,Simovski:2012,Shekhar:2014}. MSs with strong SD are specially engineered so that their nonlocal response allows the manipulation of electromagnetic field spatial distributions and desired functionalities~\cite{Asadchy:2018,Asadchy:2017}. 

\subsection*{Current status of research}

Currently, physics and applications of MMs and MSs with weak SD is a mature field. In particular, there are well-established means to create and engineer resonant chirality for control and optimization of optical activity dichroism. In addition to the use of chiral shapes, these means include also the use of arrays of negligibly thin composite layers that are non-chiral in the geometric sense (\emph{pseudochirality}~\cite{Serdyukov:2001,Asadchy:2018,Asadchy:2017}, also called \emph{extrinsic chirality}). If an electrically thin metasurface does not contain natural magnetic materials (like ferrites, for example), the only mean to use it for full control of reflection and transmission is to invoke spatial dispersion. This allows us to complement electric polarization of thin arrays with artificial magnetic response of spatially dispersive structures and realize non-reflecting (Huygens’) and more general arrays for control of reflection and transmission~\cite{Glybovski:2016,Asadchy:2018,Asadchy:2017}. Omega coupling plays the key role in realizing perfect anomalous refraction in metasurfaces~\cite{Asadchy:2018}. Also, all electrically thin absorbers covering highly reflecting bodies are spatially dispersive (omega) layers. 
Strong SD effects in MMs and MSs have been also studied recently. It was found that strong SD in MMs offers long-distance transmission of subwavelength-resolution images, super-Planckian thermal emission and absorption and super-Planckian radiative heat transfer (see in Refs.~\cite{Simovski:2012,Shekhar:2014}). For MSs, strong SD grants strong coupling of propagating and evanescent waves and offers a unique opportunity to engineer diffraction patterns~\cite{Asadchy:2017}. Furthermore, full control of reflection to realize anomalous reflection or focusing by MSs requires strong SD. Most recent studies begin exploring new electromagnetic phenomena in time-varying media with spatial dispersion. For example, phenomena at time interfaces in chiral media were studied in Ref.~\cite{Mostafa:2023}, and effects in time-varying media with omega coupling are discussed in Ref.~\cite{Mostafa:2024}. 
 
\subsection*{Challenges and opportunities}

To design and optimize metamaterials and metasurfaces, appropriate modeling tools are needed. As already discussed above, if SD is weak, there are well developed effective-medium models based on the bianisotropic material relations. However, even the second-order effects are modeled by these relations only partially, in form of artificial magnetism. Other second-order terms as well as all higher-order terms enter the material equations with spatial derivatives of the macroscopic fields~\cite{Goffi:2021,Venkitakrishnan:2023}. Solving boundary problems for MMs and MSs with strong SD is even more difficult. In this case, one needs to derive additional boundary conditions (ABCs) for each particular MM microstructure~\cite{Simovski:2018}. In Ref.~\cite{Goffi:2021}, ABCs were derived within a higher-order effective-medium model. In Refs.~\cite{Maslovski:2010,Gorlach:2020}, ABCs were derived for hyperbolic MMs. For media with strong SD difficulties arise not only in solving boundary problems. While for low-loss bianisotropic media with artificial magnetism there is a general formula for the energy density (see e.g. in Refs.~\cite{Capolino:2009,Serdyukov:2001}), for media and MSs with strong SD such general formulas are absent. In Ref.~\cite{Silveirinha:2012}, the energy density in wire media (one of two known classes of hyperbolic MMs) was deduced from microscopic consideration (telegrapher equations), and the results were shown to be consistent with the effective-medium model ($\boldsymbol{k}$-dependent permittivity tensor). The main goals of introducing effective medium models are to enable simple solutions of extremely involved electromagnetic problems of fields in complex composite media and, most importantly, get physical insights into the computed or measured results. Unfortunately, higher-order effective material models become more and more complex, so that both goals become difficult to reach. In view of these difficulties, it appears that direct numerical optimization of metamaterial and metasurface structures remains the only available tool for development of advanced metadevices.
Significant efforts have been devoted recently to development of tunable and reconfigurable MSs. One of the main envisaged applications is in future wireless communication technologies, where tunable MS can be potentially used to optimize propagation channels in wireless communications. As discussed above, only spatially dispersive MSs can offer full control of anomalous reflection and transmission, which means that we need to develop effective and practical means to tune non-local responses of MAs and their arrays or clusters electrically or optically.  

\subsection*{Future development to address challenges}

At present, research efforts in this field are focused on the design and optimization of MSs for control of reflection and transmission, including focusing. In the microwave and terahertz ranges the main envisaged applications are in future wireless communications (reconfigurable intelligent surfaces). In optics, researchers envisage, for example, extremely thin MS lenses. As discussed above, for any advanced control of waves, MSs must exhibit SD. Moreover, realizations of the optimal performance often demand specially engineered strong SD. For MSs, this usually means careful control of excitation and surface distribution of evanescent (surface) waves. Their fields exhibit fast variations at the wavelength scale, which means that MA sizes should be smaller than the wavelength scale. To realize such designs, technologies to create small, simple, and cheap MSs need to be developed. To make reconfigurable and adaptable devices, the properties of these MAs should be electrically or optically tunable in such ways that will allow optimization of meta-atom couplings. To optimize parameters, it is necessary to further advance numerical optimization methods, because fully reconfigurable devices cannot be realized as periodic arrays, so that a global optimization of huge sets of interacting meta-atoms will be needed. One possible route can be creation of standardized units (for each frequency range) and pre-computation of basic electromagnetic parameters (such as impedance matrices and effective antenna heights), after which fast purely arithmetic optimizations become possible~\cite{Vuyyuru:2023}. Perhaps, some lessons from recent fast developments of artificial intelligence can be used in this field. Here, concerted efforts of experts in electromagnetics and computer science will be needed. 
To advance understanding and use of time-modulated (4D) MMs and MSs, very significant theoretic research is needed. So far, only initial studies of time-varying media have been made, mostly completely neglecting frequency and spatial dispersion. Studies of time-varying spatially dispersive media considered only weak spatial dispersion, where it was possible to use the bianisotropic effective medium model~\cite{Mostafa:2024}. Moreover, always-present frequency dispersion of spatially dispersive materials was so far modeled only in the low-frequency approximation. Furthermore, no experimental studies of spatially dispersive media whose properties quickly change in time have been made so far. Here, we would like to remind that a huge majority of practically relevant MSs have significant SD! Many challenges are expected on this way: First, time-varying spatially dispersive media need to be studied in time domain. While the frequency-domain methods are well developed, this is not the case of the time-domain ones. For example, to solve for fields at time interfaces, time boundary conditions are needed, and to establish them, one needs to consider the microstructures of spatially dispersive materials. This means that no universal boundary condition (like continuity of fields in the frequency-domain theories) can be found. 
In summary, we expect seeing interesting and very much needed results of researchers working in this area.

\section[Nonlocal hyperbolic metamaterials and metasurfaces (Pakniyat \& Gómez-Díaz)]{Nonlocal hyperbolic metamaterials and metasurfaces}

\label{sec:Pakniyat}

\author{Samaneh Pakniyat\,\orcidlink{0000-0001-9621-4676} \& J. Sebastián Gómez-Díaz\,\orcidlink{0000-0002-6852-8317}}

\subsection*{Overview}

Hyperbolic media, including hyperbolic metamaterials (HMMs) and metasurfaces (HMSs), have driven groundbreaking advancements in photonics as well as in related fields like acoustics and thermodynamics. Hyperbolic media displays remarkable electromagnetic properties arising from its extreme anisotropy, hyperbolic dispersion, and field confinement. Such exciting properties also entail that the media response depends on the properties of the waves that it is interacting with, leading to larger nonlocal effects than the ones found in other media. Nonlocal responses grow stronger with the confinement of the supported modes and appear due to the granularity of the artificial media and the spatial dispersion of the composing materials – for instance, the finite velocity of the electrons flowing in metals. To date, nonlocality has been mostly considered a hindrance in hyperbolic media, and thus it has not been exploited to enable new functionalities or to enhance device performance. This roadmap explores the interplay of nonlocality and hyperbolic dispersion, highlighting both challenges and untapped opportunities. The limitations of conventional local effective medium theory are reviewed, emphasizing the need for sophisticated numerical tools able to solve general electromagnetic structures composed of nonlocal metallic elements. These tools will guide the development of hyperbolic devices in which nonlocality will be tailored to become an extra degree of freedom for wave control. Future directions include merging drift-current nonreciprocal plasmonic, moiré physics, and reconfigurability to realize tunable, nonlocal, and hyperbolic platforms with exciting applications in multidisciplinary fields, ranging from sensing, signal processing and computing to acoustics and thermal management. 

\subsection*{Current status}

\begin{figure}[htb]
    \centering
    \includegraphics[width=0.99\linewidth]{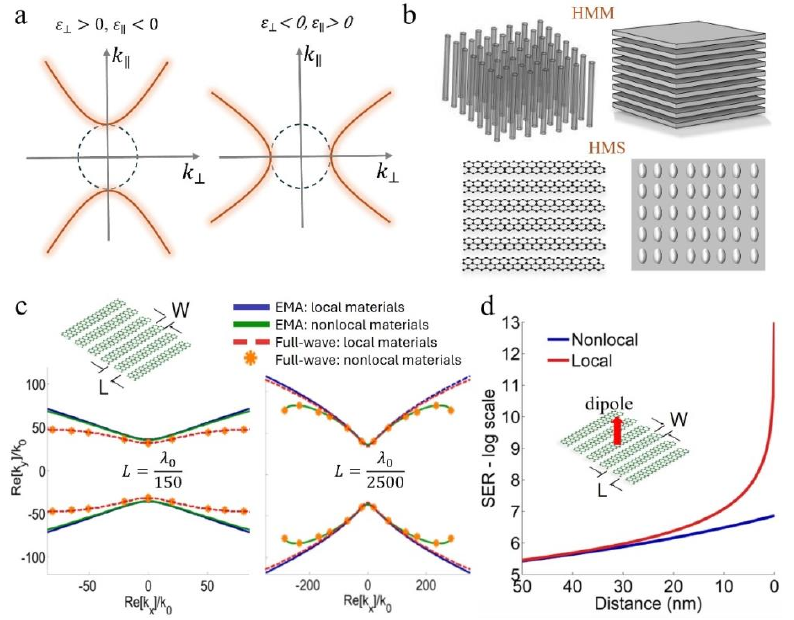}
    \caption{Hyperbolic media and nonlocal effects. \pnl{a} Isofrequency contour (i.e., dispersion relation at a fixed $\omega$) found in a local, electric, and hyperbolic media type I (left) and II (right). $\varepsilon_\perp(k_\perp)$ and $\varepsilon_\parallel (k_\parallel)$ are the out-of-plane and in-plane effective permittivity (supported mode's wave-vector) components. \pnl{b} Examples of hyperbolic metamaterials (top) and metasurfaces (bottom) constructed by properly arranging subwavelength elements. \pnl{c} Isofrequency contour of the modes supported by an array of graphene strips that behaves as a hyperbolic surface~\cite{Correas-Serrano:2015}. Results are calculated using local EMT and rigorous full-wave numerical simulations considering local and nonlocal models of graphene. Left panel shows how granularity becomes dominant when the periodicity of the unit-cells is not much smaller than the wavelength of the supported modes. Right panel illustrates how the finite velocity of electrons traveling in graphene dominates the response when granularity is negligible. \pnl{d} Spontaneous emission rate of a $z$-oriented infinitesimal dipole versus its distance above the graphene-based hyperbolic surface~\cite{Correas-Serrano:2015}. Reproduced panels \pnl{c} and \pnl{d} with permission from Ref.~\cite{Correas-Serrano:2015} (Copyright~\textcopyright~2015 Optica Publishing Group).}
    \label{fig:Pakniyat-Fig1}
\end{figure}

Hyperbolic media~\cite{Lobet:2023,Guo:2020,Vazquez-Lozano:2024} exhibits extreme anisotropy enabled by a material response that changes sign as a function of the polarization of the electric and/or magnetic field and offers unique hyperbolic dispersion together with an ideally infinite local density of states and wave confinement [Fig.~\ref{fig:Pakniyat-Fig1}\pnl{a}]. Such features have driven advancements in ultra-high-resolution imaging, negative refraction, beam shaping, optical sensing, and spontaneous or coherent emission enhancement, among other applications. The electromagnetic response of hyperbolic media is limited in practice by two mechanisms, namely the presence of loss and nonlocality, which close their otherwise open hyperbolic dispersion relation and impose a wavenumber cutoff to the supported modes. The lossy nature of frequency-dispersive materials is a direct consequence of causality, and thus unavoidable in passive devices~\cite{Landau:1984}. Nonlocality appears due to the granularity of artificial structures and to the intrinsic spatial dispersion of the materials that compose them~\cite{Yan:2012a,Correas-Serrano:2015}, for instance, the finite velocity of electrons moving within metallic elements. Nonlocal media is usually described as a wavenumber dependent and thus the material response depends on the properties of the waves that are interacting with it. In addition to capturing the fundamental response of hyperbolic media and imposing upper limits to the field confinement and local density of states, nonlocal effects must be accounted for in the design of realistic hyperbolic devices and exploited to enhance their performance. 

Hyperbolic responses appear naturally in certain materials and crystals, such as silicon nitrate, hexagonal boron nitride, $\alpha$-phase molybdenum trioxide (MoO$_3$) bilayers, or $\beta$-phase gallium trioxide (Ga$_2$O$_3$) crystals, among others~\cite{Narimanov:2015,Hu:2020b} and, as illustrated in Fig.~\ref{fig:Pakniyat-Fig1}\pnl{b}, can also be implemented by arranging subwavelength unit-cells taking advantage of metamaterial concepts~\cite{Lobet:2023,Guo:2020}. Hyperbolic media is usually characterized using local effective medium theory (EMT) based on the Maxwell-Garnett approximation coupled with a linear response model of the composing materials~\cite{Lobet:2023,Guo:2020}. EMT describes the hyperbolic structure as a homogeneous media with effective parameters and assumes that the structural elements are much smaller than the operation wavelength.

Intriguing, the response of hyperbolic media goes well beyond the predictions made by EMT. To illustrate nonlocal mechanisms, let us consider that a hyperbolic surface composed of graphene strips with periodicity $L$ supports electromagnetic modes characterized with a momentum $\boldsymbol{k}$ and a wavelength $\lambda$. First, EMT remains valid only when the artificial structure can be considered homogeneous, i.e., when the condition $L\ll\lambda$ is satisfied. As the supported modes becomes more confined, their momenta $\boldsymbol{k}$ increases and their wavelength $\lambda$ decreases to a point in which they are comparable to the strips size ($L\approx \lambda$). There, as shown in the left panel of Fig.~\ref{fig:Pakniyat-Fig1}\pnl{c}, granularity becomes the main nonlocal mechanism that closes the isofrequency contour of the supported modes and imposes a cutoff wavenumber at $k_c\approx \pi/L$. Second, the response of metallic materials is usually modeled with local models (i.e., Drude local response approximation) that do not account for the finite speed of electrons. As the wavenumber of the supported modes increases, electrons might not be able to travel fast enough to follow spatial fields variations. This imposes a wavenumber cutoff at $k_c\approx (c/v_F) k_0$, where $v_F$ is the Fermi velocity of electrons in the metal, $c$ is the speed of light, and $k_0$ is the free-space wavenumber. The right panel of Fig.~\ref{fig:Pakniyat-Fig1}\pnl{c} illustrates this scenario, confirming that the intrinsic nonlocal response of graphene becomes dominant and closes the isofrequency contour when granularity is negligible. In a more general case, the interplay between these two nonlocal mechanisms will tailor the media response. In general, EMT is just an approximation and thus local predictions regarding infinite local density of states, wave confinement, and spontaneous emission rate are incorrect. Figure~\ref{fig:Pakniyat-Fig1}\pnl{d} illustrates how the presence of nonlocality limits the maximum spontaneous emission rate of an emitter located near the graphene-based hyperbolic surface. It should also be mentioned that EMT might not be accurate to describe certain type of artificial hyperbolic media, for instance in terms of predicting the number of supported modes and their properties or accounting for polarization-dependent inter-element coupling among the unit-cells that compose the structure.

\subsection*{Challenges and opportunities}

Accounting for nonlocal effects in hyperbolic media is not an easy task. To accurately consider the influence of the elements that compose the media, the most straightforward approach is to solve Maxwell's equations using full-wave simulations that account for both geometrical details and the optical response of the materials involved. This powerful approach faces important complications, including \emph{i)} inability to incorporate nonlocal material behavior in most scenarios; \emph{ii)} lack of physical insight and analytical models that guide device design; and \emph{iii)} requirement of large computational resources. For these reasons, dedicated formulations have been developed to model canonical nonlocal hyperbolic structures, including multilayered~\cite{Orlov:2011,Li:2016a} and wire~\cite{Hanson:2013} bulk configurations and planar metasurfaces~\cite{Guo:2020,Li:2020}. Even though these efficient approaches lead to closed-form expressions in certain scenarios, they are valid only under well-defined operation conditions and cannot be generalized for arbitrary structures. 

Another challenge is to develop nonlocal optical theories to accurately describe the material response. These wavenumber-dependent models include phenomena such as quantum effects, electronic transitions, and induced charge diffusion kinetics, among other, and have been applied to metals~\cite{Mortensen:2014}, and particularized for materials such as graphene~\cite{Lovat:2013}, and black-phosphorus~\cite{Correas-Serrano:2016}. In this context, the hydrodynamic Drude model within the Thomas--Fermi approximation~\cite{Yan:2012a,Pakniyat:2022} is a common nonlocal approach to characterize thin metals by employing a combination of longitudinal and transverse waves that account for the electron dynamics. 

An exciting opportunity in the field of nonlocal hyperbolic media is the development of theoretical frameworks able to simultaneously account for the media granularity and the intrinsic nonlocal response of its constitutive materials. The main challenge lies in incorporating complex wavenumber-dependent material responses within electromagnetic approaches. Even though some studies have already been put forward in this area~\cite{Correas-Serrano:2015,Correas-Serrano:2019} tools able to predict, model, and tailor hyperbolic wave and plasmonic propagation in realistic structures would drastically accelerate the application of this technology in practice. 
To date, hyperbolic devices have been mostly implemented by relegating nonlocality to a mere hindrance. This strategy not only misses the rich potential for tailored wave-control offered by nonlocality in hyperbolic media but leads to complex implementations with sub-optimal responses. 

\begin{figure}[htb]
    \centering
    \includegraphics[width=0.99\linewidth]{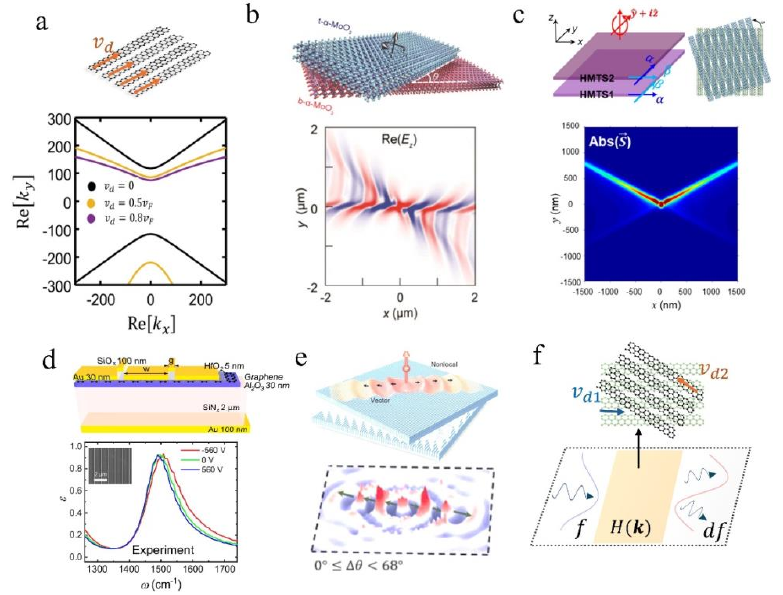}
    \caption{Applications of nonlocal hyperbolic metasurfaces. \pnl{a} Graphene-based drift-biased metasurfaces~\cite{Correas-Serrano:2019}. Drifting electrons break the symmetry of the supported modes, leading to nonreciprocal hyperbolic plasmons and immunity to backscattering. \pnl{b} Stacked moiré van der Waals crystals (top) and electric field distribution of a supported mode~\cite{Zheng:2020}. Reprinted (adapted) with permission from Ref.~\cite{Zheng:2020} (Copyright~\textcopyright~2020 American Chemical Society). \pnl{c} Moiré hyperbolic metasurfaces coupled through evanescent fields~\cite{Hu:2020a}. Spin-orbit interactions allow a rotating dipole to excite a directional beam. Reprinted (adapted) with permission from Ref.~\cite{Hu:2020a} (Copyright~\textcopyright~2020 American Chemical Society). \pnl{d} Thermal emission with hyperbolic devices. Reproduced with permission from Ref.~\cite{Siegel:2024} (Copyright~\textcopyright~2024 Nature Springer). \pnl{e} Nonlocal acoustic hyperbolic metasurfaces. Reproduced with permission from Ref.~\cite{Han:2024} (Copyright~\textcopyright~2024 Wiley). \pnl{f} In-plane optical computing with tunable nonlocal hyperbolic metasurfaces.}
    \label{fig:Pakniyat-Fig2}
\end{figure}

\subsection*{Future developments to address challenges}

Reconfigurable hyperbolic media has already been demonstrated in the local regime using phase-change and correlated materials, revealing exciting potential for programmable nanophotonic applications~\cite{Aghamiri:2022}. Tunable nonlocality has been explored in the context of current-driven plasmonics, implemented using materials such as graphene~\cite{Morgado:2017} or ultrathin metals~\cite{Bliokh:2018}, in which nonlocality appears because the supported waves effectively experience a different media when propagating along or against the drifting electrons. Exploiting this phenomena, drift-biased hyperbolic metasurfaces have been theoretically proposed~\cite{Correas-Serrano:2019}. In these structures, the interplay between nonlocality and hyperbolic dispersion breaks the symmetry of the supported modes in the $k$-space [Fig.~\ref{fig:Pakniyat-Fig2}\pnl{a}], leading to strong isolation and immunity to backscattering. The experimental demonstration of such platform would provide nonlocal, tunable, and nonreciprocal responses useful to control, collimate, and route waves with a performance well beyond to the one found in common hyperbolic structures.

Recent developments in the field of twistronics and moiré physics have also enriched the possibilities offered by hyperbolic media. For instance, it is possible to manipulate the twist angle between stacked van der Waals crystal [Fig.~\ref{fig:Pakniyat-Fig2}\pnl{b}] to control their photonic and electronic properties and engineer hyperbolic responses~\cite{Zheng:2020}. At the mesoscopic scale, such response can be obtained by rotating hyperbolic metasurfaces that are located close to each other and evanescently coupled [Fig.~\ref{fig:Pakniyat-Fig2}\pnl{c}]~\cite{Zheng:2020}. Even though these platforms have been considered mostly in the local regime, nonlocality controls and bounds their fundamental response. Merging drift-current media with moiré physics will open new and unexpected opportunities for tunable, nonlocal, nonreciprocal, and hyperbolic wave control in which tailored nonlocal responses can be spatially and temporally manipulated to enable applications in sensing, communication systems, and computing, among others. Additionally, it would also open the door to the generation of high $k$-waves~\cite{Wang:2023} and enhanced nonlinear responses~\cite{Kolkowski:2023} -- enabled by relaxed phase-matching conditions provided by nonlocality. These developments are inherently multidisciplinary and can be translated to other fields. For instance, thermally engineered devices [Fig.~\ref{fig:Pakniyat-Fig2}\pnl{d}] and nonlocal acoustic moiré hyperbolic metasurfaces [Fig.~\ref{fig:Pakniyat-Fig2}\pnl{e}] have recently been demonstrated \cite{Siegel:2024,Han:2024,Nolen:2024}. Arguably, the next step in these areas is the tunable manipulation of nonlocality in both space and time, which would significantly expand the possibilities to radiate, receive, and process acoustic and thermal waves.

In a related context, nonlocal metasurfaces have become transformative in the field of optical computing, leveraging their ability to manipulate optical wavefronts in the momentum domain with unparalleled efficiency~\cite{Shastri:2023}. Nonlocal metasurfaces have been mostly designed to interact with light propagating in free-space and to provide tailored transfer functions in the momentum domain (i.e., different responses versus the angle of incidence of the incoming waves) that perform complex mathematical operations such as spatial differentiation, edge detection, or Fourier filtering. A paradigm shift would be to exploit high $k$-waves (evanescent spectrum) for optical computing. In such platform, illustrated in Fig.~\ref{fig:Pakniyat-Fig2}\pnl{f}, input signals will be provided by near-field emitters such as quantum dots or by capturing light coming from the far-field using scatters and/or gratings. Nonlocal hyperbolic media provides unprecedented possibilities to tailor their in-plane transfer function versus the incident wavenumber. This paradigm offers new and rich possibilities for optical computing and multi-functional devices enabled by nonlocal processing of high $k$-waves that are not accessible in metasurfaces interacting with light propagating in free-space. Many technological challenges remain to be solved before this technology becomes a reality, mostly in terms of demonstrating tunable and nonlocal hyperbolic field control. Moving beyond, the integration of nonlocal hyperbolic media with optical neural networks and neuromorphic computing could pave the wave for advanced and ultra-efficient signal processing capabilities over miniaturized devices.

\subsection*{Concluding remarks}

While nonlocality has usually imposed constraints on features such as wave propagation and the local density of states offered by hyperbolic media, it also provides an opportunity to develop innovative strategies for tailored wave control. By advancing theoretical frameworks, improving nanofabrication techniques, and leveraging emerging concepts like drift-current media and moiré physics, the interplay between nonlocality and hyperbolic dispersion can be exploited to unlock the full potential of hyperbolic media and construct nonreciprocal and programmable transfer functions in the momentum domain. These advancements will address existing challenges but will also pave the way for novel applications in a variety of fields, with emphasis on optical computation using high $k$-waves, sensing, acoustics, and thermal engineering. Exploiting nonlocality as an extra degree of freedom in hyperbolic designs marks a critical step toward realizing next-generation photonic and multidisciplinary technologies with unprecedented capabilities.

\section[Nonlocal effects in transdimensional plasmonics (Bondarev \emph{et al.})]{Nonlocal effects in transdimensional plasmonics}

\label{sec:Bondarev}

\author{Igor V. Bondarev\,\orcidlink{0000-0003-0739-210X}, Svend-Age Biehs\,\orcidlink{0000-0002-5101-191X}, Alexandra~Boltasseva\,\orcidlink{0000-0001-8905-2605} \& Vladimir M. Shalaev\,\orcidlink{0000-0001-8976-1102}}

\subsection*{Overview}

Plasmonic transdimensional (TD) materials are atomically thin metal, semimetal or doped semiconductor films of precisely controlled countable number of monolayers~\cite{Boltasseva:2019}. Due to current progress in nanofabrication techniques~\cite{Pan:2024,Shah:2022,Abd-El-Fattah:2019,Das:2024}, such materials can be reproducibly grown and offer high tailorability of their electronic and optical properties not only by altering their chemical and/or electronic composition (stoichiometry, doping) or strain but also by merely varying their thickness. While ultrathin films have been extensively studied, the optical properties of TD materials remain somewhat underexplored. So far, the focus has largely been on either purely two-dimensional (2D) structures including metal-dielectric interfaces and novel 2D materials~\cite{Mak:2016}, or on conventional bulk materials, where the dimensionality and composition are the primary factors influencing the optoelectronic response. Recently it was proposed that ultrathin TD nanostructures provide a new regime‒transdimensional, in between 3D and 2D, turning into 2D as the film thickness tends to zero~\cite{Shah:2022,Abd-El-Fattah:2019,Das:2024}. In this regime, the strong vertical quantum confinement makes the linear electromagnetic (EM) response of the film nonlocal (spatially dispersive), and the degree of nonlocality can be controlled by the film thickness~\cite{Bondarev:2017,Bondarev:2020,Bondarev:2023a}. This makes plasmonic TD materials indispensable for studies of the nonlocal light-matter interactions at the nanoscale~\cite{Bondarev:2020,Bondarev:2023a,Zundel:2022,Pugh:2024,Biehs:2023,Salihoglu:2023,Bondarev:2023b,Rodriguez-Lopez:2024}, where they exhibit extraordinary tailorability including the capabilities of active tuning of their EM response and thus enabling new and unique light-matter coupling phenomena~\cite{Rivera:2016}. In addition to strong dependencies on structural parameters such as the type of substrate/superstrate and strain~\cite{Qi:2023}, plasmonic properties of ultrathin TD structures show high sensitivity to external optical and electrical stimuli~\cite{Lu:2017,Liu:2017,Karaman:2024}, which could facilitate the realization of dynamically tunable ultrathin plasmonic devices. The remarkable opportunities for tuning their highly confined in-plane plasma modes open access to novel quantum, nonlocal and nonlinear optical effects~\cite{Pugh:2024,Biehs:2023,Salihoglu:2023,Bondarev:2023b,Rodriguez-Lopez:2024,Rivera:2016,Qi:2023,Lu:2017,Liu:2017,Karaman:2024} in both the near-infrared and the visible ranges~\cite{Lu:2017,Biehs:2024}. Over the past years, TD plasmonics has grown into a promising research direction aiming to greatly enhance plasmonic functionalities through unparalleled tailorability of confinement-induced nonlocal EM response effects~\cite{Das:2024,Boltasseva:2025}.

\subsection*{Current status}

\begin{figure}[hb]
    \centering
    \includegraphics[width=0.8\linewidth]{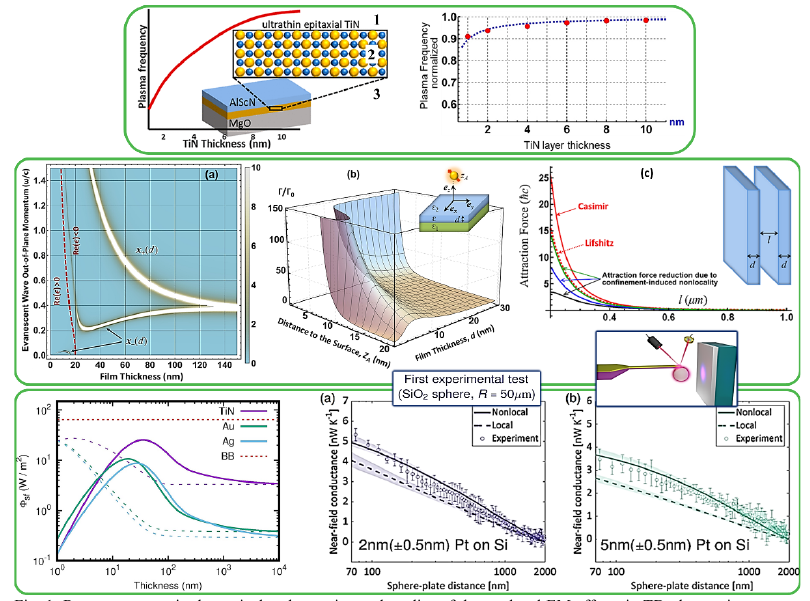}
    \caption{Recent progress in theoretical and experimental studies of the nonlocal EM effects in TD plasmonic systems. Top: Experimental demonstration of the theoretically predicted thickness dependence of the Drude plasma frequency in TD TiN films. Reprinted (adapted) with permission from Ref.~\cite{Shah:2022} (Copyright~\textcopyright~2022 American Chemical Society). Middle: \pnl{a}~Thickness-dependent epsilon-near-zero modes of TiN film with variable thickness. Reproduced with permission from Ref.~\cite{Bondarev:2020} (Copyright~\textcopyright~2022 American Physical Society); \pnl{b}~near-surface distance-/thickness-dependent spontaneous emission rate relative to vacuum~\cite{Bondarev:2023a} for a quantum dipole emitter coupled to the TD film plasma modes in \pnl{a}. Reproduced with permission from Ref.~\cite{Bondarev:2023a} (Copyright~\textcopyright~2023 Wiley); \pnl{c} Casimir attraction force calculated using the nonlocal KR model for two free-standing metallic slabs of 10 (black), 20 (blue) and 200\,nm (green) in thickness. Reproduced with permission from Ref.~\cite{Bondarev:2023b} (Copyright~\textcopyright~2023 Royal Society of Chemistry). Bottom left: Far-field heat fluxes predicted by local Drude (dashed lines) and nonlocal KR (solid lines) model for various TD metals as compared to black body (BB) radiation~\cite{Biehs:2023}. Bottom \pnl{a} and \pnl{b}: Near-field heat conductance measured for ultrathin Pt films (inset) as compared to local Drude and nonlocal KR model predictions~\cite{Salihoglu:2023}.}
    \label{fig:Bondarev-Fig1}
\end{figure}

The nonlocal optical properties of TD plasmonic films can be understood in terms of the confinement-induced nonlocal EM theory built using the Keldysh--Rytova (KR) electron interaction potential~\cite{Bondarev:2017,Bondarev:2020}. In optically dense thin films of metals, semimetals and doped semiconductors, the electrostatic Coulomb field produced by vertically confined remote charge carriers outside of their confinement region starts playing a perceptible role with the confinement size reduction~\cite{Keldysh:1979,Rytova:1967}. The Coulomb interaction of such confined charges is typically stronger than that in a homogeneous medium with the same dielectric permittivity. This is due to the increased field contribution from outside dielectric environment with lower dielectric permittivity. In the case of TD plasmonic films in a typical "sandwich" geometry [Fig.~\ref{fig:Bondarev-Fig1}\pnl{top} with titanium nitride (TiN) between magnesium oxide (MgO) and aluminium scandium nitride (AlScN)], where the film of thickness $d$ in region 2 is surrounded by semi-infinite dielectrics of constant permittivities $\varepsilon_1$ (top region 1) and $\varepsilon_3$ (bottom region 3), this turns the electron-electron Coulomb repulsion potential in the film into a much stronger $d$-dependent KR interaction potential~\cite{Keldysh:1979,Rytova:1967}. This in turn leads to the $d$-dependent electron plasma oscillation frequency [see Fig.~\ref{fig:Bondarev-Fig1}\pnl{top}], resulting in the nonlocal Drude-like, in-plane EM response of the TD film~\cite{Bondarev:2017}:
\begin{equation}
        \varepsilon_2(\omega,k)= \varepsilon_b \left[1-\frac{\omega_p^2(k)}{\omega(\omega+i \Gamma_D)}\right],\quad \omega_p(k) = \frac{\omega_p^\textrm{3D}}{\sqrt{1+1/(\tilde{\varepsilon} k d)}},\quad \tilde{\varepsilon}=\frac{\varepsilon_b}{\varepsilon_1+\varepsilon_3}
        \label{eq:Bondarev-Eq1}
\end{equation}
Here, $\varepsilon_b (> \varepsilon_1, \varepsilon_3)$ is the nearly constant permittivity due to bound electrons, $\Gamma_D$ is the damping constant, and $\omega_p(k)$ is the spatially dispersive (nonlocal) plasma frequency as a function of the in-plane momentum $k$ and film thickness $d$; $\omega_p^\mathrm{3D}$ is the bulk plasma frequency for the material the TD film is made of.

The nonlocal theoretical model for Drude-like EM response (aka the nonlocal KR model~\cite{Biehs:2023}) is verified experimentally in a variety of settings~\cite{Shah:2022,Das:2024,Salihoglu:2023,Vertchenko:2019} (Fig.~\ref{fig:Bondarev-Fig1}). The theory accounts for the vertical electron confinement due to the presence of the substrate and superstrate of dielectric permittivities, which are smaller than that of the TD film. Importantly, the thickness of the film becomes a parameter that controls its generalized linear magneto-optical response~\cite{Bondarev:2018}. The nonlocal KR model covers not only ultrathin films with thickness smaller than the half-wavelength of the incoming light but also conventional films with the thickness exceeding the optical wavelengths~\cite{Bondarev:2020}. The nonlocal EM response of TD plasmonic systems has been shown to enable a variety of new effects, such as thickness-controlled plasma frequency redshift~\cite{Shah:2022}, low-temperature plasma frequency drop-off~\cite{Vertchenko:2019}, plasma mode degeneracy lifting~\cite{Bondarev:2020}, and other quantum-optical~\cite{Bondarev:2023a,Pugh:2024}, nonlocal magneto-optical effects~\cite{Bondarev:2018}, thermal and vacuum field fluctuation effects responsible for the near-/far-field radiative heat transfer~\cite{Biehs:2023,Salihoglu:2023} and Casimir interaction phenomena~\cite{Bondarev:2023b,Rodriguez-Lopez:2024} (Fig.~\ref{fig:Bondarev-Fig1}).

\begin{figure}[hb]
    \centering
    \includegraphics[width=0.99\linewidth]{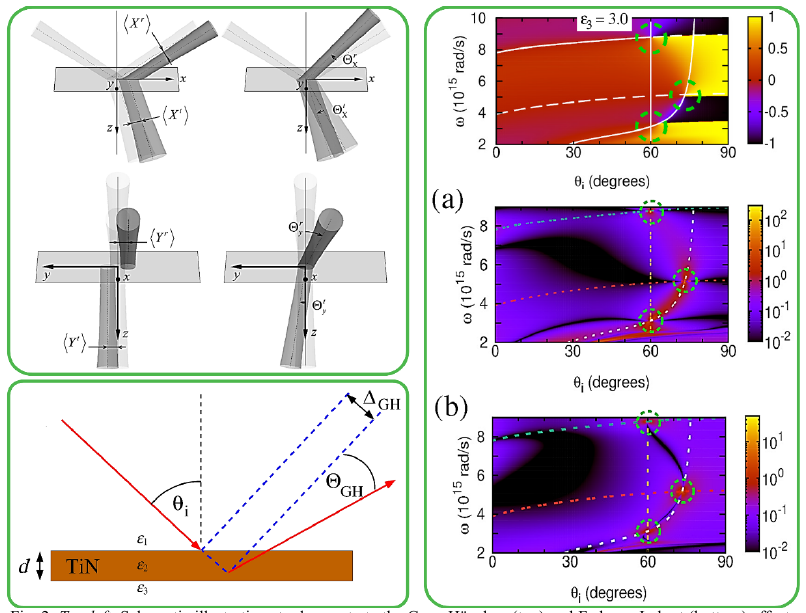}
    \caption{Top left: Schematic illustrations to demonstrate the Goos--H{\"a}nchen (top) and Fedorov--Imbert (bottom) effects for the reflection/refraction of the finite cross-section light beam as it is shifted/deflected in the plane of incidence and out of the plane of incidence, respectively. Reproduced with permission from Ref.~\cite{Bliokh:2013} (Copyright~\textcopyright~2013 IOP Publishing). Bottom left: Sketch of the lateral and angular GH shifts of the light beam incident on a TiN plasmonic slab of broken in-plane reflection symmetry. Right panel -- top, \pnl{a} and \pnl{b}: Reflection phase $\phi_p/\pi$ and absolute values of the lateral and angular GH shifts, respectively, calculated for a 40\,nm thick TiN slab with $\varepsilon_1=1$ and $\varepsilon_3=3$ (MgO substrate) neglecting dissipation. Lines show the eigenmodes available in the system. Circles indicate the topological singularity points at the eigenmode intersections. See Ref.~\cite{Biehs:2024} for details.}
    \label{fig:Bondarev-Fig2}
\end{figure}

\subsection*{Challenges and opportunities}

The confinement-induced thickness-dependent nonlocal EM response [Eq.~\eqref{eq:Bondarev-Eq1}] of the TD plasmonic systems provides the $k$-infinitesimal (first) order nonlocality type in the ultrathin regime. It is dominant in the region of small in-plane momenta $k \leq k_c$, where $k_c$ is the in-plane plasmon cut-off wave vector, as compared to the much smaller standard $k^2$-infinitesimal (second) order type nonlocality. The latter is due to the degenerated electron gas pressure in bulk metals described by hydrodynamical Drude plasma models~\cite{Raza:2011}. The remarkable opportunity to tune the first-order confinement-induced nonlocality through thickness-dependent plasma frequency $\omega_p(k)$ in Eq.~\eqref{eq:Bondarev-Eq1} is advantageous for applications of TD plasmonics in the fields of surface and quantum optics and with regard to the development of extra sensitive optical sensor technologies. To illustrate this, we refer to the TD-plasmonic-film-enhanced Goos--H{\"a}nchen (GH) and Fedorov--Imbert~(FI) shifts of a light beam incident on a metallic surface~\cite{Bliokh:2013} [see Fig.~\ref{fig:Bondarev-Fig2}\pnl{top left}]. Both effects are known to depend on the incident beam polarization~\cite{Bliokh:2013}. The GH and FI shifts are representative of TM/TE ($p$/$s$) linearly polarized and circularly (or elliptically) polarized incident waves, respectively. Below we discuss the GH shift in more detail~\cite{Biehs:2024}.
Originating from the spatial dispersion of the reflection or transmission coefficients due to the finite transverse size of the beam (and so nonlocal), the GH effect was observed in a variety of systems including plasmonic metamaterials~\cite{Shadrivov:2003} and graphene~\cite{Beenakker:2009}. In the reflection configuration [Fig.~\ref{fig:Bondarev-Fig2}\pnl{bottom left}], the lateral $\Delta_\textrm{GH}$ and angular $\Theta_\textrm{GH}$ GH shifts of TM polarized incoming waves are given by~\cite{Biehs:2024}

\begin{equation}
    \Delta_\textrm{GH}= n_1 \cos\theta_i \frac{\partial\phi_p}{\partial k}, \quad \Theta_\textrm{GH}= - \frac{\theta_0^2}{2}k_0n_1 \frac{\cos\theta_i}{\left\vert R_p \right\vert} \frac{\partial \left\vert R_p \right\vert}{\partial k}
    \label{eq:Bondarev-Eq2}
\end{equation}
Here, $k_0 = \omega/c$, $n_1$ is the refractive index for the medium that the finite cross-section (Gaussian) light beam comes from; $\theta_0 = 2/(k_0w_0n_1)$ with $w_0$ representing the beam waist; and the $p$-wave reflection coefficient is written as $R_p = \left\vert R_p \right\vert\exp(i\phi_p)$ in the complex exponential form. It can be seen that the spatial dispersion makes $\Delta_\textrm{GH}$ sensitive to reflectivity phase jumps and $\Theta_\textrm{GH}$ to zero reflection itself so that large effects are highly likely for zero-reflection modes in the system. Phase jumps and singularities make the phase ill-defined, and the reflection coefficient’s absolute value must be zero as required by the causality.

The GH shifts in Eq.~\eqref{eq:Bondarev-Eq2} are normally thought of as being due to the spatial dispersion of the incoming light beam itself while the material EM response nonlocality is usually ignored. For bulk materials or macroscopically thick films this is indeed the case. However, it is not the case for the ultrathin TD plasmonic films where the strong vertical confinement makes the in-plane EM response nonlocal as described by Eq.~\eqref{eq:Bondarev-Eq1} of the KR model. We use TD titanium nitride (TiN) as an example here, which is known for its exceptional plasmonic properties and high crystallinity down to thickness as small as 1\,nm~\cite{Shah:2022,Boltasseva:2011}. Then, there is an additional contribution to be taken into account in Eq.~\eqref{eq:Bondarev-Eq2} that is proportional to
\begin{equation}
    \frac{\partial \varepsilon_\textrm{TiN}(\omega,k)}{\partial k} = - \frac{\varepsilon_b \omega_p^2(k)}{k (1+\tilde{\varepsilon} k d) \omega(\omega+ i \Gamma_D)}= \left[\varepsilon_\textrm{TiN}(\omega,k)-\varepsilon_b\right] \frac{\tilde{\varepsilon} d}{\left(1+\tilde{\varepsilon} k d\right)^2}
    \label{eq:Bondarev-Eq3}
\end{equation}
which is not only nonzero at finite d but can also be both positive and negative, depending on the light frequency. It disappears when $d$ goes to infinity–as it should to make the EM response of thick films local in accord with the standard Drude model.

It was recently shown theoretically~\cite{Biehs:2024} that the confinement-induced nonlocality of the TD films leads to topologically protected singularities of the nonlocal reflection coefficient. Such singularities are shown to result in giant lateral and angular GH shifts in the millimeter and milliradian ranges, respectively, which exceed greatly those reported for light beams of finite transverse extent with no material-induced nonlocality~\cite{Bliokh:2013,Shadrivov:2003,Beenakker:2009,Boltasseva:2011}. The singularities appear in the TD film systems with broken in-plane reflection symmetry (substrate and superstrate of different dielectric permittivities), where due to the strong vertical confinement the eigenmode degeneracy is lifted creating the points of the topological darkness in the visible range that do not exist in the usual (local) Drude materials. Fig.~\ref{fig:Bondarev-Fig2}\pnl{right} shows the topological darkness points (green circles) that appear at the intersections of the non-degenerate EM eigenmodes of the TD film system. The full analysis and the classification of the singularity points can be found in Ref.~\cite{Biehs:2024}. The remarkable opportunity to bring the GH effect to the visible range comes from the fact that the nonlocal plasma frequency in Eq.~\eqref{eq:Bondarev-Eq1} can be red-shifted not only by thickness but also by in-plane momentum reduction through the incident angle change. Lateral and angular GH shifts as large as $\sim$0.4\,mm and $\sim$40\,mrad, respectively, are predicted theoretically for typical 40\,nm thick TD plasmonic films, using for example the conventional helium-neon (He-Ne) laser light [Fig.~\ref{fig:Bondarev-Fig2}\pnl{right}].

\subsection*{Future developments to address challenges}

While the experimental progress in TD materials for nanophotonics has earlier been impeded by challenges in producing atomically thin films of noble metals, TD films of emerging plasmonic materials such as transition metal nitrides (TiN, ZrN, HfN, etc.~\cite{Boltasseva:2011}) can be grown as epitaxial-quality films with thicknesses down to 1--2\,nm (5--10 atomic layers)~\cite{Shah:2022,Das:2024}. A variety of transition metal nitrides, their ability to grow as high-quality, ultrathin epitaxial films and the sensitivity of their optical and electronic properties to the material/structural/geometrical parameters provide a rich playground for the realization of the confinement-induced nonlocal effects. This includes using transition metal nitrides in TD plasmonics to control strong electron correlations by precise variation of the film thickness to study fundamental solid-state physics phenomena such 2D electron Wigner crystallization~\cite{Boltasseva:2025} and metal-insulator transitions~\cite{Das:2024}. Another important avenue to tailor the nonlocal optical properties of plasmonic TD films includes strain engineering. Below a critical thickness, an epitaxial thin film is expected to retain the strain induced by the substrate. This can be achieved experimentally by growing strained ultrathin films on lattice-mismatched substrates. It has been theoretically demonstrated that by varying the in-plane lattice parameter of an ultrathin film its nonlocal optical response can be tuned~\cite{Qi:2023}. In conjunction with the thickness dependence of the nonlocal EM response discussed above, the strain engineering offers an additional way to tailor the optical response of TD plasmonic materials.

\subsection*{Concluding Remarks}

Plasmonic TD materials hold great potential for enabling a new quantum material platform that utilizes strong confinement and nonlocal effects and offers new opportunities for quantum optics, quantum computing and sensing applications. A crucial step for such development is the realization of high quality, ultrathin metallic films of precisely controlled thickness with reduced surface roughness that exhibit new material functionalities and unique light-matter interactions. The quantum effects that arise in metallic TD materials along with their strong tunability may pave the way to new optical phenomena and novel, atomically thin, dynamically tunable nanophotonic devices.

\section[Light management using structural nonlocality in nanorod metamaterials\\ (Krasavin \& Zayats)]{Light management using structural nonlocality in nanorod metamaterials} 

\label{sec:Krasavin}

\author{Alexey V. Krasavin\,\orcidlink{0000-0003-2522-5735} \& Anatoly V. Zayats\,\orcidlink{0000-0003-0566-4087}}

\subsection*{Overview}

Engineering the shape of elemental metamaterial components, meta-atoms, as well as their arrangement on a subwavelength scale results in artificial materials with pre-designed optical responses both in linear and nonlinear regimes. The optical properties of such discrete media can be understood, in principle, through conventional local effective medium models. Comparing these models with full-vectorial numerical simulations, however, shows intricate differences, which are relatively small in general, but which become crucial for certain metamaterial parameters, for which a local EMT breaks down, even in the range of its applicability. This behavior is related to the so-called nonlocal, spatial dispersion effects which are not taken into account in conventional EMT models (numerical simulations of a discrete structure of a metamaterial however includes these effects automatically).

Nonlocality is a very peculiar property of optical materials, when the optical response of the medium at a given point depends not only on the driving electromagnetic field at this location, but also on the values of the field in other points of the surrounding region. This property is not so common, and it is clear why. For natural condensed non-metallic materials, liquids or solids, the optical response of an atom (or a molecule) is influenced by the optical responses of its counterparts located only in its close vicinity, particularly at the distances of the order of a lattice constant (or a molecular size), which is much smaller than the wavelength of light. Consequently, there is no variation of the field over this region and it is a valid assumption that this optical response of the atom is defined by the value of the field at its location. This can be applied to any atom of the material, and therefore the optical response of the material is local. The situation changes when larger quasiparticles are considered, such as excitons in semiconductors, resulting in spatial dispersion which strongly influences the optical response at low temperatures and leads to the appearance of additional waves~\cite{Agranovich:1984}. The nonlocality of metals important for the nanostructures with the size below few nanometers, associated with electron spill-out effects and wavevector-dependent dielectric permittivity, is discussed in 
Parts I and II of this Roadmap, including Secs.~\ref{sec:Fernandez-Dominguez}, \ref{sec:Khurgin}, \ref{sec:Shahbazyan},  \ref{sec:Wegner}, and \ref{sec:Hu}. A different situation can occur in metamaterials, in which both the size of the meta-atoms and their spacing, although being smaller than the wavelength in order to satisfy an effective medium requirements are noticeable fractions of it. This results in \emph{i)} field variation across the meta-atom, may lead to the excitation of higher-order multipoles, thus affecting its optical response, and \emph{ii)} the field at the positions of other meta-atoms affecting the optical response in a given point starts to depend on their positions~\cite{Simovski:2009}. Thus, the optical response of a metamaterial at a given location is noticeably affected by the field at other points, naturally producing a physical mechanism of the metamaterial structural nonlocality. In the reciprocal-space representation, this means that the optical parameters of the metamaterial depend on the wave vector, i.e. the material is spatially dispersive.

With subwavelength structuring and the consequent absence of the diffractive effects, a metamaterial can be considered as a uniform medium, which optical properties are defined by the metamaterial design and can be described using effective medium theories (EMTs). In the first approximation the metamaterial optical response can be considered to be local, but for its precise description development of nonlocal EMTs is needed. This is especially important for optical effects depending on the local fields inside metamaterial, such as photoluminescence or nonlinearity, as nonlocality can create high nonuniformity of the local fields due to the introduction of an additional wave. 
The development of nonlocal EMTs is not an easy task. Generally, it requires deriving microscopic optical response of the metamaterial solving the electromagnetic problem for the exact nanostructured design. Then, within two common approaches, the effective optical parameters, depending on the wave vector, can be found by averaging the derived microscopic fields over the metamaterial unit cell, or setting trial (or Taylor-expanded in the $k$-space) expressions for the effective permittivity and/or permeability and matching the dispersion of the corresponding optical modes to that found in the microscopic description~\cite{Davidovich:2019}.

\begin{figure}
    \centering
    \includegraphics[width=0.95\linewidth]{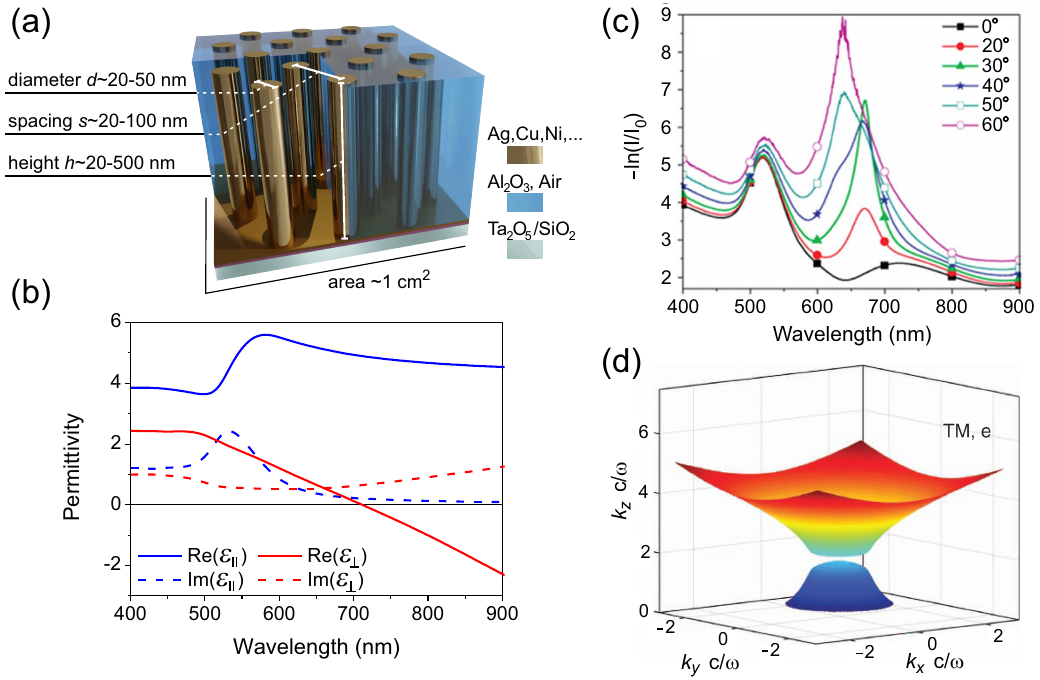}
    \caption{\pnl{a}~Schematics of a nanorod metamaterial with indicated characteristic geometrical parameters and materials. \pnl{b}~Spectral dependencies of the components of the metamaterial permittivity tensor ($d = 25$\,nm, $s = 60$\,nm) calculated using a local EMT~\cite{Roth:2024}. \pnl{c}~Experimental extinction spectra of a 300\,nm-thick nanorod metamaterial layer [Au/Al$_2$O$_3$, geometrical parameters the same as in \pnl{b}] under TM-polarized illumination as functions of the angle of incidence. Reproduced with permission from Ref.~\cite{Pollard:2009} (Copyright~\textcopyright~2009 American Chemical Society). \pnl{d}~Isofrequency surfaces of two hybridized TM-polarized waves plotted in the elliptical dispersion regime ($\lambda = 550$\,nm, $d = 50$\,nm, $s = 100$\,nm). Reproduced with permission from Ref.~\cite{Ginzburg:2017} (Copyright~\textcopyright~2017 Nature Springer).}
    \label{fig:Krasavin-Fig1}
\end{figure}

\subsection*{Current status}

There are numerous experimental demonstrations of the optical phenomena, in which the nonlocal response of a metamaterial played the crucial role. As a prominent example, a metamaterial formed by an array of aligned plasmonic nanorods represents a uniaxial medium possessing record-high optical anisotropy, hyperbolic dispersion related to different signs of dielectric permittivity along the optical axes, and an epsilon-near-zero (ENZ) response related to $\varepsilon_\parallel \simeq 0$ (permittivity component along the optical axis parallel to the nanorods)~\cite{Roth:2024} [Fig.~\ref{fig:Krasavin-Fig1}\pnl{a,b}]. The presence of structural nonlocality, which can be derived using various EMT approaches and approximations~\cite{Davidovich:2019,Pollard:2009,Wells:2014} fundamentally modifies these characteristics. In the local approximation, optical extinction of the nanorod metamaterial layer features two peaks [located at around 500 and 650\,nm in Fig.~\ref{fig:Krasavin-Fig1}\pnl{c}]. The short-wavelength peak, existing for both TE and TM polarized illumination is related to the excitation of the transverse plasmonic resonance of the nanorods, slightly shifted due to inter-rod coupling. Its long-wavelength counterpart, existing only for TM polarized illumination, corresponds to the metamaterial opacity region near the ENZ spectral point for $\varepsilon_\parallel$. The magnitude of this peak increases with the angle of incidence, but in the local approximation its spectral position does not change with the latter. However, for low ohmic losses of plasmonic metal, the behavior of the ENZ peak reveals a more intricate angle dependence~\cite{Pollard:2009} [Fig.~\ref{fig:Krasavin-Fig1}\pnl{c}]. Instead of a single peak, two ENZ extinction peaks showing an angular excitation dynamics characteristic to anti-crossing of optical modes are revealed. They are related to two TM-polarized modes, which result from the hybridization of the ordinary TM mode with an additional longitudinal mode supported solely due to the metamaterial structural nonlocality~\cite{Pollard:2009}. The latter, due to its dispersion characteristics and field distribution is optically accessible only in the ENZ region. The presence of two TM-polarized waves in the nonlocal case can be clearly seen in the plots of isofrequency surfaces [Fig.~\ref{fig:Krasavin-Fig1}\pnl{d}]~\cite{Ginzburg:2017}. The nature of the nonlocality here can be traced to the coupled cylindrical surface plasmon polaritons supported by the nanorods~\cite{Wells:2014}.
The interaction of the hybridized TM modes and, therefore, the extinction of the metamaterial in the ENZ region have been shown to be highly dependent on the optical properties of the constituting materials. 
Active control of losses in plasmonic nanorods, e.g., through excitation of hot electrons upon light absorption or temperature effects~\cite{Wu:2023} results in effective switching the nonlocal response on/off by variation of the metal optical losses, leading to ultrafast all-optical switching~\cite{Wurtz:2011} and polarisation control~\cite{Nicholls:2017} in the nonlocal regime.

\begin{figure}[htb]
    \centering
    \includegraphics[width=0.89\linewidth]{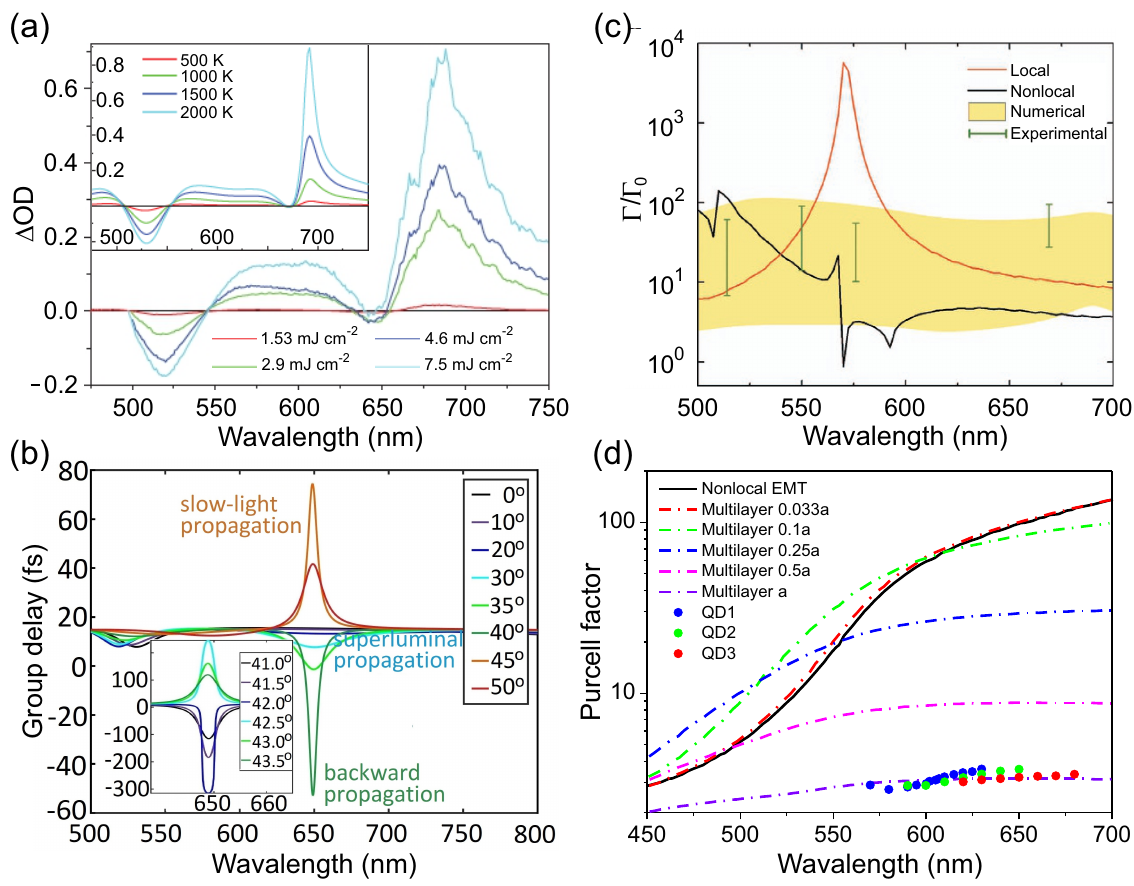}
     \caption{\pnl{a}~Transient extinction spectra of a plasmonic nanorod metamaterial (Au nanorods in alumina template, $d = 20$\,nm, $s = 70$\,nm, $h = 400$\,nm) at $20^\circ$ incident angle for various pump fluences at $\lambda_p = 465$\,nm and $\Delta t = 130$\,fs. Reproduced with permission from Ref.~\cite{Wurtz:2011} (Copyright~\textcopyright~2011 Nature Springer). \pnl{b}~Group delay experienced by an optical pulse while propagating through a 525\,nm thick metamaterial layer ($d = 25$\,nm, $s = 60$\,nm). Reproduced with permission from Ref.~\cite{Stefaniuk:2023} (Copyright~\textcopyright~2023 Wiley). \pnl{c}~Enhancement of spontaneous emission rate of quantum emitters inside a nanorod metamaterial ($d = 50$\,nm, $s = 100$\,nm): \pnl{i}~experimentally measured and \pnl{ii}~numerically simulated (with the exact microscopic geometry) for a 250\,nm thick metamaterial layer (yellow band shows the range of the values for various locations inside the metamaterial layer), and calculated using \pnl{iii}~local and \pnl{iv}~nonlocal EMTs for bulk metamaterial. Reproduced with permission from Ref.~\cite{Ginzburg:2017} (Copyright~\textcopyright~2017 Nature Springer). \pnl{d}~Purcell factor of a dipole with averaged orientation located 10\,nm above a multilayer metamaterial ($a_1 = 12$\,nm Ag, $a_2 = 83$\,nm SiO$_2$) surface calculated using nonlocal EMT (black solid line) and finite element numerical simulations (color lines, for various metamaterial unit cell size $a = a_1 +a_2$, keeping the same ratio between Ag and SiO$_2$ thicknesses $a_1/a_2$), together with experimental counterpart measured using three types od quantum dots (scatterer plots). Reprinted (adapted) with permission from Ref.~\cite{Li:2017} (Copyright~\textcopyright~2017 American Chemical Society).}
    \label{fig:Krasavin-Fig2}
\end{figure}

The hybridization of the main and additional modes, the latter associated with nonlocal response, in the ENZ region creates a very large effective frequency dispersion for optical pulses transmitted through a metamaterial layer, which has been used to manage their propagation and temporal characteristics~\cite{Stefaniuk:2023}. Particularly, sub- and super-luminal, as well as backward propagation of pulses have been demonstrated, with the switching of the group dispersion achieved in the same metamaterial by just changing the angle of illumination [Fig.~\ref{fig:Krasavin-Fig2}\pnl{a}]. 
Combining this with nonlinear functionalities discussed above, one can potentially realize all-optical control of the pulse characteristics.

Crucially, the structural nonlocality has a strong effect of the emission of quantum emitters placed inside the metamaterial. The nonlocal description has been shown to provide a more adequate description of the emitter spontaneous decay rate, contrary to its almost a singular enhancement predicted by the local EMT~\cite{Ginzburg:2017} [Fig.~\ref{fig:Krasavin-Fig2}\pnl{b}]. Experiments, together with the microscopic numerical simulations have shown, however, that considering exact metamaterial structure (as opposed to implementation of EMT theories) and particularly exact position of the emitters with respect to the meta-atoms is important for correct understanding of the emission processes.

Structural nonlocality has been also studied in multilayer metamaterials, produced by alternating metallic and dielectric layers with nanoscale thicknesses, which share major optical characteristics with the nanorod metamaterial, such as extremely high anisotropy, ENZ behavior and hyperbolic dispersion. Nonlocality in these metamaterials is underlined by strong variation of the fields across single layers (acting in this case as meta-atoms)~\cite{Elser:2007}, and is also affected by the inter-meta-atom effects. It affects the metamaterial transmission/reflection properties, spectral position of the ENZ region and the emission of quantum emitters placed in the vicinity or inside the metamaterial. Particularly, Fig.~\ref{fig:Krasavin-Fig2}\pnl{d} shows that the Purcell factor predicted by a nonlocal EMT matches the numerically simulated counterpart in the limit of very small unit cell sizes, while the numerical predictions agree well with the experimentally measured Purcell factor values~\cite{Li:2017}.

\subsection*{Challenges and opportunities}

From the theoretical perspective, various approaches have been developed to derive nonlocal EMTs for metamaterials, starting from that assuming ideally conducting wires (relevant for microwave frequencies) and wires having large negative real part of the permittivity (appropriate for wavelength in mid and far-IR) to the models with no specific restriction made on the epsilon of the nanorods (apart from the implicit assumption that they are metallic~\cite{Pollard:2009,Wells:2014}). Commonly, they return the same general dependence, featuring local effective $\varepsilon_\perp$ components and nonlocal $\varepsilon_\parallel$ counterparts, depending on the wave vector along the optical axis $k_\parallel$. At the same time, frequently models derived under similar assumptions results in different expressions for the nonlocal effective permittivity, which motivates for comparison and generalization of the nonlocal EMT models~\cite{Maslovski:2009,Davidovich:2019}. 
A related question here is determining an additional boundary condition required for the additional wave.
Substantial benefit would be the derivation of new or application of existing models to define the range of metamaterial geometrical parameters which maximize the influence of structural nonlocality and/or maximize the magnitude of nonlocality-assisted phenomena, which would provide a valuable guidance for the experimental research. Of course, all the above research directions are very challenging, as the derivation of any broadly applicable nonlocal EMT, which should ideally satisfy several important conditions: \emph{i)} provide correct dispersion relation for the metamaterial modes, \emph{ii)} give correct boundary conditions (note that in the case of nonlocality an additional boundary condition is needed to take into account the existence of the additional optical mode~\cite{Wells:2014}), and \emph{iii)} be valid beyond the case of plane wave excitation~\cite{Fietz:2009}. At the same time, this means that there is a large scope for the future theoretical research.

Experimentally, since the nonlocal response depends on losses of the plasmonic component in plasmonic metamaterials, the realization of low loss material platforms (including 2D materials, e.g. graphene) together with broader range of geometrical parameters will open opportunities for realization and exploitation of stronger spatial dispersion contributions in a desired spectral range for controlling molecular emission with potential applications in bio-imaging and quantum technologies, and enhancement of optical nonlinearities, important for laser technologies and optical information processing. 

\subsection*{Future developments to address challenges}

Nonlocal metamaterials have been shown to provide multiple functionalities, bio and chemical sensing being among the most prominent. Usually, the sensitivity of resonant excitation of guided modes supported by a metamaterial layer to the changes of the dielectric environment is used for this purpose~\cite{Roth:2024}. At the same time, nonlocal effects can also make the extinction of the metamaterial in the ENZ region highly sensitive to such changes, through the same physical mechanism as was considered above for the realization of optical switching. To demonstrate this, samples with maximized structural nonlocality, achieved by engineering the design and reducing the optical losses in metal, may be needed.

Apart from incoherent Kerr-type nonlinearity which exploit optically-induced heating of the electron gas discussed above, nanorod metamaterials possess its coherent counterpart based on a complex electron dynamics coherent with the excitation wave~\cite{Krasavin:2018}. Particularly, it has been shown that engineering of the modal structure of a nanorod metamaterial layer can enhance second harmonic generation~\cite{Marino:2018}, while alternative approaches based on back-propagating waves have also been proposed~\cite{Popov:2012}. In this respect, developing nonlinear EMT including phenomena related to structural nonlocality, e.g. along the lines presented in Ref.~\cite{Gorlach:2016}, would provide a valuable guidance to the experimental research. Possessing an extreme anisotropy and nonlocality which affect only certain components of the effective permittivity tensor, nanorod and multilayer metamaterials present a particular interest for engineering light-matter interaction with complex vectorial beams. An example here can be the study of interaction of the nonlocal $\varepsilon_\parallel$ component in the nanorod metamaterials with co-directed longitudinal fields present in radially polarized cylindrical beams of various orders. Another research direction could be the study of chiroptical, which have been shown to be inherent to the artificial media with effective structural spatial dispersion~\cite{Gompf:2011}.

\subsection*{Concluding Remarks}

Spatial dispersion in metamaterials arises from sub-wavelength structuring and results in nonlocal effects with electromagnetic fields depending not only on the field at a given point but also on its variations across several unit cells. It is essential for accurately describing wave propagation in complex media, including ultrashort pulses, and in particular the phenomena depending on local fields inside metamaterial. It has much stronger impact on optical properties of metamaterials than in semiconductors where it is related to the excitonic excitation and observable at low temperatures. Recognizing spatial dispersion additional flexibility in the design of optical response, it is important in advancing nanophotonic technologies such as, high-resolution imaging, ultrashort pulse control, nonlinear optics, optical information processing and quantum technologies.

\section[Nonlocal wave phenomena in spatio-temporal metamaterials (Al{\`u})]{Nonlocal wave phenomena in spatio-temporal metamaterials}

\label{sec:Alu}

\author{Andrea Al{\`u}\,\orcidlink{0000-0002-4297-5274}}

\subsection*{Current status}

The last few years have witnessed impressive developments in the context of metamaterials and metasurfaces, with remarkable progress in controlling electromagnetic wavefronts, and in the application of these phenomena within a wide range of technologically relevant contexts. One important breakthrough, detailed in several sections of this Roadmap, has been the discovery that nonlocal wave phenomena are not necessarily just a nuisance, a correction over well-understood phenomena in light-matter interactions, or a modeling headache, but rather they can be controlled, enhanced and engineered in metamaterial platforms, offering new degrees of freedom for enhanced wave control both in real and in reciprocal space. In this context, the emerging field of nonlocal metasurfaces~\cite{Kwon:2018} has been flourishing, with remarkable demonstrations in three relevant directions: \emph{i)} spatial shaping of lattice resonances for enhanced metasurface control~\cite{Overvig:2022}, \emph{ii)} analog-based image processing and optical computing at the nanoscale~\cite{Zhou:2020,Cordaro:2023}, and \emph{iii)} enhanced coherence control over light generation~\cite{Nolen:2024}.

Beyond nonlocality, in a parallel research direction the dimension of time has recently offered new opportunities to enhance the degrees of freedom available for wave control in metamaterials. Beyond the three spatial dimensions, by engineering the arrow of time through tailored temporal variations of the optical properties of engineered materials it is possible to realize advanced functionalities and new wave phenomena, such as time interfaces, time scattering and time crystals. In turn, suitable combinations of spatial and temporal structuring of a material have been enabling space-time- or four-dimensional metamaterials with enhanced degrees of freedom for wave control~\cite{Engheta:2023}. 

These seemingly disconnected research lines are likely to offer tremendous opportunities when combined, with interesting prospects for fundamental research breakthroughs and for applications across different scales. Indeed, nonlocalities broadly arise in natural and artificial materials, both in space and time, and are particularly relevant in the context of resonant wave phenomena, at the basis of metamaterials and metasurfaces. For instance, nonlocality engineering in time has been considered in the context of temporal metamaterials~\cite{Rizza:2022}, and  space-time nonlocality engineering in metasurfaces has been recently explored~\cite{Esfahani:2024} to enable analog image processing in space and time. By engineering the spatial and temporal dispersion of a passive metasurface [Fig.~\ref{fig:Alu-Fig1}\pnl{a}], it is possible to perform space-time differentiation of an incoming image stream, enabling neuromorphic event detection through an ultrathin, passive metasurface~\cite{Esfahani:2024}. This initial proposal leaves open several questions: to what extent is it possible to engineer nonlocalities both in space and time, leading to superior control over spatial and frequency dispersion in metamaterials? What range of applications can benefit from this control? What are the fundamental limits in engineering space-time nonlocalities? What role do nonlocalities play in the emerging field of space-time metamaterials?

\subsection*{Challenges and opportunities}

\begin{figure}[htb]
    \centering
    \includegraphics[width=0.99\linewidth]{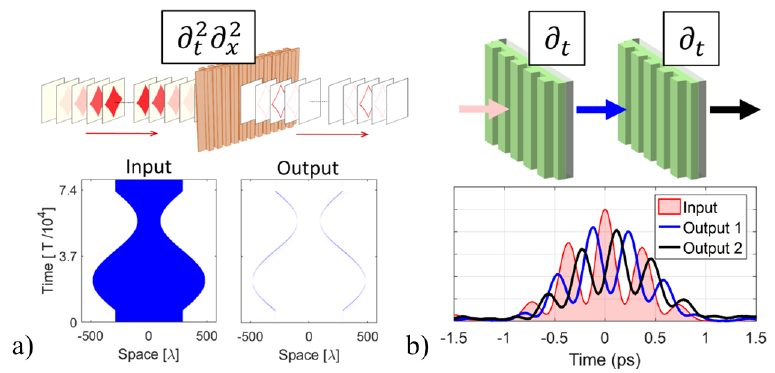}
    \caption{\pnl{a}~A space-time nonlocal metasurface performing derivatives both in space and time, with application as an analog-domain neuromorphic event detection device: transmission is only allowed for the edges of objects that are changing in time. Adapted with permission from Ref.~\cite{Esfahani:2024} (Copyright~\textcopyright~2024 American Physical Society). \pnl{b}~Experimental demonstration of a cascade of time-derivative metasurfaces, each of them independently transmits the time-derivative of the input wave. As a bi-layer, the measured output is the second-derivative in time of the incoming signal. Adapted with permission from Ref.~\cite{Cotrufo:2024a} (Copyright~\textcopyright~2024 Nature Springer).}
    \label{fig:Alu-Fig1}
\end{figure}

A first step towards the experimental demonstration of space-time nonlocal metasurfaces was recently shown in Ref.~\cite{Cotrufo:2024a}, in which passive metasurfaces performing the time-derivatives of the incoming signal were demonstrated by tailoring their frequency dispersion. By cascading two of such metasurfaces performing the first derivative of the incoming signal [Fig~\ref{fig:Alu-Fig1}\pnl{b}], a second-derivative device was also demonstrated, showing opportunities to realize more complex operations by simply stacking multiple devices. While a metasurface showcasing engineered nonlocalities in both space and time has not been demonstrated, this initial experiment has highlighted a few challenges: the limited spatial extent over which a metasurface engages with the wave in the propagation direction makes the control over nonlocalities inherently difficult. In the devices demonstrated in Ref.~\cite{Cotrufo:2024a}, the event detection response is limited to temporal variations in the picosecond range, which may be of interest for ultrafast imaging, but limits other applications. For significantly slower temporal dynamics the demonstrated metasurfaces operate at much reduced efficiency, and better responses can be obtained only at the price of significantly increased resonance $Q$ factors, which is challenging for free-space operation and limits the overall operational bandwidth. Another related challenge is reconfigurability: even though these nonlocalities control both spatial and temporal dispersion, the underlying devices are static in nature, which is appealing in terms of scalability and ease of implementation, but it implies a non-reconfigurable response, which limits the application space. Moreover, the demonstrated responses have so far been limited to linear operations, implying that the resulting application is limited to Fourier-based operations. Extensions towards nonlinear space-time operations may largely broaden the degree of applicability. 

Engineered space-time nonlocality opens interesting opportunities not only in the field of analog-based image processing, but also in other contexts. Metasurfaces with engineered nonlocalities both in space and time can shape the coherence profile of their own thermal emission and photoluminescence~\cite{Nolen:2024}, converting oscillations of matter driven by incoherent processes to efficient radiation with large temporal and spatial coherence, in the range of hundreds of oscillation periods and free-space wavelengths. By combining these nonlocal phenomena with local structuring in space of the underlying resonances, it is also possible to tailor the polarization and wavefront profile at will~\cite{Overvig:2021}, offering a new platform for compact, efficient and cheap custom light generation. Yet, several challenges remain: taming incoherent processes is difficult, making these metasurfaces sensitive to fabrication defects and disorder. The enhanced coherence is fundamentally limited by these practical factors, as well as by the extent of the area over which these devices can be reliably fabricated. 
Space-time nonlocality emerges as a very important element in the accurate modeling of time and space-time metamaterials. Similar to how spatial nonlocalities play an important role in capturing the response of metamaterials~\cite{Silveirinha:2005}, temporal nonlocalities cannot be neglected in the proper modeling of the wave scattering at temporal interfaces~\cite{Galiffi:2024,Moussa:2023}. As a result, the microscopic implementations of time- and space-time metamaterials, and the associated nonlocalities, are expected to play a crucial role in determining the overall wave response, and their engineering may be leveraged to enhance its control. Theoretical progress in the understanding of these interactions in space-time metamaterials, and experimental progress on the implementation of space-time nonlocality engineering will be fundamental to enable this vision.

\subsection*{Future developments to address challenges}

The growing interest in engineered nonlocalities is driving impactful research that holds the promise to address the challenges mentioned in the previous section, and leverage the many available opportunities. The possibility of reconfiguring spatial nonlocalities has been recently demonstrated in proof-of-concept metasurface devices relying on phase-change materials~\cite{Cotrufo:2024b,Yang:2025}, which may be extended to reconfigurable space-time nonlocality. Electro-optical or all-optical modulation may extend the degrees of freedom that can be reconfigured, opening dramatically the design and application. Space. Similarly, the limitation of resonance quality-factors of free-space metasurfaces can be addressed with tailored designs and improved fabrication techniques. Recent efforts have enabled demonstrations with quality-factors comparable to integrated photonic platforms~\cite{Fang:2024}, opening many opportunities in the context of space-time nonlocality engineering. 
Similar to diffractive nonlocal metasurfaces~\cite{Overvig:2022}, which combine local wavefront engineering in the spatial domain with nonlocality engineering in momentum space, future developments may consider space-time local and nonlocal responses coming together in the metasurface platform, offering opportunities to tailor dynamically space-time dispersion of complex wavefronts, further expanding the degree of control over waves. Finally, enhanced nonlinearities in metasurfaces, relying on strong light-matter interactions, as well as on material engineering as in the case of multiple quantum wells~\cite{Lee:2014}, hold the promise to further broaden the opportunities and impact of space-time nonlocalities by extending it to the control of nonlinearly-generated signals through wave mixing, as well as to perform nonlinear image processing in space and time. 

\subsection*{Concluding remarks}

The fields of nonlocal metasurfaces and of space-time metamaterials have been thriving in recent years, and the initial attempts of merging these research areas have been showcasing tremendous opportunities for the future of metamaterials and metasurfaces, providing enhanced wavefront control, signal processing and neuromorphic computing, and coherence control in space and time over thermally generated light or photoluminescence. Once mature, space-time nonlocality engineering holds the promise to be another exciting toolbox for photonic engineers and metamaterial scientists.

\section[Spectral and angular control of light with nonlocal metasurfaces\\ (Song \& Brongersma)]{Spectral and angular control of light with nonlocal metasurfaces}

\label{sec:Song}

\author{Jung-Hwan Song\,\orcidlink{0000-0001-9502-5718} and Mark L. Brongersma\,\orcidlink{0000-0003-1777-8970}}

\subsection*{Overview}

Many emerging imaging, sensing, communication, display, and non-linear optics applications require compact optical elements that can selectively manipulate light waves at very well-defined wavelengths and/or angles of incidence. In this Roadmap article we argue that nonlocal metasurfaces are ideally suited for this purpose. We further highlight that a careful spectral engineering of the optical materials loss in such optical elements opens a unique opportunity to fully-decouple the optical functions at different illumination wavelengths.

\subsection*{Current status}

The field of metasurface flat-optics has opened a myriad of new ways to control the flow of light beyond the capabilities of polished pieces of glass and geometrically-shaped metallic mirrors. Metasurface optical elements can be created by judiciously nanostructuring thin films of metal or semiconductor materials. They effectively harness light scattering and optical interference processes to achieve a very wide range of valuable optical functions~\cite{Kuznetsov:2024,Schulz:2024}. Low-cost and large-area nanofabrication techniques as well as advanced computational tools are now propelling these optical elements into real, commercial applications~\cite{Brongersma:2025} and highly-integrated device technologies~\cite{Zheludev:2012,Ha:2024}. 
Metasurfaces can broadly be divided into two distinct classes based on their operational principle. There are local metasurfaces for which the optical interaction between neighboring nanostructures or "meta-atoms" is weak and nonlocal metasurfaces in which the coupling is substantial and can extend over a long range. Recent work provides helpful insights into the ways that the meta-atom geometry can be engineered to transition between local and nonlocal behaviors~\cite{Liang:2024}. Each type has their unique application spaces, advantages, and limitations. The design of local metasurfaces is conceptually very simple as one can think of each nanostructure imparting independent, local, spatially-varying amplitude, phase, and polarization changes on an optical wavefront. These optical manipulations can be controlled through judicious choices of the nanostructure sizes, shapes, orientations and spatial arrangements. A growing intuition about the operation of local metasurfaces together with powerful rapid-design software tools based on topological optimization, inverse design and deep-learning principles has facilitated rapid progress in their development~\cite{Jensen:2011,Lalau-Keraly:2013,Jiang:2021a}. At the same time, it is worth pointing out that there is always some degree of optical coupling between meta-atoms present and this can result in notable performance limitations in terms of the achievable diffraction efficiency~\cite{Gigli:2021}. In contrast, nonlocal metasurfaces strategically take advantage of the optical coupling and collective responses of nanostructures to boost efficiency. 
Based on their distinct nature, local and nonlocal metasurfaces display notably different spectral and angular responses. For local metasurfaces these properties are intimately connected to those of the meta-atom building blocks they are constructed from. They can be either non-resonant, truncated waveguides~\cite{Lalanne:1998} or employ optically-resonant plasmonic~\cite{Schuller:2010,Brongersma:2015} or Mie~\cite{Kuznetsov:2016} resonators with low optical quality factor ($Q$) resonances. As a result, the nanostructures deliver display light scattering responses that are broad in terms of their spectral and angular characteristics. As the nanostructures in a local metasurface more-or-less act independently, these traits are naturally transferred to the entire optical element. This is advantageous when the goal is to realize optical elements that can operate across a broad spectral range and need to deliver a high numerical aperture. However, this is detrimental in applications that require the selective manipulation of light waves at well-defined frequencies and/or angles incidence. This is an area where nonlocal metasurfaces can shine. 

\subsection*{Challenges and opportunities}

\begin{figure}
    \centering
    \includegraphics[width=0.99\linewidth]{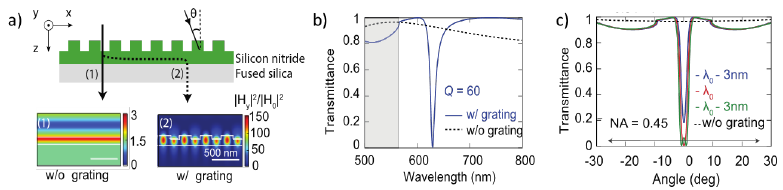}
    \caption{\pnl{a}~A cross sectional schematic of a basic nonlocal metasurface~\cite{Ji:2022}. Incident transverse magnetic (TM) polarized light can take a direct (1) and indirect, resonant pathway (2) through this optical element. The corresponding field distributions are shown underneath the schematic. \pnl{b}~The simulated transmittance spectrum of the metasurface at normal incidence shows a pronounced dip that results from destructive interference between the direct and indirect pathways on resonance. \pnl{c}~Simulated angular transmittance spectra of the metasurface on and near the resonant wavelength $\lambda_0$. The case without the surface relief grating on the nitride layer is shown for reference. Reproduced with permission from Ref.~\cite{Ji:2022} (Copyright~\textcopyright~2022 Nature Springer).}
    \label{fig:Song-Fig1}
\end{figure}

Many modern day optics applications require extreme control over the spectral and angular properties of light. For example, in imaging and microscopy spectral control helps to mitigate aberrations and produce sharper images. Spectroscopy and sensing tools require spectrally-selective optical excitation and analysis of light to analyze materials properties, chemical composition, and molecular structure. In optical communication systems, spectral and angular control enables the precise management of light signals for efficient data transmission, multiplexing, and de-multiplexing. Local metasurfaces have displayed challenges in delivering the required spectral and angular control, but nonlocal metasurfaces can capitalize on the optical interaction between the building blocks to achieve such functions in an integrated form factor. 
To understand why nonlocal metasurfaces can provide superior control over the angular and spectral properties of light, we investigate how they are designed and operate at a conceptual level. In the design of most local metasurface structures, one arranges the meta-atoms to achieve a transmission function $t(x,y)$ that is dependent locally on the spatial coordinates $x$ and $y$. For example, in a metalens the transmission function is designed to achieve a local space-variant phase shift that shapes the wavefront in the same fashion as a polished lens~\cite{Lin:2014}. In contrast, a nonlocal metasurface is designed to deliver a transmission function $t(k_x, k_y)$ that is dependent on the wave-vectors $k_x$ and $k_y$ of the incident light~\cite{Kwon:2018,Guo:2018} (see also Section~\ref{sec:Levy} of this Roadmap). In other words, this type of metasurface will differently manipulate light waves with distinct incident angles. Fig.~\ref{fig:Song-Fig1} illustrates in a tutorial example how the spectral- and angle-dependent behaviors can be engineered. Figure~\ref{fig:Song-Fig1}\pnl{a} shows a very basic nonlocal metasurface comprised of a 240\,nm thick single-mode silicon nitride (Si$_3$N$_4$) slab waveguide with a surface-relief grating etched into its top surface~\cite{Ji:2022}. Light incident on this optical element can follow two distinct pathways. Most commonly, the incident light waves follow a direct transmission pathway through the patterned nitride layer. The spectral transmission behavior for this pathway mimics the transmission for a regular, low quality-factor ($Q \sim 2$) Fabry--P{\'e}rot resonator and only a gentle buildup of field can be seen in the optical simulation. However, at selected wavelength and incident-angle combinations, the grating can resonantly couple free-space light waves to quasi-guided modes. Such coupling is facilitated by the power of constructive interference, where all of the small grating teeth can work in concert to scatter and effectively redirect the incident light waves into the nitride waveguide. For the discussion that follows, it is important to note that this can lead to a notable energy storage in the waveguide and a very substantial increase in the electric field in the waveguide. This field build up can be seen in the optical simulation for the second pathway. In the same way the grating can couple the light into the waveguide, it also causes the guided light to slowly leak back out of the structure. In the forward direction the leaked waves interfere with the directly transmitted light. On the grating coupling resonance, this interference is destructive and gives rise to a pronounced dip in the spectral transmission spectra [Fig.~\ref{fig:Song-Fig1}\pnl{b}]. The amplitude of the grating elements determines the coupling efficiency in/out of the waveguide and thus also the radiative quality factor ($Q_r$) for the resonances. The period controls the wavelength at which the light is coupled.  In this example, a grating period of 390 nm places the resonance at 630\,nm for normally-incident light with a transverse magnetic (TM) polarization (the magnetic-field is pointing along the grooves of the grating). The grating depth was chosen to be 90\,nm to produce a resonance with a quality factor $Q = 60$. 

\begin{figure}[htb]
    \centering
    \includegraphics[width=0.99\linewidth]{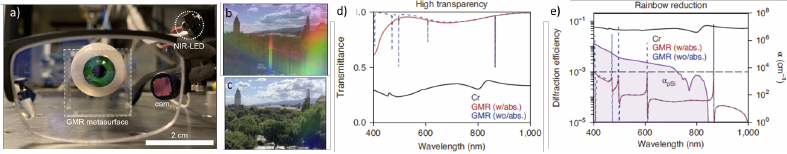}
    \caption{\pnl{a}~Optical image of an eyetracking prototype showing the highly transparent nonlocal metasurface in front of the eye~\cite{Song:2021}. \pnl{b}~Photograph taken through a glass surface with a 200\,nm thick Si$_3$N$_4$ anti-reflection coating patterned with a 30\,nm thick Cr grating on top showing multiple strong rainbows. \pnl{c}~Photograph taken through a nonlocal metasurface with a 3\,nm thick \emph{p}Si grating showing a strongly suppressed rainbow and a high, color balanced transmission. \pnl{d}~Simulated zeroth-order transmittance spectra of the Cr grating and optimized GMR metasurface under normally-incident planewave illumination. The dashed blue curve shows the situation where the materials absorption in the Si is artificially set to zero. The polarization direction is in parallel to the grooves of the grating (TE). \pnl{e}~Simulated diffraction efficiency into the +1 diffracted-order transmission for the 30\,nm thick Cr grating (black solid curve) and the nonlocal metasurface under normally-incident TE polarized planewave incidence (red solid curve). Both structures reach $\sim10\%$ efficiency at the target wavelength of 870\,nm. The dashed blue curve shows the situation where the materials absorption in the Si is artificially set to zero. The horizontal dashed line indicates a $0.1\%$ upper limit for visible light diffraction to avoid perceptible rainbows. The purple shaded region shows the absorption of \emph{p}Si. Reproduced with permission from Ref.~\cite{Song:2021} (Copyright~\textcopyright~2021 Nature Springer).}
    \label{fig:Song-Fig2}
\end{figure}

Figure~\ref{fig:Song-Fig1}\pnl{c} shows the simulated angle-dependent transmittance on resonance ($\lambda_0 = 630$\,nm) and for illumination wavelengths 3\,nm above and below it. It highlights that the sharp transmission feature also naturally comes with a narrow angular selectivity. The dispersion relation of the nitride waveguide mode governs for what pairs of wavelengths and angles light coupling can be achieved. For analog optical computing, image processing, and the compression of free-space in optical systems, it is critical to have nonlocal metasurfaces with well-defined dispersive behaviors that can be engineered through their geometric parameters. In other applications, it can be desirable to fully decouple the optical functions at different wavelengths and angles of incidence. Eye tracking is an illustrative application and we discuss it below. 
Figure~\ref{fig:Song-Fig2}\pnl{a} shows a photograph of a basic eye tracker prototype where we patterned a large-area high-$Q$, nonlocal metasurface on a pair of glasses [18]. A near-infrared (NIR) light-emitting diode emitting at 870\,nm illuminates the eye and the metasurface redirects the scattered light from the eye toward a miniature camera that is attached to one of the arms. The metasurface should provide the person wearing the glasses with an unperturbed view of the outside world, while a high-speed camera can image and follow the eye’s motion. This can in principle be accomplished with other compact optical elements (e.g. gratings, miniature prisms, local metasurfaces, and waveguides) placed directly in front of the eye, but these dispersive components inevitably produce unwanted rainbows [Fig.~\ref{fig:Song-Fig2}\pnl{b}]. This is very distracting when used in environments with bright optical sources, such as the sun. A high-$Q$, nonlocal metasurface can provide a solution to this problem by performing different, independent operations on light waves at different frequencies and incident angles. In our example, all light waves in the visible (VIS) spectrum should be allowed to pass completely unperturbed, independent of the incident angle. However, in the NIR we aim to only redirect light across a narrow band of wavelengths and a small range of angles to the camera for imaging. 

To see the benefits of our metasurface design, we first consider the properties of a conventional grating comprised of 30\,nm thick chromium (Cr) strips whose dimensions were optimized to redirect 10\% of a normally-incident NIR beam through the first diffracted order. Figure~\ref{fig:Song-Fig2}\pnl{b} shows the low transmission of this element and the intense rainbows that it creates. These aspects are shown quantitatively as the black curves in Figure~\ref{fig:Song-Fig2}\pnl{d,e}. The rainbow is formed because of the high efficiency for light redirection into the first diffracted order. Figure~\ref{fig:Song-Fig2}\pnl{c} shows the same scene as viewed through a high-$Q$, nonlocal metasurface that has 3\,nm thick poly-crystalline silicon (\emph{p}Si) strips placed on a Si$_3$N$_4$ waveguide. This nonlocal metasurface shows a very high transmissivity at normal incidence across the visible range (89\% on average). This can be expected as the grating is at least 2 orders of magnitude thinner than the absorption depth in \emph{p}Si. Moreover, with proper design the Si$_3$N$_4$ waveguide and silica cover can also serve as a high-performance, double-layer antireflection coating. The transmission spectrum for the nonlocal metasurface shows a sharp spectral feature near 870\,nm that can be attributed to the excitation of the guided resonance. At this resonance, a second indirect pathway for the light through the structure. On resonance, about 40\% of the light is taken out of the incident beam and redirected into four first-order diffracted beams ($+1$ and $-1$ in forward and backward directions). Based on the high diffraction efficiency seen at 870\,nm, one would naturally expect to see high diffraction-efficiencies in the visible range as well. However, the guide mode resonances in the VIS (at 608\,nm, 496\,nm, and 470\,nm) display a very low diffraction-efficiency ($< 0.07\%$), explaining the virtual absence of rainbows as shown by the solid red curve in Fig.~\ref{fig:Song-Fig2}\pnl{e}. The difference in behavior in the VIS and NIR lies in the spectral absorption properties of the 3-nm-thin \emph{p}Si metasurface elements. The significant suppression of diffractions in the VIS takes place when the absorption coefficient of \emph{p}Si is above 1\,cm$^{-1}$ [purple shaded region in Fig.~\ref{fig:Song-Fig2}\pnl{e}]. This can be seen by analyzing the dashed curves in Fig.~\ref{fig:Song-Fig2}\pnl{d,e} that show the transmission and diffraction efficiency for the case that the materials absorption in \emph{p}Si is artificially set to zero. They highlight how the materials absorption in the metasurface elements can control the flow of light through the device on resonance; In the presence/absence of materials absorption, the diffracted orders can selectively be turned off/on without notably impacting the response off-resonance.
One key insight is that all of the light that is redirected by the nonlocal metasurface has to flow through the waveguide first (i.e. follows the resonant channel). Because of the high $Q$ of the resonance, the light-matter interaction in the waveguide is very strong. As such, even very small amounts of materials absorption in the waveguide can dramatically impact the flow of light through the structure. When the absorption efficiency of the metasurface elements (\emph{p}Si strips) exceeds their scattering efficiency, the guided mode is damped and all of the light will necessarily follow the direct path only. As \emph{p}Si strongly absorbs light in the visible, the diffraction channels and rainbows are effectively suppressed. When the absorption is weak (for \emph{p}Si in the infrared), the light will couple to the guided mode and be decoupled from the waveguide to the camera before it is absorbed. At a high level, this example shows that diffractive optical functions can be deactivated by introducing a little bit of materials absorption.

\subsection*{Future developments and concluding remarks}

The use of materials absorption in nonlocal metasurfaces brings the valuable opportunity to decouple optical functions at different wavelengths and angles. In the eye tracking example, this was utilized to achieve different behaviors in the visible (transparency) and infrared regime (imaging of the eye). By further engineering the spectral absorption behaviors, one can imagine achieving highly-selective transmission/reflection of light waves at certain frequencies and/or incident angles. Through the introduction of materials with dynamically tunable absorption features, it may also be possible to modulate such transmission speeds that are ultimately only limited by the quality factor of the nonlocal metasurface. Such optical elements will have a range of applications in optical imaging, computation, communication, and sensing.

\section[Nonlocal metalenses (Levy)]{Nonlocal metalenses}

\label{sec:Levy}

\author{Uriel Levy\,\orcidlink{0000-0002-5918-1876}}

\subsection*{Overview}

Metasurfaces are artificially engineered, subwavelength-thickness and subwavelength patterned optical surfaces that can control the amplitude, phase, and polarization of light. They can also enhance light-matter interactions at the nanoscale~\cite{Bomzon:2002,Levy:2004,Lalanne:1999,Overvig:2019,Lin:2014,Kuznetsov:2024,Kildishev:2013}.  
Such metasurfaces can be classified into local or nonlocal metasurfaces, depending on the interaction of light with the nanoscale meta-atoms of the metasurface. If the metasurface is based on the interaction of light with an individual meta-atom and long-range interactions does not play a role, the metasurface is coined "local metasurface". If on the other hand, the effect depends not only on the local phase control arising from the independent response of each meta-atom but also on the nonlocal (collective) resonance excitation originating from the interactions between neighboring meta-atoms, the metasurface is coined "nonlocal metasurface". The latter offers opportunities related to high $Q$ resonance~\cite{Fang:2024,Huang:2023}, wavelength selectivity and tunability~\cite{Chen:2025,Damgaard-Carstensen:2023,Klopfer:2022,Li:2019}. 
One of the most dominant types of metasurfaces is the metalens, sometimes defined also as a metasurface lens. The metalens is a flat optical device that performs similar operations as conventional lenses. However, its phase is controlled by the interaction of light with the meta-atom, and the diffraction angle is also dictated by the local periodicity of the metalens, which is typically chirped away of its center, to facilities larger diffraction angles that are needed to focus all incident parallel light rays to the same focal spot~\cite{Lalanne:2017,Khorasaninejad:2017}. In essence, the metalens is an advanced version of the diffractive lens, and it offers several advantages over diffractive lens, the such as polarization control, continuous phase control using a binary structure, CMOS compatibility, and reduced thickness to name a few~\cite{Engelberg:2020}. Being based on diffraction, metalenses are inherently dispersive, featuring strong chromatic aberrations~\cite{Engelberg:2021}. This is perhaps one of the main, if not the major drawback of metalenses, limiting their applicability to narrow band use cases. However, for broadband illumination, it would be ideal if the metalens can support large bandwidth, or alternatively provide high transmission only at the wavelength which it is designed for, while not affecting other wavelengths to avoid blurred image. To do so, the metalens needs to be highly spectral selective, which is typically achieved using a high $Q$ resonance mechanism. This is where nonlocal metalenses come into play. 

\subsection*{Current status}

In recent years, several nonlocal metalenses have been demonstrated. Malek \emph{et al.}~\cite{Malek:2022} demonstrated a nonlocal metalenses based on nonlocal dielectric metasurfaces that offer both spatial and spectral control of light. The demonstrated metalens was focusing light exclusively over a narrowband resonance while leaving off-resonant frequencies unaffected. Meta-atoms were designed based on the concept of quasi bound state in the continuum (q-BIC), encoded with a spatially varying geometric phase. They achieved effective transmission (polarization conversion efficiency) of 8\%, lower than the theoretical limit of 25\%, and a $Q$ factor of 86. It is mentioned that conversion efficiency can be further increased by better design, better fabrication and higher out of plane symmetry breaking. Yao \emph{et al.}~\cite{Yao:2024} also demonstrated a nonlocal high-$Q$ Huygens’ metalens operating under circular polarized illumination and explored its application in wavelength-selective imaging. To do so, an integrated-resonant unit (IRU) was designed to achieve a balance between $Q$-factor, efficiency, and the robustness of meta-atom’s orientation. By introducing structural asymmetry in the configuration, they could excite the leaky q-BIC mode, resulting in a high $Q$-factor. Each IRU consisted of a crescent nanopillar made of amorphous silicon placed on the silica substrate. Three different designs were fabricated, with $Q$ factor as high as $\sim$100 and very high polarization conversion efficiency of $\sim$70\%. The high efficiency, effectively exceeding the theoretical limit of 25\%, is attributed to varying the structural parameters to modulate the resonant wavelengths of magnetic dipole resonance and the q-BIC mode. 
Basically, subwavelength-thick local metalenses perform well only over a limited input angular range, whereas nonlocal metalenses can achieve high efficiency and diffraction-limited characteristics over a much wider field of view (FOV). Yet, such nonlocal metalenses are typically thicker, as they need a minimal thickness to provide the nonlocality. Since a large FOV requires an angle-dependent response (i.e., angular dispersion), a spatially localized incident wave must spread as it propagates through the metalens under space-angle Fourier transform. Thus, increasing the angular diversity can be facilitated by increasing the degree of nonlocality. Based on this, Li and Hsu~\cite{Li:2022c} discussed the thickness bound for nonlocal wide field of view (FOV) metalenses. They were able to show that there is an intrinsic trade-off between achieving a desired broad-angle response and reducing the thickness of the metalens. This thickness bound originates from the Fourier transform duality between space and angle. When applied to wide-FOV lenses, their approach predicted the minimal thickness as a function of the FOV, metalens diameter, and numerical aperture. Their formalism can provide guidance for the design of nonlocal metasurfaces in general, and nonlocal metalenses in particular, with the understanding that mitigating high FOV by a metalens of negligible thickness might be an extremely challenging task. In another paper~\cite{Li:2023}, the same authors introduced transmission efficiency limit for nonlocal metalenses. Using the same concept of describing the device as a transmission matrix that relates the input to the output, they were able to develop a term that links the efficiency to various metalens parameters. Specifically, it was shown that if the input and the output apertures are equal, the efficiency drops with the increase of the numerical aperture ($\mathrm{NA}$) following the relation $\eta\propto \sqrt{1-\mathrm{NA}^2}$, where $\eta$ is the efficiency. However, if the input aperture is set to an optimal value, an increased transmission efficiency bound is obtained significantly surpassing this relation.  

\subsection*{Challenges and opportunities}

One of the major opportunities of nonlocal metalenses is the ability to control the amplitude, phase and polarization response as a function of angle (see also Section~\ref{sec:Song} of this Roadmap). This may provide an opportunity to address the challenge of reducing aberration in large FOV applications, typically encountered in modern imaging. For example, the main camera of a smartphone can now support FOV as high as $\sim$800, and the ultra-wide camera supports an even higher FOV. This imposes a great challenge in the design process of the metalens. Clearly, a slightly thickener metalens which is designed to take advantage of the nonlocal phase response may cope better with the goal of supporting a higher FOV. Here, focusing on narrow band illumination may become helpful, because supporting simultaneously large bandwidth, large FOV and large Fresnel number (for high light collection efficiency) is perhaps beyond the capability of a wavelength or subwavelength-thick metalens. Another challenge of the nonlocal metalens is related to nanofabrication. While simulations show that high $Q$ nonlocal metalenses are feasible, it is extremely difficult to achieve such high $Q$ values in practice. Any small deviation from the design parameters due to fabrication imperfection may result in a significant decrease in the $Q$ factor. This is particularly relevant for nonlocal metalens, relying on a collective response of the meta-atoms. In such a case, a small perturbation in one or a few meta-atoms will affect the response of the entire device. Considering the need for large apertures to support high light collection efficiency and high $Q$ factor in collective modes, the material from which metalens is made is crucial. While for the near infrared (beyond $\sim$1 micron wavelength) amorphous silicon is an excellent material choice, this is not the case for shorter wavelength, where the absorption of amorphous silicon is a limiting factor. Materials such as titanium dioxide are not CMOS compatible and are very difficult to handle over a large area. A promising direction for the implementation of nonlocal metasurfaces and particularly nonlocal metalenses operating in the visible range is to use silicon rich nitride (SRN). SRN films can be easily applied on almost any substrate using a relatively simple process, e.g. using plasma enhanced chemical vapor deposition (PECVD). The refractive index of the SRN film can be tuned by controlling the atomic ratio between silicon and nitride. We have recently implemented metasurfaces based on SRN for digital holography and structural colors applications, and a similar material platform can be applied for nonlocal metalenses~\cite{Goldberg:2024,Goldberg:2025}.  
As mentioned before, one of the major opportunities of nonlocal metalenses is the capability to achieve high $Q$ factor. On the one hand, this requires a delicate design that combines the local phase response with the collective response of many meta atoms. On the other hand, as discussed before, the interference between local modes and nonlocal modes is helpful in overcoming transmission limitations. The high $Q$ factor implies that the design is sensitive to fabrication imperfection and environmental effects such as temperature drifts. Yet, it allows large tunability using e.g. the Pockels or the Kerr effect, the plasma dispersion effect in semiconductors, the thermo optic effect. Tunability can also be achieved by structural modifications (e.g. MEMS based device~\cite{Han:2022,Meng:2024}, and the use of external medium such as liquid~\cite{Li:2022d} and even quantum vapor~\cite{Bar-David:2017}. Real time tunability of nonlocal metalenses is probably one of the major research topics in the field. 

\subsection*{Future developments to address challenges}

\begin{figure}[hbt!]
    \centering
    \includegraphics[width=0.8\linewidth]{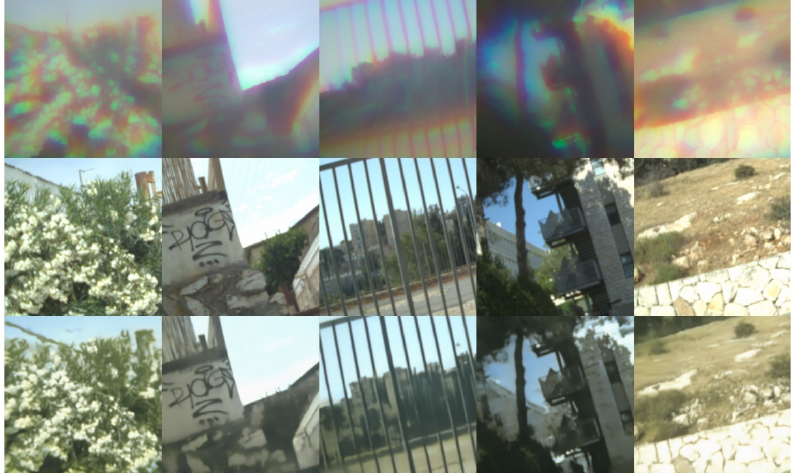}
    \caption{Top row -- 5 aberrated test-set images. Middle row -- The corresponding ground truth images. Bottom row -- The reconstructed images. Reprinted (adapted) with permission from Ref.~\cite{Maman:2023} (Copyright~\textcopyright~2023 American Chemical Society).}
    \label{fig:Levy-Fig1}
\end{figure}

The previous sections have been dealing with the status of nonlocal metalenses and highlighted opportunities related to nonlocality. Yet, a single nonlocal metalens, while holding a great promise, will probably be unable to cope with the grand challenge of simultaneously providing diffraction limited imaging with ultra-high resolution over the entire visible range, with large Fresnel number and large FOV.  One possible way to overcome this limitation is to consider a hybrid refractive-metasurface design, where most of the optical power will be delivered by the refractive optical system, whereas the metasurface/metalens will be used for improving the imaging quality by providing some sort of aberration correction. The metasurface/metalens can also be used to simplify the requirements from the refractive optical system. Another possible avenue is to use a multi-layer nonlocal metalens, providing additional degrees of freedom and a slightly larger thickness to improve the quality of the metalens. In our opinion, an even more promising direction is to use the hybrid concept of combining the nonlocal metalens with computational imaging approach. This has been recently demonstrated for a local flat lens~\cite{Maman:2023}. As shown in Fig.~\ref{fig:Levy-Fig1}, the flat lens produced a highly chromatic aberrated image. Yet, by applying a deep-learning-based approach, and a proper training procedure it was possible to overcome chromatic aberrations of flat lens-based imaging systems operating outdoors under normal ambient illumination conditions.  This approach, combined with advanced nonlocal metalens design, may be used to achieve high quality imaging under different scenarios, ranging from ambient light conditions to multi-color LED illumination.

\subsection*{Concluding remarks}

The use of nonlocal metalenses in imaging systems may provide an enhanced tool for the realization of high quality imaging systems based on flat optics technology. The capability to control amplitude, phase and polarization as a function of incident angle and wavelength of illumination may pave the way for advanced imaging functionalities. Nonlocal metalenses may also provide multi-functionality, such as imaging and spectral filtering, imaging and FOV manipulations, and polarimetric imaging to name a few. While this chapter was focused on the visible and the near infrared spectral regime, nonlocal metalenses can also be used for imaging applications at longer wavelengths, ranging from the mid infrared, through the thermal domain, and all the way to the terahertz. Finally, it is important to understand that while a single nonlocal metalens can be very useful for many use cases, it cannot cope with the grand challenge of performing diffraction limited imaging over large bandwidth, large FOV and high Fresnel number simultaneously. For such a task, applying hybrid approaches e.g. refractive lens and nonlocal metalens, or nonlocal metalens with additional computation imaging effort is a key for mitigating the stringent requirements of modern imaging systems. 

\section[Nonlocal metasurfaces for space compression (Long \emph{et al.})]{Nonlocal metasurfaces for space compression}

\label{sec:Long}

\author{Olivia Y. Long\,\orcidlink{0000-0003-4636-9463}, Cheng Guo\,\orcidlink{0000-0003-4913-8150} \& Shanhui Fan\,\orcidlink{0000-0002-0081-9732}}

\subsection*{Current status}

Miniaturization has been a common trend in the development of optical technologies, driven by the goal of improved efficiency, compactness, and portability of devices. With the rapid rise of artificial intelligence, there is an accompanying need for fast processing of large data inputs in a scalable way, further fueling the endeavor for smaller components~\cite{Wetzstein:2020}. The bulk of conventional optical setups consists of free-space, including space filled with uniform dielectric materials~\cite{Goodman:1996}.
In recent decades, metasurfaces with subwavelength scatterers have attracted increasing attention for their ability to manipulate the phase, amplitude, and polarization of light, paralleling advances in nanofabrication techniques~\cite{Chen:2016}. However, most research efforts have been toward designing more compact optical components such as lenses, bandpass filters, quarter-wave plates, and polarization converters~\cite{Yu:2014}. 
Here, we focus on a complementary effort: the compression of free-space.

\begin{figure}[ht!]
\centering
\includegraphics[width=0.7\textwidth]{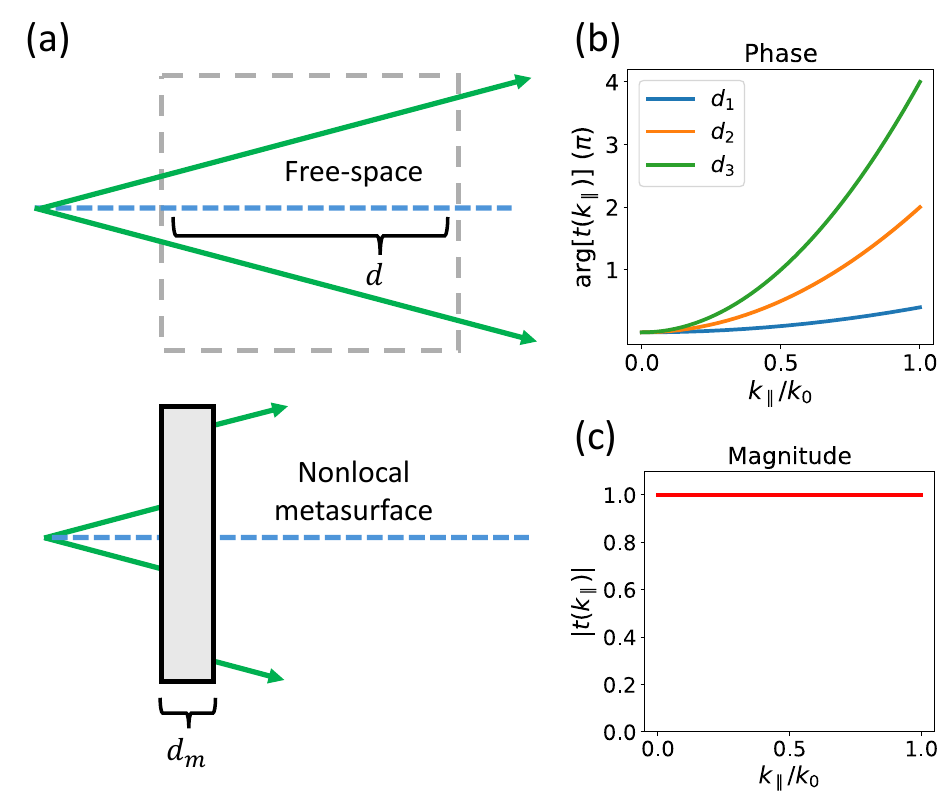}
\caption{\pnl{a} Schematic of nonlocal metasurface for space compression. Top inset shows the equivalent free-space propagation through a distance $d$ mimicked by transmission through the nonlocal metasurface with thickness $d_m < d$ (bottom inset). \pnl{b}~Ideal phase dependence of metasurface transmission coefficient on the transverse wavevector $k_\parallel$ for designs with different compression ratios.
Here, the effective free-space propagation distances of the curves are related by $d_{1} < d_{ 2} < d_{3}$ due to the different quadratic curvatures. \pnl{c}~Ideal magnitude of transmission coefficient, allowing unity transmission of light at all angles of incidence.}
\label{fig:Long-Fig1} 
\end{figure}

The propagation of light in free space over a distance $d$ in the paraxial regime is described by a quadratic phase transfer function~\cite{Goodman:1996}: 
\begin{align}\label{eq:Long-Eq1}
    H(k_{\parallel}) = e^{-ik_0 d} e^{i\frac{\lambda_0 d}{4\pi} k_{\parallel}^2 }
\end{align} 
where $k_0$, $\lambda_0$, and $k_{\parallel}$ are the free-space wavevector, free-space wavelength, and transverse component of the wavevector, respectively. To compress free space, one aims to construct a device that achieves the same phase transfer function of Eq.~\eqref{eq:Long-Eq1}, but with a physical thickness that is much smaller than the distance $d$. To achieve this aim, the dependence on the transverse wavevector component $k_{\parallel}$ naturally suggests the use of nonlocal metasurfaces, which manipulate light in the wavevector space. To effectively compress space, the designed metasurface, with a thickness $d_m$, should impart a phase factor $e^{i\phi k_{\parallel}^2 }$, with the equivalent free-space propagation distance $d=4\pi \phi / \lambda_0 > d_m$. 
The performance of the metasurface can thus be quantified by a compression ratio $R = d/d_m$. Moreover, the amplitude of each $k_{\parallel}$-component of the incident light should ideally be preserved over a wide frequency bandwidth and range of incident angles. Fig.~\ref{fig:Long-Fig1}\pnl{a} schematically shows the free-space propagation distance $d$ (top inset) that is replaced by transmission through a nonlocal metasurface (bottom inset). The compression ratio is determined by the quadratic dependence of the transmission phase on $k_\parallel$. A comparison of three ideal transmission phase profiles with different compression ratios is shown in Fig.~\ref{fig:Long-Fig1}\pnl{b}, and the ideal transmission amplitude as a function of $k_\parallel$ is depicted in Fig.~\ref{fig:Long-Fig1}\pnl{c}, where the magnitude remains unity for all angles of incidence. 

\begin{figure}[ht!]
\centering
\includegraphics[width=0.9\textwidth]{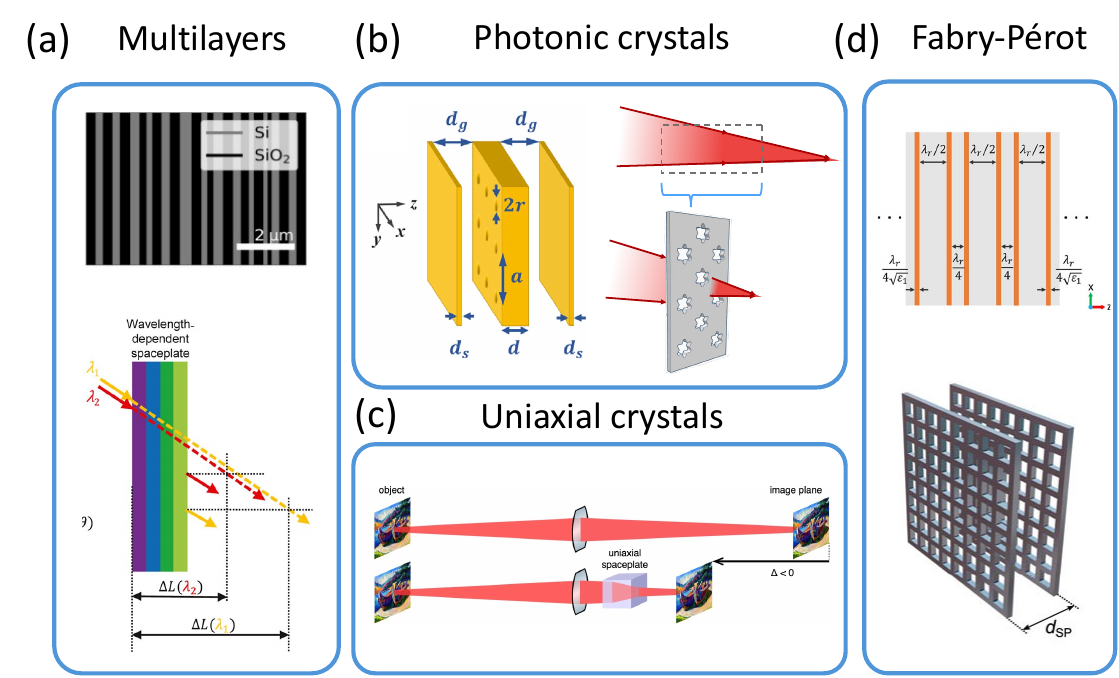}
\caption{Nonlocal metasurface designs for space compression. \pnl{a} Optimized multilayer designs. Top panel is reproduced with permission from Ref.~\cite{Page:2022} (Copyright~\textcopyright~2022 Optica Publishing Group) while bottom panel is reproduced with permission from Ref.~\cite{Shao:2024} (Copyright~\textcopyright~2024 American Chemical Society). \pnl{b} Photonic crystal slab designs. Left panel is reproduced with permission from Ref.~\cite{Cheng:2020} (Copyright~\textcopyright~2020 Optica Publishing Group), while right panel is reproduced with permission from Ref.~\cite{Long:2022} (Copyright~\textcopyright~2022 American Physical Society). \pnl{c} Uniaxial crystal spaceplate. Reproduced with permission from Ref.~\cite{Reshif:2021} (Copyright~\textcopyright~2021 Nature Springer). \pnl{d} Coupled multi-resonator design~ \cite{Chen:2021} and single Fabry--P\'erot cavity design from Ref.~\cite{Mrnka:2022}. Top panel is reprinted (adapted) with permission from Ref.~\cite{Chen:2021} (Copyright~\textcopyright~2021 American Chemical Society), while lower panel is reproduced with permission from Ref.~\cite{Mrnka:2022} (Copyright~\textcopyright~2022 American Institute of Physics).}
\label{fig:Long-Fig2}
\end{figure}

Pioneering theoretical proposals for space compression harnessed guided mode resonances of photonic crystal slabs~\cite{Cheng:2020}  and multilayer structures~\cite{Reshif:2021}, while the first experimental demonstrations employed a uniaxial crystal and low-index homogeneous media~\cite{Reshif:2021} to achieve the desired transfer function [Eq.~\eqref{eq:Long-Eq1}].
To date, other nonlocal metasurface designs for space compression (also referred to as "spaceplates") include Fabry--P\'erot resonators~\cite{Chen:2021,Mrnka:2022}, $4f$ lens-based approaches~\cite{Sorensen:2023}, and Huygens' metasurfaces~\cite{Diaz-Fernandez:2024}, as well as photonic crystal and multilayer designs with additional functionalities and optimizations~\cite{Long:2022,Shao:2024,Page:2022}. A visual summary of many of these design approaches is provided in Fig.~\ref{fig:Long-Fig2}.

Single guided-mode resonance designs have achieved compression ratios ranging from 5 to 144 with numerical apertures (NA) of 0.32 and 0.01, respectively, often with polarization dependencies and limited bandwidths of operation~\cite{Cheng:2020, Long:2022, Chen:2021}.
Multiple coupled resonances have been shown to yield wider operating bandwidths~\cite{Chen:2021}, and optimized multilayers have achieved larger compression ratios of 340 at the expense of a relatively small NA (0.017)~\cite{Page:2022}. 
Thus, specific platforms may be more favorable depending on the specific use case. Ongoing research strives to achieve the ideal structure with simultaneously high compression ratio, large NA, wide frequency bandwidth, and polarization independence. In this article, we focus on the approach of nonlocal metasurfaces for space compression.

\subsection*{Challenges and opportunities}

In general designs for space compression, one would like to have strong quadratic dependence of the phase $\phi(k_{\parallel})$ on the transverse wavevector $k_{\parallel}$ to achieve higher space compression ratios. In the nonlocal metasurface approach, one uses resonances to achieve such quadratic dependence. However, since the strongest dependencies on $k_{\parallel}$ occur near resonances, such designs are limited by a narrow frequency bandwidth of operation. Thus, one challenge in this approach to space compression is the inherent tradeoff between the frequency bandwidth and the transverse wavevector range of operation (numerical aperture). Proposed approaches to achieve wider bandwidths to date include using multiple resonances~\cite{Chen:2021}.

Polarization-independent operation would also be advantageous for integration into existing optical systems. One drawback of many recent designs, including both designs using nonlocal metasurfaces and designs based on multilayer films, is that the optical response of the device differs for TM and TE polarized light. Recent work has proposed polarization-independent space compression using a photonic crystal slab structure that features bands with similar curvatures, resulting in similar resonance responses for each polarization~\cite{Long:2022}. It is of interest to achieve polarization independence as well as a large compression ratio over a sufficiently large numerical aperture.

Since nonlocal metasurfaces for space compression have been mostly applied to imaging applications thus far and are often used in conjunction with lenses, it is desirable to achieve designs with tunable compression ratios for the purpose of reconfigurable focal distances. Thus, another opportunity for future development is in improving the programmability of such nonlocal nanophotonic structures.
Efforts in the area of active nonlocal metasurfaces have included electro-optic, mechanical, and thermo-optic tuning~\cite{Malek:2021}. Challenges in such approaches include operation efficiencies, speed of reconfigurability, and maintaining a pure phase response throughout the tuning process.

There is ample opportunity to explore integration of space compression with other functionalities, with the goal of expanding the domain of applications beyond conventional imaging. For example, the narrow bandwidth and angular ranges of current designs could be harnessed to achieve simultaneous polarization conversion or diffraction suppression at other operating wavelengths, which can be useful in augmented reality and eye-tracking technologies~\cite{Song:2021, Overvig:2020}. Nonlocal spaceplates can also be co-designed with existing lenses and local metasurfaces for functionalities such as aberration correction, as demonstrated in Ref.~\cite{Shao:2024}.

Finally, in the design of nonlocal metasurface for space compression, a theoretical challenge is to understand the tradeoff between various performance metrics, such as frequency bandwidth, numerical aperture, polarization independence, and compression ratio. For multilayer and single-resonance structures, the tradeoffs between frequency bandwidth and numerical aperture, and between frequency bandwidth and compression ratio, have been examined in Ref.~\cite{Shastri:2022}. There have also been works on the fundamental constraints of the thickness of optical devices in general \cite{Miller:2023}. It would be of interest to apply and further develop the formalisms in these papers to understand the design tradeoff in nonlocal metasurfaces.

\subsection*{Future developments to address challenges}

One of the challenges in achieving widespread adoption is designing nonlocal metasurfaces with the target performance metrics such as numerical aperture, bandwidth, and efficiency. Thus, future development in the understanding of the spaceplate design space would help accelerate the integration of such devices in optical systems.
Exploration of the design space of space-compressing nonlocal structures has thus far included gradient-based optimization~\cite{Shao:2024} and exploiting topological properties in the structural parameter space~\cite{Long:2023}. With the rise of artificial intelligence, we anticipate further optimization of existing designs using machine learning techniques such as those based on neural networks.

The development of new techniques for fast programmability of the compression ratio would help increase the range of applications for space-compressing nonlocal metasurfaces. Technological advances in areas such as electrically and optically tunable materials would contribute to greater reconfigurability in both nonlocal and local metasurface designs~\cite{Abdelraouf:2022}. The simulation of metasurfaces designed with such materials would also be an interesting theoretical opportunity to explore.

Given that nonlocal metasurfaces are a more recent development compared to local metasurfaces, future integration of the two could help address the challenge of increasing multifunctionality. By harnessing the momentum-space control of nonlocal metasurfaces and the local wavefront control of local metasurfaces, new emergent functionalities will enable next-generation optical systems. For instance, a recent work demonstrates the use of spaceplates in correcting both chromatic and spherical aberrations from refractive and local metasurface lenses~\cite{Shao:2024}. Building upon this work, the development of adaptive, tunable devices for automatic aberration correction represents an exciting frontier. Such multifunctional capabilities further expand the potential applications of nonlocal metasurfaces. Furthermore, the co-design of nonlocal metasurfaces with a computational backend may help overcome some of the limitations of the metasurface alone, as demonstrated recently for local metasurfaces~\cite{Colburn:2018}.

Beyond single-resonance-based designs, multiple coupled resonances offer promising opportunities for improved performance. The key challenge lies in optimizing the collective arrangement of resonance poles to achieve both unity transmission amplitude and the desired phase response across a broader bandwidth. This complex optimization problem is particularly well-suited for machine learning and AI methods, which can efficiently explore the vast design space.

While our discussion has focused on compression of free space, similar principles could be extended to guided waves. An intriguing question emerges: can we compress the propagation length in waveguide arrays or in multimode waveguides? This concept could significantly reduce the footprint of photonic links in integrated photonic circuits, opening new possibilities for compact optical interconnects.

Looking beyond electromagnetic waves, these space compression principles could be extended to other wave varieties, including electron, spin, and acoustic waves. Particularly promising is the application to electron optics, where compressed free-space propagation could enhance electron microscopy capabilities. Similarly, the manipulation of spin and acoustic waves through space compression could enable novel functionalities in materials science and device engineering.

\subsection*{Concluding remarks}

With further advancements in key metrics such as numerical aperture, bandwidth, and compression ratios, the integration of nonlocal metasurfaces for space compression into everyday optical devices such as cameras could become a reality in the near future. 
Moreover, the functionality of nonlocal space compression, in combination with those of local metasurfaces, highlights the rapidly advancing techniques in the precise manipulation of light at an increasingly smaller scale. The demonstration of space compression points to the potential for many more possibilities in the area of nonlocal metasurfaces, which may expand beyond imaging.

\section[Electro-optic spatiotemporal nonlocal metasurfaces (Bozhevolnyi)]{Electro-optic spatiotemporal nonlocal metasurfaces}

\label{sec:Bozhevolnyi}

\author{Sergey I. Bozhevolnyi\,\orcidlink{0000-0002-0393-4859}}


\subsection*{Overview}

Spatiotemporal light control with dynamic optical metasurfaces has been high on the research agenda, promising attractive solutions and new avenues for modern highly integrated optics and nanophotonics. Due to ultrafast responses of electro-optic materials, various electro-optic metasurface configurations have been scrutinized with the aim of realizing both efficient and ultrafast dynamic metasurfaces. Given fundamentally nm-thin metasurfaces and very weak electro-optic responses, advances towards this ambitious goal have been facing enormous challenges without any clear way through at hand. Here, several promising electro-optic metasurface configurations are presented that exploit nonlocal interactions between incident radiation and grating-excited waveguide modes propagating inside nm-thin electro-optic films. Challenges and opportunities for spatiotemporal light control with electro-optic nonlocal metasurfaces are discussed, outlining also future developments towards tackling the most serious challenges and exploiting the most exciting opportunities.

\subsection*{Current status}

Optical metasurfaces formed by surface arrays of nanoscale resonant elements constitute one of the most vibrant and explosively developing fields in modern optics and photonics, primarily due to their ability to effectuate complete control over the transmitted/reflected fields enabling thereby diverse optical functionalities, including those not realizable with conventional bulky optics~\cite{Ding:2018}. Most metasurfaces are however static in nature with their optical responses being determined at the design stage and set in the process of fabrication. Recent years have therefore seen increasing efforts to realize dynamically controlled (tunable) optical metasurfaces, rapidly progressing towards faster operation and realization of spatiotemporal light control, which is immensely attractive for many fascinating phenomena waiting to be realized and exploited in highly integrated optics and nanophotonics~\cite{Shaltout:2019}. Among various types of dynamically controlled metasurfaces, electrically tunable optical metasurfaces have shown great promise due to their fast response time, low power consumption, and compatibility with existing electronic control systems, offering unique possibilities for dynamic tunability of light–matter interactions via electrical modulation. It should be noted that to exercise truly spatiotemporal light control, the electrical operation bandwidth should exceed, preferably by orders of magnitude, the optical bandwidth of the radiation to be controlled, requiring thereby ultrafast modulation~\cite{Shaltout:2019}. The linear electro-optic (Pockels) effect, which is found in several crystalline (ordered) media without center of symmetry, ensures inherently fast electrical control of material birefringence with femto- or even attosecond-short response times. The main challenge when utilizing the electro-optic materials in optical metasurfaces is associated with inherently small refractive index variations accessible, a circumstance that, in combination with fundamentally thin nature of metasurfaces, results in enormous difficulties in their practical realization.

Electro-optic metasurface configurations, whether based on electro-optic polymers~\cite{Benea-Chelmus:2022} or lithium niobate (LN) films~\cite{Weiss:2022,Damgaard-Carstensen:2023}, are designed to have electro-optic thin films sandwiched between bottom and top electrodes that can modify the film refractive index and thereby control the metasurface response~\cite{Benea-Chelmus:2022,Weiss:2022,Damgaard-Carstensen:2023}. Since the film thickness is typically subwavelength and the electro-optically induced index changes are very small ($< 10^{-3}$), one should exploit resonant
configurations that would increase the effective interaction length either by multiple reflections across the metasurface layer (i.e., electro-optic thin film) in Fabry--P{\'e}rot resonators~\cite{Weiss:2022} or by the nonlocal interaction~\cite{Shahbazyan:2023} of incident radiation with waveguide modes propagating along the metasurface layer~\cite{Benea-Chelmus:2022,Damgaard-Carstensen:2023}. The latter approach is somewhat alike to that based on the nonlocal interaction with surface lattice resonance (SLR) modes, typically enhanced by simultaneous excitation of localized nanoparticle Mie~\cite{Benea-Chelmus:2022} or plasmon~\cite{Weiss:2022} resonances. The excitation of quasi-bound states in the continuum (qBIC) has also been exploited~\cite{Benea-Chelmus:2022}. Considering the experimental characteristics demonstrated for these configurations~\cite{Benea-Chelmus:2022,Weiss:2022,Damgaard-Carstensen:2023}, while those based on electro-optic polymers exhibited superior efficiency and bandwidth~\cite{Benea-Chelmus:2022}, LN-based electro-optic metasurfaces~\cite{Weiss:2022,Damgaard-Carstensen:2023} offer potentially similar bandwidth along with the excellent environmental (temperature) stability, a feature that is in a stark contrast with electro-optic polymers exhibiting typically a glass temperature below $100^\circ$C.

\subsection*{Challenges and opportunities}

The prior research into electro-optic metasurfaces has convincingly demonstrated that the best performance is expected from the configurations involving the nonlocal interaction of incident radiation with waveguide modes propagating along the electro-optic layer. The design optimization goal is then to identify the physical mechanism along with the modes involved that would result in ultra-narrow transmission/reflection minima/maxima, since those would most efficiently be influenced even with very weak, electro-optically induced, refractive index changes. The guided mode excitation by gratings is governed by phase-matching that can result in very narrow transmission/reflection spectral features, often referred to as guided-mode resonances (GMRs), for very wide gratings and nearly parallel incident light~\cite{Wang:1990}. The latter requirement is not compatible with designing compact and robust practical devices and components. One can circumvent this challenge by making the excited waveguide modes to bounce back and forth in a distributed Bragg resonator (DBR), thus effectively extending the interaction between guided modes and grating couplers. Fortunately, this possibility seems being naturally present for the waveguide mode excitation by normal light incidence~\cite{Yezekyan:2024}.

\begin{figure}[hb]
    \centering
    \includegraphics[width=0.5\linewidth]{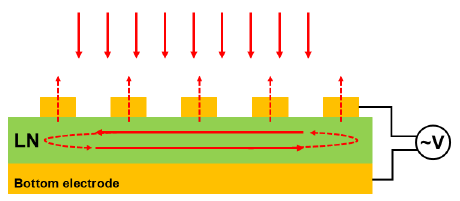}
    \caption{Schematic of an LN-based electro-optic metasurface operating under normal light incidence that, under the phase-matching condition, results in the excitation of counter-propagating waveguide modes. These modes, upon interaction with the electrode grating structure, couple out of the waveguide and Bragg reflect into each other.}
    \label{fig:Bozhevolnyi-Fig1}
\end{figure}

Let us consider a generic LN-based configuration of electro-optic metasurfaces illuminated at normal light incidence (Fig.~\ref{fig:Bozhevolnyi-Fig1}). The first-order phase matching between the incident light and counter-propagating waveguide modes requires that the grating period, $\Lambda$, matches the waveguide mode wavelength, $\lambda_W = \lambda/N_\mathrm{eff}$, where $N_\mathrm{eff}$ is the effective mode index and $\lambda$ is the wavelength in vacuum. Once this condition is satisfied, the waveguide modes experience the Bragg reflection in the second-order diffraction by the grating~\cite{Yezekyan:2024}. Thus, the resonant excitation of waveguide modes under normal light incidence is intrinsically linked to the DBR formation. The latter is associated with the occurrence of the photonic band gap reflected in the corresponding BIC and qBIC dispersion branches. The qBIC resonance is accessible at normal incidence, resulting typically in ultrahigh-$Q$ resonances~\cite{Huang:2023}. Note that the excitation of qBIC resonances in electro-optic metasurfaces, that cannot be avoided in the grating-assisted
coupling to waveguide modes at normal incidence, has already been exploited both explicitly~\cite{Benea-Chelmus:2022} and implicitly~\cite{Damgaard-Carstensen:2023}. The former (based on the electro-optic polymer) exhibited high modulation efficiency (60\%) and broad bandwidth (3\,GHz) when driven by $\pm 100$\,V~\cite{Benea-Chelmus:2022}. The latter (utilizing environmentally stable LN films) did not yet reach the same level of performance, but several avenues for the improvement are clearly in sight~\cite{Damgaard-Carstensen:2023}. One of the matters to be considered is that strong gratings enhance not only the Bragg reflection, improving thereby the DBR performance, but also the out-coupling of waveguide modes, thus increasing the energy dissipation in the system. The challenge of optimally balancing the metasurface lateral extension and DBR loss to radiation out-coupling by the grating and absorption by electrodes is yet to be carefully considered from the viewpoint of reaching the utmost (efficient and fast) performance of electro-optic metasurfaces.

\subsection*{Future developments to address challenges}

The general design direction can be formulated as realizing ultrahigh-$Q$ resonant transmission/reflection with the relevant electromagnetic fields efficiently overlapping with the electro-optic layer and controlling (electrostatic) fields (assuming that the electrostatic field is oriented to make use of strong components of the electro-optic susceptibility tensor~\cite{Huang:2023}). This requires minimizing the radiation losses, both by absorption and out-coupling. The former is unavoidable because of the presence of control electrodes (Fig.~\ref{fig:Bozhevolnyi-Fig1}), but can be decreased by exploiting transparent conductive materials, e.g., indium tin oxide (ITO)~\cite{Weiss:2022}, and/or increasing the layer thickness. It should be noted though that electrical characteristics of transparent oxides are significantly worse than noble metals, and their use might jeopardize the speed of operation, i.e., the modulation bandwidth. Increasing the dielectric layer thickness would also decrease the electrode capacitance and thereby increase the modulation bandwidth, but the driving voltage would proportionally increase as well, resulting in the overall increase of the electrical energy consumed by the device.
Decreasing the radiation loss requires reaching a delicate balance between the strength of coupling between incident free propagating fields and counter-propagating waveguide modes and that of Bragg reflection of waveguide modes. Both scattering channels originate in the diffraction of waveguide modes by the grating, corresponding to the first- and second-order diffraction, respectively. For the configuration shown in Fig.~\ref{fig:Bozhevolnyi-Fig1}, the design freedom is limited, since enhancing the scattering strength of individual grating elements (e.g., by making the metal ridges thicker and wider) would increase the corresponding diffraction efficiencies in a similar way. In that respect, recently introduced asymmetric (dielectric) gratings, in which a grating supercell contains several nonidentical grating elements, promise new avenues for reaching ultrahigh-$Q$ coupling (when considering only radiative losses) by decreasing the grating asymmetry~\cite{Yezekyan:2024,Huang:2023} and thereby boosting up the performance of electro-optic metasurfaces. Thus, for example, one can realize vanishingly weak in- and out-coupling diffraction channels, while maintaining very strong Bragg reflection and thereby forming a high-$Q$ DBR for waveguide modes~\cite{Yezekyan:2024}. It should be emphasized that very strong Bragg reflection implies the possibility of drastically reducing the grating lateral extension by effectively folding the waveguide propagation region, while maintaining the same overall resonance quality. This folding has immediate consequences for the electrode capacitance, which is proportional to the electrode system area, allowing thereby to drastically increase the operation bandwidth. Concomitantly, the area of individually addressed metasurface elements making up electro-optic spatiotemporal metasurfaces for dynamic molding of electromagnetic wavefronts could also be significantly reduced without loss of efficiency, increasing thus the spatial resolution available for spatiotemporal wave shaping~\cite{Sisler:2024}.
Perfecting the operation of electro-optic metasurfaces both as a single array and as an ensemble of individually addressed metasurface arrays requires different optimization strategies and targets different application domains. Thus, achieving complete light extinction with the electro-optic metasurface, and thereby 100\% modulation efficiency, would allow one
to realize dynamic phase contrast imaging using the same approach as that demonstrated recently with static non-local angle-selective metasurfaces~\cite{Ji:2022}. Dynamic functionality would in turn open other application directions, such as adjustable contouring of images features by gradually turning on (when applying the electrical control signal) the phase contrast. Combining active temporal and spatial field modulation with individually addressed arrays would enable unique optical functionalities, such as frequency mixing, harmonic beam steering and molding, as well as non-magnetic nonreciprocity among other things~\cite{Sisler:2024}.

\subsection*{Concluding remarks}

The considered configuration of electro-optic nonlocal metasurfaces (Fig.~\ref{fig:Bozhevolnyi-Fig1}) is deceitfully simple, leaving one wondering about its potential. Indeed, DBRs and associated bandgap dispersion branches are well known since the 1980s, when being extensively investigated and exploited for mode selection in distributed feedback semiconductor lasers~\cite{Kazarinov:1985}. However, the realization of waveguide grating couplers with metal (lossy) electrodes~\cite{Damgaard-Carstensen:2023}, addition of asymmetry into grating configurations~\cite{Yezekyan:2024,Huang:2023}, making beneficial use of nonlocality in electromagnetic interactions~\cite{Damgaard-Carstensen:2023,Shastri:2023} and yet unexplored finite-size effects add up to resulting in the configuration that is surprisingly reach in physical phenomena and potential applications as discussed above. It should be noted that the presented here considerations are all concerned the case of normal light incidence. Deviations from normal incidence would result in the occurrence of two transmission/reflection minima/maxima (instead of one at normal incidence) corresponding to the coupling at two dispersion branches~\cite{Yezekyan:2024}. This would enable certain tunability of the operation wavelength by controlling the incident angle close to normal incidence. This tunability, although limited to the DBR bandgap, is a very interesting (and yet unexplored) feature of the considered electro-optic nonlocal metasurfaces that should broad and enhance their application potential. Overall, I believe that many exciting and unexpected developments in this area are just around the corner, waiting to be discovered and explored.

\section[Cascaded diffractive nonlocal metasurfaces for highly multifunctional meta-optics (Overvig)]{Cascaded diffractive nonlocal metasurfaces for highly multifunctional meta-optics}

\label{sec:Overvig}

\author{Adam Overvig\,\orcidlink{0000-0002-7912-4027}} 

\subsection*{Current status}

Metasurface optics leveraging nonlocal scattering effects (e.g., interelement coupling) are emerging as a new frontier for compact optical systems with high information density. Taking advantage symmetry-breaking design principles and the frequency- and angle-selective responses of quasi-bound states in the continuum, highly multifunctional meta-optics may be achieved by cascading several metasurfaces~\cite{Overvig:2022,Malek:2022}. While local metasurface optics often aims for broadband and angle-insensitive operation, such an approach encodes comparatively negligible information in the electromagnetic wave compared to having distinct encodings at each frequency and angle. Nonlocal metasurfaces therefore offer a platform for vastly increasing our command of light within compact form factors. Most notably, the selectivity and mutual transparency of several resonant metasurface layers simplifies the design of a cascaded meta-optic; in contrast, cascaded local meta-optics require redesigning all layers upon the introduction of each new layer. Hence, cascaded nonlocal meta-optics offers a greatly simplified design space and approach for highly multifunctional devices.
Nonlocal metasurface optics may be distinguished into two categories~\cite{Shastri:2023}: periodic devices that manipulate many plane waves (operating solely in momentum space), and aperiodic devices that generalize plane wave selectivity to spatial selectivity (operating in both momentum and real space). The latter are referred to as "diffractive nonlocal metasurfaces", a generalized category of device encompassing both periodic nonlocal metasurface or local metasurfaces~\cite{Overvig:2022}. Here, we emphasize the potential of a cascaded set of these devices to encode distinct functionalities across many discrete angles and/or frequencies. Their nonlocality is engineered through high quality-factor ($Q$-factor) guided resonances, called quasi-bound states in the continuum, whose dispersion yields a sharp dependence on incident momentum (encompassing a combination of both angular and frequency selectivity). Meanwhile, the local scattering of these delocalized modes is controlled similarly to local metasurfaces: by leveraging the selection rules governing the leakage of these states to free space, tailored geometric perturbations applied pointwise to each unit cell customizes the response beyond plane waves, i.e., to any spatial profile of $Q$-factor, phase, and polarization state. Since the resulting devices would be periodic but for the applied perturbation, only resonant states interact strongly with the patterning; non-resonant states only weakly interact. This mechanism results in high wavelength and frequency selectivity, or "confidentiality", of the response: outside of resonances, the metasurface behave as unpatterned thin films. By simultaneously engineering the perturbation and the non-resonant (broadband) response, these devices can operate with resonant efficiency approaching unity for a reflection peak within a transmission window (i.e., a stopband filter in which the rejected band is reflected and spatially shaped)~\cite{Malek:2022} or a transmission peak within a reflection window (i.e., a passband filter with broadband rejection and a spatially shaped passed band)~\cite{Zhou:2023a}. With polarization manipulation, a converted state may pass in transmission while the broadband response is also transmissive, but the maximal efficiency is 25\%~\cite{Overvig:2022,Malek:2022}. 
A closely related class of metasurfaces employs the resonant phase of localized high $Q$-factor resonators to pattern diffracted light~\cite{Lawrence:2020,Hail:2023}. In these devices, the goal is frequency selectivity without angular selectivity, conferring high numerical aperture performance. While there inevitably exists some degree of nonlocality due to the long lifetime of the modes, flat dispersion can yield a moderate to strong degree of localization. This behavior is compatible with multi-frequency operation, but not multi-angle operation. In contrast, certain metasurface approaches have achieved modest angular-dependent wavefront transformation~\cite{Kamali:2017,Cheng:2017}. When rationally designed, these devices are primarily treated as if their unit cell response has angular dependence, but without explicit introduction of large interelement coupling~\cite{Kamali:2017}; when computationally designed~\cite{Cheng:2017}, interelement coupling associated with resonances often arises during optimization. This suggests that rational control of interelement coupling is a crucial tool for highly multifunctional optics of this kind.

\subsection*{Challenges and opportunities}

\begin{figure}[hb!]
    \centering
    \includegraphics[width=0.9\columnwidth]{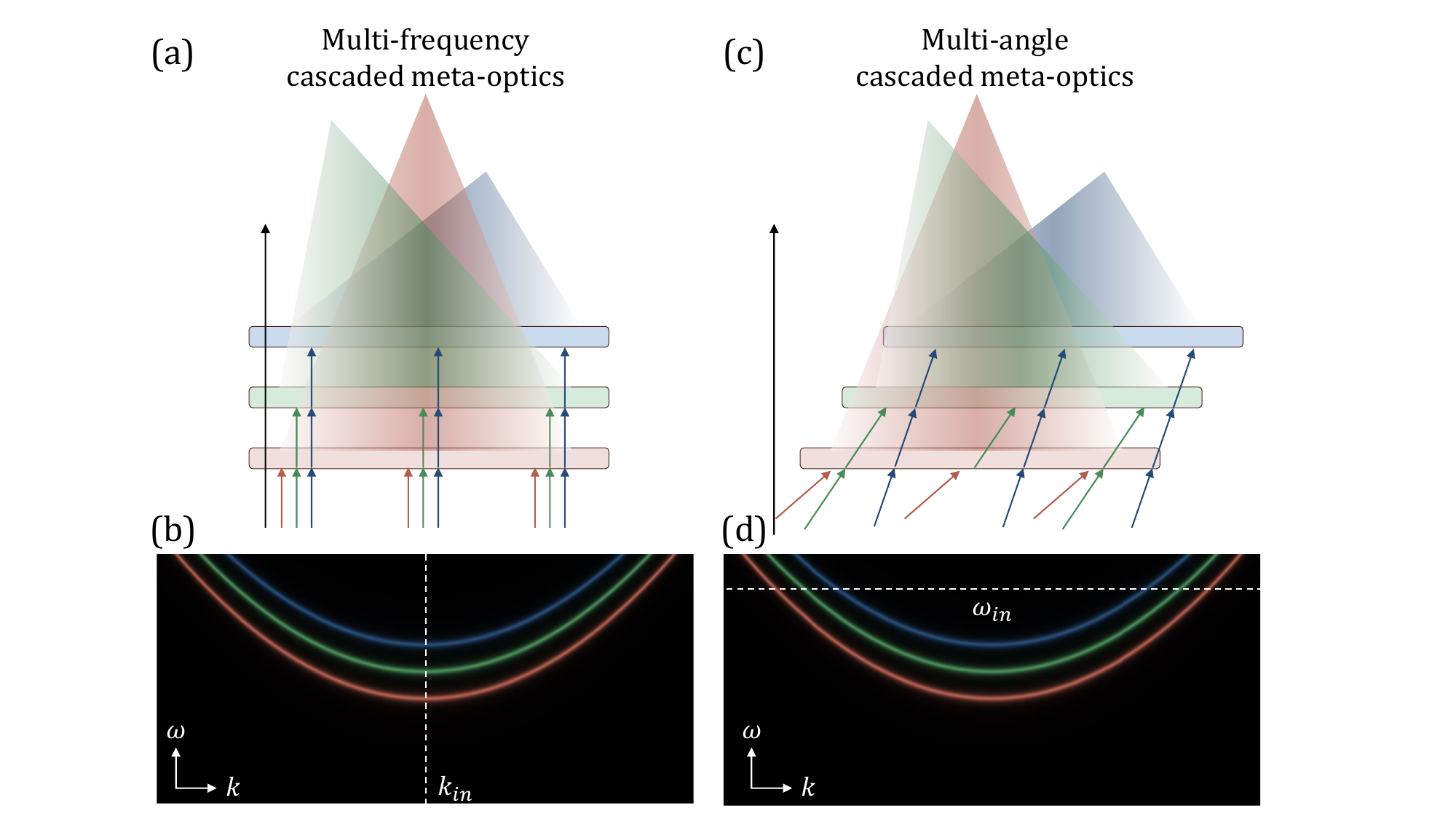}
    \caption{Cascaded nonlocal meta-optics. \pnl{a} For light incident at a fixed incident angle, cascaded resonant metasurfaces may achieve multi-wavelength operation. \pnl{b}~Example use of three dispersive metasurfaces (marked as red, green, and blue) for multi-wavelength operation \pnl{c} For light incident at a fixed frequency, cascaded nonlocal metasurface may achieve multi-momentum operation. \pnl{d} Example use of three dispersive metasurfaces for multi-angle operation. In \pnl{a} and \pnl{c}, the black arrow indicates non-resonant waves, which simply pass through the system.}
    \label{fig:Overvig-Fig1}
\end{figure}

By leveraging several modes within each metasurface and/or several mutually transparent metasurfaces cascaded together~\cite{Malek:2022}, the future of highly multifunctional cascaded meta-optics looks highly promising. Here, we suggest a picture for how to consider the potential for these devices as multi-frequency devices, multi-angle devices, or both; we note that there are many open questions on the implementation physicality of such systems and nature of the information encoding in the following, and hope that this viewpoint will motivate efforts to clarify bounds and constraints on these systems. Figure~\ref{fig:Overvig-Fig1}\pnl{a} depicts a cascaded meta-optic with multi-frequency operation. At three frequencies, independent wavefront modulation is imparted at each device for light incident with a lateral momentum $k_\mathrm{in}$ near 0 (near normal incidence). Three example dispersion profiles are shown in Fig.~\ref{fig:Overvig-Fig1}\pnl{b}, showing three shifted copies of the same mode (easily achieved by altering the thickness, duty cycle, or period of the metasurface). Each mode may be locally patterned at will, imparting independent operations to each resonant frequency~\cite{Overvig:2022,Malek:2022}. Meanwhile, the broadband response is simple transparency (black arrow), indicating the scalability of this approach to an arbitrary number of frequencies according to: the bandwidth of the mutual transparency window of the devices, the $Q$-factors of each device, the required angular range to encode the desired functionalities, and the acceptable tolerances of the system. Realization of this vision began in Ref.~\cite{Malek:2022}, which showed up to four functionalities. Beyond this, by using multi-layer nanofabrication, truly dense compound meta-optics with dozens of wavelengths controlled within a narrow band of operation are possible. Such highly multi-color meta-optics promise unprecedented capabilities for structured light, spectropolarimetry, and spatiotemporal pulse shaping. Advances could open opportunities for imaging, augmented reality, and highly customized light sources. 

On the other hand, Fig.~\ref{fig:Overvig-Fig1}\pnl{c} depicts a cascaded meta-optic with multi-angle operation at a fixed frequency $\omega_\mathrm{in}$ [Fig.~\ref{fig:Overvig-Fig1}\pnl{d}]. Compared to Figs.~\ref{fig:Overvig-Fig1}\pnl{a,b}, this approach effectively trades degrees of freedom from frequency to angle by using the intrinsic spatial dispersion of the underlying modes [i.e., the very same stack of devices could be used as in Figs.~\ref{fig:Overvig-Fig1}\pnl{a,b}]. This scheme breaks the numerical aperture of the meta-optic into a number $N_k$ of small ranges $\delta k$, each controlled by a distinct mode and/or metasurface layer. Moreover, it will be constrained by reciprocity: the output waves of each metasurface must be orthogonal to each other, or else the efficiency will necessarily suffer. Such an approach, if suitably generalized, could extend the multifunctionality of multi-angle approaches such as Refs.~\cite{Kamali:2017,Cheng:2017}, greatly increasing the customizability of a meta-optic at given frequency. 

\begin{figure}[htb]
    \centering
    \includegraphics[width=0.99\columnwidth]{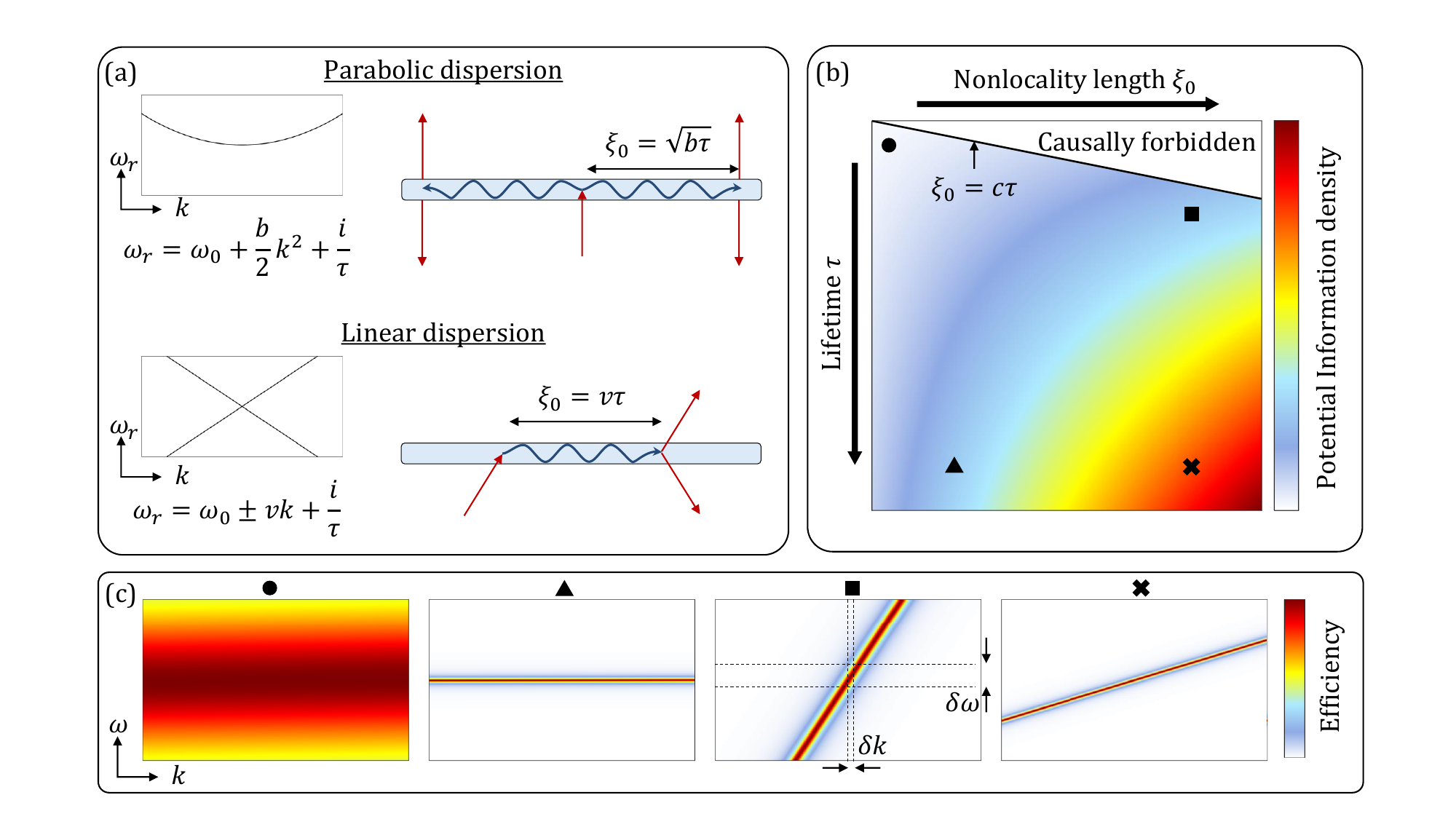}
    \caption{Spatio-temporal dispersion of nonlocal metasurfaces. \pnl{a} Parabolic dispersion (top) results in diffusive transport with nonlocality length, $\xi_0$, varying as the square root of the lifetime of the mode, $\tau$. Linear dispersion (bottom) results in transport with $\xi_0\propto \tau$. \pnl{b} Potential information density that may be encoded using both frequency and angle, as a function of $\xi_0$ and $\tau$. The upper right region is superluminal, forming an upper bound on the angular selectivity for a given spectral selectivity. \pnl{c} Four representative responses in the four extremes of \pnl{b}.}
    \label{fig:Overvig-Fig2}
\end{figure}

However, the bandwidth and information capacity of such approaches are limited by constraints such as dispersion and causality. In general, due to the dual nature of frequency and time, the number $N_\omega$ of small ranges $\delta\omega$ will vary according to the lifetime $\tau$ of the optical modes: $N_\omega \propto 1/\delta\omega\propto \tau$, reminiscent of an "uncertainty principle". Here, $\tau=Q/\omega_r$ where $Q$ is the $Q$-factor and $\omega_r$ is the resonant frequency. This is paralleled in momentum and space, namely: $N_k\propto 1/\delta k \propto \Re\xi$, where $\xi$ is the (complex) nonlocality length, and $\xi_0 = \Re\xi$ characterizes the in-plane distance a mode travels before radiatively decaying. Moreover, the allowable $N_k$ for a given bandwidth $\delta\omega$ depends on the nature of dispersion (e.g., parabolic or linear). Figure~\ref{fig:Overvig-Fig2}\pnl{a} depicts parabolic dispersion of a resonant frequency $\omega_r$ with band curvature $b$ near a band-edge mode (top), and the linear dispersion with group velocity $v$ (bottom). The distance $\xi_0$ increases with the square root of lifetime or linearly with lifetime, respectively.  Hence, the "diffusive" transport of a band-edge mode results in increased localization (i.e., decreased $\xi_0$) compared to the "advective" transport of a linear mode with the same $\tau$. In other words, for multi-angle devices: to increase bandwidth (decrease $Q$) for the same $N_k$, a linear dispersion should be used. While for multi-frequency devices: to increase localization (or range of angles) for a high-$Q$ structure, parabolic dispersion should be used. 

While the scheme in Fig.~\ref{fig:Overvig-Fig1} depicts either frequency or angle multifunctionality, the ultimate case is where both the frequency and angle ranges are independently addressable. In this case, the total number of independent functionalities grows as $N_\omega N_k \propto \xi_0\tau$, representing the potential information density of a cascaded meta-optic [Fig.~\ref{fig:Overvig-Fig2}\pnl{b}]. Figure~\ref{fig:Overvig-Fig2}\pnl{c} shows four representative dispersion relations of the resonant efficiencies for four extreme cases in Fig.~\ref{fig:Overvig-Fig2}\pnl{b}. Note that by causality, no mode can access the region of Fig.~\ref{fig:Overvig-Fig2}\pnl{b} such that the distance traveled exceeds the speed of light $c$; we require that $\xi_0 < c\tau$. A given device platform generally falls in a limited region of Fig.~\ref{fig:Overvig-Fig2}\pnl{b}; quasi-bound states have proven highly popular because they can rationally design the lifetime at will using symmetry-breaking concepts, extending their applicability in this context. A more generalized class of high $Q$-factor metasurface with both continuously tunable spectral and angular selectivity, yet locally addressable scattering phase and polarization, is highly desirable as a tool for both multi-frequency and multi-angle meta-optics. Such a platform does not exist to our knowledge. 

Importantly, we note that the implementation of such an information-dense meta-optic cannot be achieved for arbitrarily thin systems: the required thickness will grow as the required independent modes increases. In other words, intuitively, information capacity grows with volume: both physical aperture size and thickness~\cite{Miller:2023}. In this sense, cascaded diffractive nonlocal meta-optics amount to a rationally designed scheme for scaling the volume layer-by-layer according to the desired information capacity. 

\subsection*{Future developments to address challenges}

To begin to realize this potential information capacity with high efficiency, many challenges must be overcome. For instance, in contrast to the limitations of Ref.~\cite{Malek:2022}, for high efficiency as well as use of the polarization channels, a transmission mode nonlocal meta-optic should be developed that resonantly alters light upon transmission within a broadband transparent response. Such a response has not been demonstrated, yet may be possible with unidirectional guided modes~\cite{Yin:2020}, motivating the incorporation of point-wise symmetry-breaking design principles with directional control of the leakage of these states. Multi-layer nonlocal metasurfaces have sufficient degrees of freedom for such a task, and should be a priority in future research. Moreover, while flat bands and linear bands are achievable within periodic flat optics, even within a single system via a rational design paradigm~\cite{Nguyen:2018}, such approaches have not been shown to be compatible with spatially aperiodic functions of phase and polarization. A "knob" to tune the spatial dispersion that does not conflict with the knobs that spatially control the phase and polarization is highly desirable to create flexible cascaded meta-optics over a wide range of angular selectivity responses.
While the primary message here has been to point to the ceiling of potential information capacity, full access to the full information space is not realistic in the near term; practical choices should be made to make tangible progress. We believe that the potential information capacity of current high-$Q$ and nonlocal metasurfaces is often too high as a starting point, and as a result the implementation is not sufficiently flexible; the sensitivities to experimental and design errors may inhibit progress. That is, we argue that we should first seek individual nonlocal metasurfaces associated with less information capacity (i.e., smaller $\xi_0$ and $\tau$) for use in cascaded meta-optics, and then increase the capacity of each layer in the future as processes mature. For instance, while linear dispersion is optimal for extreme values of $N_k$ for a given bandwidth, the moderate localization of a band-edge mode affords tolerances to experimental and design errors that a linear dispersion does not. Hence, parabolic bands are more practical for low to moderate $N_k$. Unfortunately, if the band edge mode of a system is limited to normal incidence, only one such mode may be used for a given operating frequency. This motivates the incorporation of custom band-edge angles~\cite{Overvig:2023} with spatially aperiodic perturbations. Moreover, while high $Q$-factors are attractive for information density, practical experimental considerations often limit $Q$-factors to moderate values of less than $10^3$ or suffer drastically from low efficiency. Ideally, for exploration of proofs-of-principle and for robust, high-efficiency operation over moderate bandwidths, the radiative $Q$-factor would be lower than 100; it is well-known how to increase it later. That is, an arbitrarily high $Q$-factor may be designed using quasi-bound state concepts, yet the symmetry-breaking approach of quasi-bound states often has a practical lower bound in $Q$-factor. This can be simply understood as follows: larger symmetry breaking perturbations yield lower $Q$-factor, but a geometric perturbation has an upper bound before geometric features overlap each other. Future efforts lowering the $Q$ in diffractive nonlocal metasurfaces will offer increased reliability and flexibility in the design space of Fig.~\ref{fig:Overvig-Fig2}\pnl{b} while maintaining the symmetry-based pointwise control of light afforded by the selection rules.

\subsection*{Concluding remarks}

While local meta-optics have been ushering in unprecedented information density encoded at will into flat optical devices, the new frontier of cascaded diffractive nonlocal meta-optics reveals that the local approach only begins to scratch the surface of what is possible. In nonlocal metasurfaces, selectivity to incident frequency and/or angle is possible. In diffractive nonlocal metasurfaces, the spatial customization of amplitude, phase, and polarization may be retained in addition to this selectivity. Here, we have emphasized that in cascaded diffractive nonlocal meta-optics, the non-resonant transparency of each layer renders each selectively encoded functionality mutually transparent (orthogonal) to one another, establishing a scalable approach to highly multifunctional meta-optics. Such an approach amounts to volumetric metamaterials composed of many two-dimensional layers that are rationally designed to achieve unprecedented customizability of light. Success in these endeavors promises advances across optics, including spectrospatial structuring of light, optical computing, and dense holographic data storage. Beyond linear optics, the strong light-matter interactions of the customized high $Q$-factor responses promise new pathways for highly efficient and multifunctional reconfigurable, nonlinear, and quantum volumetric metamaterials that are designable and manufacturable with standard procedures.

\usection{\emph{Part V} --- Nonlocality, topology and nonreciprocity, extreme geometries and fundamental limits}
\label{roadmap:Part5}

\section[Role of nonlocality in topological metamaterials (Prud{\^e}ncio \& Silveirinha)]{Role of nonlocality in topological metamaterials}

\label{sec:Prudencio}

\author{Filipa R. Prud{\^e}ncio\,\orcidlink{0000-0002-7073-0987} \& M{\'a}rio G. Silveirinha\,\orcidlink{0000-0002-3730-1689}}

\subsection*{Current status}

\begin{figure}[hbt]
    \centering
    \includegraphics[width=0.99\linewidth]{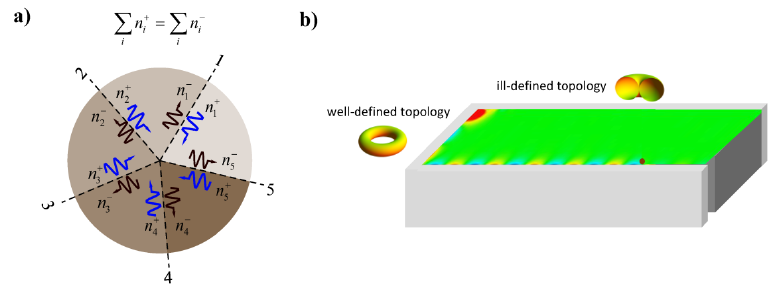}
    \caption{\pnl{a} Junction between different bulk materials that share a common band-gap. In a topological system, the number of incoming edge channels and the number of outgoing edge channels is exactly the same. \pnl{b}~A system with an ill-defined topology is not bound by the constraint in panel \pnl{a}. The ill-defined topology can be used to abruptly halt the wave propagation at a topological singularity with a massive field enhancement. Adapted with permission from Ref.~\cite{Fernandes:2022} (Copyright~\textcopyright~2022 SPIE).}
    \label{fig:Prudencio-Fig1}
\end{figure}

Topological photonics is an exciting and rapidly advancing field that brings the mathematical elegance of topology into the realm of electromagnetic waves and photonic materials~\cite{Ozawa:2019,Silveirinha:2023}. At its heart lies the principle of bulk-edge correspondence, a profound result that serves as the cornerstone of the field’s theoretical framework and practical applications. This principle states that the topological invariants, which are properties of the bulk band structure, dictate the existence and robustness of edge states that propagate along the material boundary~\cite{Silveirinha:2019}.
From a practical standpoint, the bulk-edge correspondence manifests as a remarkable conservation law: at the junction of different topological materials, all of which share a band gap, the number of edge states propagating toward the junction must precisely equal the number of edge states leaving it [Fig.~\ref{fig:Prudencio-Fig1}\pnl{a}]~\cite{Silveirinha:2019}. This conservation law ensures that waves traveling along such boundaries follow well-defined pathways, dictated by the topological invariants of the materials. If this correspondence were violated—if, for instance, there were more states entering the junction than leaving—it would imply the presence of a singular behavior in conservative systems~\cite{Silveirinha:2019,HassaniGangaraj:2020a,Fernandes:2022}. Such a scenario might correspond to the material acting as an electromagnetic sink, where light could enter but not escape [Fig.~\ref{fig:Prudencio-Fig1}\pnl{b}]. The absence of such singularities is a fundamental feature of nature’s physical laws and reflects the robustness of topological protection in these systems.

However, this robustness relies on subtle physical mechanisms, and it is here that nonlocality plays a critical role. Nonlocality refers to a material’s response that depends not only on the electromagnetic field at a given point but also on its spatial variations. In plasmonics, a well-known example of nonlocality is found in the drift-diffusion (hydrodynamic) model, where electron-electron repulsive interactions effectively establish a minimum localization length, thereby preventing extreme wave localization. Thus, diffusion effects inherently counteract forms of field localization that might otherwise compromise the integrity of the bulk-edge correspondence.
In the context of topological metamaterials, nonlocality becomes more than a secondary effect: it is an essential feature. Nonlocal effects act as the key mechanism that ensures singularities, such as electromagnetic sinks, are naturally avoided. By mediating how the response to fast varying fields tailors the material’s photonic structure, nonlocality enforces the constraints required for bulk-edge correspondence to hold universally.

\subsection*{Challenges and opportunities}

The Chern theorem applies strictly to closed manifolds, which initially suggests that electromagnetic continua—systems without intrinsic periodicity—are unsuitable for nontrivial photonic topologies. For example, early studies revealed a fundamental challenge: conventional local models of materials, such as magnetically biased plasmas or ferrites, fail to produce a well-defined topological classification. This deficiency is attributed to the noncompact nature of the wavevector space in electromagnetic continua, which is identified with the Euclidean plane~\cite{Silveirinha:2015}.
However, recent advances reveal that nonlocality can relax these constraints, serving as a foundational approach to resolve such issues~\cite{Silveirinha:2015,Jin:2016,Pakniyat:2022}. Specifically, by introducing a dependence of the material response on the wavevector $\boldsymbol{k}$, nonlocal models can effectively suppress nonreciprocal responses at large $\boldsymbol{k}$\cite{Silveirinha:2015}. This suppression eliminates the problematic contributions from high-$\boldsymbol{k}$ modes, ensuring that the band gaps in electromagnetic continua have well-defined topological invariants.

Interestingly, the importance of nonlocal effects extends beyond electromagnetic continua to periodic systems such as photonic crystals~\cite{Prudencio:2022}. In these systems, the wavevector space is isomorphic to a torus and thus naturally satisfies the conditions of the Chern theorem. Nonetheless, the topology of photonic crystals can also become ill-defined when the nonlocal response of the materials is disregarded. This issue arises from the bosonic nature of electromagnetic waves, which gives photonic band structures properties that differ fundamentally from those of electronic band structures.
Unlike electronic systems, which have a well-defined ground state (a finite number of bands below the Fermi level), photonic systems exhibit spectra that are unbounded from below. This property stems from the real symmetry of the electromagnetic field, which enforces a particle-hole symmetry in the spectrum: for every positive frequency mode, there is a corresponding negative frequency partner. Consequently, photonic systems always have an infinite number of bands below any given bandgap. This feature places the Chern theorem in a precarious position, as it traditionally relies on having a finite number of bands below the gap to compute the topological invariants.

This issue can be vividly illustrated with an example. Consider a system with an infinite number of isolated bands below a gap, each contributing alternately $+1$ or $-1$ to the Chern number~\cite{Prudencio:2022 ,Camara:2024}. The gap Chern number in this case is given by the non-convergent series $\mathscr{C}_\mathrm{gap}=1+(-1)+1+(-1) + \ldots$, resulting in an ill-defined topology. Additionally, material dispersion can exacerbate the problem. For example, the combination of a flat band in a material and the band folding induced by periodicity can result in an accumulation of an infinite number of bands at certain resonance frequencies. This accumulation creates further challenges, as the corresponding contributions to the Chern number may again form a divergent series~\cite{Prudencio:2022}.

As is well-known, an infinite sum of integers converges only if all but a finite number of terms are zero. Without a mechanism to suppress contributions from high-$\boldsymbol{k}$ modes or to regulate the effects of band folding and dispersion, the topology of photonic systems cannot be reliably defined. Here again, nonlocality offers an elegant solution. By introducing a wavevector cutoff that suppresses nonreciprocal responses at large $\boldsymbol{k}$, nonlocal effects provide the regularization needed to resolve these ambiguities. This ensures that the band gaps in realistic dispersive photonic crystals have well-defined topological invariants, allowing them to serve as robust platforms for topological photonics~\cite{Prudencio:2022}.

\subsection*{Future developments to address challenges}

Despite the important role of nonlocal effects in ensuring well-defined topologies in photonic systems, several challenges remain to be addressed. One critical issue is the limited knowledge of the material response at high wavevectors $\boldsymbol{k}$, as most physical models are developed for small $\boldsymbol{k}$, corresponding to long-wavelength physics. The behavior of materials at high $\boldsymbol{k}$, which governs short-wavelength phenomena, is often poorly characterized, leading to uncertainties in the topological classification of materials.

\begin{figure}[htb]
    \centering
    \includegraphics[width=0.99\linewidth]{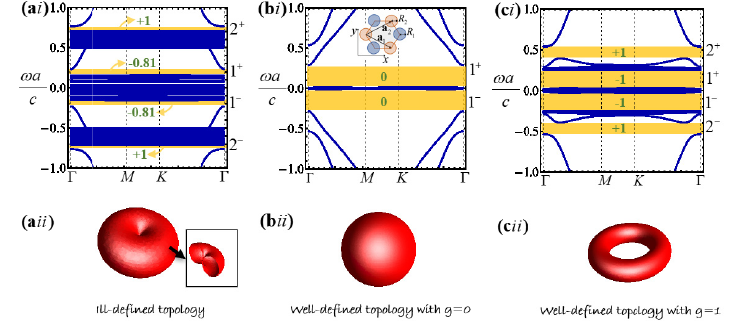}
    \caption{Top row: Photonic band structures of dispersive photonic crystals formed by a hexagonal array of air rods embedded in a magnetized electric plasma [geometry of the unit cell is sketched in the inset of panel \pnl{b\emph{i}}]. The green insets give the numerically calculated gap Chern numbers. The dispersive host material is characterized by \pnl{a} local model, \pnl{b} hydrodynamic model, and \pnl{c} a full cutoff model. Bottom row: Geometrical illustration of the concepts of ill-defined topology and topology regularization. Similar to the torus with vanishing inner radius, the topology of the photonic crystal depends on the regularization procedure. Adapted with permission from Ref.~\cite{Prudencio:2022} (Copyright~\textcopyright~2022 American Physical Society).}
    \label{fig:Prudencio-Fig2}
\end{figure}

It has been shown that the topology of a material critically depends on its high-$\boldsymbol{k}$ response and that different types of nonlocality can lead to entirely different topological phases~\cite{Pakniyat:2022,Prudencio:2022}. This concept can be geometrically visualized using an analogy with the theory of surfaces. Imagine a torus with a vanishing inner radius, which forms a cusp at its center, as illustrated in Fig.~\ref{fig:Prudencio-Fig2}\pnl{a\emph{ii}}. Mathematically, the cusp introduces a nondifferentiable point, rendering the topology ill-defined. This issue is resolved by an infinitesimal deformation: either making the inner radius finite, creating a torus with genus one, or severing the cusp, yielding a sphere with genus zero.

This analogy underscores that different perturbations can regularize an ill-defined topology, yielding different topological invariants. In photonic systems, starting with an ill-defined topology (e.g., a local material model), various types of nonlocality can similarly create distinct topological phases (Fig.~\ref{fig:Prudencio-Fig2}). For example, mechanisms like diffusion, odd viscosity, or full wavevector cutoffs have all been proposed to regularize the topology of materials~\cite{Silveirinha:2015,Pakniyat:2022,Souslov:2019}. However, this raises a critical issue: the exact nature of the nonlocal response—and therefore the resulting topology—often depends on assumptions about the high-$\boldsymbol{k}$ behavior, which is not always experimentally or theoretically well-constrained.

Importantly, similar to the topological properties, the number of unidirectional edge state channels also depends on the specific form of nonlocality controlling the material response~\cite{HassaniGangaraj:2019a,Buddhiraju:2020,Silveirinha:2016,Serra:2025}. The number of unidirectional edge states consistently aligns with the predictions of the bulk-edge correspondence~\cite{Silveirinha:2016,Serra:2025}. While this may seem problematic, such ambiguity often affects only very short-wavelength waves. Fortunately, dissipation tends to suppress contributions from these short wavelengths, ensuring that in practical scenarios the physical response remains consistent across different nonlocal models. Nevertheless, a deeper understanding of the mechanisms that determine the nonlocal responses at high-$\boldsymbol{k}$ in weakly dissipative systems remains an open problem.

Another significant challenge lies in extending the bulk-edge correspondence to non-Hermitian systems, such as those with loss and gain. Recent studies have demonstrated that in active systems, the bulk-edge correspondence can fail due to the non-Hermitian skin effect. In such systems, the bulk spectrum can change dramatically when switching from periodic to open boundary conditions, due to wave localization along specific boundaries. An open question is whether it is possible to identify a subclass of non-Hermitian systems where bulk-edge correspondence can be preserved. For instance, could the correspondence be constrained to dissipative systems (with only loss) or stable systems (with the spectrum in the lower-half complex frequency plane)? Understanding these conditions could open new avenues for topological photonics, particularly in active and non-equilibrium platforms. Nonlocality, with its ability to mediate high-$\boldsymbol{k}$ responses, may play a crucial role in addressing these challenges, potentially enabling robust bulk-edge correspondence even in non-Hermitian contexts.

\subsection*{Concluding Remarks}

Understanding and incorporating nonlocality into the design and analysis of photonic topological metamaterials is not merely an academic exercise but a practical necessity. It provides the physical mechanism that underpins the robustness of topological edge states, ensuring that light retains its predictable and protected pathways. Nonlocality, in this sense, serves as the bridge between the abstract mathematical elegance of topology and the physical reality of photonic materials, safeguarding the interplay of edge states and topological invariants that makes this field so vibrant.

\section[Nonlocality in nonreciprocal plasmonics (Hassani Gangaraj \& Argyropoulos)]{Nonlocality in nonreciprocal plasmonics}

\label{sec:HassaniGangaraj}

\author{S. Ali Hassani Gangaraj\,\orcidlink{0000-0003-1818-3215} \& Christos Argyropoulos\,\orcidlink{0000-0002-8481-8648}}

\subsection*{Current status}
\begin{figure}[hbt]
    \centering
    \includegraphics[width=0.68\linewidth]{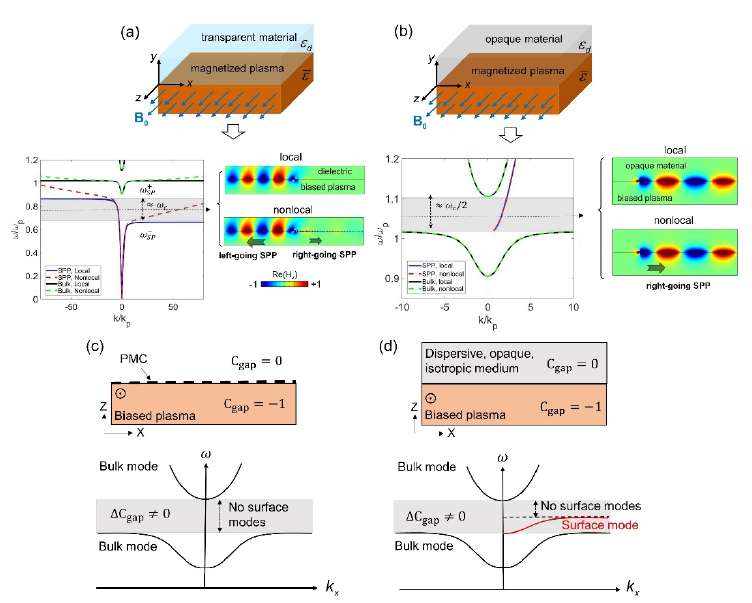}
    \caption{\pnl{a}~Typical band diagrams for nonreciprocal bulk modes and SPPs formed inside or at the interface of biased plasmonic material (InSb) and transparent dielectric (Si) under both local and nonlocal InSb material properties~\cite{HassaniGangaraj:2019a}. Solid black and blue lines represent the local bulk and SPP dispersion, while dashed lines represent the nonlocal case, with nonlocality of InSb accounted for by a hydrodynamic approach~\cite{HassaniGangaraj:2019a}. The applied bias on InSb is $\omega_c/\omega_p=0.2$, where $\omega_c$ is the cyclotron frequency proportional to the bias strength and $\omega_p$ is the plasma frequency. In the nonlocal case, bulk mode asymptotes bend upward, closing the bandgap~\cite{HassaniGangaraj:2019a}. The gray strip indicates the nonreciprocal SPP operation region where $\omega_\textrm{SP}$ shifts to $\omega_\textrm{SP}^{\pm}=\omega_\textrm{SP}\pm \omega_c/2$, opening an asymmetric frequency window with a bandwidth of $\omega_c$~\cite{HassaniGangaraj:2019a}. The solid blue line represents the local nonreciprocal response with two asymptotes at $\omega_\textrm{SP}^{\pm}$; while the dashed red line represents the nonlocal response, where the asymptotes bend upward, leading to bidirectional but asymmetric SPP propagation. The field profiles for local and nonlocal cases are represented in the inset~\cite{HassaniGangaraj:2019a}, where it is clearly depicted that in the local case SPPs are strictly one-way but including nonlocality leads to a weak backward propagation. \pnl{b}~Similar results but for opaque material at the interface leading to topological SPPs within the bulk mode bandgap, where one-way propagation stems from the nontrivial topological properties of the biased plasmonic material. In this case, as is clear from the field profiles, the inclusion of nonlocality does not alter the nonreciprocal SPP propagation~\cite{HassaniGangaraj:2019a}. \pnl{c,d}~Typical dispersion diagrams for the two classes of bulk-edge correspondence principle violations~\cite{HassaniGangaraj:2020a}. Solid black and red curves indicate bulk and surface modes, respectively, and gray areas highlight the bulk-mode bandgap. Nonreciprocal SPPs exist only in \pnl{d} which only covers the bulk band gap partially in the local case. Reproduced with permission from Ref.~\cite{HassaniGangaraj:2019a} (Copyright~\textcopyright~2019 Optica Publishing Group).}
    \label{fig:HassaniGangaraj-Fig1}
\end{figure}

Surface-plasmon polaritons (SPPs) can be excited along metal-dielectric interfaces giving rise to the emerging field of plasmonics. While most plasmonic systems are reciprocal, unidirectional (nonreciprocal) SPPs can be realized by biasing metallic media with a quantity that is odd under time reversal symmetry, typically a static magnetic field. For instance, one-way SPPs can be induced along the surface of magnetized highly doped semiconductors (acting as metals). Interestingly, the realization of nonreciprocal SPPs may lead to extreme wave physics phenomena which, however, cannot be characterized correctly unless nonlocality (a.k.a., spatial dispersion) is included in the analysis~\cite{Buddhiraju:2020}. One method to achieve nonreciprocal SPP propagation is by introducing asymmetry in the conventional SPP flat dispersion band at the surface plasmon resonance. This is done by interfacing a magnetized plasmonic material with a transparent dielectric, leading to asymmetrical SPP dispersion bands.
However, recently~\cite{Buddhiraju:2020}, it was argued that the presence of a flat asymptote band leads to a thermodynamic paradox, i.e., infinite photonic states, $k\in (k_\textrm{min},\infty)$ within the unidirectional SPP frequency range $\omega\in (\omega_\textrm{SP}^{-}, \omega_\textrm{SP}^{+})$, resulting in infinite energy at finite temperatures. This issue arises from the locality assumption implying that the material stays polarized even near asymptotes where SPP modes have diverging wavenumbers. Interestingly, the inclusion of nonlocality corrects this unphysical behavior since it bends the dispersion upwards at large wavenumbers and close the nonreciprocal window, as depicted in Fig.~\ref{fig:HassaniGangaraj-Fig1}\pnl{a}, leading to asymmetric SPP propagation but in all directions~\cite{HassaniGangaraj:2019a}.
Another type of unidirectional SPP is realized by interfacing a magnetized plasma with an opaque material. Here, one-way propagation stems from asymmetric cutoff at $\boldsymbol{k}=\boldsymbol{0}$ for counter-propagating modes in the plasmonic bandgap, rather than asymmetric asymptotes described before, as shown in Fig.~\ref{fig:HassaniGangaraj-Fig1}\pnl{b}. This type of nonreciprocal SPPs remain unaffected by the strong nonlocality in biased semiconductors, such as indium antimonide (InSb), mainly due to the topological properties of the resulted magnetized plasma~\cite{HassaniGangaraj:2019a}. 
While topological protection suggests robustness against nonlocality, there are instances where nonlocality violates bulk-edge correspondence in topological photonics. One violation occurs when a magnetic wall on a magnetized semiconductor "short-circuits" the tangential SPP magnetic field [Fig.~\ref{fig:HassaniGangaraj-Fig1}\pnl{c}], preventing SPP existence despite a non-zero gap Chern number difference. However, it was proven that SPP theoretically exists only at diverging wavenumbers~\cite{HassaniGangaraj:2020a}. Incorporating nonlocality keeps the SPP wavenumber large but finite, formally restoring bulk-edge correspondence. Nonetheless, SPPs with large wavenumbers are hard to excite and suffer attenuation due to Landau damping, even when lossless materials are considered~\cite{HassaniGangaraj:2020a}.
The second violation of bulk-edge correspondence occurs when SPP dispersion does not span the entire bandgap due to an opaque material with a plasma frequency near the upper bandgap edge, causing an asymptote within the bandgap [Fig.~\ref{fig:HassaniGangaraj-Fig1}\pnl{d}]. While the lower bandgap supports a topological state, no edge mode exists in the upper portion, violating the principle. Including nonlocality in InSb response bends the SPP band asymptote upward, leading to SPP leakage into the bulk of InSb~\cite{HassaniGangaraj:2020a}. This nonlocality-induced radiation loss quickly attenuates SPP power. Thus, while nonlocality restores bulk-edge correspondence, it is violated practically as SPPs are immediately attenuated, even in lossless materials. Hence, nonlocality's role should be treated cautiously in nonreciprocal plasmonic systems. The results in Ref.~\cite{HassaniGangaraj:2020a} clarify that although nonlocality does not break the unidirectionality of topological SPPs, we still need to consider the material spatial dispersion in our analysis to prevent paradoxical observations such as bulk-edge correspondence violation. 
The above discussion highlights the role of nonlocality in nonreciprocal plasmonics and paves the way to future experimental observations. By leveraging nonreciprocal SPPs via realistic materials, one can develop highly efficient and more compact, directionally controlled novel light transport systems, which promise to be fundamental components of envisioned advanced classical and quantum photonic technologies. Hence, nonreciprocal plasmonics offers a versatile tool for fine-tuning the nanophotonic device performance with applications in different technological frontiers, such as communication and computing.

\subsection*{Challenges and opportunities}

The investigation of nonlocal effects in nonreciprocal plasmonics is still in its infancy. The discussion presented above indicates that accounting for material nonlocality gives rise to complex wave phenomena, playing a crucial role in the nonreciprocity of the induced surface modes. While these findings highlight the potential of nonreciprocity to control light propagation in unprecedented ways, they also underscore the growing need to practically realize this effect by using new photonic systems. Currently, the predominant method for achieving such response remains heavily reliant on magneto-optical effects that require static magnetic field bias. The inherent weaknesses of magneto-optical effects, along with the sensitivity of one-way SPP propagation achieved by this method to nonlocality, raise major concerns about the practical applicability of nonreciprocal plasmonics.
Drifting electrons resulting in current bias is also an odd quantity under time reversal symmetry, similar to magnetic field. Therefore, instead of magnetic bias, one may use currents in both bulk conducting media (metals, semiconductors, and plasmas) and emerging two-dimensional (2D) materials such as graphene. The origin of this nonreciprocal effect is rooted in the Doppler effect due to the electrons’ movement along the medium, which produces a Doppler frequency shift ($\omega \rightarrow \omega -\boldsymbol{k}\cdot \boldsymbol{v}_d$) in the material optical response, where $\boldsymbol{k}$ is the wavenumber and $\boldsymbol{v}_d$ is the drifting velocity~\cite{Bliokh:2018}. The wavenumber inclusion in the material response indicates a form of nonlocality that can lead to nonreciprocity. 
Although the inclusion of material’s natural nonlocality challenges the nonreciprocal response of plasmonics, paradoxically, induced nonlocality via current bias provides opportunities to achieve this response. The nonlocality-induced nonreciprocity can be intuitively explained as follows: SPPs involve collective electron-photon oscillations. When these modes are dragged or opposed by the drifting electrons, SPPs "see" different media when propagating along or against the current flow, i.e., $\omega(\boldsymbol{k}) \neq \omega(-\boldsymbol{k})$.
Different platforms can be used to demonstrate nonlocality-induced nonreciprocity. Recently, nonreciprocal SPPs propagation along a gold nanowire carrying a small electric current with a drift velocity around $v_d/c_0 \approx 10^{-6}$ was demonstrated~\cite{Bliokh:2018}. Another work studied n-type InSb which allowed for even larger drift velocities up to $v_d/c_0 \approx 10^{-3}$~\cite{HassaniGangaraj:2022a}. The latter study was focused not only on nonreciprocal SPP propagation, but also on the excitation of slow light, steerable SPPs, and exceptional point-like mode transitions at dispersion inflection points. Such nonlocality-induced nonreciprocity can be utilized in thermal photonics, specifically to enable nonreciprocal near-field radiative heat transfer between two planar bodies.
Graphene has also emerged as a promising material for nonlocality-induced nonreciprocal plasmonics. Recent experiments demonstrated nonreciprocity with electrons drifting at speeds around its Fermi velocity~\cite{Dong:2021}. Interestingly, a relevant theoretical work~\cite{HassaniGangaraj:2023} highlighted the potential of nonlocality-induced nonreciprocity in enhancing nonlinear light-matter interactions in graphene plasmonics. By utilizing drifting electrons induced by current on a voltage-biased graphene sheet placed on a periodically corrugated silicon (Si) grating, it is possible to create asymmetric SPP modes with field enhancement shown in Fig.~\ref{fig:HassaniGangaraj-Fig2}\pnl{a}. Blocking the path of nonreciprocal SPPs enables the generation of dramatically enhanced and localized broadband electric field hotspots near the termination, since the radiation cannot be reflected back, with the intensity of these hotspots controlled by the speed of the drifting electrons. The resulted energy accumulation can significantly enhance third-order nonlinear optical effects in graphene, with predicted conversion efficiencies of up to 0.3\% around the plasmon resonance frequency~\cite{HassaniGangaraj:2023}.

\subsection*{Future developments to address challenges}

Plasmonic systems exhibit unique properties, such as subwavelength confinement and strong field enhancement, making them ideal solutions to realize nanophotonic technologies. One of their key features is the extreme enhancement of nonlinear light-matter interactions~\cite{Krause:2022}. In this context, two general strategies are employed traditionally to generate strong electric field enhancement: \emph{i)} localized resonances and \emph{ii)} slow-light effects with adiabatic impedance matching. However, these methods suffer from sensitivity to material absorption losses, limited bandwidth, and large device footprint due to the very long adiabatically tapered structures.

\begin{figure}[htb]
    \centering
    \includegraphics[width=0.8\linewidth]{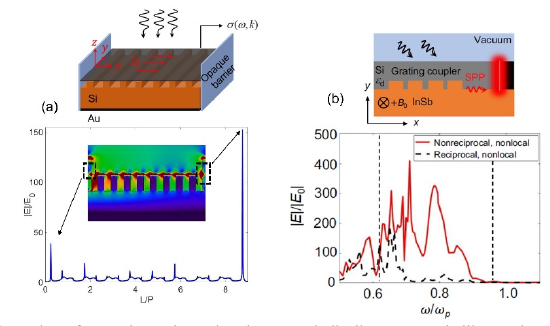}
    \caption{\pnl{a} Illustration of a graphene sheet placed on a periodically corrugated Si substrate. Current bias is applied to graphene shown by the red arrows. The realized nonreciprocal structure is terminated at the two ends by opaque barriers. The line graph shows the distribution of the electric field enhancement along the graphene sheet at the absorption peak frequency. Two strong asymmetric hot spots exist at the two ends of the graphene waveguide. The inset shows the spatial distribution of the electric field enhancement. Reprinted (adapted) with permission from Ref.~\cite{HassaniGangaraj:2023} (Copyright~\textcopyright~2023 American Chemical Society). \pnl{b}~Illustration of the terminated one-way waveguide geometry consisting of a Si grating on top of a magnetic biased InSb. The structure is terminated at one end by an opaque material. The line graph shows the realized field enhancement as a function of frequency. Vertical dashed lines indicate the broadband nonreciprocal operation which is achieved even in the presence of nonlocality. Reproduced with permission from Ref.~\cite{HassaniGangaraj:2020b} (Copyright~\textcopyright~2020 American Physical Society).}
    \label{fig:HassaniGangaraj-Fig2}
\end{figure}

An alternative strategy to address these technological challenges is the use of terminated nonreciprocal plasmonic waveguides to enhance weak optical nonlinear effects. Towards this end, strong electric field hot spots were demonstrated~\cite{HassaniGangaraj:2020b} when the unidirectional SPP propagation is blocked suitably in a nonreciprocal plasmonic waveguide made of Si and magnetically biased InSb with results depicted in Fig.~\ref{fig:HassaniGangaraj-Fig2}\pnl{b}. The obtained large and broadband electric field enhancement is ideal to boost the efficiency of various optical nonlinear processes. Additionally, impedance matching is automatically guaranteed in this nonreciprocal waveguide configuration due to the absence of a backward mode, eliminating the need for long adiabatic structures, hence, substantially reducing the device footprint. While plasmonic effects are typically constrained by material loss, in the limit of negligible material loss, other effects — such as nonlocality-induced surface-to-bulk mode coupling — can become a primary limitation on field localization. However, recent studies have shown that electric field hot spots achieved based on this method can lead to substantial second~\cite{Mann:2021} and third~\cite{HassaniGangaraj:2020b} harmonic generation enhancement.
As previously outlined, the weak magnetic response of materials at optical frequencies limits the effectiveness of magnetic bias for this purpose. These technological challenges necessitate concerted efforts to advance the underlying platforms. Drifting electrons offer an alternative method to break time reversal symmetry. Achieving such nonlocality-induced nonreciprocity requires high current bias values, which, although difficult, are not experimentally unattainable, as was recently confirmed in Ref.~\cite{Dong:2021} through the demonstration of Fizeau drag in graphene.
Nonlocality-induced nonreciprocity holds great promise for classical nanophotonic technologies, offering enhancement in both linear and nonlinear light-matter interactions. Similarly, in the realm of quantum technologies, nonreciprocal nanophotonic media can provide benefits, such as enhancing interactions between two-level emitters (e.g., atoms, molecules, quantum dots). This improved interaction between two-level emitters can serve as a key mechanism for boosting inter-atomic energy transport efficiency~\cite{HassaniGangaraj:2022b} and entanglement~\cite{Berres:2023}. Such advances could, in turn, enable the generation and mediation of entanglement in multipartite systems, consisting a fundamental objective in quantum technology.
Although much of the theory behind nonlocal nonreciprocal systems has been developed, offering a glimpse to various potential groundbreaking applications, experiments of the presented effects are still lagging. The experimental demonstration of nonreciprocity-induced field hotspots in any optical system, either drift current-biased or magneto-optical, is predicted to be an ongoing active research area in the broad fields of photonics and optics. Efficient operation is required for these nonreciprocal effects to become practical which has not been experimentally demonstrated yet.

\subsection*{Concluding Remarks}

The above discussion provides a glimpse into how nonlocality and its effects on nonreciprocal plasmonics are crucial not only to advance classical photonic processes, such as enhancing linear and nonlinear light-matter interactions and one-way waveguiding, but also in quantum optics, where induced nonlocality may enable more efficient interactions in multipartite emitter systems. Nonlocal effects are essential for correcting the unphysical behavior arising in nonreciprocal plasmonic systems, particularly when one-way SPPs rely on asymmetric dispersion asymptotes that could otherwise lead to thermodynamic paradoxes. However, this resolution may come at the cost of SPP unidirectionality, unless the one-way behavior originates from the nontrivial topological properties of the underlying platform, such as in the case of magnetized plasma-like materials. Although nonlocality can challenge nonreciprocal plasmonics in extreme cases, it also offers opportunities to achieve one-way propagation through nonlocality-induced effects. Introducing nonlocality via drifting electrons presents a promising alternative for creating a one-way path for edge states, particularly in semiconductors (n-type InSb) or emerging 2D conductive materials such as graphene. Future work should focus on optimizing these nonlocal effects, exploring innovative natural or artificial materials with strong nonreciprocal properties, and incorporating these advances into practical photonic devices for both classical and quantum optical applications.

\section[Nonlocal effects in singular plasmonic geometries unveiled by transformation optics (Huidobro \emph{et al.})]{Nonlocal effects in singular plasmonic geometries unveiled by transformation optics}

\label{sec:Huidobro}

\author{Paloma A. Huidobro\,\orcidlink{0000-0002-7968-5158}, Emanuele Galiffi\,\orcidlink{0000-0003-3839-8547}, Fan Yang\,\orcidlink{0000-0002-8648-1858} \& John B. Pendry\,\orcidlink{0000-0001-5145-5441}}

\subsection*{Current status}

\begin{figure}[hb]
    \centering
    \includegraphics[width=0.9\linewidth]{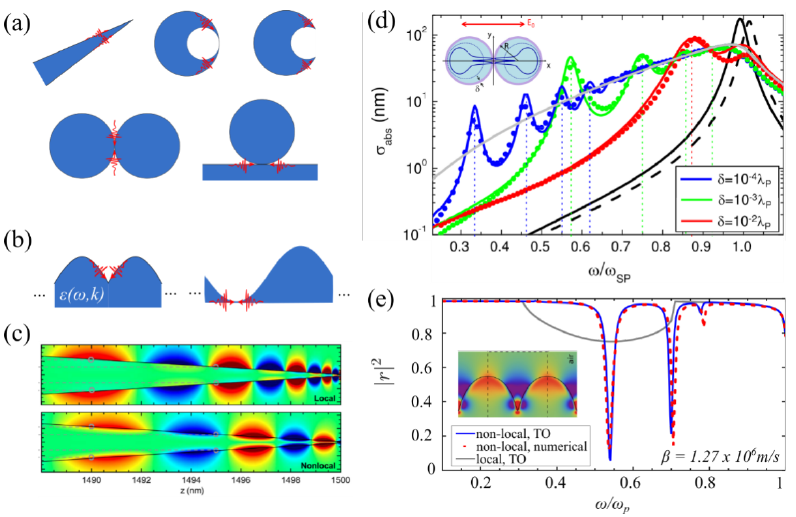}
    \caption{Plasmonic singularities unveiled by TO. \pnl{a} Gallery of singular plasmonic structures that localize the EM fields on a single hot spot: tip, crescents, touching nanoparticles and nanoparticle-on-mirror structures. All these structures transform from the same infinite "mother" structure, a planar waveguide, and thus share the same continuous spectrum in the local limit. \pnl{b} Two plasmonic gratings with periodic singularities that localize the EM fields at a series of spots. \pnl{c} Nonlocal TO calculation of a plasmonic nanotip. Spatial dispersion effectively blunts the tip and limits field enhancement at singular points. Reprinted (adapted) with permission from Ref.~\cite{Wiener:2012} (Copyright~\textcopyright~2012 American Chemical Society). \pnl{d} The scattering cross section of two touching 10\,nm nanowires shows a discrete set of resonance peaks due to nonlocality (solid lines: TO, dots: FEM simulations). Reproduced with permission from Ref.~\cite{Fernandez-Dominguez:2012} (Copyright~\textcopyright~2012 American Physical Society). \pnl{e} The reflectivity spectrum of a silver surface with a periodic array of sharp grooves displays dips at a discrete set of modes when nonlocality is taken into account, as opposed to the continuous spectrum of a local description (grey line). Reproduced with permission from Ref.~\cite{Yang:2019b} (Copyright~\textcopyright~2019 American Physical Society).}
    \label{fig:Huidobro-Fig1}
\end{figure}

Transformation optics (TO) has played a pivotal role in advancing metamaterials research~\cite{Zhang:2019}. This theoretical framework exploits the form-invariance of Maxwell’s equations under coordinate transformations to prescribe the spatial tailoring of electromagnetic constitutive parameters that have to be implemented in a metamaterial in order to obtain a desired optical effect. Being a very general theory framework, TO has also been used as a design tool for nanoplasmonic structures. 
By concentrating light within subwavelength volumes, plasmonic nanostructures act as nanoscale light harvesters. At optical frequencies, free electrons in plasmonic materials interact with incident electromagnetic fields forming surface plasmon polaritons that oscillate with much shorter wavelengths than incident radiation. These nanostructures concentrate light at nanometer length scales, with the strongest localization occurring in singular plasmonic structures, that feature sharp edges, nanometer-sized gaps, or touching points, towards which the surface plasmon polaritons carry electromagnetic energy. Examples include nanotips, nanocrescents and touching spheres, shown in Fig.~\ref{fig:Huidobro-Fig1}\pnl{a}, as well as gratings with sharp edges or almost touching points, shown in Fig.~\ref{fig:Huidobro-Fig1}\pnl{b}. As the surface plasmon travels towards, e.g., the edge of a nanotip, its wavelength shrinks and the energy density is localized [see Fig.~\ref{fig:Huidobro-Fig1}\pnl{c}]. In an ideal, lossless system, this generates singular hot spots where the intensity of electromagnetic fields diverges.

TO brings further insight into this process by revealing "hidden symmetries" by relating a structure with no apparent symmetry to another more symmetrical one. For example: a cylinder can be transformed into a plane which explains degeneracy of the plasmonic modes of a cylinder. According to this framework, all the singular structures shown in Fig.~\ref{fig:Huidobro-Fig1} can be transformed into an extended and nonsingular one: a planar waveguide for the geometries in panel \pnl{a} and the smooth grating in \pnl{b}, and a waveguide array for the grating with sharp singularities in \pnl{c}. Additionally, TO is not limited to the strict quasistatic limit, and radiative corrections can be incorporated in the theory to accurately deal with larger structures.

We can thus exploit this connection to predict the optical response of a structure of interest by studying the modes in a highly symmetric structure, which allows for an analytical treatment~\cite{Luo:2010}. For instance, in a planar waveguide (or waveguide array), surface plasmons excited at the origin travel away towards infinity. Singular coordinate transformations map infinity into singular points in the frame of a localized or grating structure. Consequently, all singular structures share a broadband optical spectrum associated with the energy build up at the singular points. Interestingly, in the case of gratings, this can be understood as the compacting of one of the dimensions of the non-singular structure into the singularities~\cite{Pendry:2017}. Huge concentrations of electromagnetic energy at nanoscale volumes can then be exploited for highly sensitive sensing or improved photovoltaic devices. 

However, this immense energy concentration has limits. Fields cannot be confined beyond the electronic screening length, where nonlocal effects come into play and impose a limit to the achievable field enhancement with these light harvesting structures. The dynamics of the free carriers in plasmonic media, which mediate a response to incident electromagnetic fields over finite distances, explain the failure of the local response approximation of classical electrodynamics. 

\subsection*{Challenges and opportunities}

The diverging field concentrations in singular (or near-singular) plasmonic nanostructures, as well as the optically small length scales involved in the nonlocal response, make most theoretical approaches challenging or even inapplicable. For instance, numerical methods are limited by the size of the finite elements used to spatially discretize a structure. These must be smaller than the length scales involved in the problem, which prevents from realizing a singular point. On the contrary, the analytical nature of TO makes it an ideal tool for dealing with these singular structures, by exploiting the fact that they can be mapped to extended, non-singular structures. This capability, together with the insight provided by TO, represents a unique opportunity to understand the key role played by nonlocal effects in singular plasmonic geometries. 

Reference~\cite{Fernandez-Dominguez:2012} showed that nonlocal effects can be incorporated into the TO framework by considering a hydrodynamic description of the free carriers. Within this model, electron-electron interactions result in a wave-vector dependent, spatially dispersive, dielectric function for longitudinal modes. This reflects the fact that incident electromagnetic fields can excite longitudinal plasmon oscillations that are not described in the common local response approximation. As a result, surface charges accumulate on a finite-length layer with a non-vanishing decay length [see purple area in the sketch of Fig.~\ref{fig:Huidobro-Fig1}\pnl{d}, representing two touching nanowires]. The existence of this layer smooths out any singularity and has critical implications on the field enhancement capabilities and the optical response of these structures. 

An example of field localization calculated by means of TO is shown in Fig.~\ref{fig:Huidobro-Fig1}\pnl{c} for the case of a plasmonic nanotip~\cite{Wiener:2012}. Under a local-response approximation (top panel), the wavelength of the surface plasmons shrinks as they travel towards the singularity, a point that is never reached as it corresponds to infinity in the other frame. In contrast, when spatial dispersion in the metal is included, the tip effectively blunts, quantizing the modes, and limiting the smallest achievable wavelength for the surface plasmons (bottom panel).

Figure~\ref{fig:Huidobro-Fig1}\pnl{d} displays the optical response (absorption cross section) of two touching nanowires calculated by means of TO. In a local approximation (grey line), the singular nature of this structure results in a broad absorption peak due to its continuous spectrum. On the other hand, mode quantization due to nonlocality results in a series of discrete absorption peaks (blue, green and red lines for different values of the nonlocal decay length). In sharp contrast, for the case of a single nanowire (black lines), nonlocality only results in a blueshift (dashed line).

TO also presents opportunities to understand singularities in plasmonic gratings. Subwavelength arrays of sharp grooves etched into metallic surfaces, such as gold (Au) or silver (Ag), convert these materials from efficient reflectors into good broadband absorbers. TO reveals how the surface plasmons localized at the two-dimensional (2D) surface have a three-dimensional (3D) nature with an additional wave-vector inherited from the transformed structure compacted into the singularities~\cite{Pendry:2017}. The missing selection rule for this extra wave-vector produces a broadband absorption spectrum, unlike conventional gratings with discrete absorption lines. Figure~\ref{fig:Huidobro-Fig1}\pnl{e} shows the broadband spectral response of a local singular silver metasurface with a 10\,nm period calculated analytically with TO: a small decrease in reflectivity is seen for a broad band. By contrast, when nonlocality is considered, the spectrum splits into a discrete series of peaks with very low reflectivity, thus high absorption~\cite{Yang:2019b}. 

\subsection*{Future developments to address challenges}

So far, most approaches to nonlocality using TO have been limited to the hard wall approximation and the hydrodynamic model. However, this fails to account for spill-out of the electronic density away from the metal. These nonclassical spill-out effects can be incorporated to the TO framework by combining Feibelman $d$-parameters~\cite{Yang:2022}. This relies on introducing the electronic scale length through the modification of the classical boundary conditions with the Feibelman $d$-parameters (see Sec.~\ref{sec:Christensen} of this Roadmap). These are a set of mesoscopic complex surface-response functions which play a role analogous to the local bulk permittivity, but for interfaces between two materials~\cite{Yang:2019a}. 

\begin{figure}[hbt]
    \centering
    \includegraphics[width=0.99\linewidth]{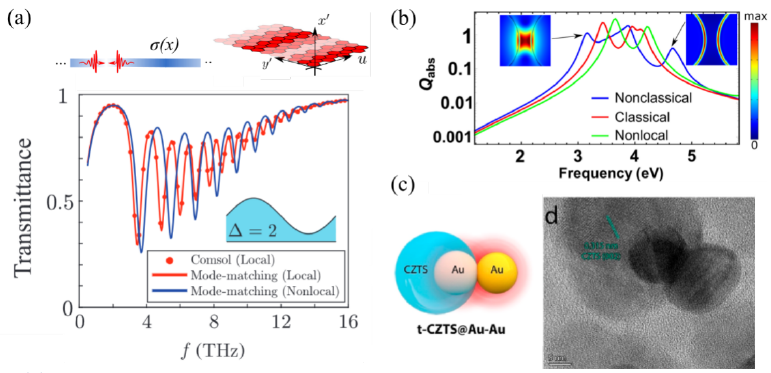}
    \caption{\pnl{a} Nonlocality leads to saturation of the density of states in a singular graphene metasurface. The optical response of a graphene conductivity grating with closely vanishing values at periodic points is a balance between a continuous spectrum due to the singularity and a discrete set of resonances due to nonlocality. Reproduced with permission from Ref.~\cite{Galiffi:2020} (Copyright~\textcopyright~2020 De Gruyter). \pnl{b} Absorption efficiency spectrum of two almost touching nanowires under three TO approaches: classical and local, classical and nonlocal (hydrodynamic model), and nonclassical (mesoscopic approach). Reproduced with permission from Ref.~\cite{Yang:2020b} (Copyright~\textcopyright~2020 American Physical Society). \pnl{c} Sketch (left) and TEM image (right) of a Au--Au touching dimer utilized for photocatalysis. The average size of Au nanoparticles is 18\,nm. Reprinted (adapted) with permission from Ref.~\cite{Lin:2023} (Copyright~\textcopyright~2023 American Chemical Society).}
    \label{fig:Huidobro-Fig2}
\end{figure}

The approach of Ref.~\cite{Yang:2022} combines the analytical power of TO with the capability of this mesoscopic model to accurately incorporate nonlocality, spill-out, and surface-enabled losses. Fig.~\ref{fig:Huidobro-Fig2}\pnl{b} shows the absorption efficiency spectra of two almost touching nanowires. While the hard-wall hydrodynamic model predicts a blueshift of the first resonance peak, the nonclassical, mesoscopic calculation predicts a redshift and the appearance of the Bennet mode (see inset panel), showing how this approach successfully captures both electron spill-out and nonlocal effects. 

On the other hand, tunneling and size quantization are not contemplated within this model. Therefore, an outstanding challenge would be to incorporate these effects into the TO approach. This would be relevant for optical nanostructures with the sharpest singularities, since these effects are important for feature sizes below about 1\,nm~\cite{Yang:2019a}.

Furthermore, there are plasmonic materials beyond metals, and TO can be extended to treat them~\cite{Huidobro:2016}. For instance, graphene, the paradigmatic example of a two-dimensional material, hosts surface plasmons when doped. Interestingly, singular plasmonic gratings can be realized on graphene by having periodic conductivity modulations that approach zero at the grating minima, see Fig.~\ref{fig:Huidobro-Fig2}\pnl{a}. Under THz illumination, this monoatomic structure shows a broad absorption band. The stronger the suppression of conductivity at the grating valleys, the more broadband the absorption spectrum is~\cite{Pendry:2017}. Ref.~\cite{Galiffi:2020} showed how, under a semiclassical description of graphene’s conductivity (a local analogue model), these conductivity gratings could probe the nonlocal response of graphene by far field measurements only. However, quantum calculations have found that plasmon propagation is damped by strong modulations of the local Fermi energy~\cite{Brey:2020}, pointing to the need to implement fully quantum conductivity models within optical calculations, where TO can be of help. 

Finally, if TO is to be employed for devising applications, further advances are needed to treat more complex structures. For instance, TO-inspired singular plasmonic nanospheres have been proposed for catalysis applications at metal-semiconductor interfaces~\cite{Lin:2023}. Three-dimensional hybrid nanostructures with plasmonic singularities, see Fig.~\ref{fig:Huidobro-Fig2}\pnl{c}, were shown to provide broad and strong light harvesting at the active site of the semiconductor, facilitating a chemical reaction. However, these and other fully three-dimensional structures are challenging to model with TO, with only a few examples of nanosphere dimers having been developed in the literature~\cite{Li:2016b}. 

\subsection*{Concluding remarks}

Combining the TO framework with nonlocal hydrodynamic or nonclassical mesoscopic models offers a powerful tool to unveil the effect of spatial dispersion, electron spill out or surface-enabled losses in nanoplasmonic structures. These effects are especially important in singular or near singular geometries where TO is particularly interesting, as it can treat structures that are commonly very challenging to simulate. Additionally, TO is able to uncover the geometrical basis of their optical response, by linking hotspots of field concentration in these systems with points at infinity in partner structures, to which surface plasmons travel. In conclusion, TO enables analytical calculations in complex geometries, in the presence of nonlocal and retarded effects. This makes TO a critical tool to understand nonlocal effects in plasmonics.

\section[The concept of overlapping nonlocality and the ultimate thickness limits of optics (Miller)]{The concept of overlapping nonlocality and the ultimate thickness limits of optics}

\label{Sec:Miller}

\author{David A. B. Miller\,\orcidlink{0000-0002-3633-7479}}

\subsection*{Current status}

\begin{figure}[hb]
    \centering
    \includegraphics[width=0.85\linewidth]{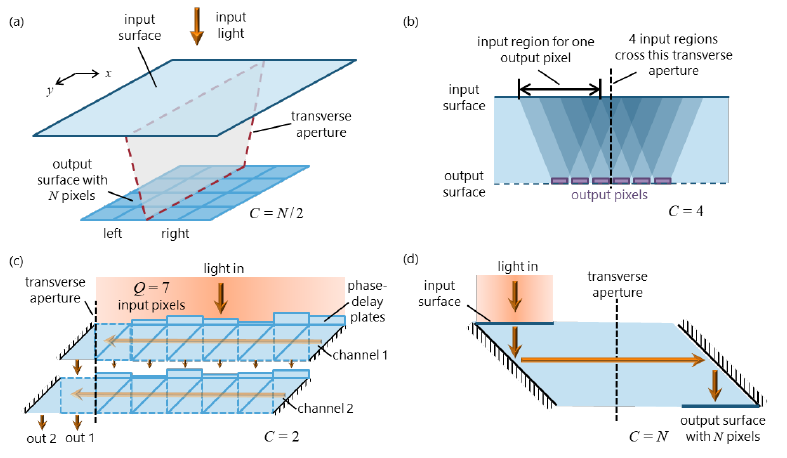}
    \caption{Illustrations (after Ref.~\cite{Miller:2023}) of overlapping nonlocality and corresponding number $C$ of required channels that must flow through a "transverse aperture" inside the device. \pnl{a} Input and output surfaces for an $N$-pixel imager, with a transverse aperture defined by cutting the system from top to bottom through the middle, giving $C = N/2$. \pnl{b}~Side view of a hypothetical image differentiator where each output pixel depends on the input field on four adjacent pixels, giving $C = 4$. \pnl{c}~A side view of a hypothetical arrangement of beamsplitters and phase-delay plates separating two orthogonal input functions (as given by their vectors of amplitudes on the line of $Q = 7$ input pixels) to separate outputs, with $C = 2$. \pnl{d}~Side view illustration that folding the optics of an imager can actually increase the required thickness. Now all $N$ pixels or channels must flow through the transverse aperture, so $C = N$, hence increasing the required transverse aperture size.}
    \label{fig:Miller-Fig1}
\end{figure}

In many optical and wave systems, the output at a given point or pixel depends on the input at many other points or pixels, which we could regard as a definition of nonlocal behavior~\cite{Overvig:2022,Shastri:2023}. For example, in an imager, the light on one output image pixel depends on the light over the entire input surface of the lens~\cite{Miller:2023}. However, the light on another output pixel also depends on the input light over the same entire input lens surface. So, these nonlocalities are "overlapping". Such overlapping nonlocalities~\cite{Miller:2023} can be found in many other systems, including metasurfaces for image differentiation~\cite{Wang:2022} or space compression~\cite{Chen:2021}; there, too, multiple adjacent input pixels contribute to each output pixel (Fig.~\ref{fig:Miller-Fig1}). 

To understand overlapping nonlocality, we can imagine making a cut through both the input and output surfaces, which will define a "transverse aperture" in the optics (Fig.~\ref{fig:Miller-Fig1}). The key idea here is that, for the device to do what we want, some number $C$ -- the overlapping nonlocality~\cite{Miller:2023} -- of independent channels must pass through this transverse aperture from side to side. Because we believe that physically we can only get so many propagating channels through some finite cross-section (e.g., the number of waveguide modes that could fit in that cross-section), this transverse aperture must be large enough to carry this number $C$ of channels. 

This idea of overlapping nonlocality addresses a key question that is particularly important for metastructures: can we do what we want with just a single layer of material or metamaterial? Or, more generally, just how much thickness will that structure need? This approach also gives bounds for the thickness of imagers, for example~\cite{Miller:2023}.
Note that, as in Fig.~\ref{fig:Miller-Fig1}\pnl{d}, any given channel could be collecting from a large number $Q$ of input pixels, and we could regard $Q$ as the (simple) "nonlocality". (To distinguish from nonlocality as a physical distance, we could call $Q$ the "nonlocality number" of the channel.) The overlapping nonlocality $C$ is not the same as $Q$; $C$ depends on the number of separate output channels that much be constructed from those same input pixels. 

\begin{figure}[hb]
    \centering
    \includegraphics[width=0.5\linewidth]{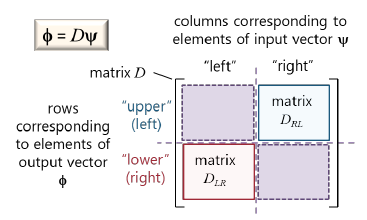}
    \caption{Illustration of the matrix multiplication $\boldsymbol{\phi}=D \boldsymbol{\psi}$ that represents the operation of the device from inputs $\boldsymbol{\psi}$ to outputs $\boldsymbol{\phi}$. The choice of the transverse aperture position on the input surface divides the matrix into "left" and "right" parts, which also correspond to the input positions to the left and the right of the transverse aperture position. The choice of the transverse aperture position on the output surface divides the matrix into "upper" and "lower" parts, which correspond to the left and right parts of the output surface (relative to the transverse aperture position). These divisions generate matrix quadrants $D_\textrm{LR}$ and $D_\textrm{RL}$. The singular value decomposition of these quadrants effectively defines numbers $C_\textrm{LR}$ and $C_\textrm{RL}$ of channels that are required to pass through the transverse aperture in the operation of the device. The overlapping nonlocality is then $C = C_\textrm{LR} + C_\textrm{RL}$.}
    \label{fig:Miller-Fig2}
\end{figure}

This overlapping nonlocality can be precisely specified mathematically. Some matrix $D$ describes the mathematical mapping between input pixels (which feed matrix columns) and output pixels (which are fed by matrix rows) (Fig.~\ref{fig:Miller-Fig2}). We can write that matrix so that all the columns for pixels on the "left" of the dividing surface are to the "left" of some vertical line through the matrix, and similarly the rows for all output pixels to the "left" of the dividing surface are "above" some horizontal line through the matrix. This defines the "quadrant" sub-matrices $D_\textrm{LR}$ for "left-inputs-to-right-outputs" and $D_\textrm{RL}$ for "right-inputs-to-left-outputs". 
We can view optical systems quite generally as mode converters~\cite{Miller:2012,Miller:2019}, mapping one set of orthogonal input functions one-by-one to orthogonal output functions; these pair of functions -- one each in the input and output spaces -- form the mode-converter (or communication mode) basis sets for the system. If we know the desired matrix $D$ for the system, we can establish these pairs of functions by singular value decomposition of $D$, a standard matrix operation. Importantly, this process, through the resulting singular values, also gives the coupling strengths of these pairs. From this process, we can establish just how many such couplings are really needed, above some coupling strength we consider negligibly weak. (Quite generally~\cite{Miller:2025}, such couplings anyway fall off rapidly past some finite number in physical wave systems.) For our matrices $D_\textrm{LR}$ and $D_\textrm{RL}$, we can therefore deduce numbers of channels $C_\textrm{LR}$ and $C_\textrm{RL}$ needed in each case. Reciprocity~\cite{Miller:2023} requires we add these numbers to get the required overlapping nonlocality $C = C_\textrm{LR} + C_\textrm{RL}$. 

The number $C$ arguably then can tell us a thickness of some optical element, based on a "maximum number of modes" conjecture that 

\begin{itemize}
    \item[] "the number of available channels cannot exceed the number of modes of a waveguide with a cross-section given by the transverse aperture and made from the largest refractive index used in the design" 
\end{itemize}

This approach~\cite{Miller:2023} has given suggested minimum thicknesses for differentiators~\cite{Wang:2022}, space plates~\cite{Chen:2021}, and imagers; optimized designs are thicker than these limits by no more than a factor of 3. It also gives limits for thickness of optical neural networks~\cite{Huang:2024,Li:2024a}; helps suggest better designs for optical processing networks~\cite{Li:2024a}; is consistent with subsequent more specific limits for metalenses~\cite{Li:2024b}, metalens system design~\cite{Wirth-Singh:2025}, and space plates in the microwave region~\cite{Mrnka:2024}; and will impose thickness limits on metastructures for aberration correction~\cite{Shao:2024}. 
Note the following about the overlapping nonlocality $C$.

\begin{itemize}
    \item[1)] $C$ is not related to the physical size of the function being performed; a small camera and a large camera with the same number of image pixels have the same overlapping nonlocality. 
    \item[2)] $C$ is an effective mathematical dimensionality of our desired input-output relation, and in that sense is independent of how we implement the device physically. In optics we may use propagating modes in dielectric structures to carry these transverse channels, but the same number of channels is required for any other physical implementation, such as plasmonics or even electrical wires in radio-frequency systems (e.g., in "reflective intelligent surfaces" with non-diagonal reflection coefficients~\cite{Bartoli:2023}).
\end{itemize}
This concept of overlapping nonlocality is a quite general property of any optical or wave device intended to implement some relation between inputs and outputs. Understanding and using this concept can arguably help us design and operate all such systems. 

\subsection*{Challenges and opportunities}

Though the overlapping nonlocality $C$ is well defined just from what we want the optical or wave system to do, there are at least two challenges in using this approach to deduce the necessary thickness of the required system. 

\begin{itemize}
    \item[1)] We need to understand whether our approach to building the system can perform "dimensional interleaving"~\cite{Miller:2023}. If we think of our system as having its input degrees of freedom in two dimensions -- $x$ and $y$ in Fig.~\ref{fig:Miller-Fig1}\pnl{a} -- we need to know if we can move those degrees of freedom between the $x$ and $y$ dimensions. Now, a photonic integrated circuit with a two-dimensional array of grating couplers can readily couple their outputs to a one-dimensional array of waveguides, for example, hence "interleaving" the input information from $x$ and $y$ degrees of freedom into a line in the $y$ direction~\cite{Miller:2023}. Many other optical systems do not, however, do this interleaving; conventional imagers and lens optics, free-space propagation, dielectric stacks, and photonic crystals all preserve the degrees of freedom in each of the $x$ and $y$ directions~\cite{Miller:2023}. If there is no dimensional interleaving, the required thickness is determined by the number of degrees of freedom in the $x$ or $y$ directions (whichever is larger) and on a one-dimensional version of channel counting in a waveguide, which generally will lead to a larger thickness than if we can perform dimensional interleaving.
\end{itemize}
We can form a "dimensional interleaving" conjecture here that 

\begin{itemize}
    \item[] "dimensional interleaving is not possible without segmenting the field or, equivalently, it requires a discontinuous mapping between input and output fields."
\end{itemize}
This conjecture remains to be proved or disproved. Note, for example, that a hypothetical array of grating couplers segments the input optical field, essentially discontinuously. Another device that could do dimensional interleaving, a photonic lantern~\cite{Birks:2015}, also segments the field as the fused fibers at one end separate into discrete fibers at the other. This author knows of no way of performing such dimensional interleaving without segmenting the field somewhere; of course, this author’s possible ignorance proves nothing.

\begin{itemize}
    \item[2)] Using the overlapping nonlocality $C$ to deduce minimum thickness of an optical system that we have not yet designed has required the "maximum number of modes" conjecture stated above. This conjecture also remains to be proved or disproved.
\end{itemize}
One possible approach to the "maximum number of modes" conjecture could be to use a multiple scattering theorem~\cite{Miller:2007a} that relates dielectric constant contrast in a volume to achievable optical functions; this has been used successfully for limits in one-dimensional structures~\cite{Miller:2007a,Miller:2007b}, though it has not so far been extended to two or three dimensions.   

A limitation of this approach so far is that it is only analyzed for one frequency. Since different frequencies are anyway orthogonal functions in that they can separately propagate through the same channel, each containing different information (e.g., amplitudes), this approach has nothing to say about additional limits or opportunities~\cite{Pahlevaninezhad:2024} in multi-frequency systems, and further such additional limits remain open questions. 

\subsection*{Future developments to address challenges}

Clear future directions for understanding and applying overlapping nonlocality are to prove or disprove the "maximum number of modes" and "dimensional interleaving" conjectures stated above. These conjectures are, of course, interesting for their direct relevance to overlapping nonlocality. Possibly, however, proving or disproving them might require moving beyond our typical ideas for understanding waves; as such, addressing them might expose new areas of opportunity.  
A final comment on future directions is that this overlapping nonlocality work emphasizes the importance of analyzing nonlocality also in real space, not just in the $k$-space "plane-wave basis" approach. Such plane waves are, of course, mathematically completely nonlocal because they have infinite lateral extent. This approach has been powerful and successful in many ways. It is, however, quite awkward mathematically in a $k$-space approach to introduce a surface that divides real space into two parts. Possibly as a result, overlapping nonlocality, though implicit in any actual design based on $k$-space approaches, may be far from obvious in that description. Such plane-wave approaches are also generally problematic as we move to small structures -- e.g., from several wavelengths to sub-wavelength sizes -- where they are arguably not a good basis and can cause us to miss physics in the problem. For small structures, physically there are really no (plane) evanescent waves, for example~\cite{Miller:2025}. To model waves in and out of such structures, spherical waves~\cite{Miller:2025,Kuang:2025} are a better fundamental basis. Unlike plane waves, such spherical waves are countable, which can help in counting channels, and by using "bounding spheres" they can implicitly include the finite size of real objects~\cite{Miller:2025,Kuang:2025}. These results remind us of the importance of thinking also in real space when analyzing wave systems and generally of being prepared to move beyond $k$-space descriptions.

\subsection*{Concluding remarks}

The idea of overlapping nonlocality is ultimately a simple one that we can deduce directly from what we want our optical system to do. Using it to predict requirements and limits in optical designs of various kinds looks to be a potential productive and useful direction. At the same time, it is exposing new conjectures and ways of thinking about wave devices that may test and expand our understanding and use of waves in general.

\usection{Closing remarks and back matter}

\section[Conclusion and outlook (Monticone \& Mortensen)]{Conclusion and outlook}
\author{Francesco Monticone\,\orcidlink{0000-0003-0457-1807} \& N.~Asger~Mortensen\,\orcidlink{0000-0001-7936-6264}}

\label{sec:Conclusion}

Assembling this roadmap has given us a unique opportunity to work with many of the leading experts in this field to explore the rapidly expanding landscape of nonlocal responses in natural and artificial materials. Through more than thirty sections, this roadmap highlights an impressive array of exciting new developments and emerging directions. It is clear that the multifaceted concept of nonlocality is playing an increasingly important role in optical materials science and technology, from advancing our understanding of anomalous and extreme forms of light-matter interactions and revealing new insights into the fundamental limits of wave physics, to unlocking new opportunities for optical devices based on nonlocality engineering. In the spirit of this roadmap, we believe that future progress in this field will not be "local" but will depend on an extended range of contributions from multiple areas. We hope that this roadmap, bringing together insights from different corners of this field, serves as a valuable guide to the current landscape of nonlocality in photonic materials and metamaterials and will inspire further innovation in expanding its frontier.

\addcontentsline{toc}{section}{\protect\numberline{}Funding}

\section*{Funding}

\begin{itemize}

    \item[Sec.~\ref{Sec:Mortensen}] Danish National Research Foundation (project No.~DNRF165). Air Force Office of Scientific Research (grant no. FA9550-22-1-0204) and Office of Naval Research (grant no. N00014-22-1-2486). 
    
     \item[Sec.~\ref{sec:Fernandez-Dominguez}] Danish National Research Foundation (project No.~DNRF165); Spanish Ministry of Science, Innovation and Universities (grant No.~PID2021-126964OB-I00); Otto M{\o}nsteds Fond (grant No.~24-12-1979).

     \item[Sec.~\ref{sec:Khurgin}] No funding to report.
     
    \item[Sec.~\ref{sec:Shahbazyan}] National Science Foundation (grants No.~DMR-2000170, No.~DMR-2301350, and No.~NSF-PREM-2423854).
    
    \item[Sec.~\ref{sec:Chaves}] Portuguese Foundation for Science and Technology (FCT) in the framework of the Strategic Funding (UIDB/04650/2020, COMPETE 2020, PORTUGAL 2020, FEDER, and through project PTDC/FIS-MAC/2045/2021); Independent Research Fund Denmark (grant No.~2032-00045B); Danish National Research Foundation (project No.~DNRF165); CNPq --Conselho Nacional de Desenvolvimento Cient\'ifico e Tecnol\'ogico (grants No.~423423/2021-5, No.~408144/2022-0, No.~315408/2021-9); FAPESP (grant No.~2022/08086-0); CAPES-PrInt scholarship.
    
    \item[Sec.~\ref{sec:Wegner}] German Research Foundation (DFG) in the framework of the Collaborative Research Centers 1375 "Nonlinear Optics down to Atomic Scales (NOA)" (Project ID 398816777 -- Project A06) as well as 1636 "Elementary Processes of Light-Driven Reactions at Nanoscale Metals" (Project ID 510943930 -- Project A06).
    
    \item[Sec.~\ref{sec:Hu}] European Innovation Council, Pathfinder project (NEHO, project No.~101046329); ICSC -- Centro Nazionale di Ricerca in High Performance Computing, Big Data and Quantum Computing, funded by European Union -- NextGenerationEU -- PNRR.
    
    \item[Sec.~\ref{sec:Aizpurua}] Spanish Ministry of Science and Innovation (project No.~PID2022-139579NB-I00); Basque Government program for groups of the University of the Basque Country (project No.~IT1526-22); German Research Foundation (DFG) in the framework of 1636 "Elementary Processes of Light-Driven Reactions at Nanoscale Metals" (Project ID 510943930 -- Project A06).
    
    \item[Sec.~\ref{sec:Zhang}] National Natural Science Foundation of China (grant No.~12274160).
    
    \item[Sec.~\ref{sec:Tserkezis}] Danish National Research Foundation (project No.~DNRF165).
    
    \item[Sec.~\ref{sec:Christensen}] VILLUM Fonden (project No.~42106); Ms.~Belinda Hung, the Asian Young Scientist Fellowship; Croucher Foundation; New Cornerstone Science Foundation through the Xplorer Prize. 
    
    \item[Sec.~\ref{sec:Hohenester}] Austrian Science Fund (FWF project No.~10.55776/P37150); German Research Foundation (DFG) in the framework of 1636 "Elementary Processes of Light-Driven Reactions at Nanoscale Metals" (Project ID 510943930 -- Project A06).
    
    \item[Sec.~\ref{sec:Wubs}] Danish National Research Foundation (project No.~DNRF147).

    \item[Sec.~\ref{sec:DeLiberato}] Leverhulme Trust (grant No.~RPG-2022-037).

    \item[Sec.~\ref{sec:Goncalves}] European Research Council (Advanced Grant 789104--eNANO); European Commission (Horizon 2020 Grants 101017720 FET-Proactive EBEAM and 964591-SMART-electron); Spanish MICINN (PID2020-112625GB-I00 and Severo Ochoa CEX2019-000910-S); Catalan CERCA Program; Fundaci\'{o}s Cellex and Mir-Puig.

    \item[Sec.~\ref{sec:Hess}] Research Ireland  via the METAQUANT Research Professorship Programme (grant No.~18/RP/6236).

    \item[Sec.~\ref{sec:Cox}] Independent Research Fund Denmark (grant No.~0165-00051B); Danish National Research Foundation (project No.~DNRF165).
     
    \item[Sec.~\ref{sec:Jelver}] Independent Research Fund Denmark (grant No.~0165-00051B); Danish National Research Foundation (project No.~DNRF165).
    
    \item[Sec.~\ref{sec:SanchezSanchez}] Ministerio de Ciencia, Innovaci{\'o}n y  Universidades (grant No.~PID2020-113164GBI00 funded by MCIN/AEI/10.13039/501100011033, PID2023-146461NB-I00); CSIC Research Platform on Quantum Technologies (PTI-001. G.G-S); Spanish Ministry of Science, Innovation and Universities  through the "Mar{\'i}a de Maeztu" Programme for Units of Excellence in R\&D (CEX2023-001316-M).
    
    \item[Sec.~\ref{sec:Tretyakov}] European Innovation Council, Pathfinder project (PULSE, project No.~101099313); Research Council of Finland, Finland-USA bilateral cooperation on Future Information Architecture for IoT (project No.~365679). 
    
    \item[Sec.~\ref{sec:Pakniyat}] Keck Foundation.
    
    \item[Sec.~\ref{sec:Bondarev}] U.S. Army Research Office (award No.~W911NF2310206); U.S. Department of Energy, Office of Basic Energy Sciences, Division of Materials Sciences and Engineering (award No.~DE-SC0017717); U.S. National Science Foundation (DMREF award No.~10002504).

    \item[Sec.\ref{sec:Krasavin}] European Research Council (Advanced Grant 789340--ICOMM); Engineering and Physical Sciences Research Council (EPSRC project EP/Y015673/1).

    \item[Sec.~\ref{sec:Alu}] Department of Defense; Simons Foundation. 

    \item[Sec.~\ref{sec:Song}] Ministry of Education, Singapore, under the Academic Research Fund Tier 1 (FY2024).
    
    \item [Sec.~\ref{sec:Levy}] The Israeli Innovation Authority; The Metamaterials and Metasurfaces consortium program. 

    \item[Sec.~\ref{sec:Long}] Stanford Graduate Fellowship; U. S. Air Force Office of Scientific Research, MURI project (grant No.~FA9550-21-1-0312)  from the U. S. Air Force Office of Scientific Research.

    \item[Sec.~\ref{sec:Bozhevolnyi}] European Innovation Council, Pathfinder project (OPTIPATH, project No.~101185769).

    \item[Sec.~\ref{sec:Overvig}] Airforce Office of Scientific Research (grant No.~FA9550-24-1-0068). 

    \item[Sec.~\ref{sec:Prudencio}] Institution of Engineering and Technology (IET); Simons Foundation (Award No.~SFI-MPS-EWP-00008530-10); Instituto de Telecomunica\c{c}{\~o}es (project No.~UIDB/50008/2020).
  
    \item[Sec.~\ref{sec:HassaniGangaraj}] National Science Foundation (grant No.~DMR-2224456).

    \item[Sec.~\ref{sec:Huidobro}] Spanish Ministry for Science, Innovation and Universities (grant No.~RYC2021-031568-I and project No.~PID2022-141036NA-I00 financed by MCIN/AEI/10.13039/501100011033 and FSE+).
    
    \item[Sec.~\ref{Sec:Miller}] Airforce Office of Scientific Research (grant No.~FA9550-21-1-0312).
   
\end{itemize}

\addcontentsline{toc}{section}{\protect\numberline{}Acknowledgments}

\section*{Acknowledgments}

The lead authors are grateful for the community's strong support in organizing this Roadmap. We also extend our thanks to the Editorial Board of the journal for inviting us to take on this project.

\addcontentsline{toc}{section}{\protect\numberline{}Disclosures}

\section*{Disclosures}

The authors declare no conflicts of interest.

\newpage

\addcontentsline{toc}{section}{\protect\numberline{}References}

\bibliography{references}

\end{document}